\documentclass[a4paper,prb,twocolumn,showpacs,floatfix]{revtex4}

\usepackage{graphicx}
\usepackage{amsmath}

\newcommand{\avec}{\boldsymbol{a}}
\newcommand{\lvec}{\boldsymbol{l}}

\newcommand{\qvec}{\boldsymbol{q}}
\newcommand{\mvec}{\boldsymbol{m}}
\newcommand{\rvec}{\boldsymbol{r}}
\newcommand{\nvec}{\boldsymbol{n}}
\newcommand{\Kvec}{\boldsymbol{K}}

\begin{document}

\title{Light and stable triplet bipolarons on square and triangular lattices}

\author{J.P. Hague}
\affiliation{Department of Physics and Astronomy, The Open University, Walton Hall, Milton Keynes, MK7 6AA, UK}

\author{P.E. Kornilovitch}
\affiliation{Hewlett-Packard Company, 1000 NE Circle Blvd, Corvallis, Oregon 97330, USA}

\begin{abstract}
We compute the properties of singlet and triplet bipolarons on two-dimensional lattices
using the continuous time quantum Monte Carlo algorithm. Properties of
the bipolaron including the total energy, inverse mass, bipolaron
radius and number of phonons associated with the bipolaron demonstrate the qualitative difference between models
of electron phonon interaction with long-range interaction (screened
Fr\"ohlich) and those with purely local (Holstein) interaction. A major result of our survey of the parameter space
is the existence of extra-light hybrid singlet bipolarons consisting of an
on-site and an off-site component on both square and triangular lattices. We also compute triplet properties of the
bipolarons and the pair dispersion. For pair momenta on the edge of the Brillouin zone of the
triangular lattice, we find that triplet states are more stable than
singlets.
\end{abstract}


\date{7th October 2009}

\maketitle

\section{Introduction}

%DISCUSS 2D EFFECTS HERE: ON SQUARE AND TRIANGULAR LATTICES. NOTE THAT
%WE'VE GOT THE BEST ALGORITHM TO LOOK AT THIS.

Layered superconductors such as the cuprates \cite{bednorz}, iron
pnictides \cite{kamihara2008a} and others such as hydrated cobaltates
(Na$_{x}$CoO$_{2}\cdot 1.3$H$_{2}$O) (Ref. \onlinecite{cobaltate})
have the potential to completely change thinking about the origins
of superconductivity. Given the importance of such a rethink, a number
of competing theories and mechanisms have been suggested, which aim to
understand superconductivity in layered materials on at
least a qualitative level
\cite{andersonbook,monthoux1992a,laughlin2006a,sashabook,hardy2009a,tranquada1995a,mishchenko2006a,carbotte1999a}. For
example, it has been suggested that one way of explaining high
temperature superconductivity in cuprates is the condensation of pre-formed
bipolarons \cite{sashabook}. A complete quantitative understanding of
bipolaron formation, especially in the intermediate coupling regime is
needed to judge the applicability of such a theory, especially when
the electron-phonon interaction is not perfectly local (i.e. when the
Holstein interaction is not appropriate) as is to be expected in real materials.

There are a number of exact numerical methods for studying bipolaron
formation in one-dimensional (1D) systems with electron-phonon
interactions such as exact diagonalization (ED) \cite{wellein1996a,
weisse2000a}, advanced variational techniques \cite{bonca2000a}, density
matrix renormalization group (DMRG) \cite{jeckelmann1999a} and various quantum
Monte-Carlo (QMC) approaches \cite{hague2009a,
hohenadler2005a,hohenadler2005b}. In two-dimensional (2D) systems,
many of those techniques are not applicable. Exact diagonalization
cannot cope with large numbers of lattice sites, DMRG schemes are not
easy to develop in 2D and both ED and DMRG can suffer from a
truncation of the phonon Hilbert space, rendering those methods
inefficient at strong electron-phonon coupling. Also, advanced
variational techniques are not easily generalized to 2D. On the other
hand, QMC techniques (and especially continuous time QMC) can cope
with large lattice sizes and treat the phonon degrees of freedom
exactly, making them well suited for computations of bipolarons on
two- and potentially even three- dimensional lattices.

% N.B. I DIDN'T FIND ANY 2D BIPOLARON WORK BY HOHENADLER, BUT SHOULD DOUBLE CHECK WITH HIM

There is a fascinating possibility relating to bipolarons in 2D. If a
lattice is made up from triangular plaquettes, and if there is a
strong Coulomb repulsion keeping bipolarons from pairing on-site,
superlight small bipolarons can form
\cite{alexandrov1996a,alexandrov2002a,hague2007a,hague2007b}. Our aim here is to comprehensively compute the properties of
bipolarons with long range interaction on both square and triangular
lattices to determine if there are any other promising ways that bipolarons can pair. 
%
%Studies of the properties of 2D bipolarons using exact
%diagonalization have been limited by large Hilbert spaces, which must
%be truncated reducing practical computations to a few lattice sites,
%even when advanced schemes are employed \cite{wellein1995a}. 
%
The Hubbard-Holstein bipolaron on the square lattice was studied by
Macridin {\it et al.} using the related diagrammatic quantum Monte
Carlo technique (DMC) \cite{macridin2004a}. Our current study goes
beyond this previous work by computing the properties of bipolarons
formed from long range electron-phonon interactions. We also compute
properties of bipolarons on triangular lattices, which have not been
comprehensively studied even for the Holstein interaction. Our
continuous-time quantum Monte Carlo (CTQMC) algorithm is particularly
efficient for this task, since phonon degrees of freedom can be
treated exactly to generate an effective retarded interaction between electrons.

In this article, we study the screened Hubbard-Fr\"ohlich model, which has the Hamiltonian,
\begin{eqnarray}
H & = & - t \sum_{\langle \nvec \nvec' \rangle\sigma}
c^{\dagger}_{\nvec'\sigma} c_{\nvec\sigma} + U\sum_{
\nvec} c^{\dagger}_{\nvec\uparrow}
c_{\nvec\uparrow}c^{\dagger}_{\nvec\downarrow} c_{\nvec\downarrow}\\&& +
\sum_{\mvec} \frac{\hat{P}^{2}_{\mvec}}{2M} + 
\sum_{\mvec} \frac{\xi^{2}_{\mvec} M\omega^2}{2} -
\sum_{\nvec\mvec\sigma} f_{\mvec}(\nvec)
c^{\dagger}_{\nvec\sigma} c_{\nvec\sigma} \xi_{\mvec}\nonumber
\: .
\label{eqn:hamiltonian}
\end{eqnarray}
Here, $c^{\dagger}_{\nvec}$ creates an electron on site $\nvec$, $t$
is the intersite hopping integral, $U$ is the Hubbard repulsion, $M$
is the ion mass, $\omega$ the ion oscillation frequency, $\zeta$ the
ion displacement and $\hat{P}$ the ion momentum. The electron-ion
force function is $f$, and has the form
$f_{\mvec}(\nvec)=\kappa\left[(\mvec-\nvec)^2+1\right]^{-3/2}\exp\left(-|\mvec-\nvec|/R_{sc}\right)$. The
screening radius $R_{sc}$ controls the length of the interaction, and
$\kappa$ is the strength of the interaction. $\mvec$ represents the position of the ion and $\nvec$ is the position of the electron.
The electron-ion interaction leads to an effective retarded electron-electron interaction characterized by the function,
\begin{equation}
\Phi_{\Delta\rvec}(\rvec(\tau),\rvec(\tau')) = \sum_{\mvec}f_{\mvec}[\rvec(\tau)]f_{\mvec+\Delta\rvec}[\rvec(\tau')]
\end{equation}
where $\Delta\rvec$ is the offset between end configurations of the
paths. By changing $R_{sc}$ it is possible to investigate the
Hubbard-Holstein model ($R_{sc}\rightarrow 0$) and the
Hubbard-Fr\"ohlich model ($R_{sc}\rightarrow\infty$). We also consider
a model similar to the nearest-neighbor model of Bon\v{c}a and Trugman
\cite{bonca2001a}, where effective electron-electron interactions are
truncated at near-neighbor sites ($\Phi(\avec)=\Phi(0)/2$ and all
other $\Phi=0$ where $\avec$ are vectors to the near-neighbor
sites). This simplified model maps directly onto a $U-V$ model in the
large phonon frequency limit, and could be particularly useful for
understanding limiting behavior
\footnote{The strict analogue of the near-neighbor model would have
ions above near-neighbor bonds, which would only interact with
electrons at the ends of those bonds. In that case,
$\Phi(\avec)=\Phi(0)/4$ for the square lattice and
$\Phi(\avec)=\Phi(0)/6$ on the triangular lattice.}.

This article is structured as follows. In section \ref{sec:formalism},
we briefly review the differences between the CTQMC algorithm in 1D
and 2D. In sections \ref{sec:singlet} and \ref{sec:triangular} we make a
comprehensive survey of the parameter space of singlet bipolarons in
2D. Triplet properties are discussed in section \ref{sec:triplet}. One
of the additional advantages of our algorithm is that bipolaron
dispersions can be computed efficiently, especially in the large
$\lambda$ regime, and these are shown in section
\ref{sec:dispersion}. Finally we summarize in section \ref{sec:summary}.

\section{Method}
\label{sec:formalism}

We use a continuous time quantum Monte Carlo algorithm based on path
integrals. We have
previously discussed our algorithm in detail with regard to computations in 1D \cite{hague2009a}, so we
do not repeat those details here. There are a few small subtleties
relating to using the algorithm in 2D, especially on a triangular
lattice, which are discussed here. Our
algorithm has been thoroughly checked against results from the $U$-$V$
model.

\subsection{Triple kink insertions on triangular lattices}

There is a subtle difference to the algorithm on triangular lattices,
or more generally on any lattice where an electron can return to its
original position in 3 hops (such as lattices with nearest and
next-nearest neighbor hopping). We illustrate this by considering an
example configuration that is permitted on such lattices: Let one of
the paths have no kinks and the other path have 3 kinks - one in each
of the nearest neighbor directions - so that the start and end of each
path lies on the same lattice site. Such a configuration is clearly
permitted, since the periodic boundary conditions in imaginary time are
satisfied. However, if only binary kink insertions are included in the
algorithm then it is not possible to update between configurations with
odd and even numbers of kinks. Thus, to ensure ergodicity, an update
with a 3 kink insertion is proposed.

Our scheme for 3 kink insertion is similar to the one that we use for
binary updates. To avoid complications, we do not weight the positions
of the kinks in imaginary time. Ternary kink insertions do not need to
be especially efficient; if path configurations can be
updated from even to odd numbers of kinks (and vice-versa) with
reasonable regularity, binary insertions can be used to sample the
remaining configurations efficiently. Our scheme is as follows,

\begin{enumerate}
\item We select a kink type from the 6 possible kinks and assign the label $\lvec_{1}$.
\item We choose kinks at 120$^{o}$ and 240$^{o}$ rotations from $\lvec_{1}$ and assign them the labels $\lvec_{2}$ and $\lvec_{3}$
\item We choose insertion or removal of kinks with equal probability $1/2$.
\item If inserting, we choose imaginary times $\tau_1$, $\tau_2$ and $\tau_3$ for the new kinks with equal probability $1/\beta$ from the interval $[1,\beta)$.
\item If removal is selected and there is not at least one of each type of kink, then abort. Otherwise select a kink of type $\lvec_1$ for removal with equal probability $1/N_{\lvec_1}$, etc
\end{enumerate}

If configuration $[D]$ has three more kinks than configuration $[C]$, then insertion takes place with probability
\begin{equation}
P[C\rightarrow D] =  \min\left\{\frac{(t\beta)^3}{N_{\lvec_1}[D]N_{\lvec_2}[D]N_{\lvec_3}[D]}e^{A[C]-A[D]},1\right\}
\end{equation}
and for removal,
\begin{equation}
P[D\rightarrow C] = \min\left\{\frac{N_{\lvec_1}[D]N_{\lvec_2}[D]N_{\lvec_3}[D]}{(t\beta)^3}e^{A[D]-A[C]},1\right\}
\end{equation}
Note that the number $N_{\lvec_1}[D]$ represents the number of kinks
of type $\lvec_1$ in configuration D. Therefore when kinks are
inserted, this is the number of kinks in the final configuration. When kinks are removed, $N_{\lvec_1}[D]$ is the number of kinks in the initial configuration.

\subsection{Path exchange in 2D}

There is a small difference between exchange in 1D and 2D, especially
on the triangular lattice. In 1D, exchange can be carried out by
inserting and removing kinks and antikinks. The
form of the exchange update in 1D is,
\begin{equation}
P(C\rightarrow D) = {\rm min}\{1,Q^{(A)}_{\lvec,-\lvec}Q^{(B)}_{-\lvec,\lvec}\exp(A[C]-A[D])\}
\end{equation}
where,
\begin{equation}
Q^{(A)}_{\lvec,-\lvec}=(t\beta)^{2n_{A}-\Delta}\frac{\min(N_{A-\lvec},\Delta)+1}{\min(N_{A\lvec}+n_{A},\Delta)+1}\frac{_{N_{A-\lvec}}P_{\Delta-n_{A}}}{_{N_{A\lvec}+n_{A}}P_{n_{A}}}
\end{equation}
here $\Delta$ is the displacement between the $\tau=\beta$ ends of
path A and path B. $_{n}P_{k}=n!/(n-k)!$ is the number of permutations. There are $N_{p}=\min(N_{A-\lvec},\Delta)+1$
possible updates that can be made by inserting $n_{A}$ kinks into path
A and removing $m_{A}$ antikinks, where $n_{A}+m_{A}=\Delta$. $m_{A}$
is chosen with equal weighting $1/N_{p}$. $N_{A\lvec}$ is the number
of kinks of type $\lvec$ on path A. The expression for path B is
similar, but the kink and anti-kink assignments are reversed to get
displacement $-\Delta$.

In 2D, two sets of kinks and antikinks in different directions are
required to exchange the ends of the paths, so the exchange update has
the form,
\begin{equation}
P(C\rightarrow D) = \min\{1,\exp(A[C]-A[D])\prod_{i=1}^{d}Q^{(A)}_{\lvec_{i},-\lvec_{i}}Q^{(B)}_{-\lvec_{i},\lvec_{i}}\}
\end{equation}
where there is a different $\Delta$ for each kink type to be inserted
(and thus a different $N_p$ and $m$ which should be chosen
independently for each kink type and path) as determined from the
number of kinks that would be required to hop from the end of one path
to the end of the other. $d$ is the number of lattice dimensions.

For the triangular lattice, the choice of the two kink directions is not unique,
since the kinks are not orthogonal. There are six nearest neighbor
vectors representing kinks. To carry
out the update, kinks representing two different directions are chosen
with equal weight from all available possibilities (that is the two chosen kinks may
not be antikinks of each other). Since an equal weight scheme is used,
the factors relating to the probability of this choice cancel on both sides of the balance equations. It
is important to choose kinks from all possible directions to improve
the efficiency of the algorithm.

\subsection{Global path shifts}

As in the 1D case, a useful update displaces a single path when the
configurations are not exchanged. In both cases, one of the paths is
chosen with equal probability, a kink direction is chosen with equal
probability and then the path is shifted through $r$ lattice sites,
where $r$ is an integer chosen randomly between $0$ and
$R_{\max}-1$. Typically $R_{\max}=50$ for lattice size 100. Failure to include global path
shifts leads to an inefficient algorithm with large correlation
between measurement, which is especially bad when the bipolaron is only just bound.
Global shifts are essential when the bipolaron radius is to be measured.

\begin{figure*}
\includegraphics[height=55mm,angle=270]{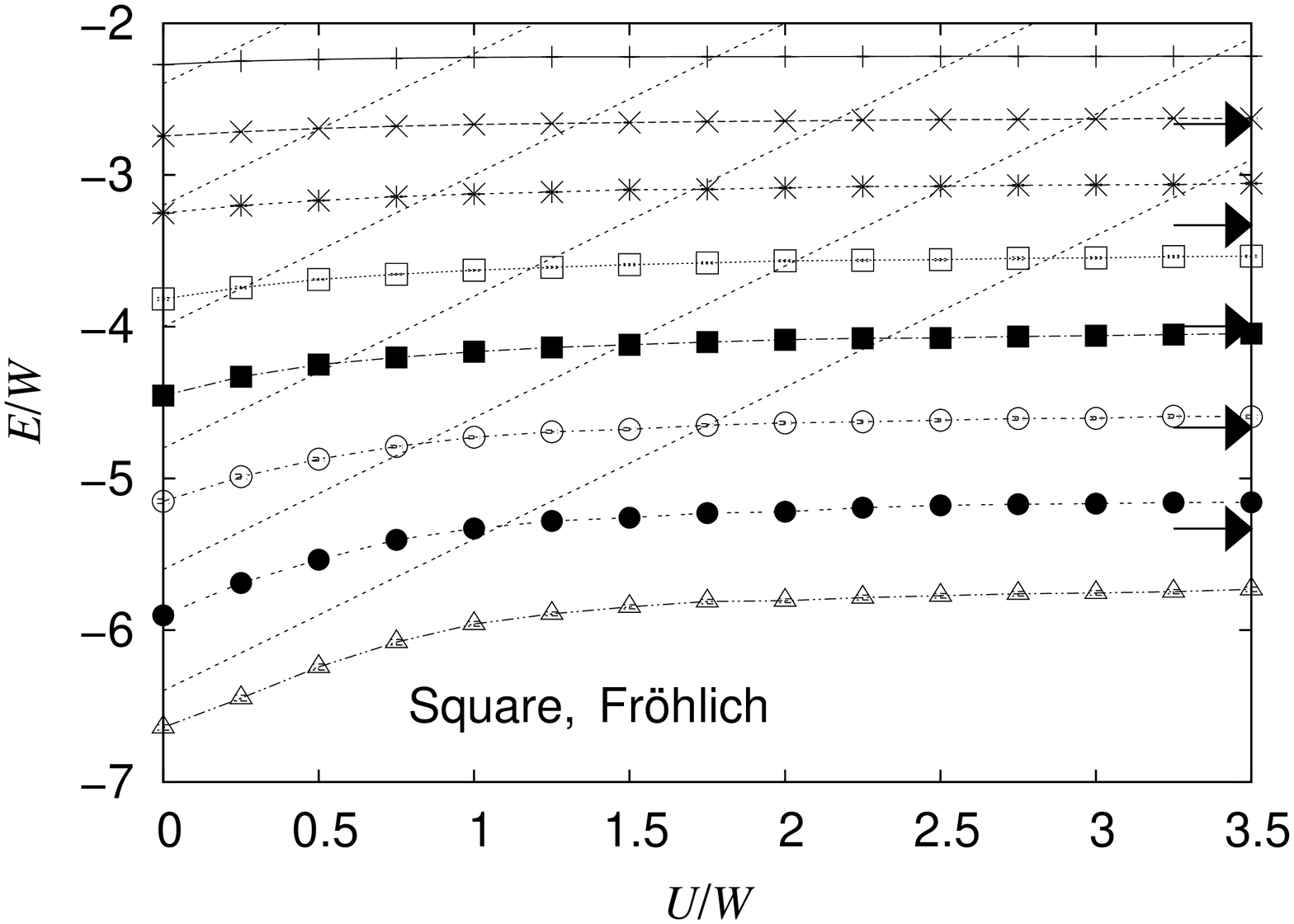}
\includegraphics[height=55mm,angle=270]{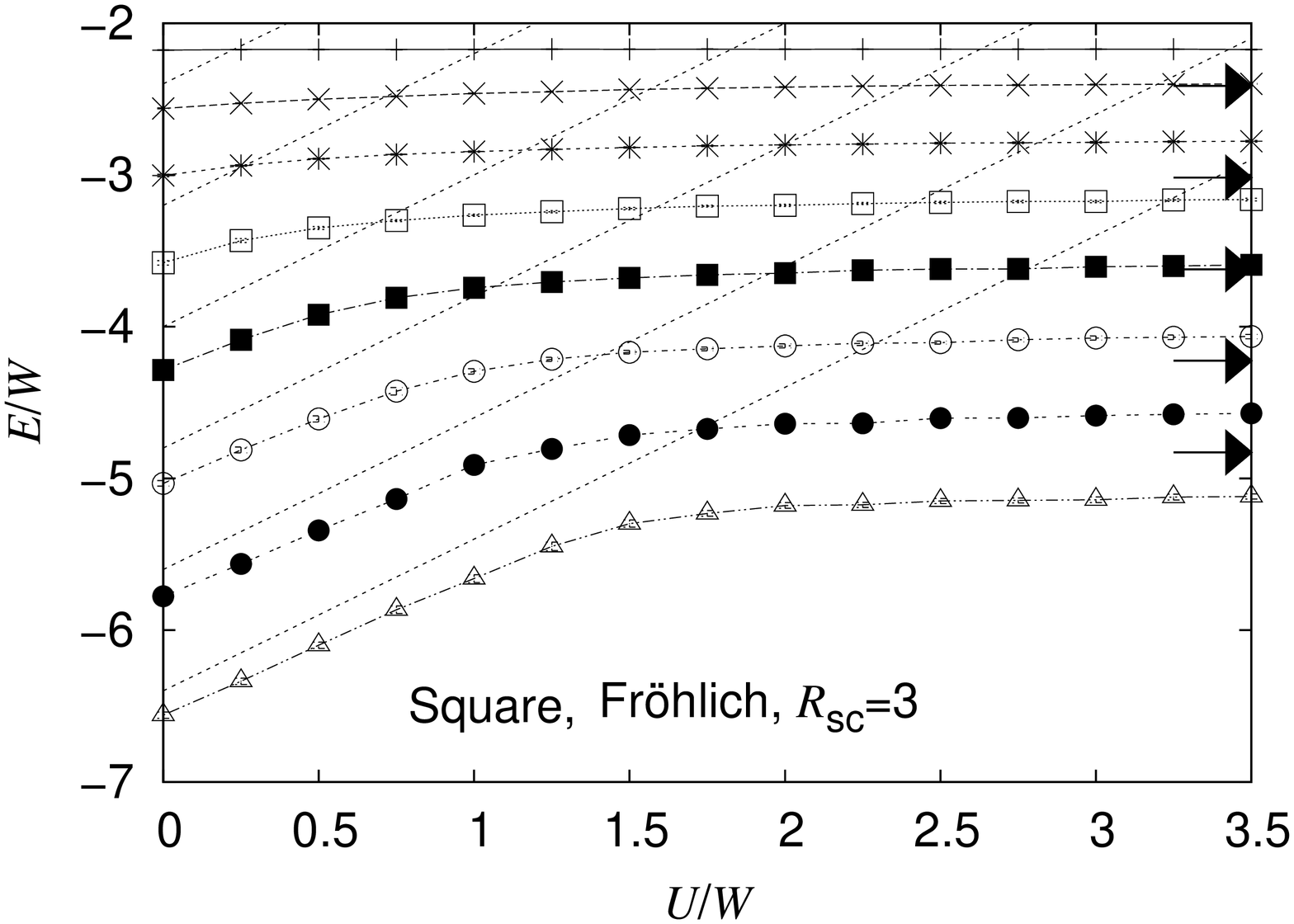}
\includegraphics[height=55mm,angle=270]{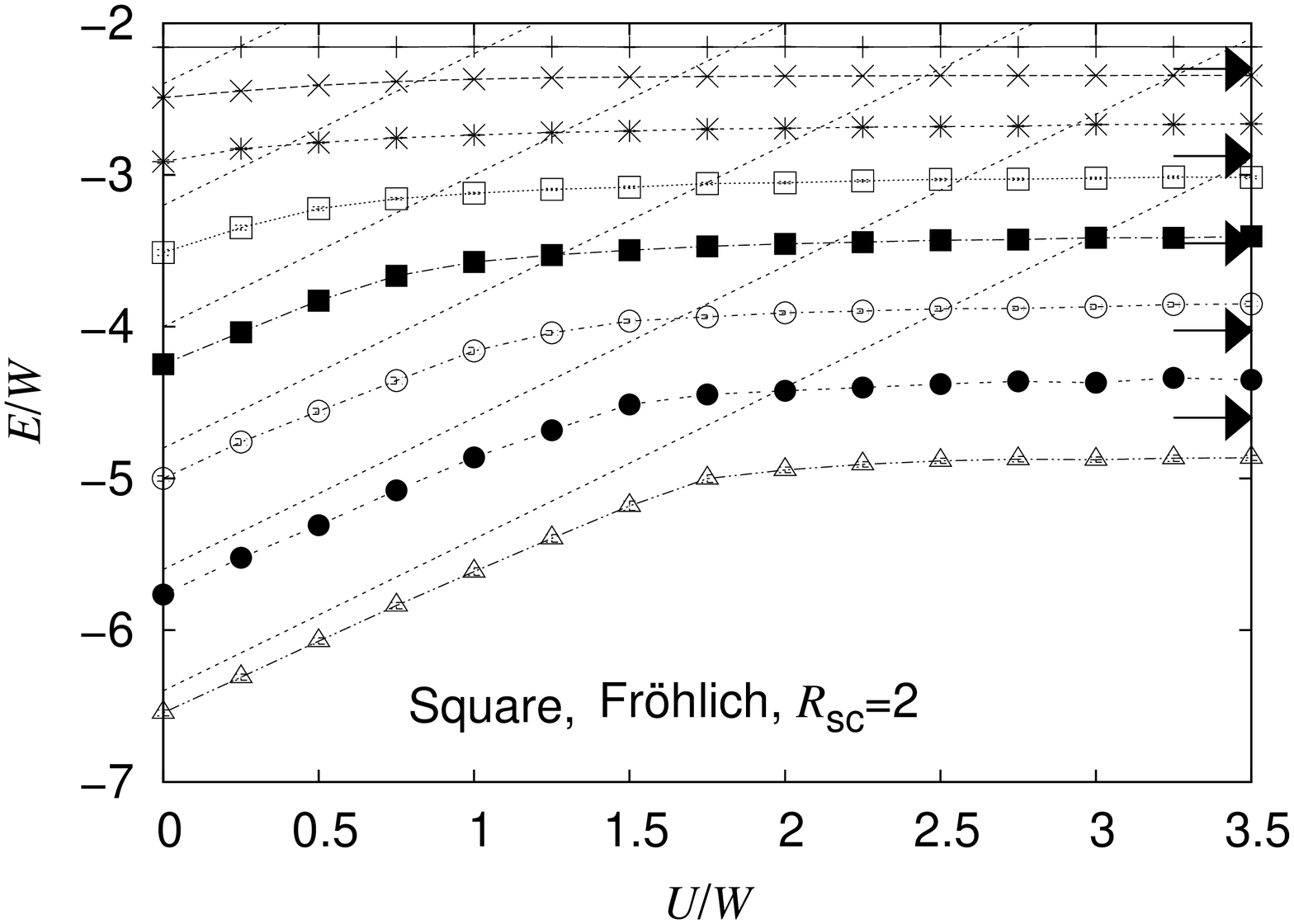}
\includegraphics[height=55mm,angle=270]{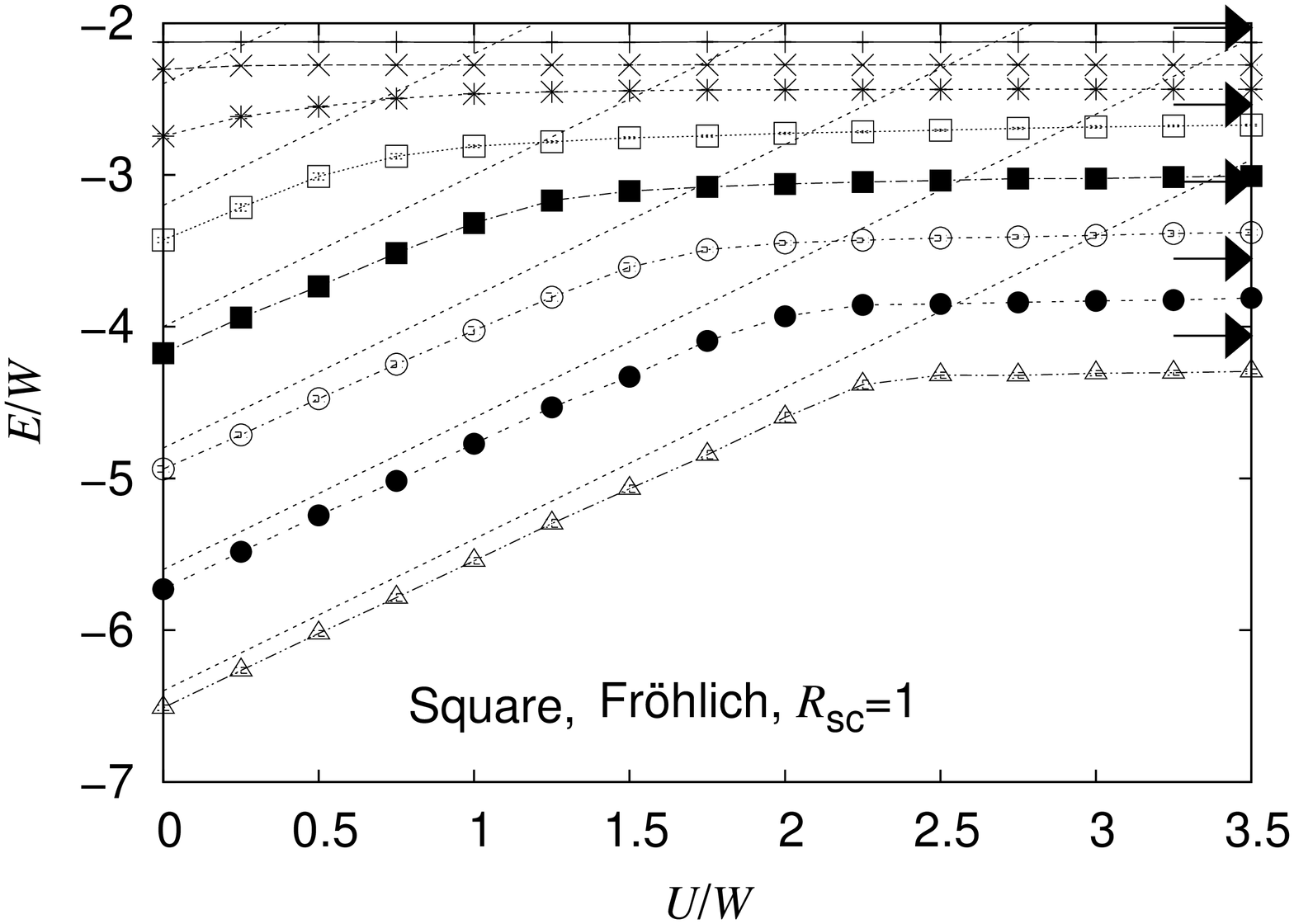}
\includegraphics[height=55mm,angle=270]{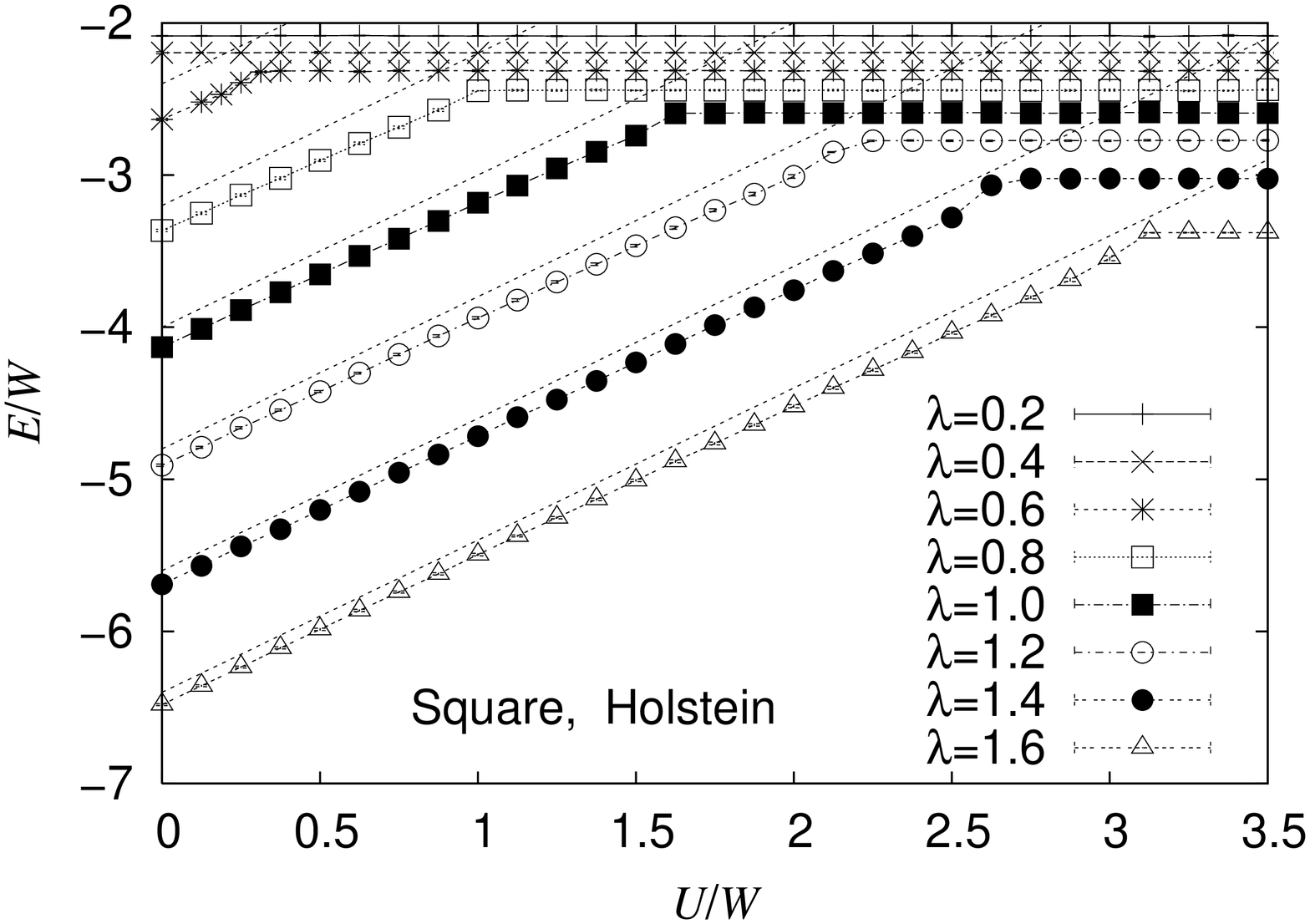}
\includegraphics[height=55mm,angle=270]{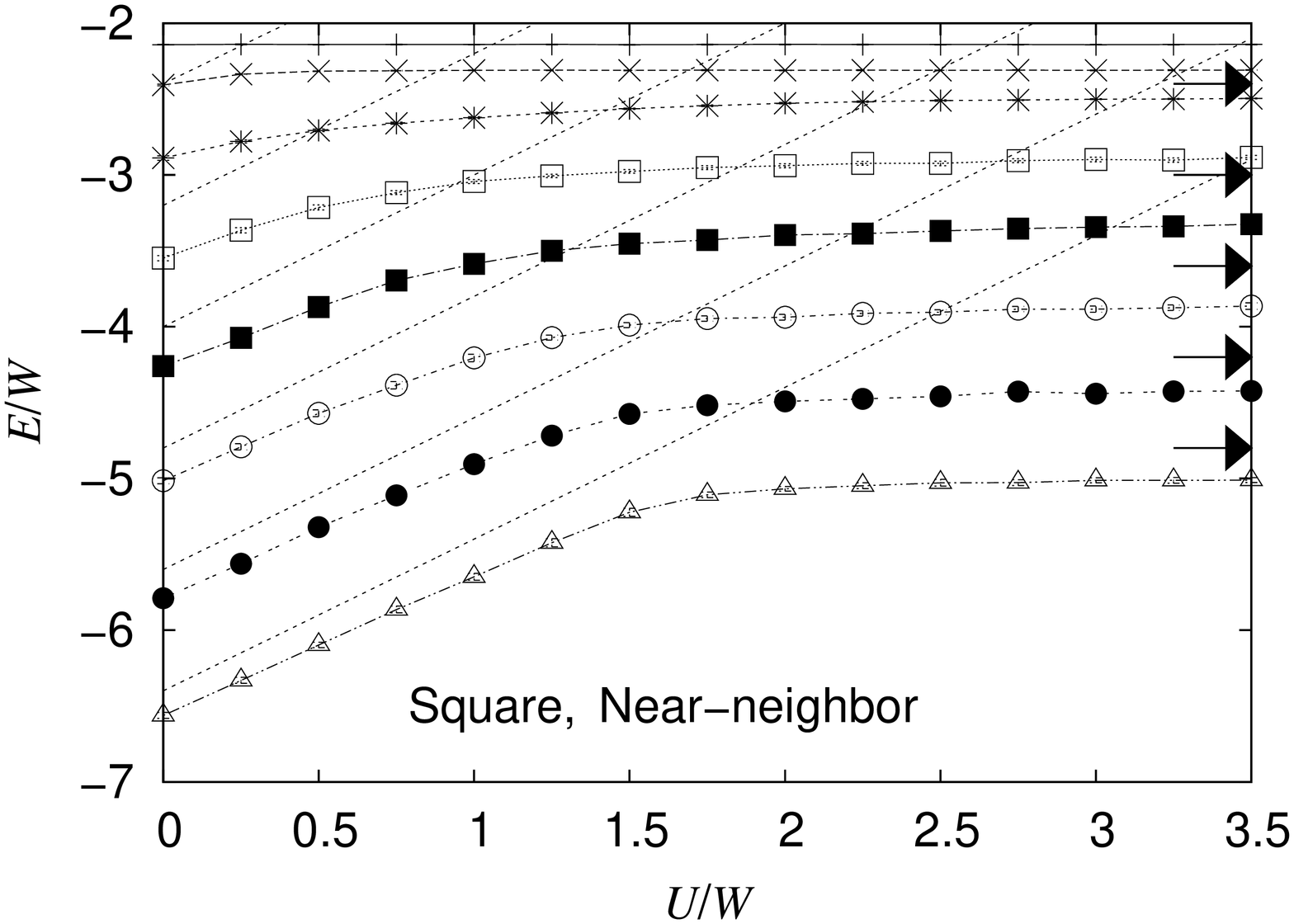}
\caption{Total energy of bipolarons on the square lattice for various
$U$, $\lambda$, $R_{sc}$, $\bar{\omega}=1$ and $\bar{\beta}=\beta/t=14$. There
is a clear qualitative difference between bipolarons with long range
interactions and those formed from the purely local Holstein
interaction. Unless otherwise indicated, error bars represent three
standard deviations. Where error bars are not visible, they are smaller than the points.}
\label{fig:energysinglet}
\end{figure*}

\begin{figure*}
\includegraphics[height=55mm,angle=270]{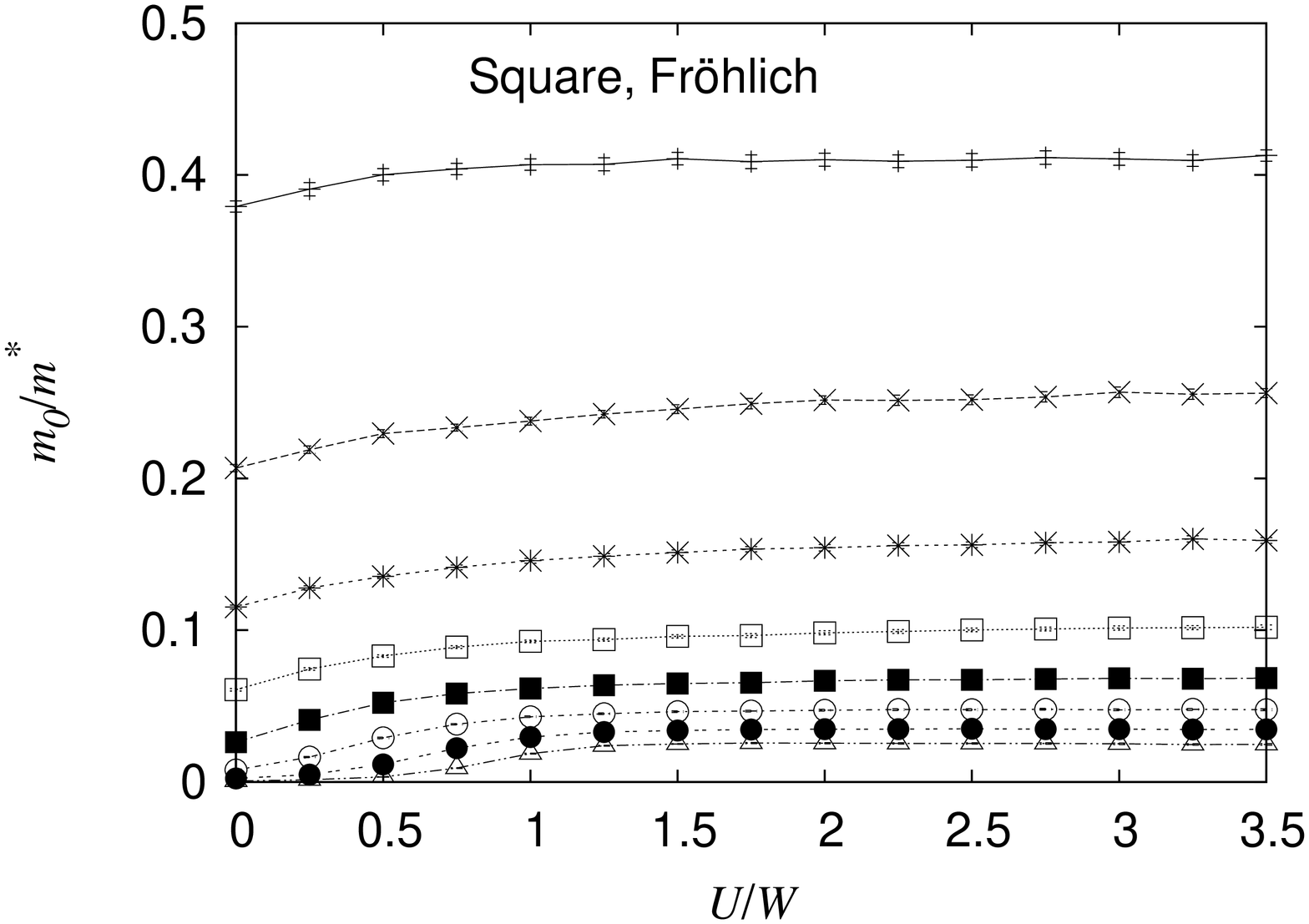}
\includegraphics[height=55mm,angle=270]{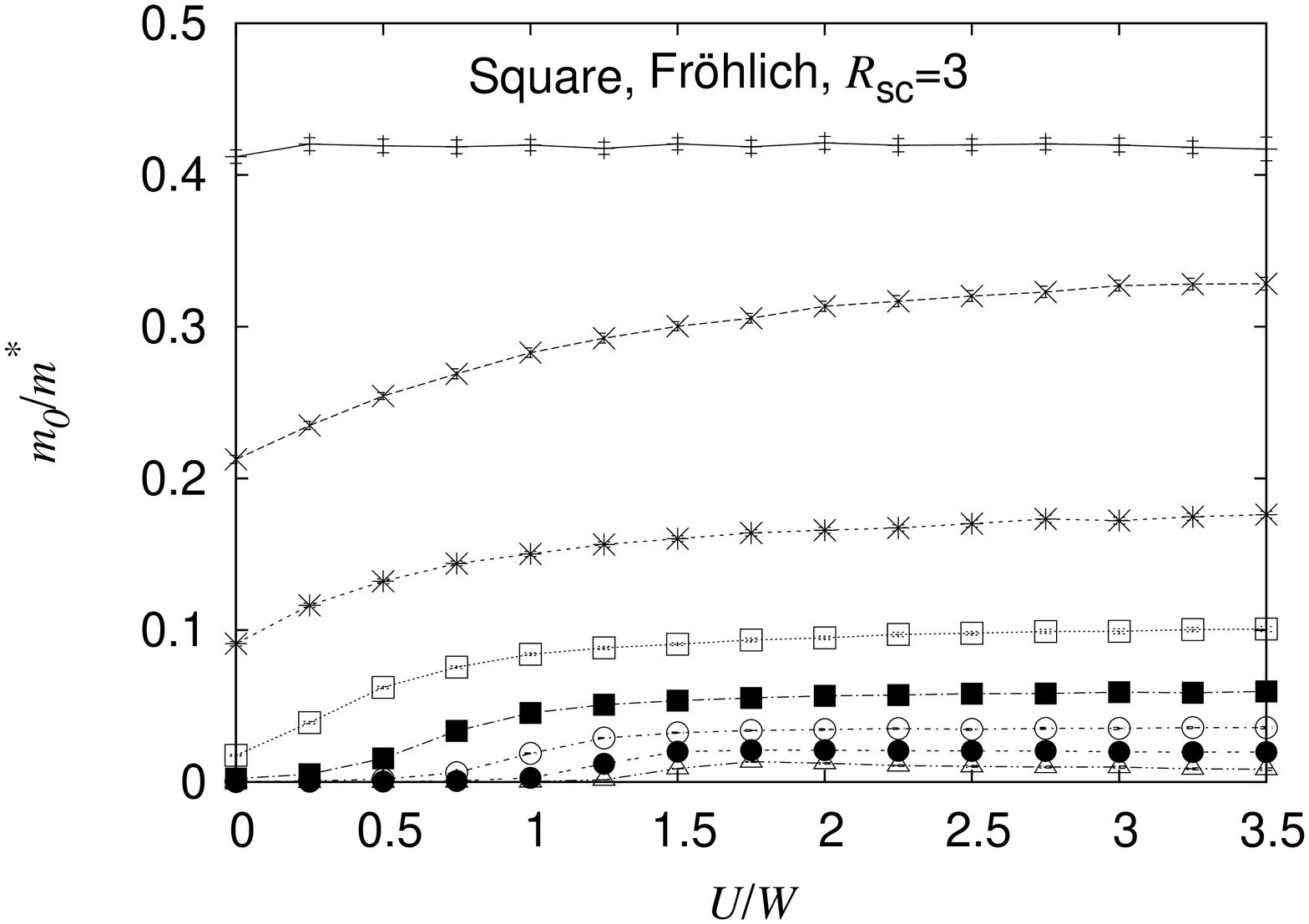}
\includegraphics[height=55mm,angle=270]{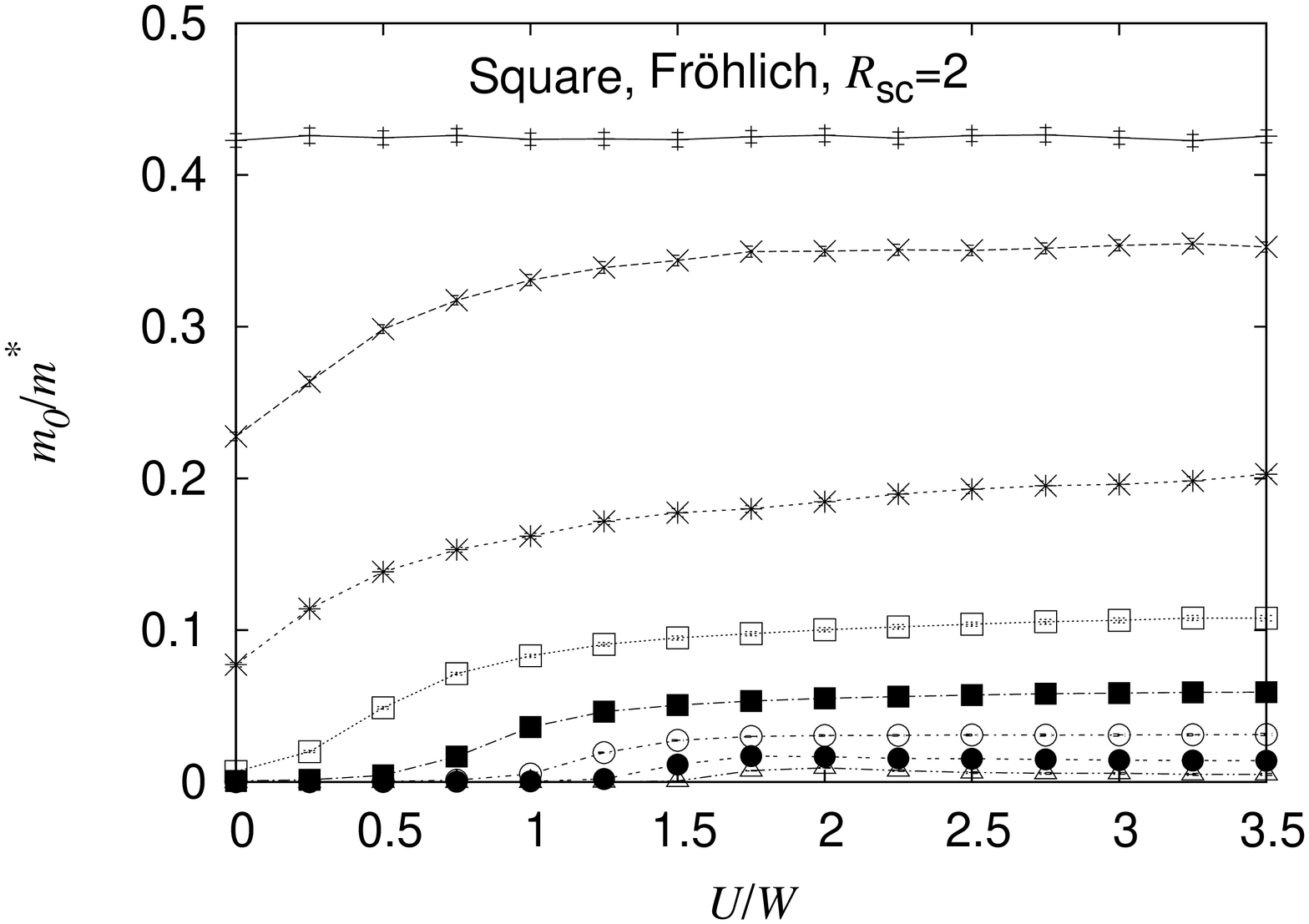}
\includegraphics[height=55mm,angle=270]{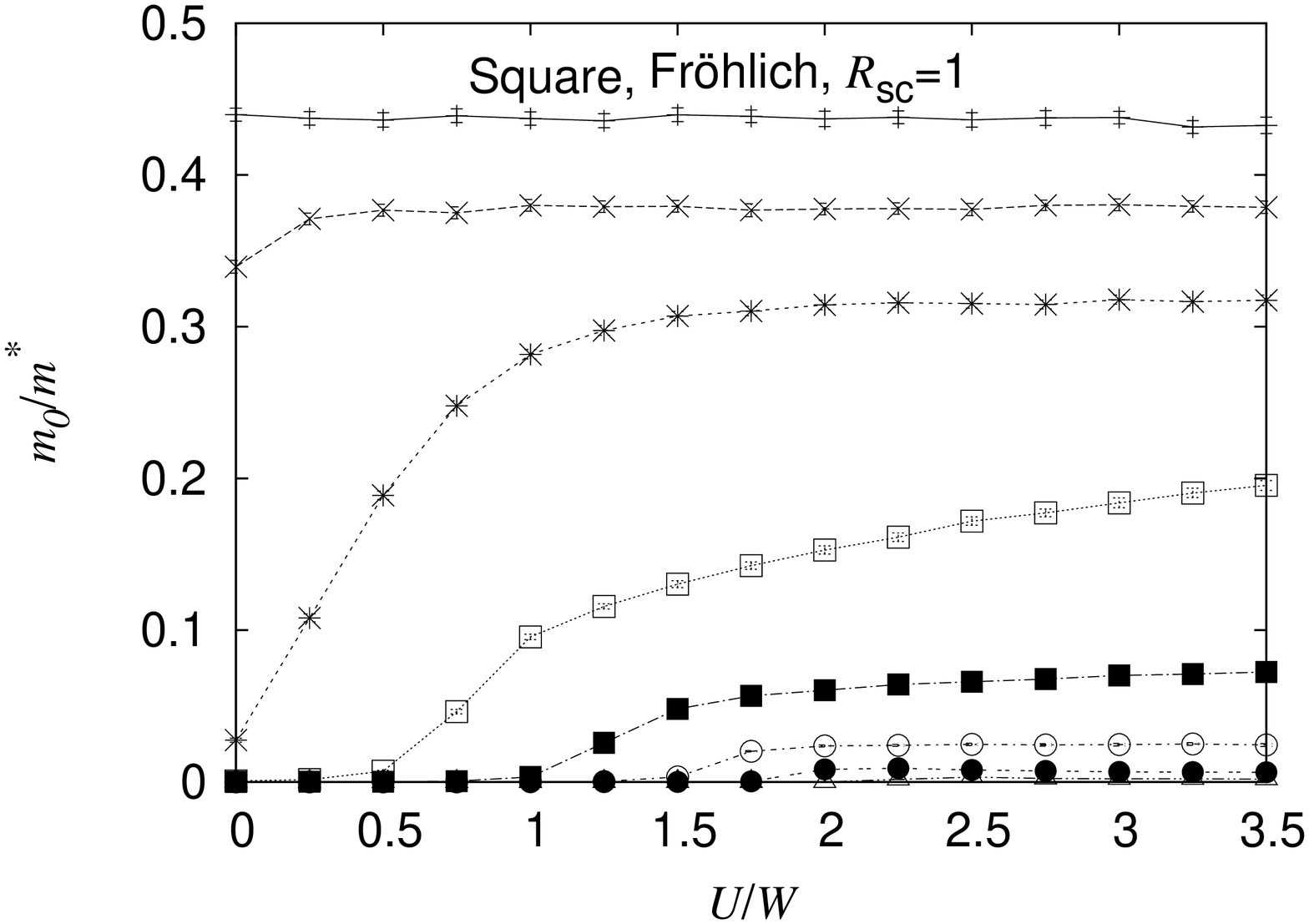}
\includegraphics[height=55mm,angle=270]{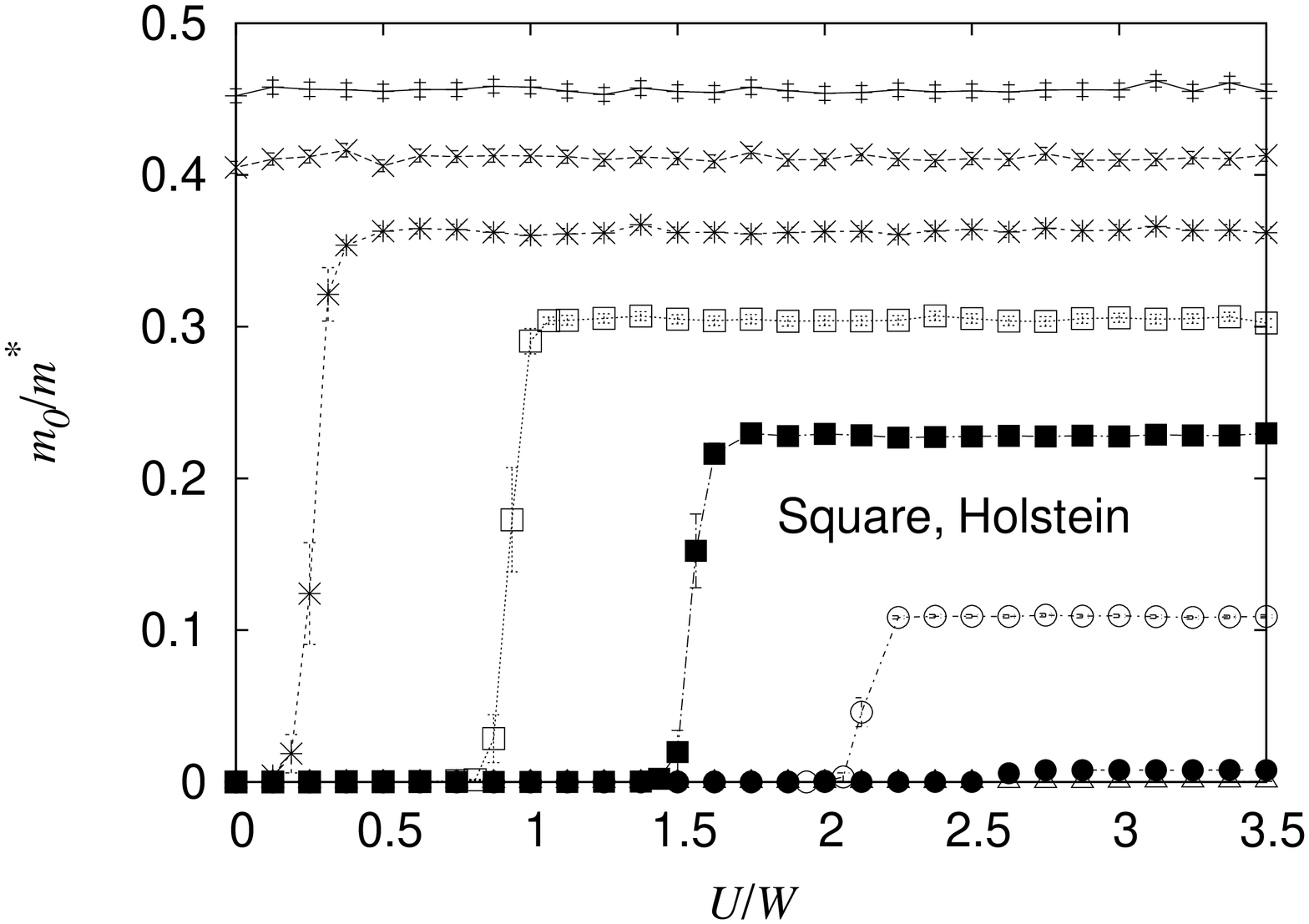}
\includegraphics[height=55mm,angle=270]{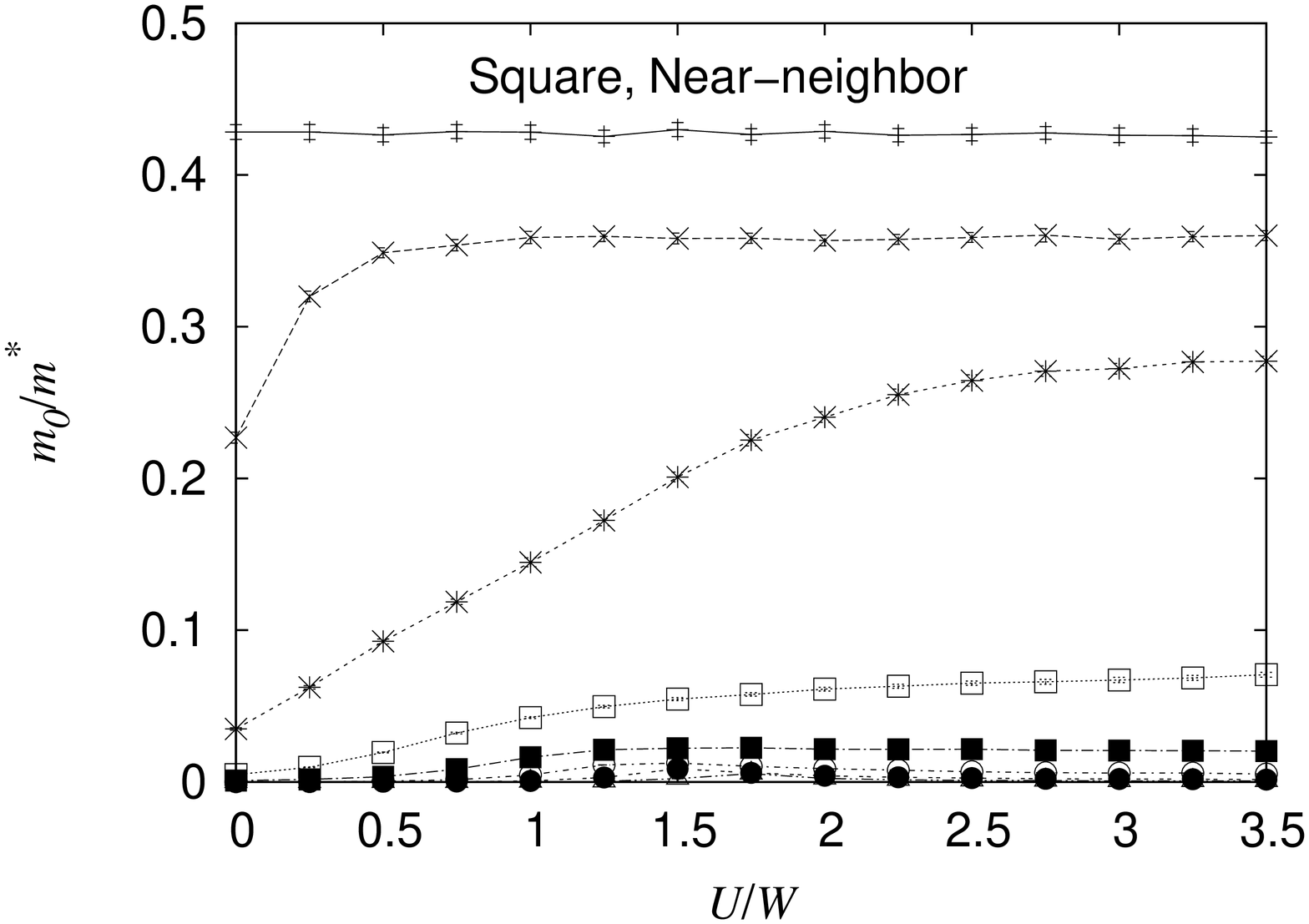}
\caption{Inverse mass of bipolarons on the square lattice. Parameters are the same as in Fig. \ref{fig:energysinglet}. Again, qualitative differences between the bipolarons formed via the purely site local Holstein interaction and the long-range interactions can be seen.}
\label{fig:imsquaresinglet}
\end{figure*}

\begin{figure}
\includegraphics[height=75mm,angle=270]{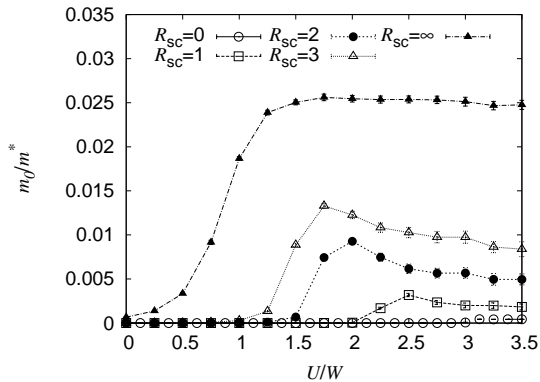}
\caption{Comparison of the inverse mass for bipolarons on the square lattice. $\lambda=1.6$ and various interaction ranges are considered. Other parameters are as in Fig. \ref{fig:energysinglet}. At intermediate $U$, the energies of the on-site (S0) and intersite (S1) bipolarons are degenerate, so a light bipolaron can be formed without the need for triangular plaquettes.}
\label{fig:imcomparelargelambda}
\end{figure}

\begin{figure*}
\includegraphics[height=55mm,angle=270]{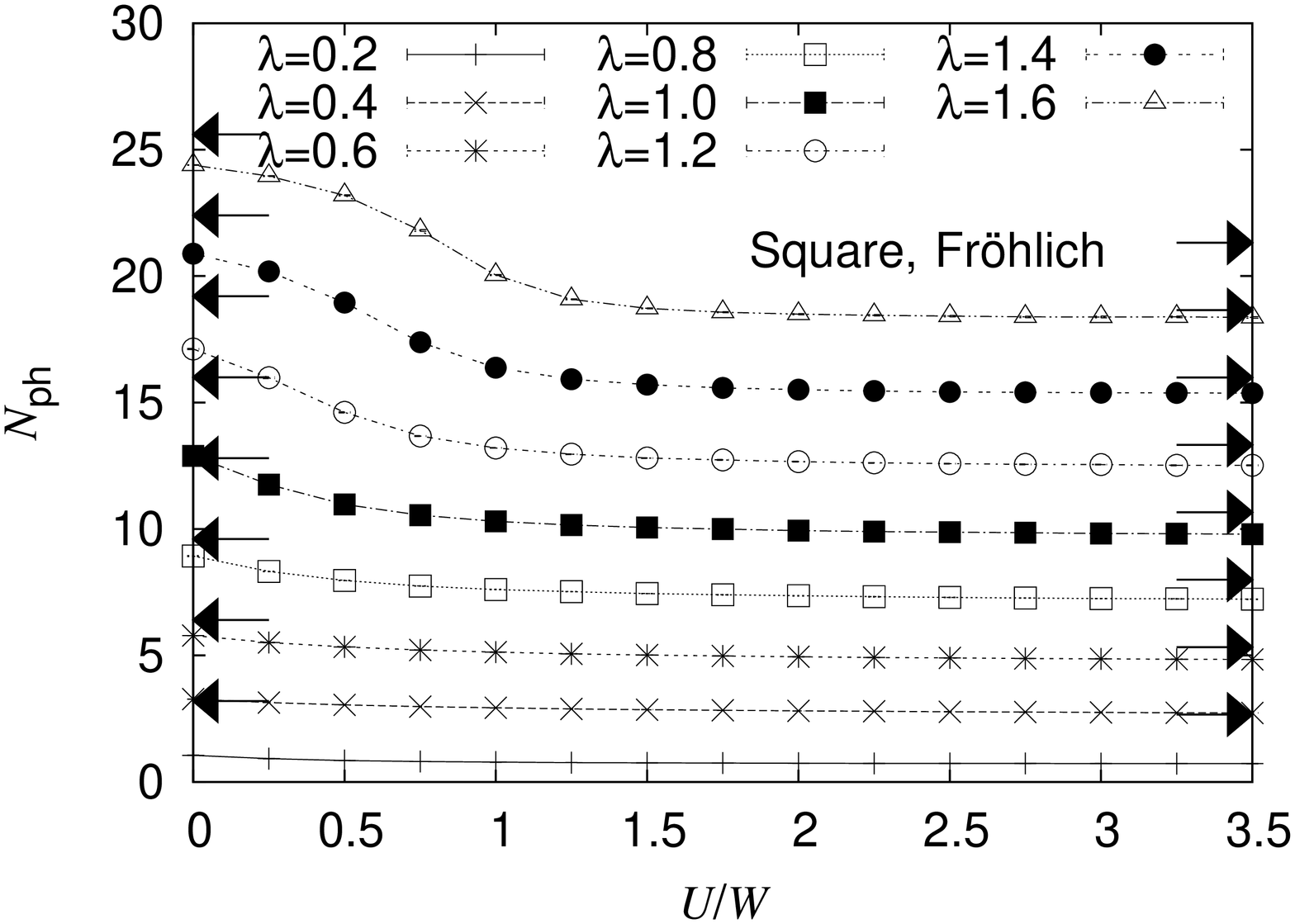}
\includegraphics[height=55mm,angle=270]{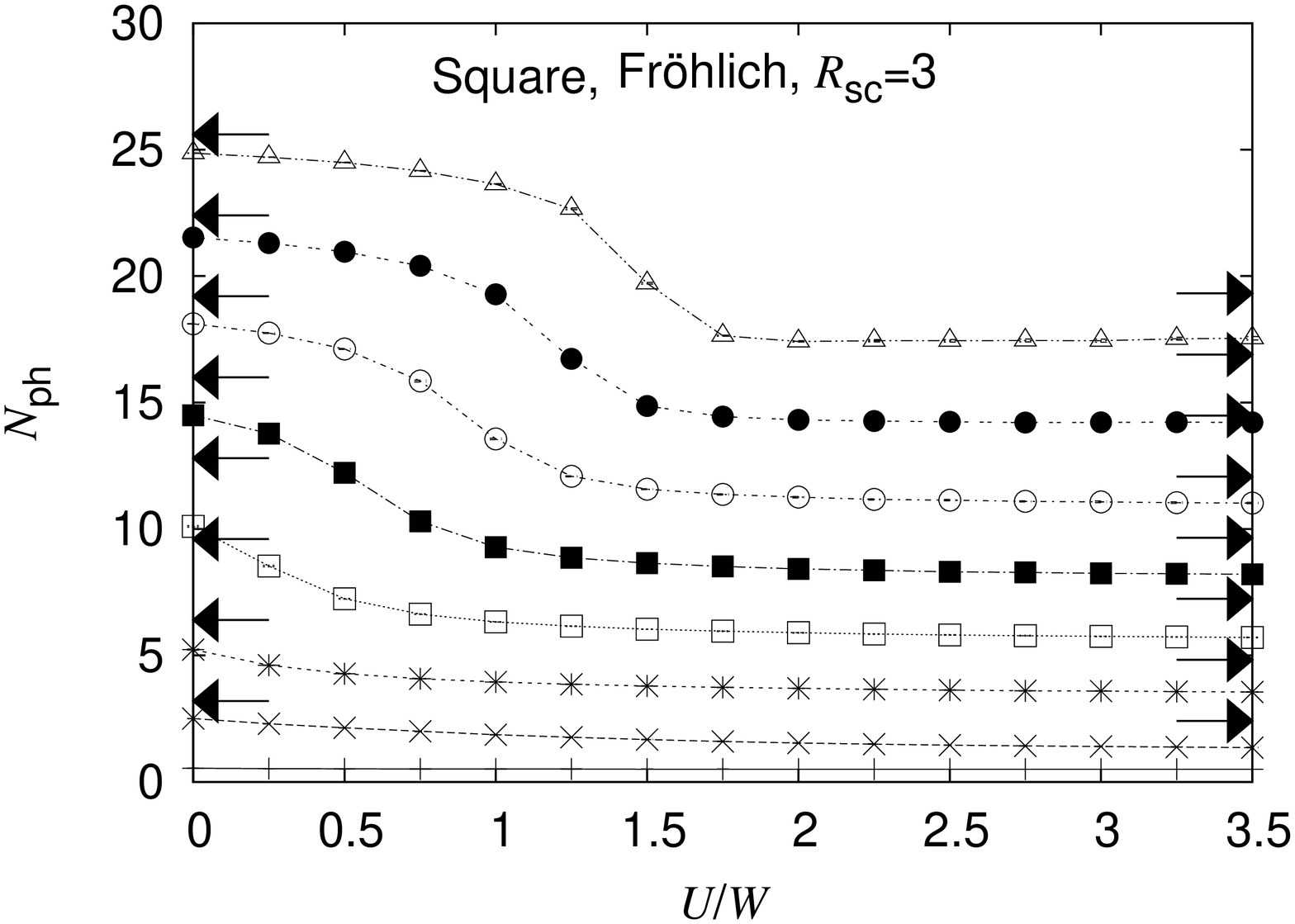}
\includegraphics[height=55mm,angle=270]{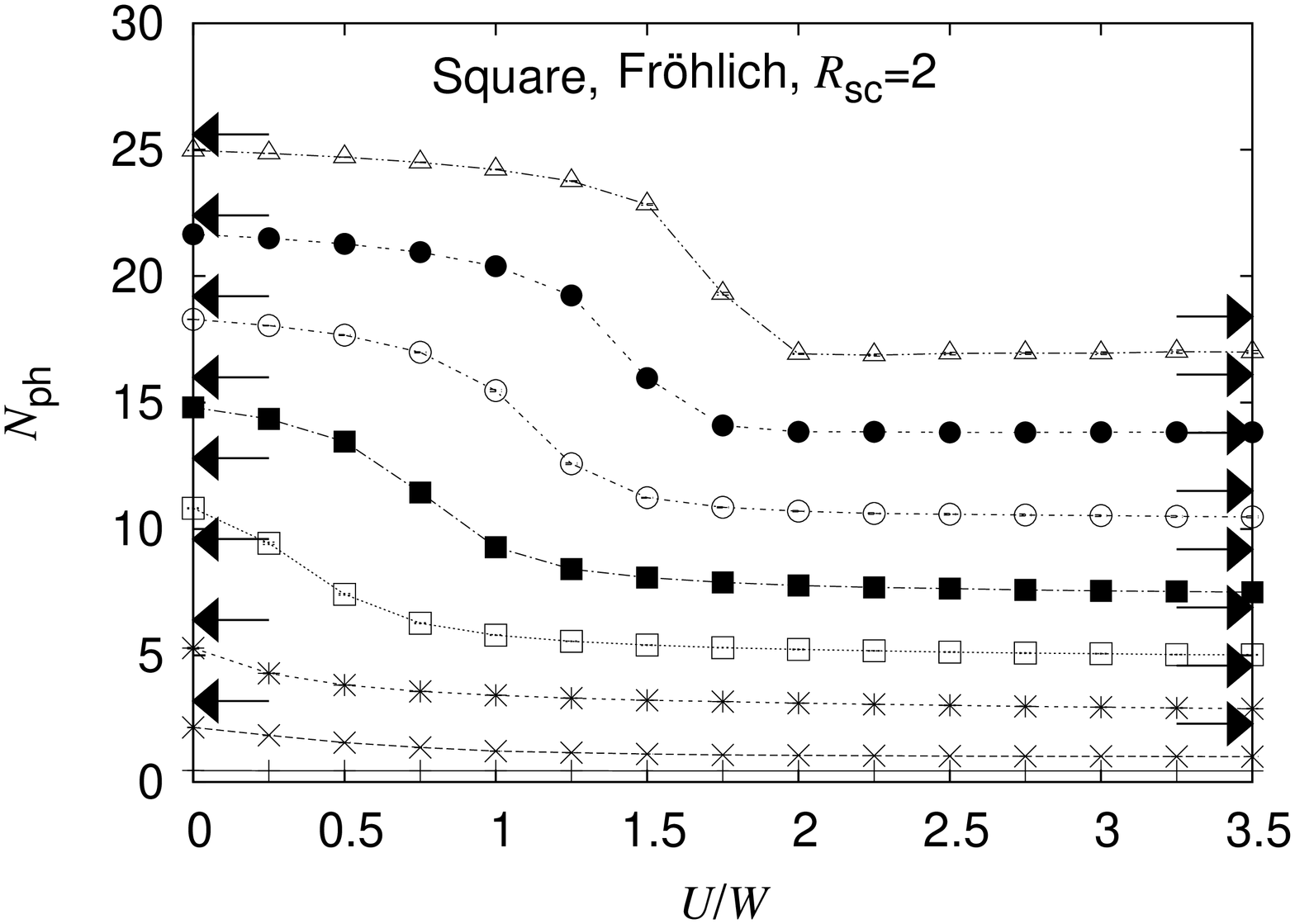}
\includegraphics[height=55mm,angle=270]{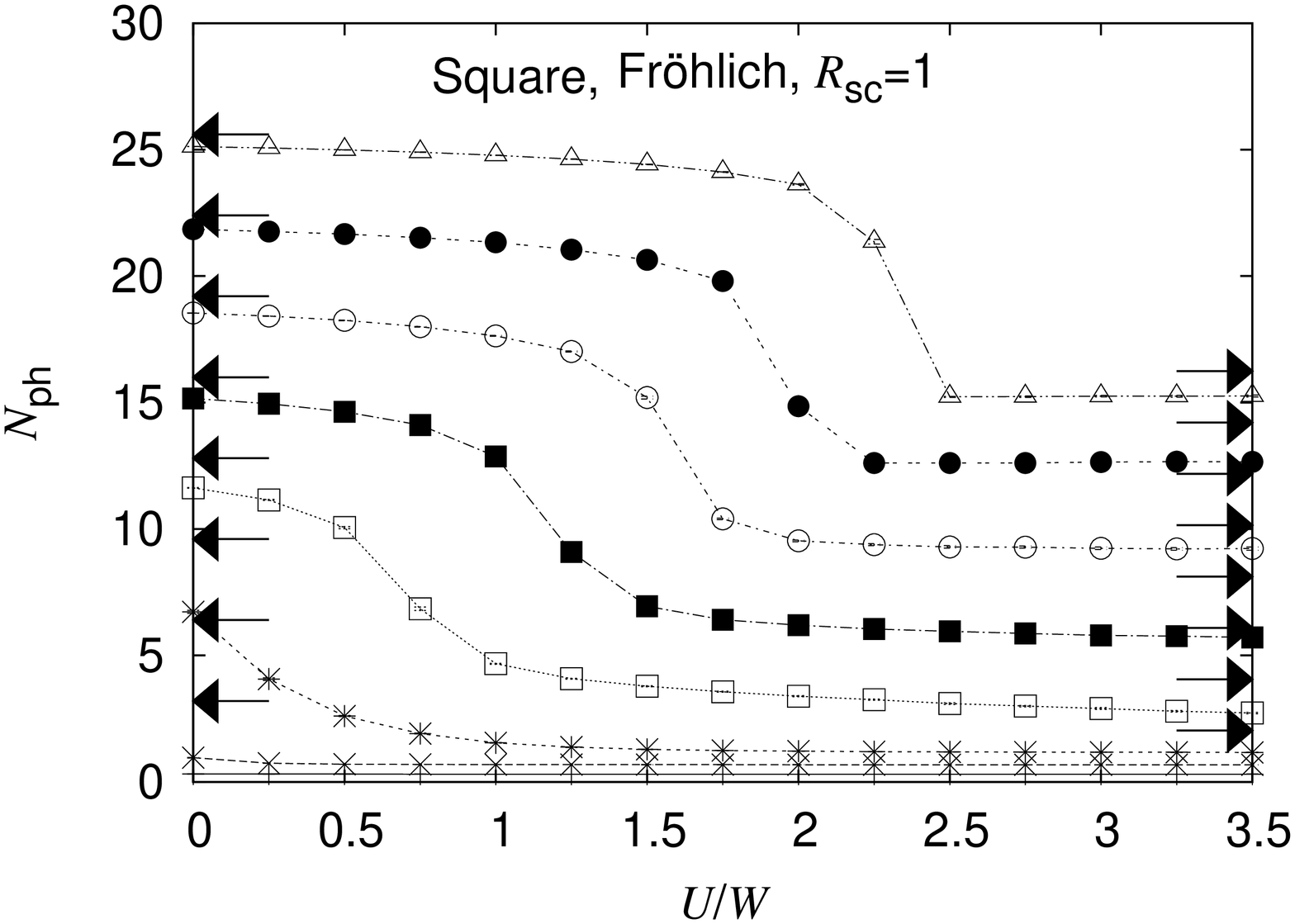}
\includegraphics[height=55mm,angle=270]{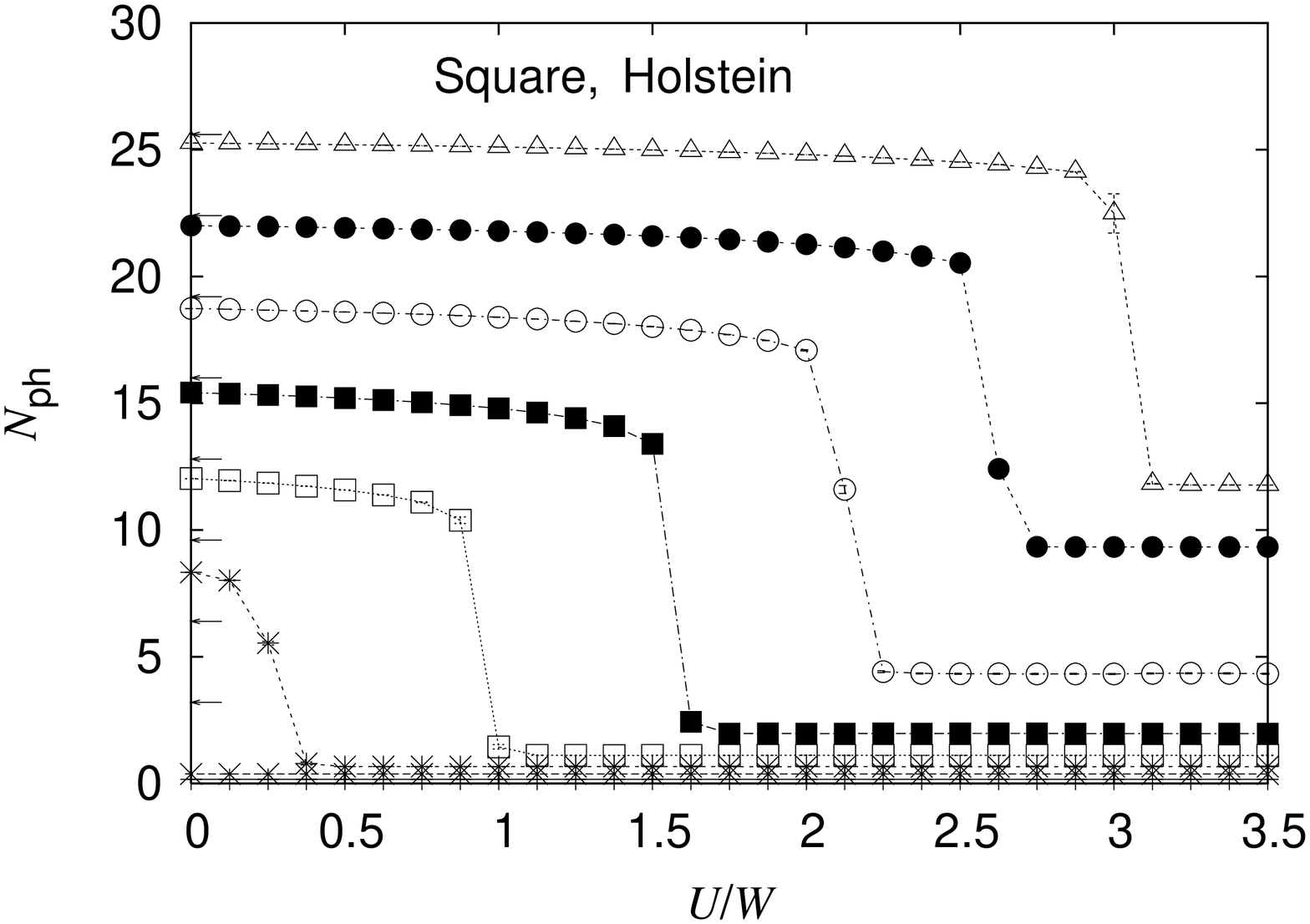}
\includegraphics[height=55mm,angle=270]{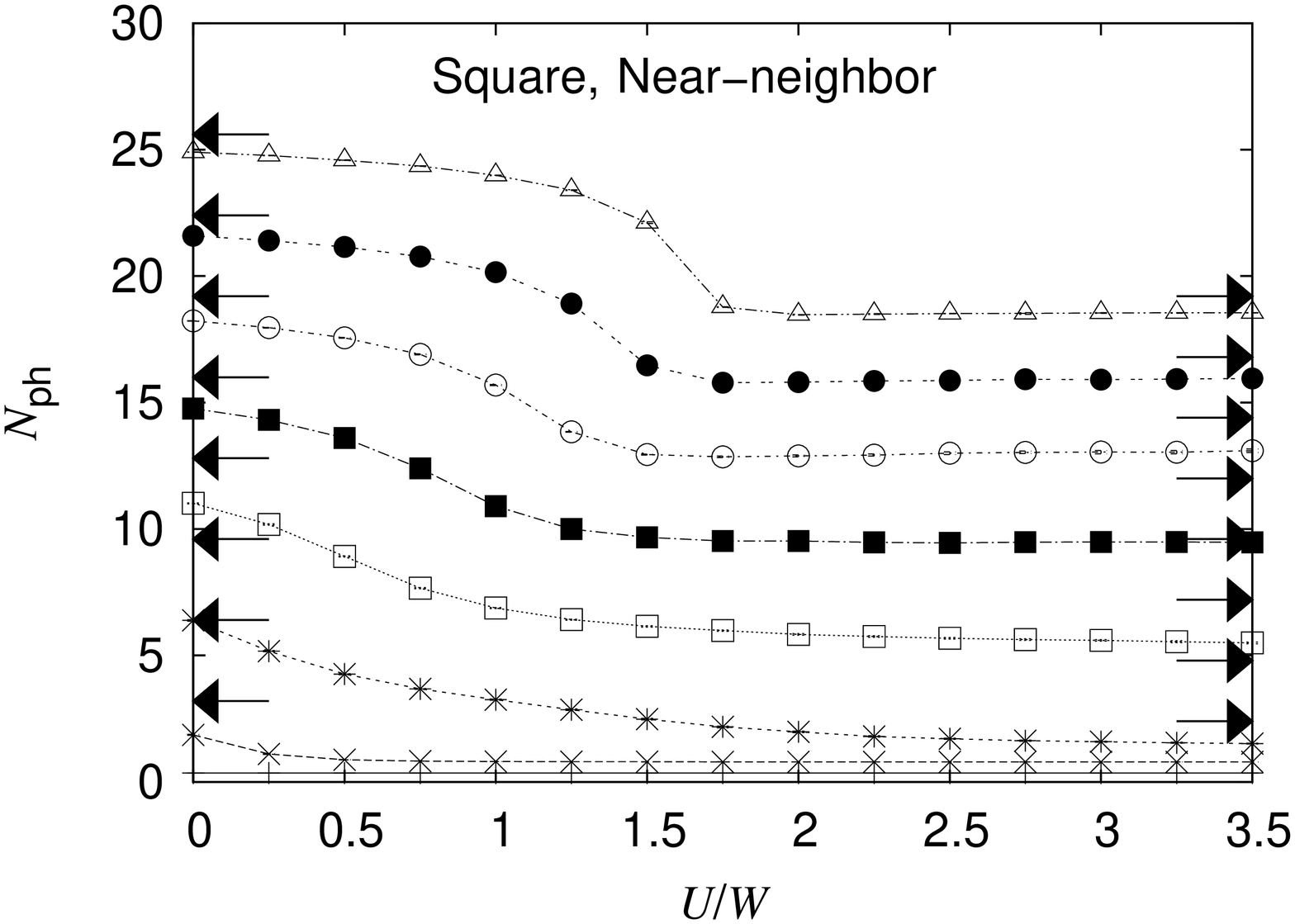}
\caption{Number of phonons associated with bipolarons on the square lattice. Parameters are identical to Fig. \ref{fig:energysinglet}. Again, the properties of the Holstein bipolaron are qualitatively different to those of the Fr\"ohlich bipolaron.}
\label{fig:nophonsinglet}
\end{figure*}

\begin{figure*}
\includegraphics[height=55mm,angle=270]{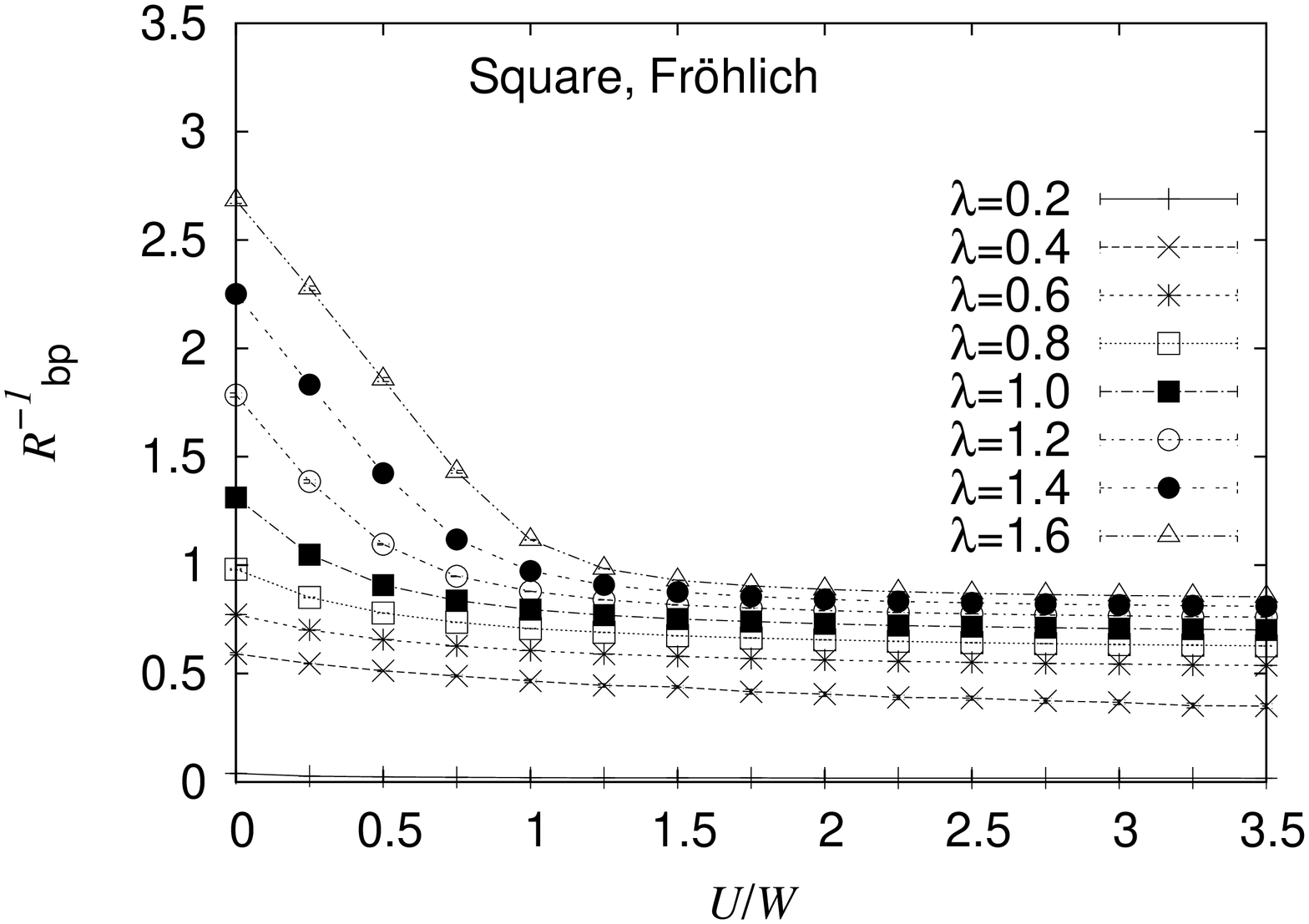}
\includegraphics[height=55mm,angle=270]{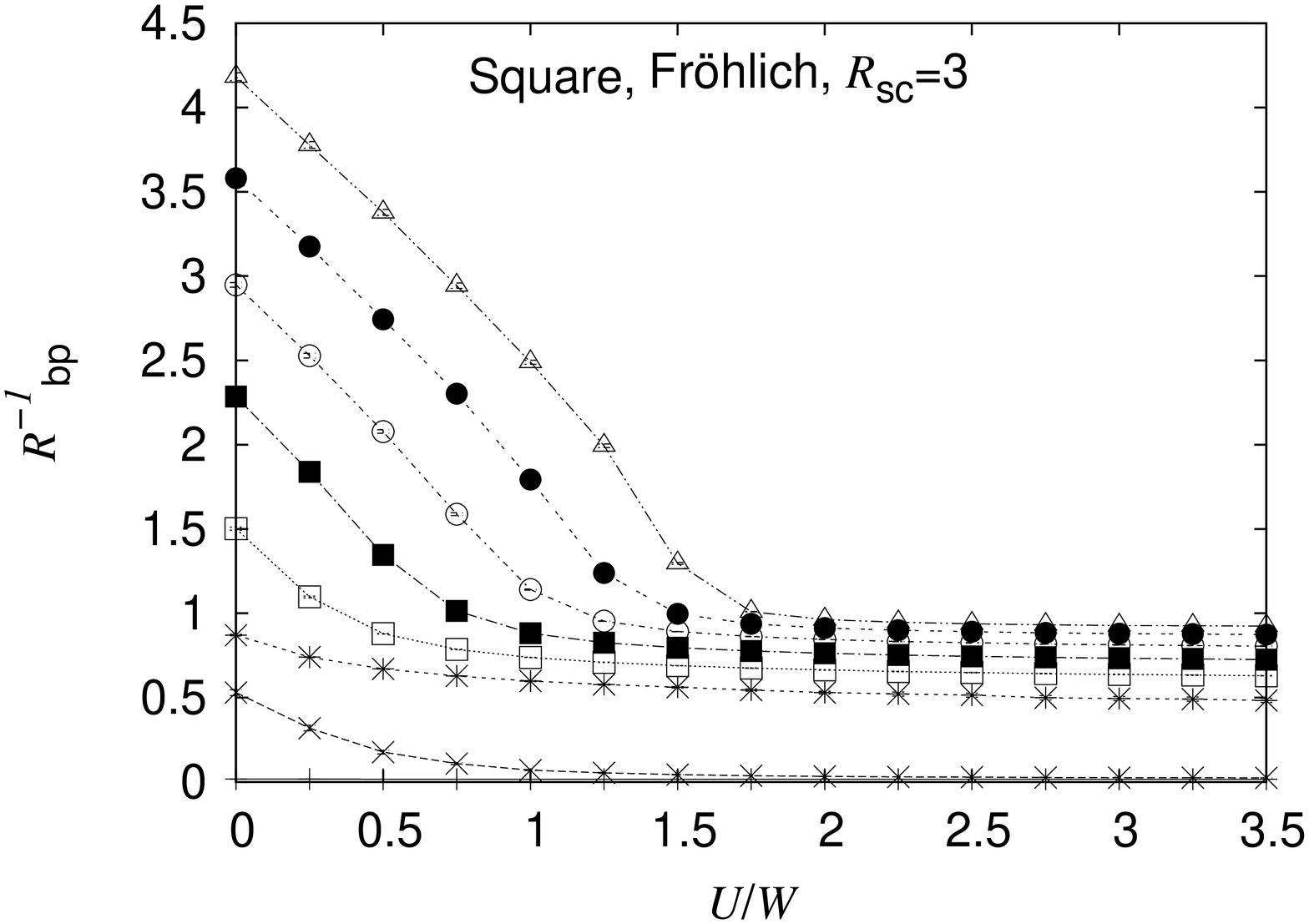}
\includegraphics[height=55mm,angle=270]{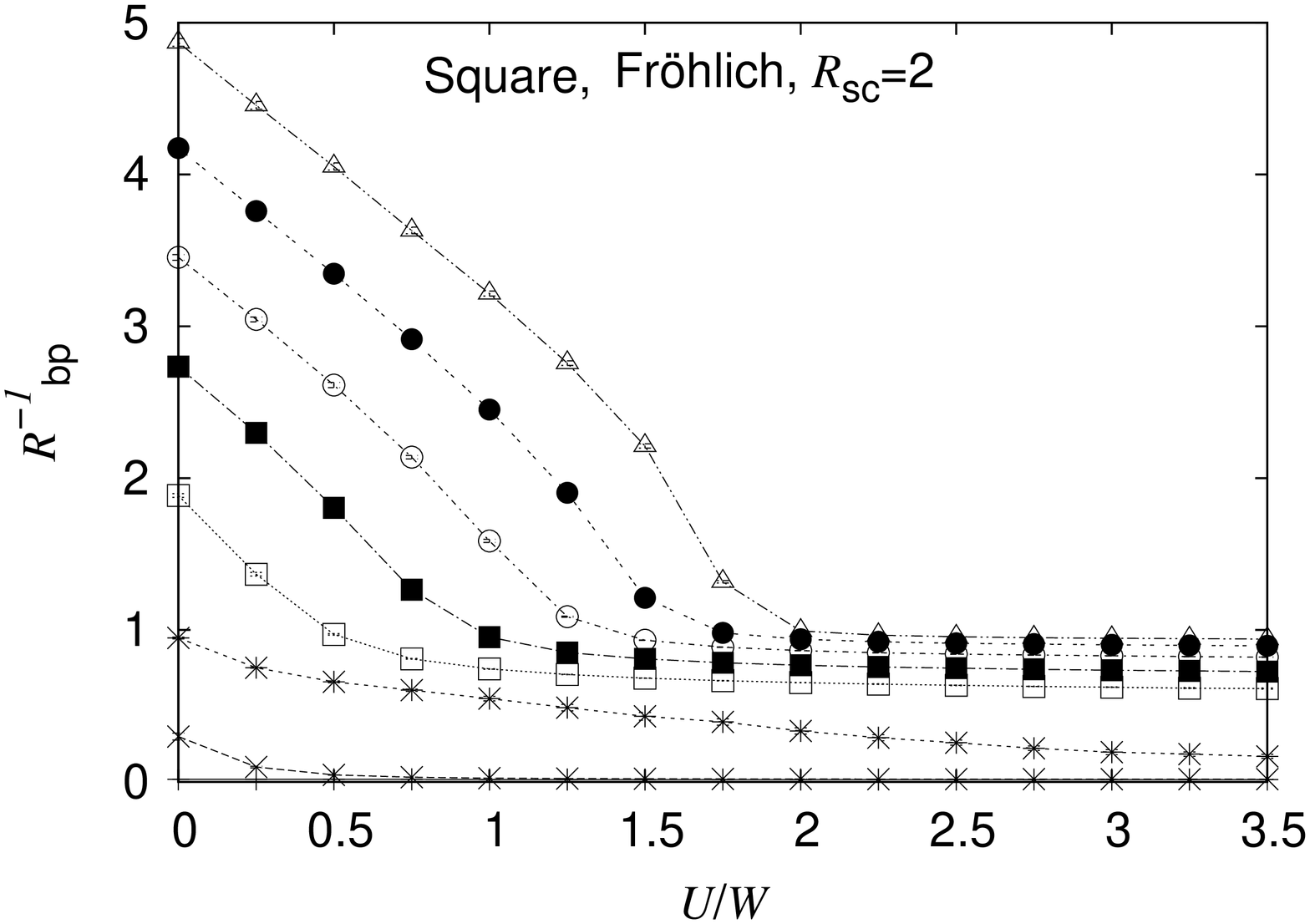}
\includegraphics[height=55mm,angle=270]{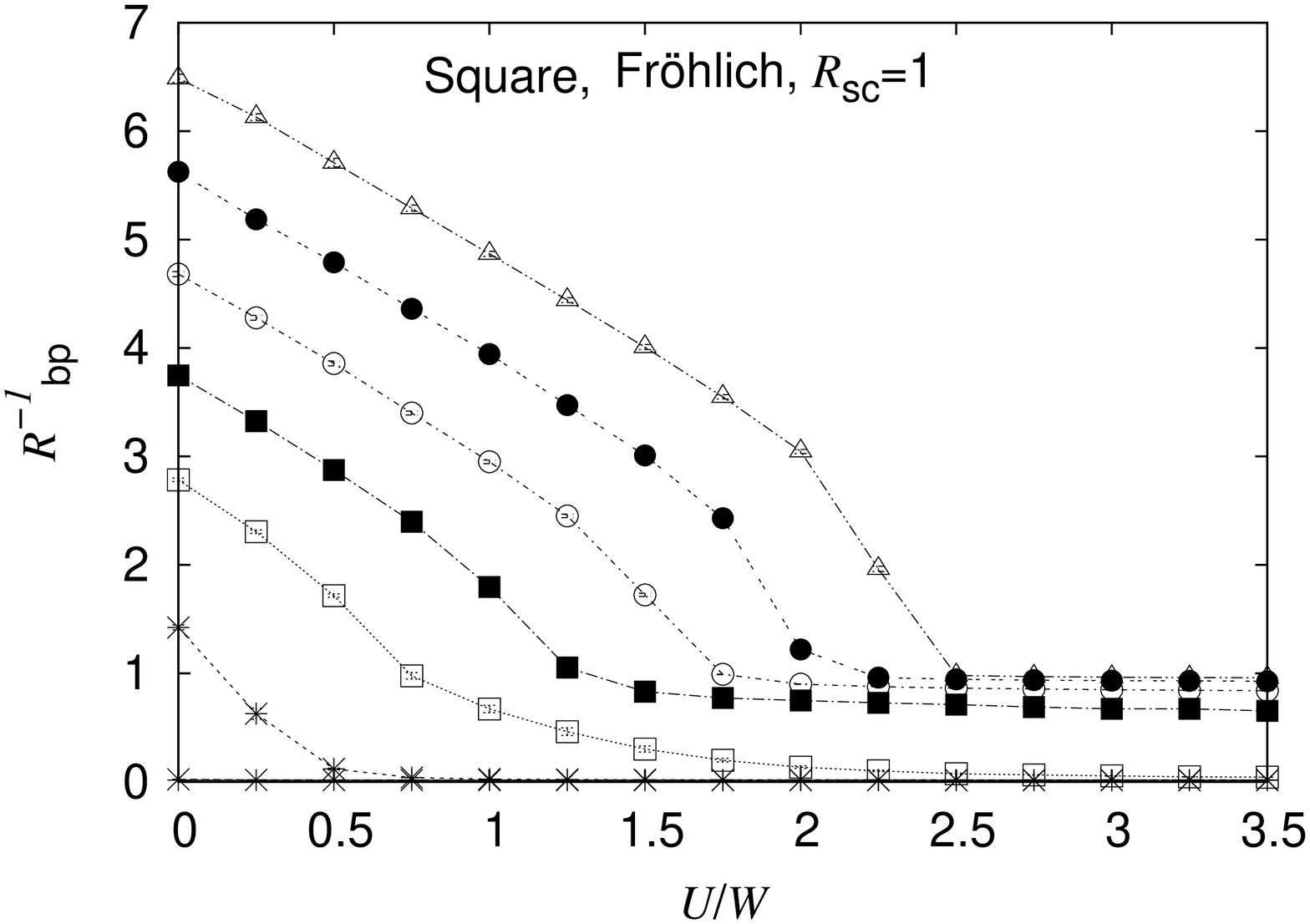}
\includegraphics[height=55mm,angle=270]{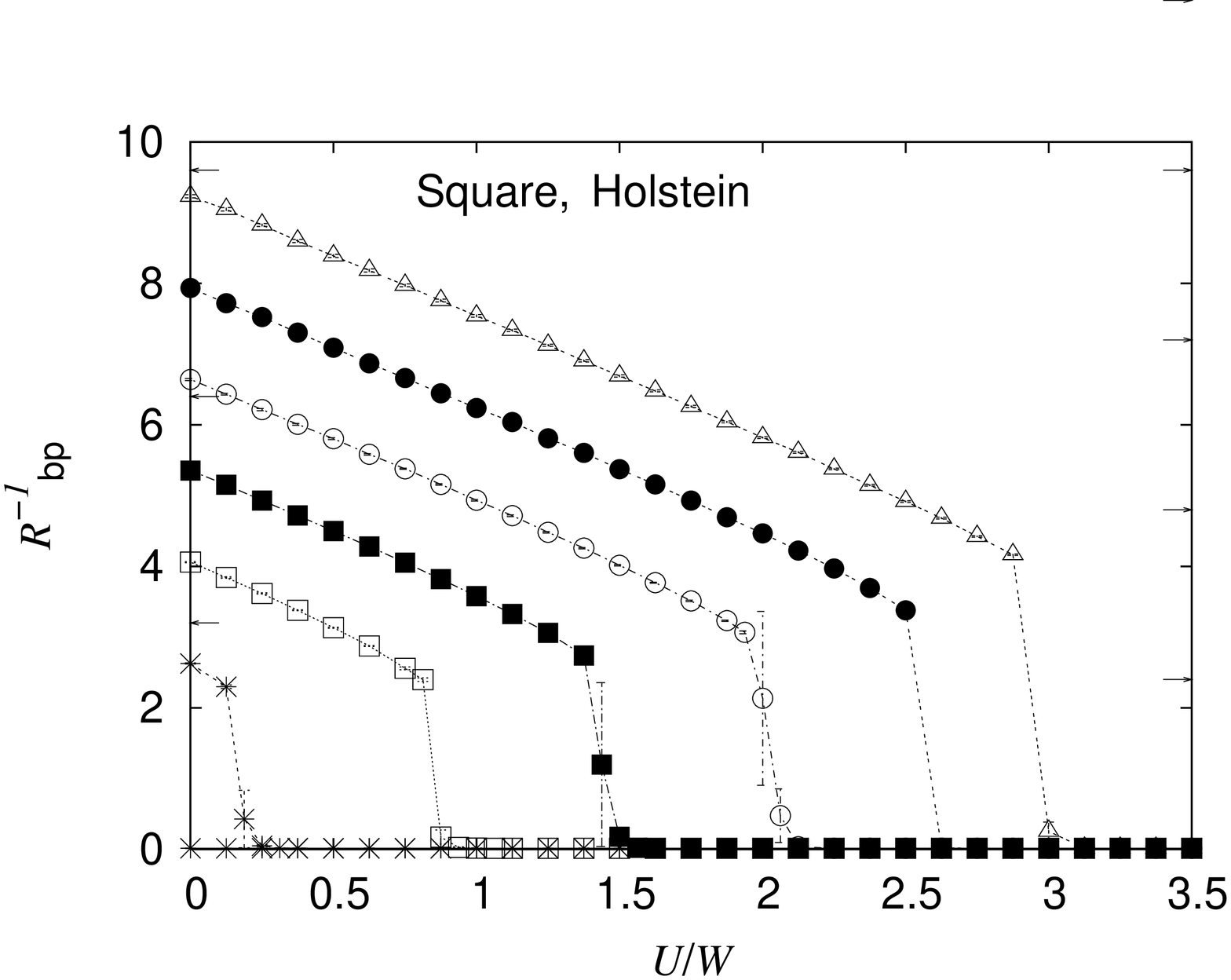}
\includegraphics[height=55mm,angle=270]{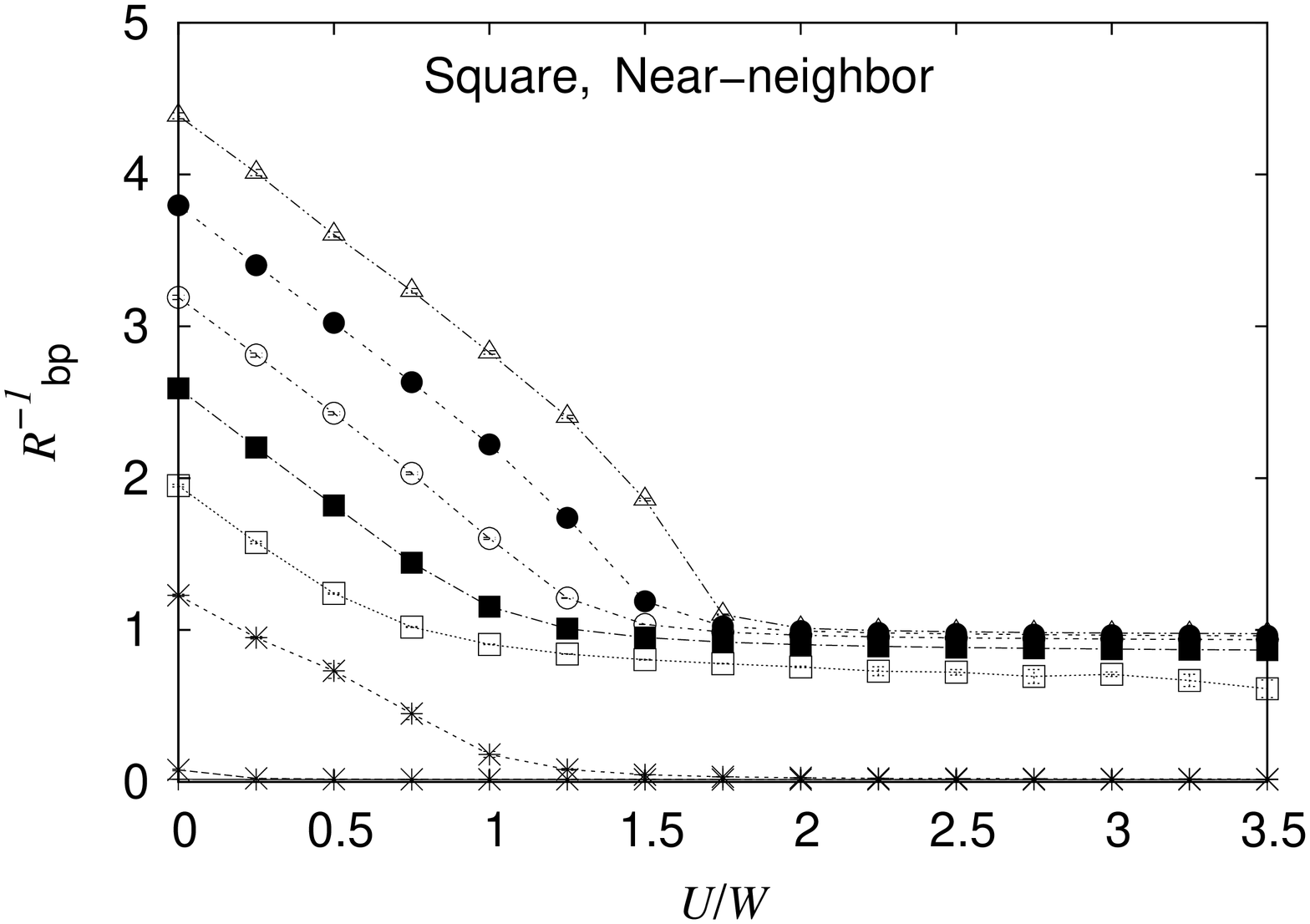}
\caption{Inverse bipolaron size associated with bipolarons on the square lattice. Parameters as in Fig. \ref{fig:energysinglet}.}
\end{figure*}

\section{Singlets on the square lattice}
\label{sec:singlet}

The survey of the bipolaron parameter space begins by considering
singlet properties of bipolarons on the square lattice. We will
examine the effects of the electron-phonon coupling, Hubbard repulsion
and interaction range on the bipolarons. In the following, we take
$\bar{\omega}=\omega/t=1$, which corresponds to $\omega/W=1/4$ which is well inside the
adiabatic regime. $W$ is the half band width or energy of a single non-interacting electron. $W=4t$ for the square lattice and $W=6t$ for the triangular lattice. We restrict paths to lie within 100 lattice
spacings of each other, with the large lattice size used to ensure that the effects of the lattice
restriction are minimal (this should be true where the bipolaron radius
is much smaller than the lattice size).  Most properties are
not strongly affected by finite size effects. We note that
the inverse radius is the most sensitive measurement, and the total energy the
least sensitive. The most severe finite size effect is that premature binding can
occur if the paths are too closely restricted.

Figure \ref{fig:energysinglet} shows the total energy of the two
polarons / bipolaron pair for a range of electron-phonon coupling
strengths, interaction screening radius and Hubbard $U$. As is the
case in 1D \cite{hague2009a}, a qualitative difference can be seen
between bipolarons formed with Holstein and screened Fr\"ohlich
interactions. At large $U$, the energy curves for the Hubbard-Holstein
bipolaron are flat, indicating that the bipolaron has unbound into two
polarons. This is true for all the $\lambda$ shown, indicating that
there is no bound singlet bipolaron in the $U\rightarrow\infty$
Hubbard-Holstein model. The significance for triplet states will be
discussed later in the article.

In contrast to the flat energy curves characterizing the large $U$,
large $\lambda$ Holstein model, the energy of the Hubbard-Fr\"ohlich
bipolaron changes on varying $U$, indicating that there are bound
singlet states even for very large $U$ (we will revisit this point
later in the article). For small $\lambda$ and $R_{sc}\neq 0$, the
energy curves are flat at large $U$. The model with only
nearest-neighbor interaction ($\Phi(\avec)=0.5$) has similar properties to the $R_{sc}=2$
and $R_{sc}=3$ models. $\Phi(\avec)=0.438$ to 3 significant figures for the model
with $R_{sc}=2$ and $\Phi(\avec)=0.509$ to 3 significant figures for the model with
$R_{sc}=3$, indicating that bipolaron properties depend
most strongly on the nearest neighbor part of the Fr\"ohlich interaction, rather than the tails.

Is is useful to quickly describe the limiting behaviors of the
bipolarons. First, it is appropriate to introduce some notation. We use S0 to define a strongly bound on-site singlet pair where the two electrons
are separated by no lattice spacings and use S1 to denote a strongly bound
inter-site pair where the electrons are separated by one lattice space \cite{macridin2004a}. As discussed for the 1D model \cite{hague2009a}, if bipolarons
are strongly bound the renormalized intersite hopping is small, and an atomic
Hamiltonian can be determined using the Lang--Firsov transformation:
\begin{equation}
\tilde{H}_{\rm at} = -\sum_{\nvec\nvec'}\frac{W\lambda\Phi_{0}(\nvec,\nvec')}{\Phi_0(0,0)}\hat{n}_{\nvec} \hat{n}_{\nvec'}+\omega\sum_{\mvec}\left(d^{\dagger}_{\mvec}d_{\mvec}+\frac{1}{2}\right)
\end{equation}
Since the electron number operator, $\hat{n}$ is unchanged under the
Lang--Firsov transformation, the effects of the Hubbard $U$ can
quickly be reintroduced. Thus, for a strongly bound onsite (S0)
bipolaron,
\begin{equation}
E=U-4W\lambda
\label{eqn:s0energy}
\end{equation}
$W$ is the non-interacting kinetic energy of a single particle (half band-width). The limiting behavior of the S0 bipolaron is shown on the plots as
dashed lines. As in the 1D case, the S0 Hubbard-Holstein bipolaron rapidly becomes strongly bound on decreasing $U$. For the longer range interactions,
it takes much more attraction to strongly bind the S0 bipolaron. We
can also compute the energy of the strongly bound S1 bipolaron which
forms at large $\lambda$ and $U\gg W\lambda$,
\begin{equation}
E=-2W\lambda[1+\Phi(\avec)/\Phi_0].
\label{eqn:s1energy}
\end{equation}
The energy of the strongly bound S1 state is plotted in
Fig. \ref{fig:energysinglet} as the arrows at $U/W=3.5$ (note that
only 5 arrows corresponding to the largest $\lambda$ are shown). The
energy of the strongly bound S1 bipolaron is higher than the simulated
energy, but the values converge as $\lambda$ increases. From the
difference in energies between the S1 limit and the simulated energy,
it appears that the bipolarons formed from heavily screened
interactions have a stronger S1 character.

The inverse mass of the bipolarons is shown in
Fig. \ref{fig:imsquaresinglet}. Again, significant differences between Hubbard-Holstein and Hubbard-Fr\"ohlich bipolarons can immediately be seen. In the
Hubbard-Holstein model, the mass changes very rapidly with $U$ when the
bipolaron binds, with a mass change of several orders of magnitude over a very
small range of $U$ values. This rapid change is not seen for the
Fr\"ohlich bipolaron, which maintains a similar mass over a wide
range of $U$. For the larger $\lambda$ values and large $U$, the
system of two particles appears to be heavier in the Fr\"ohlich case. This is because the Holstein bipolaron is not bound for those values, whereas
the Fr\"olich interaction leads to an intersite (S1) bipolaron.

In comparison to the $U$ dependence of the mass of the 1D Holstein bipolaron, which can be found in Ref. \onlinecite{hague2009a}, the change
in mass with $U$ is extremely rapid on the square lattice. In contrast, the mass relating to long
range interactions is qualitatively unchanged, with the formation of a relatively light bipolaron (even in comparison with a free electron) over a very wide range of
the parameter space. The qualitative difference between the properties of
Holstein and Fr\"ohlich bipolarons emphasizes that while the Holstein
model may be a reasonable approximation for considering polarons in
organic molecular compounds, the simple Holstein
approximation is inappropriate for many crystalline systems where
complete screening of the electron-phonon interaction is not
possible. Even if the screening length is as small as a single lattice
spacing, bipolarons have significantly modified properties
compared with those formed in the Holstein model.

In 1D, a crab like bipolaron can form in models with long range
interaction when the energies of on-site and off-site configurations
are degenerate (S1 like configurations can be changed to S0 with a
single hop, without any energy barrier to make the process second
order in $t$). This leads to a noticeable decrease in mass at intermediate
$U$. It is of interest to determine if these lighter bipolarons can
also form on the square lattice. In Fig.
\ref{fig:imcomparelargelambda}, the masses of bipolarons formed when
$\lambda=1.6$ are examined as $R_{sc}$ is changed. A peak in the inverse mass (decrease in the mass) is clearly visible at intermediate $U$ and is especially pronounced when $R_{sc}=2$. Comparing equations \ref{eqn:s0energy} and \ref{eqn:s1energy}, it can be seen that for large phonon frequencies, the peak
would be expected at $U/W=2\lambda[1-\Phi(\avec)/\Phi(0)]$, which corresponds to $U/W=1.798$ for $R_{sc}=2$ (to 4 s.f) in good agreement. The presence of this peak shows that light bipolarons can exist on square lattices without the need for triangular plaquettes. Naturally, such a state depends on a very subtle balance of parameters and could not be realized in all crystalline systems.

As in the case of the 1D bipolaron, analytical determination of the
effective mass, even for limiting values of the parameters, is difficult (and may not be possible) when there are
long range tails in the interaction. While we have not computed
limiting behaviors of the mass for the model with only near-neighbor
interaction, we expect that a similar approach to that applied to the
1D bipolaron by Bon\v{c}a and Trugman could yield results
\cite{bonca2000b}.

We also show the number of phonons associated with the bipolaron cloud
(Fig. \ref{fig:nophonsinglet}). As before, the properties of the
Hubbard-Holstein bipolaron are qualitatively different to the
bipolarons formed from long-range electron phonon interaction. The
number of phonons associated with the Holstein bipolaron shows an
abrupt increase as the bipolaron binds on decreasing $U$, whereas there is a smoother crossover from S1 to S0 behavior associated with the screened Fr\"ohlich interaction.

The total number of phonons associated with the strongly bound onsite
bipolaron (following the argument in Ref. \onlinecite{hague2007a}) is,
\begin{equation}
N_{\rm ph} = \frac{4W\lambda}{\omega} \: .
\end{equation}
This number is independent of $U$, so the number of phonons reaches a
limiting value on decreasing $U$ as seen in
Fig. \ref{fig:nophonsinglet} (the strongly bound S0 limit is represented by arrows on the left of the graph). The Holstein bipolaron with large
$\lambda$ quickly reaches the S0 limit at low $U$. Bipolarons formed
with long range interactions approach the S0 limit less rapidly since
the longer range tails lead to a shallower effective potential.

The number of phonons associated with the S1 bipolaron can also be
found,
\begin{equation}
N_{\rm ph} = \frac{2W\lambda}{\omega}\left[1+\frac{\Phi(\avec)}{\Phi_0}\right].
\end{equation}
This is represented as arrows on the large $U$ side of the
plots. The agreement improves on increasing $\lambda$. Again, it is clear that the most heavily screened interactions have the most clearly defined S1 character.

%Figure \ref{fig:isotope} shows the isotope exponent plotted for a
%range of parameters. In common with the 1D case, the bipolaron
%associated with the screened Fr\"ohlich interaction has a reduced
%isotope exponent when the hybrid S0-S1 crab is formed. The Holstein
%model shows some unusual features, with an anomalous increase in the
%exponent before a significant drop (on increasing $U$). We believe that this is due to the
%very rapid mass change as the bipolaron binds. As with all other
%properties, the Holstein interaction leads to a bipolaron with very
%different isotope exponent.

We complete this section by examining the inverse bipolaron size
associated with the square lattice. Again, qualitative differences
exist between the Holstein bipolaron and the Fr\"ohlich bipolarons,
with S1 bipolarons forming in the screened Fr\"ohlich model at large
$U$ and $\lambda$ indicated by a radius of around a lattice
spacing. Comparison with figure \ref{fig:imsquaresinglet} shows that
bipolarons on the square lattice with long range interaction may be
both small and light at intermediate and large $\lambda$ and intermediate $U$, where polarons are bound into the hybrid S0-S1 bipolaron.

\begin{figure*}
\includegraphics[height=55mm,angle=270]{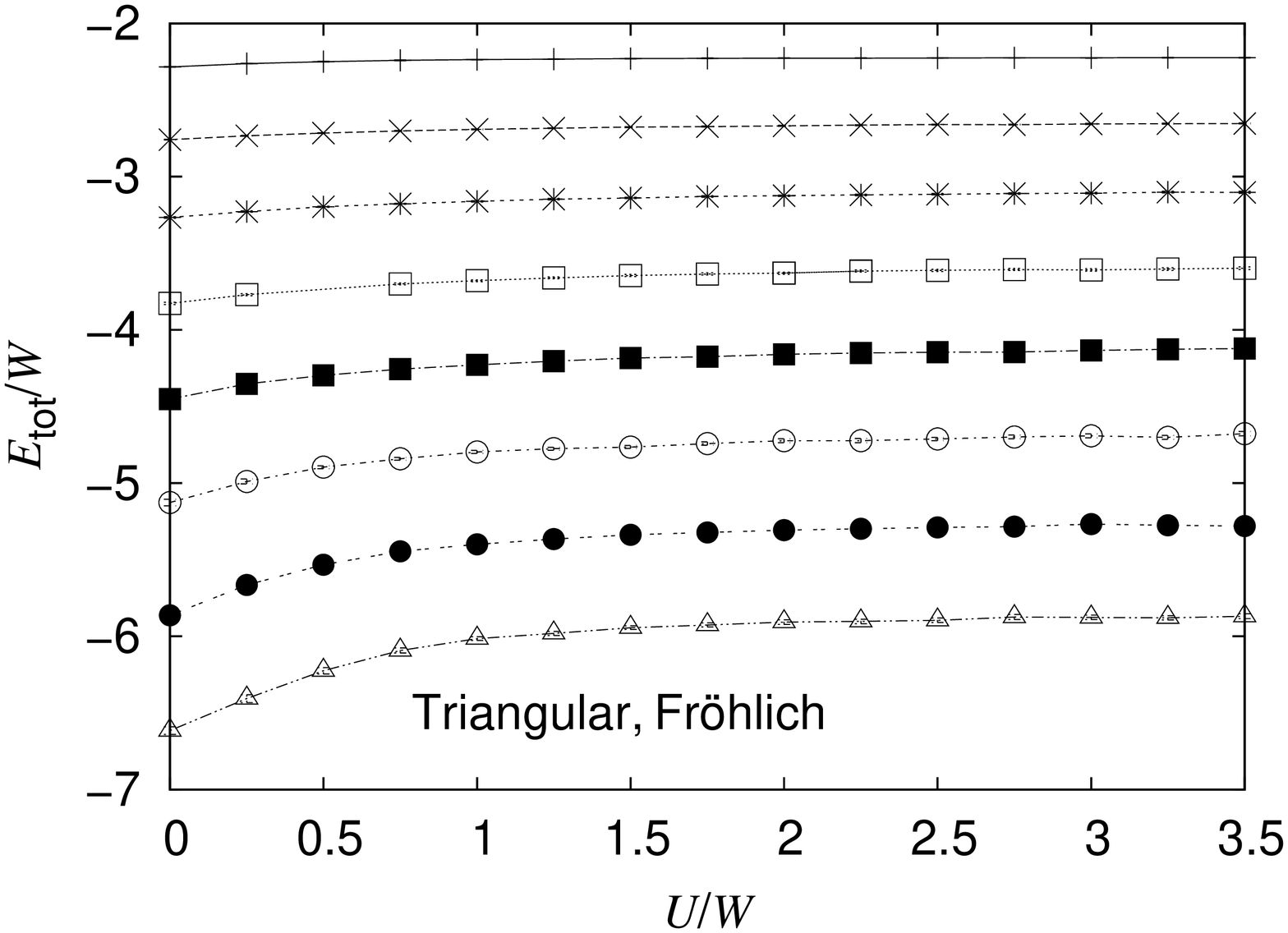}
\includegraphics[height=55mm,angle=270]{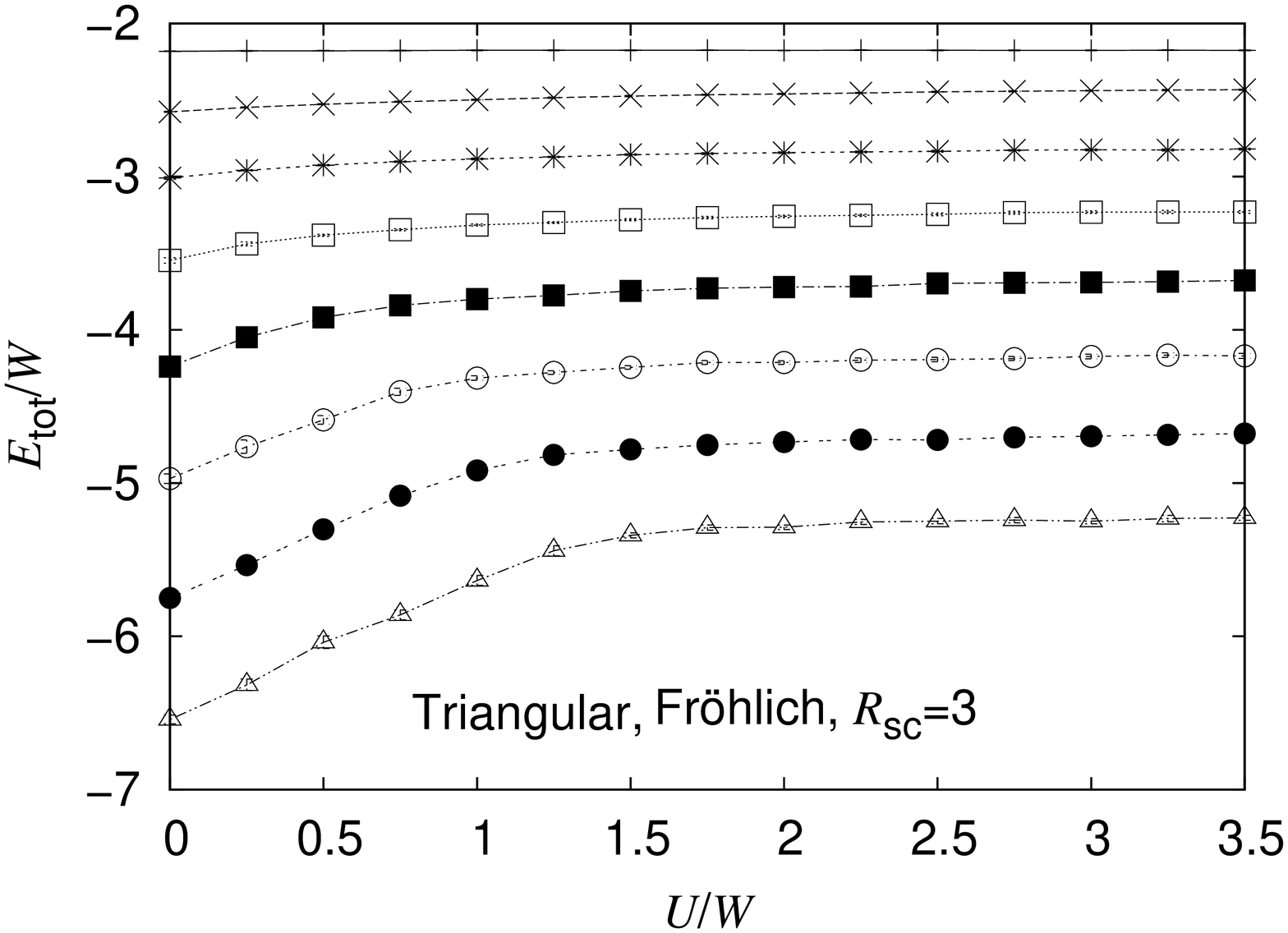}
\includegraphics[height=55mm,angle=270]{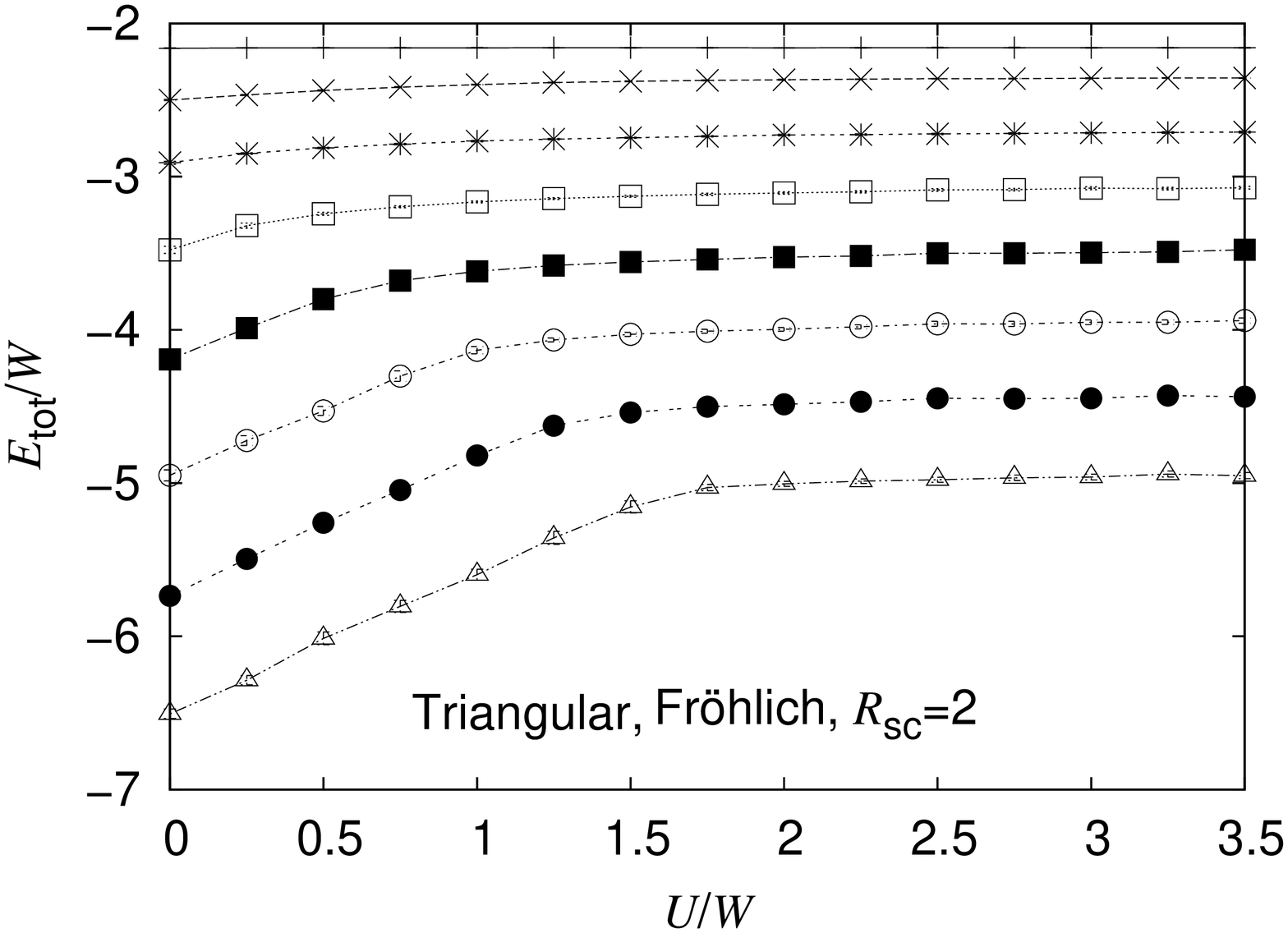}
\includegraphics[height=55mm,angle=270]{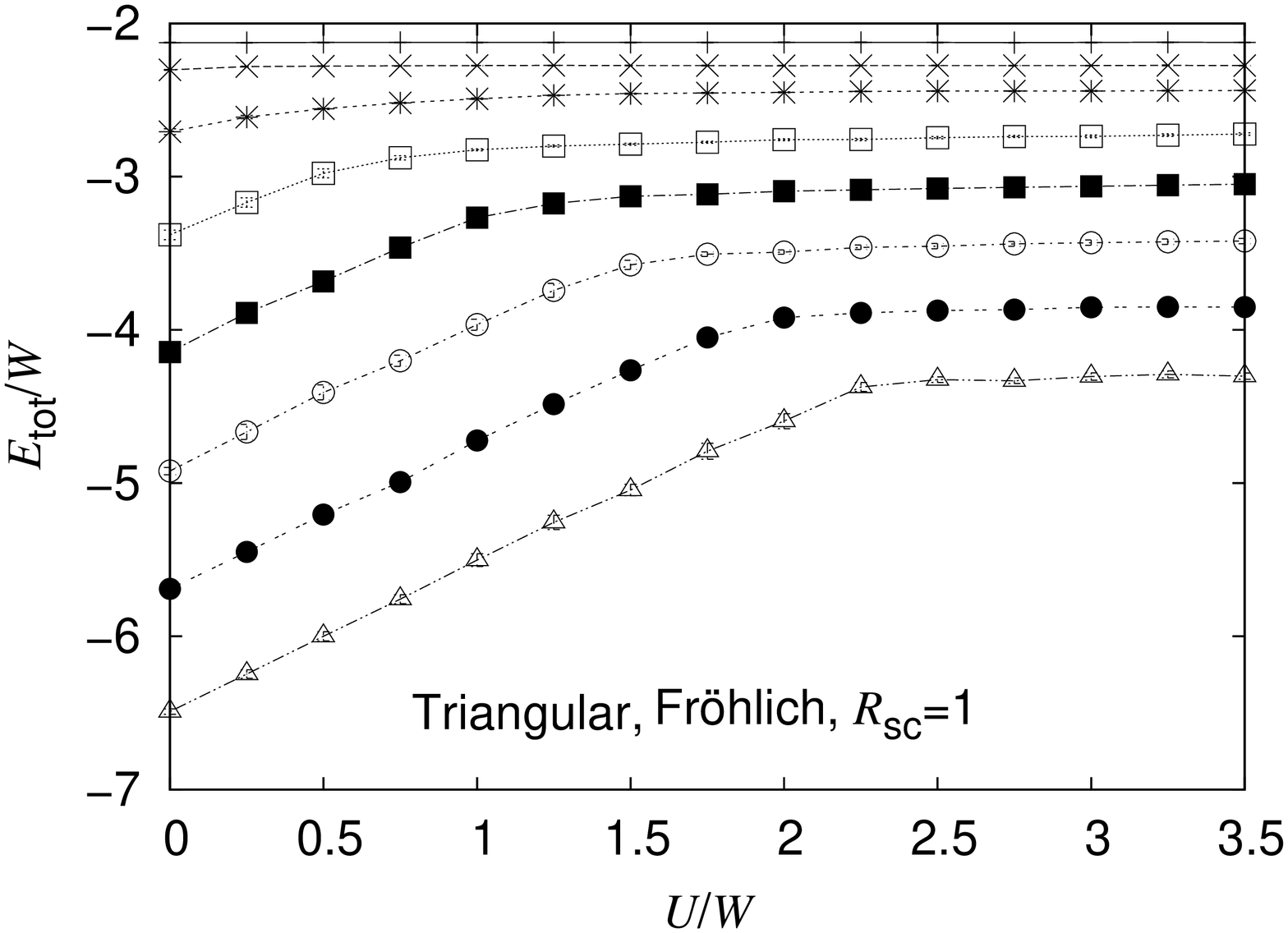}
\includegraphics[height=55mm,angle=270]{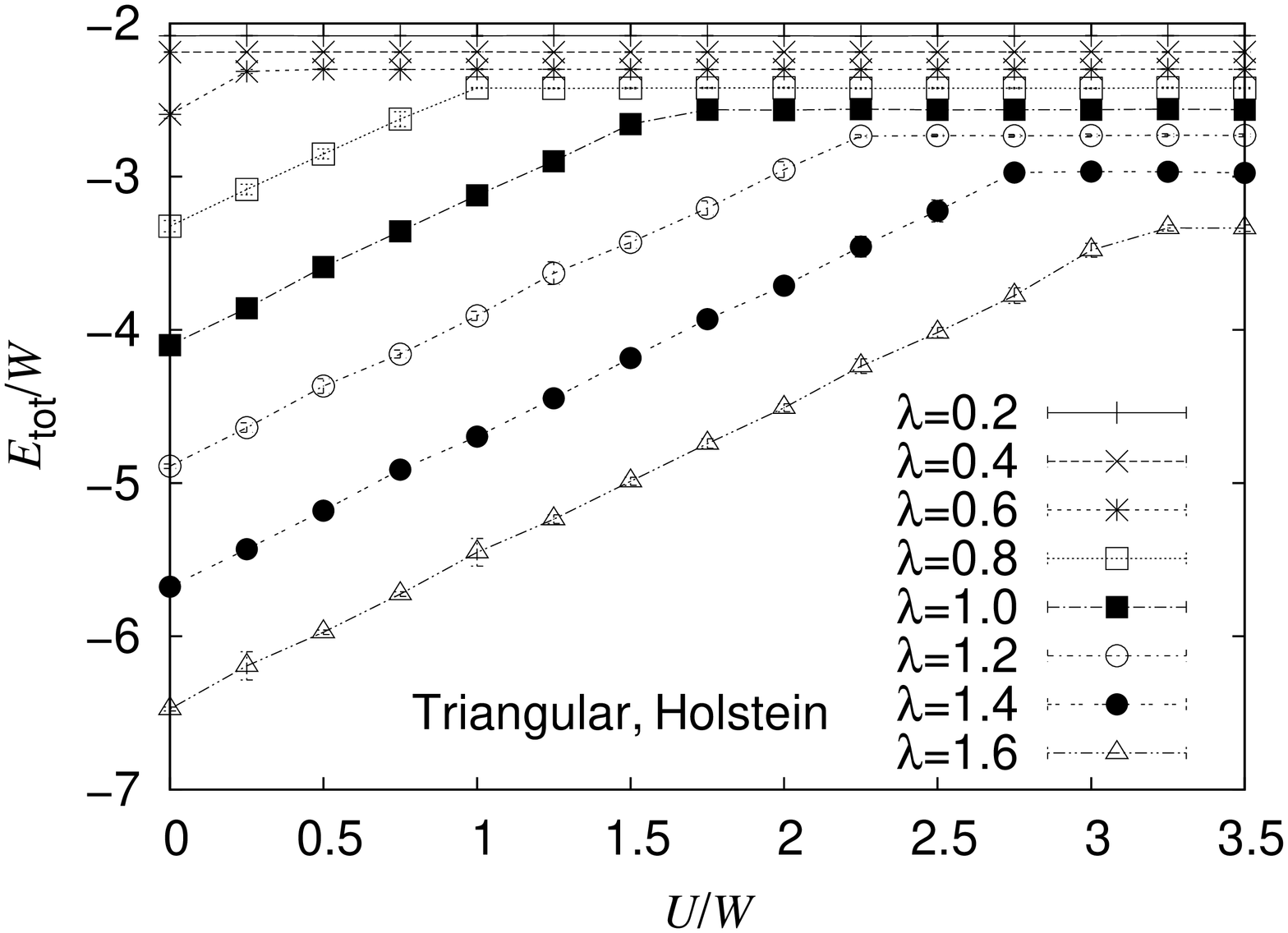}
\includegraphics[height=55mm,angle=270]{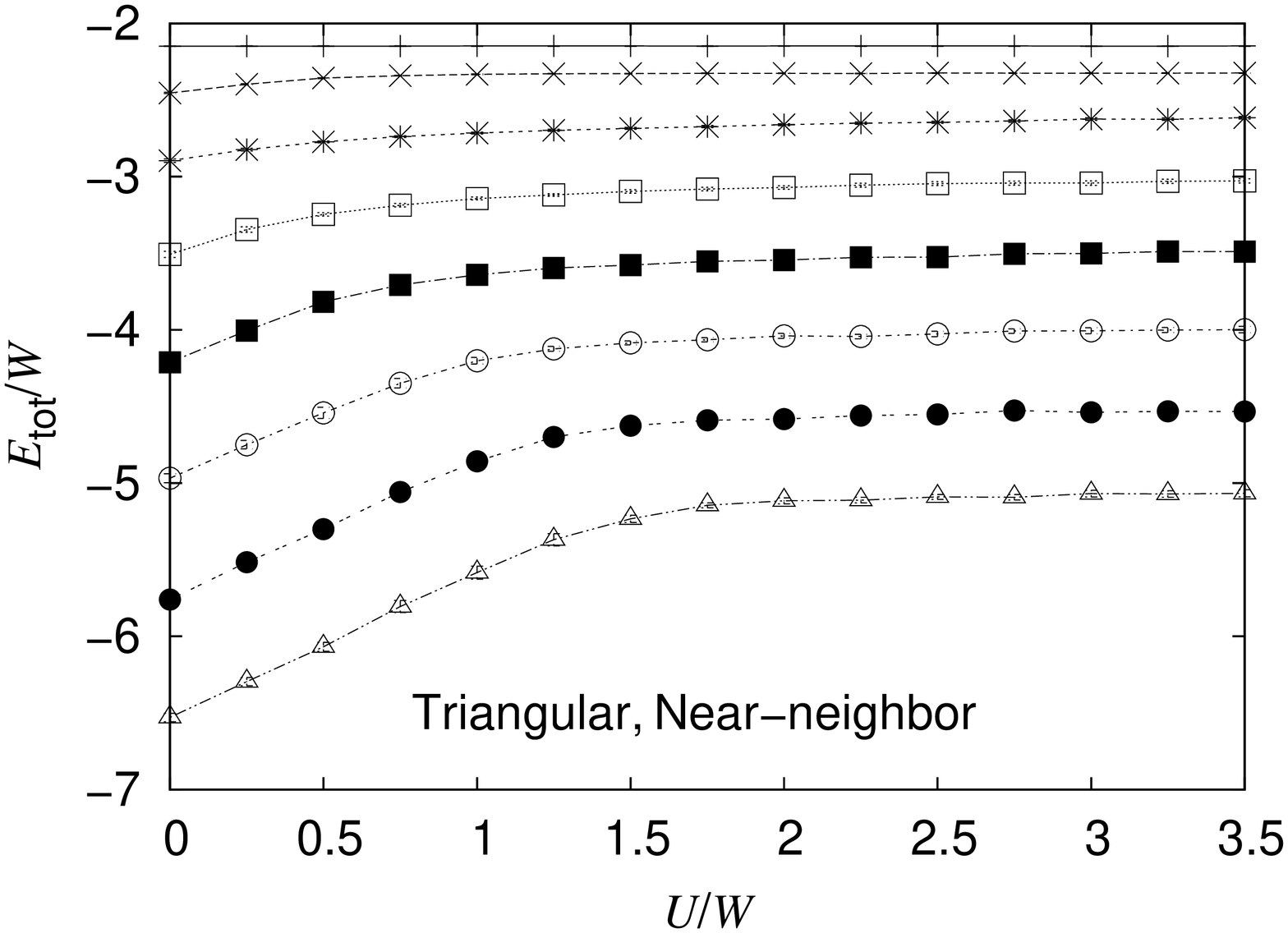}
\caption{Total energy of singlet bipolarons on the triangular
lattice. $\omega/t=1.5$, $U$, $\lambda$ and $R_{sc}$ are varied,
$\bar{\beta}=28/3$. The value of $\omega/t$ is chosen so that
$\omega/W$ is the same for both square and triangular lattices. The
energies are qualitatively similar to those associated with the square
lattice.}
\label{fig:tetriangular}
\end{figure*}

\begin{figure*}
\includegraphics[height=55mm,angle=270]{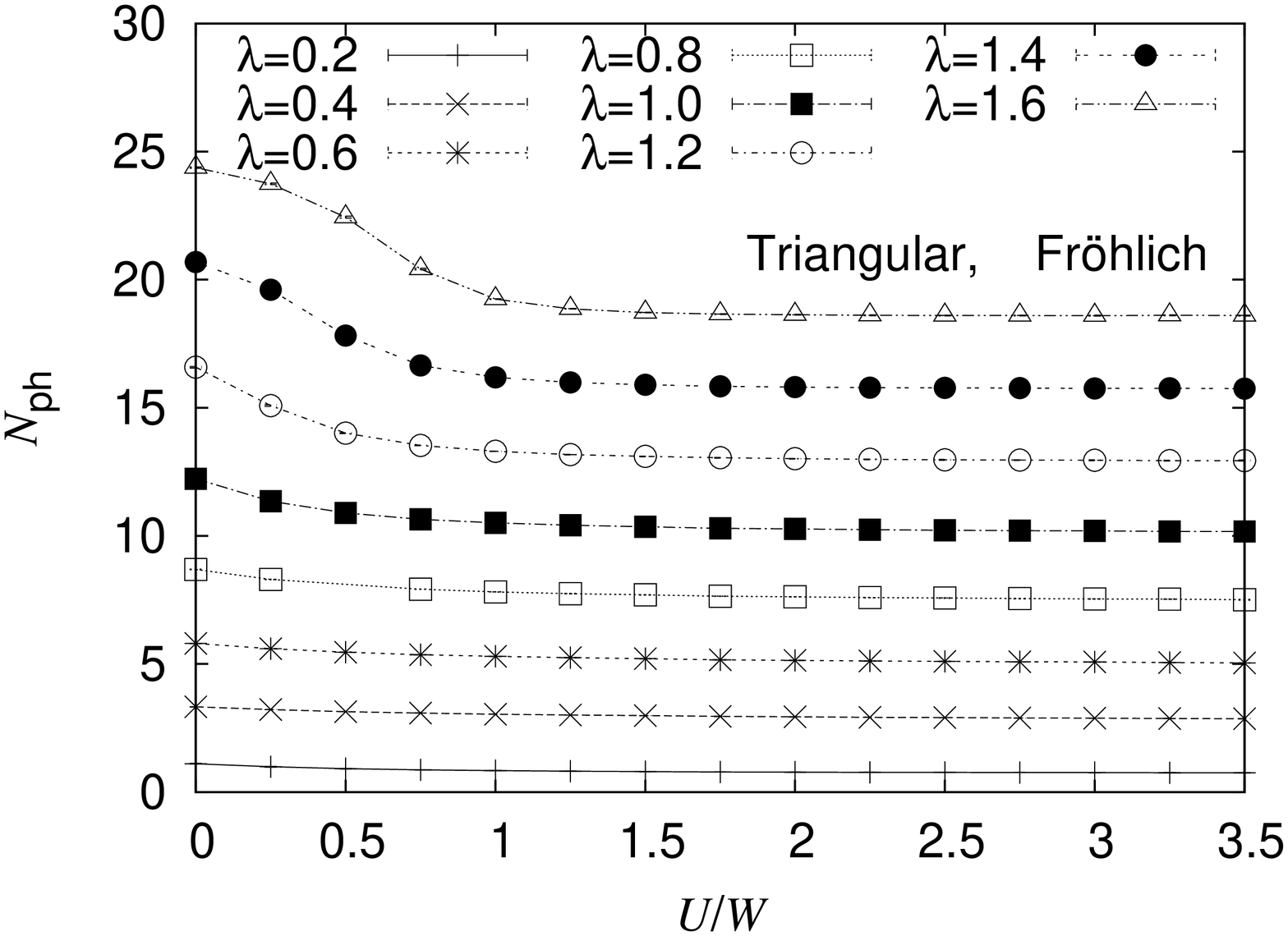}
\includegraphics[height=55mm,angle=270]{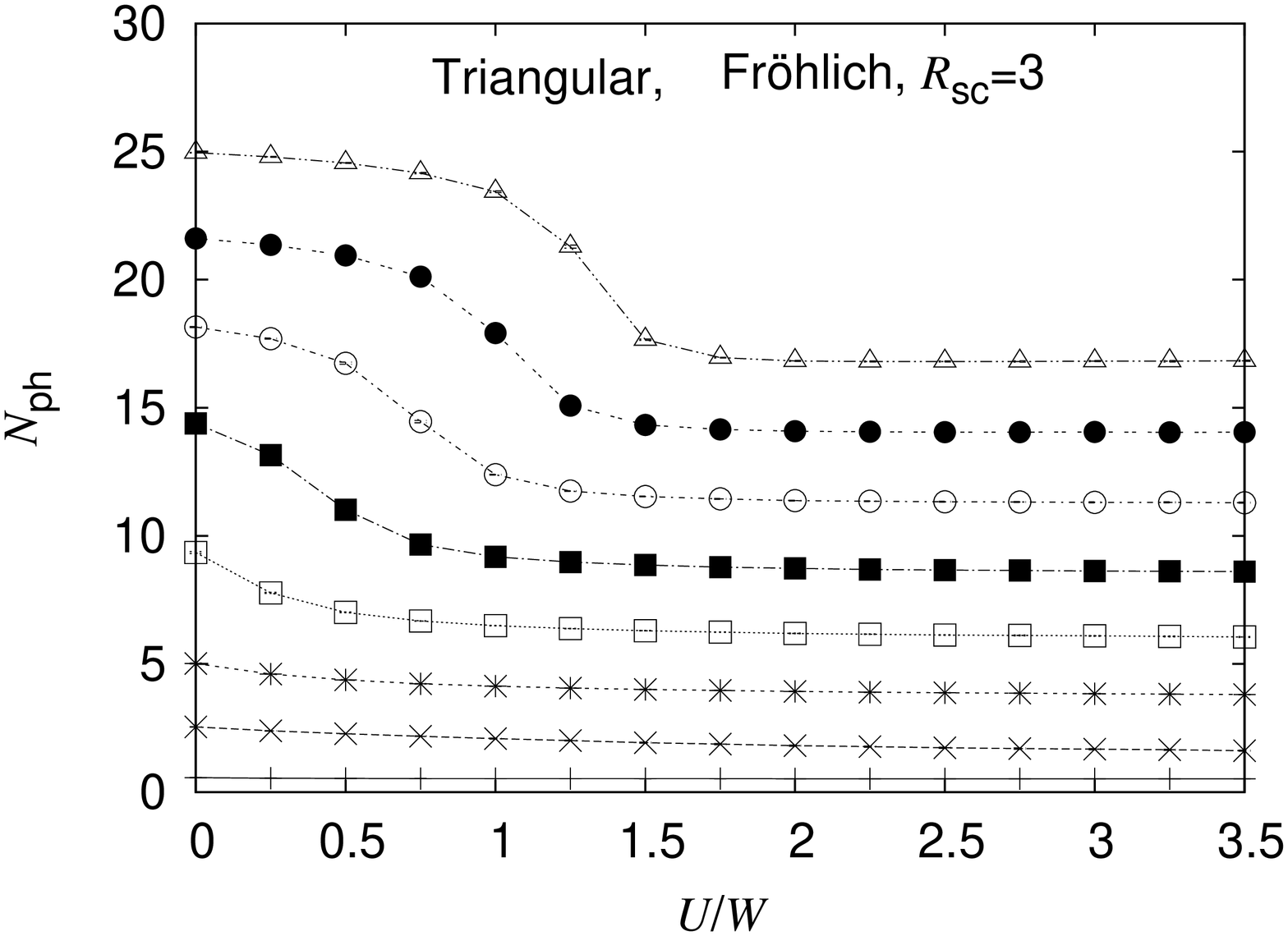}
\includegraphics[height=55mm,angle=270]{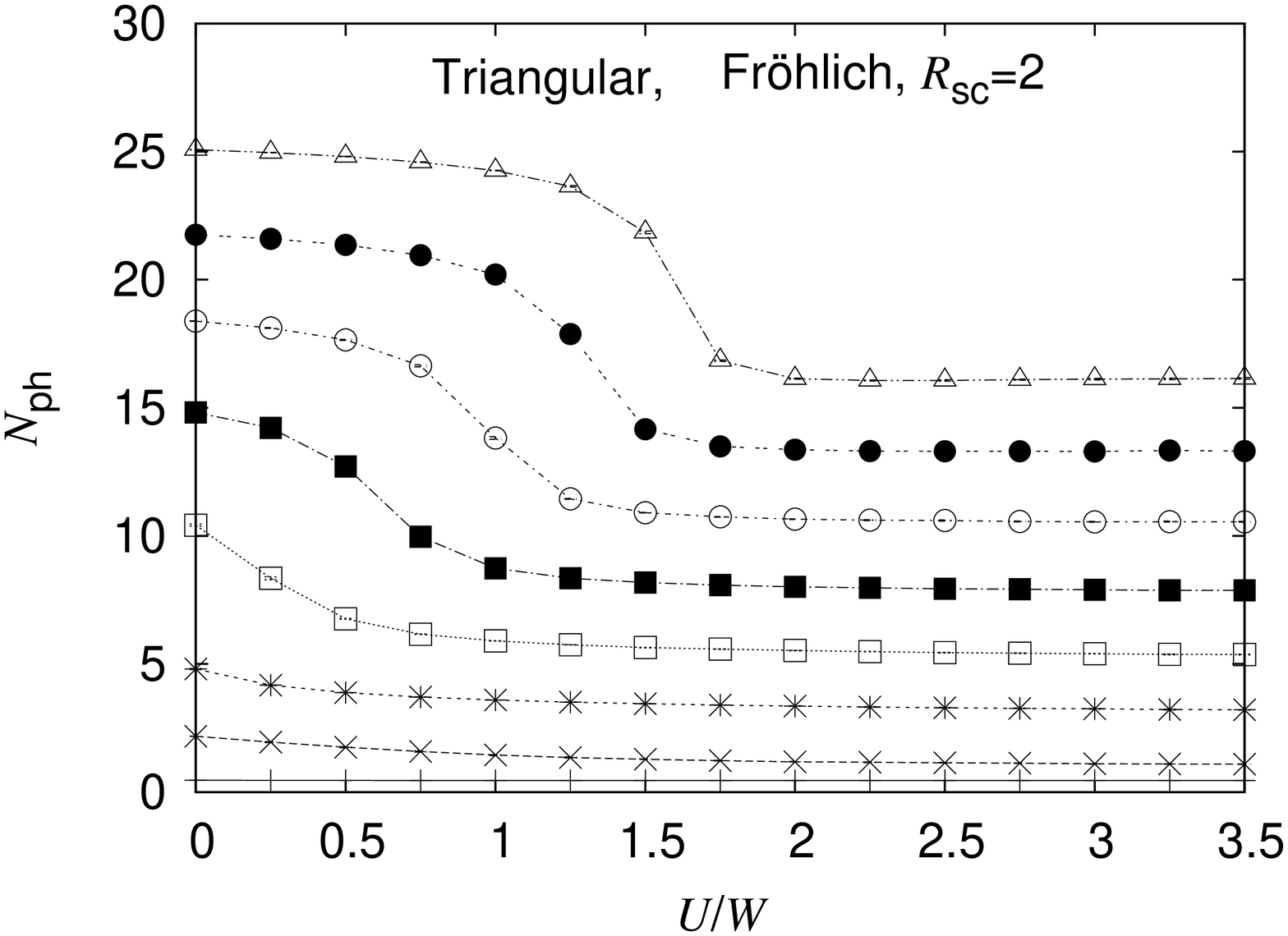}
\includegraphics[height=55mm,angle=270]{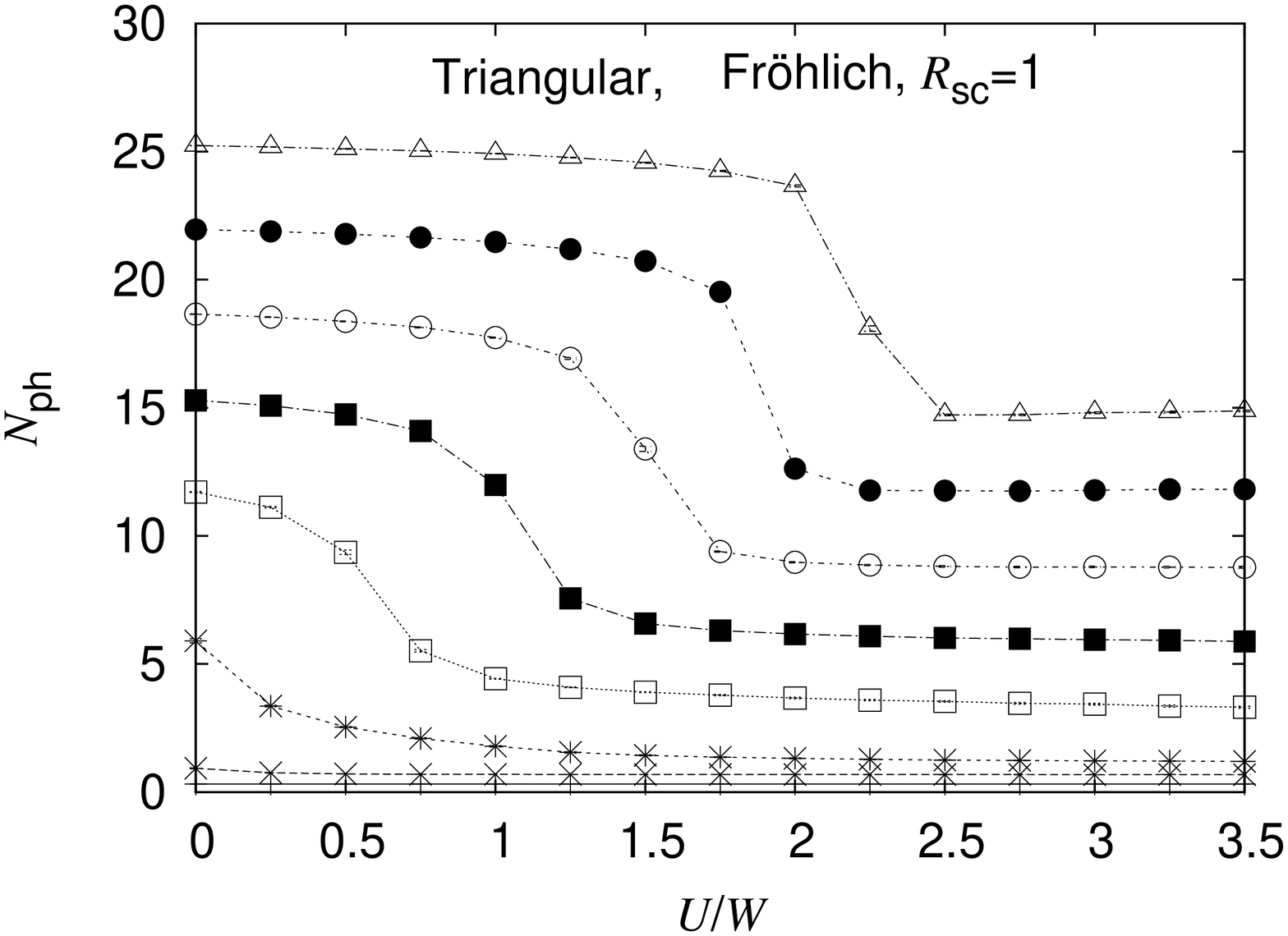}
\includegraphics[height=55mm,angle=270]{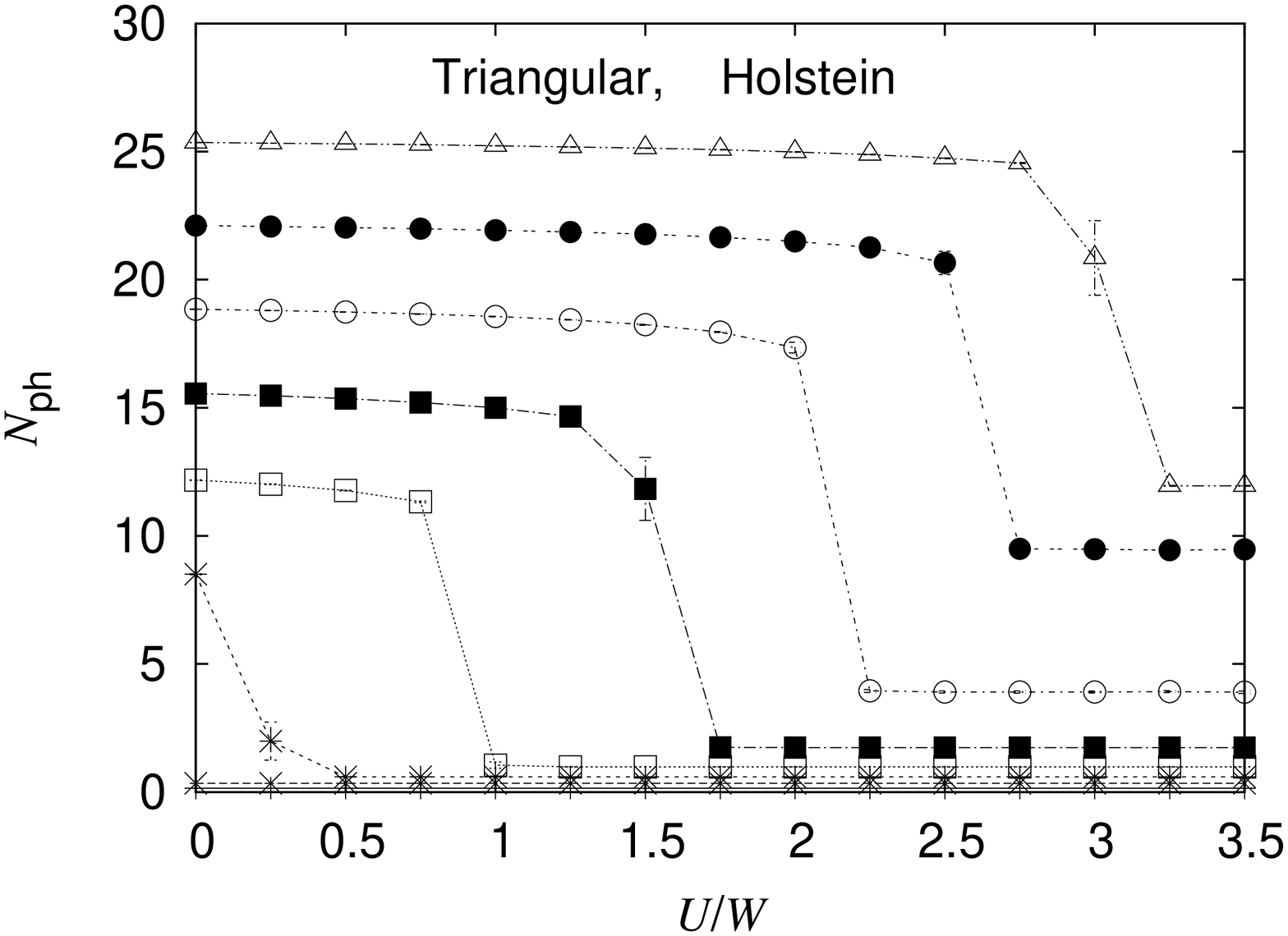}
\includegraphics[height=55mm,angle=270]{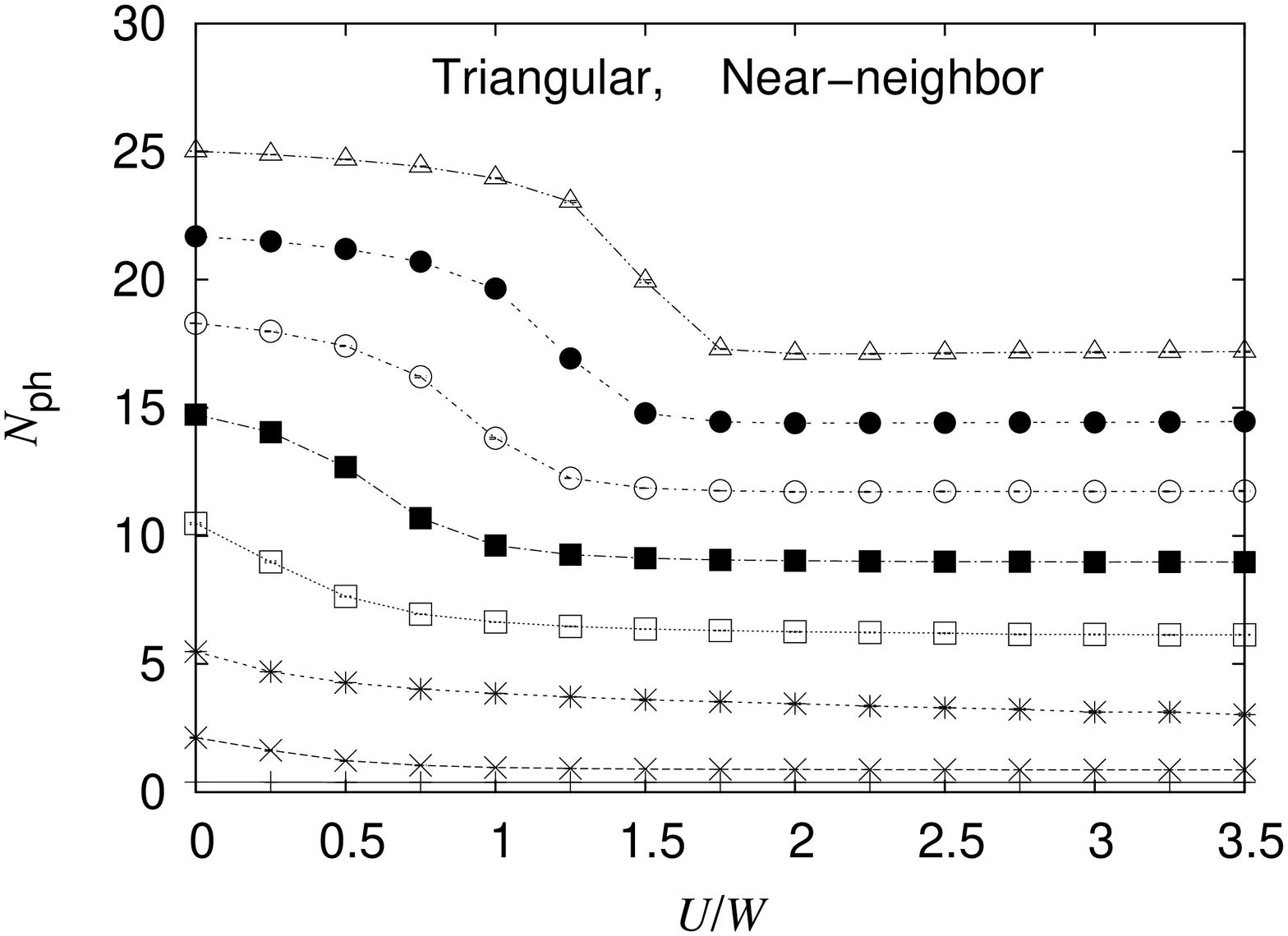}
\caption{Number of phonons associated with bipolarons on the triangular lattice. Parameters are as in Fig. \ref{fig:tetriangular}.}
\label{fig:nptriangular}
\end{figure*}

\begin{figure*}
\includegraphics[height=55mm,angle=270]{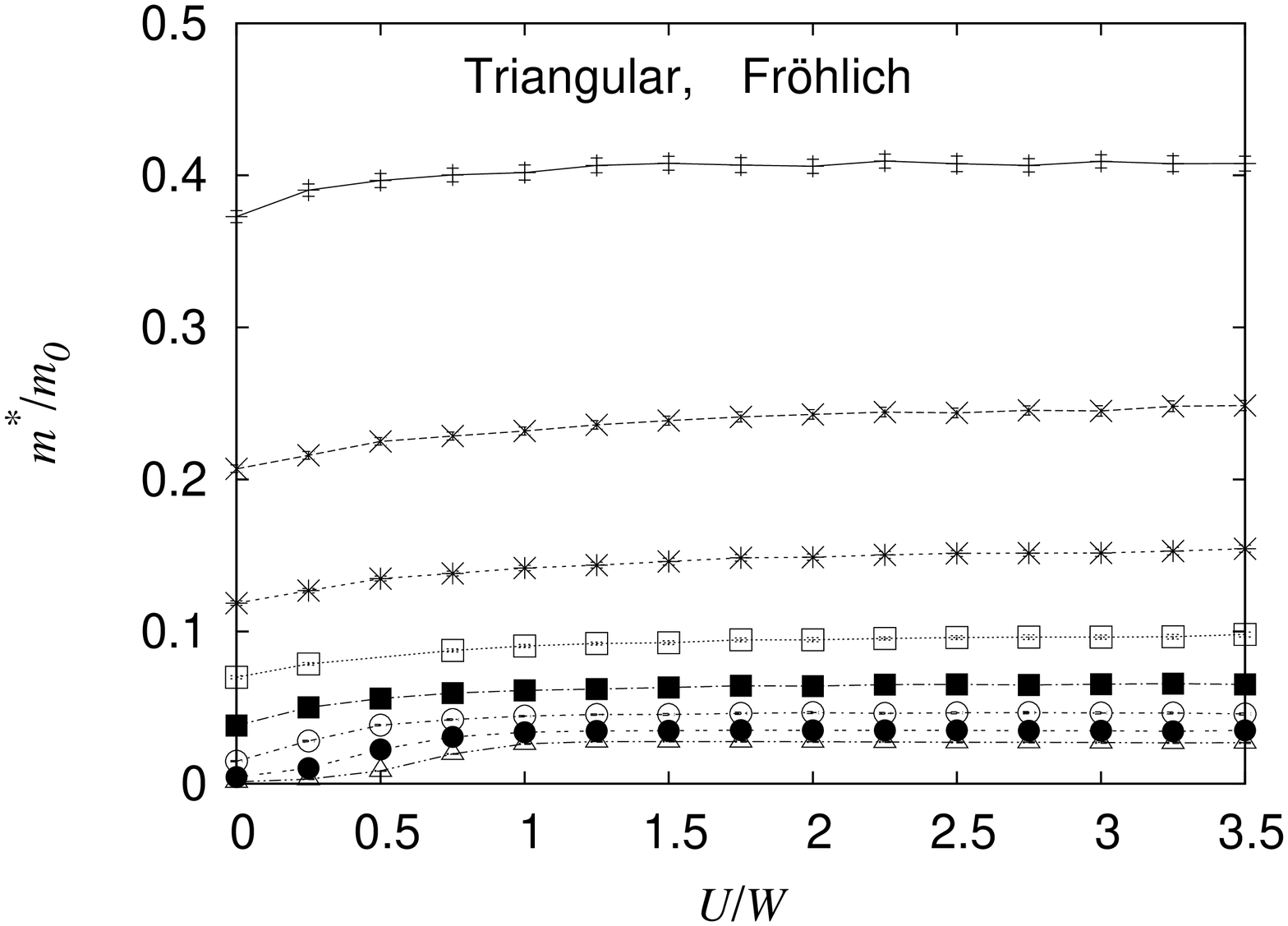}
\includegraphics[height=55mm,angle=270]{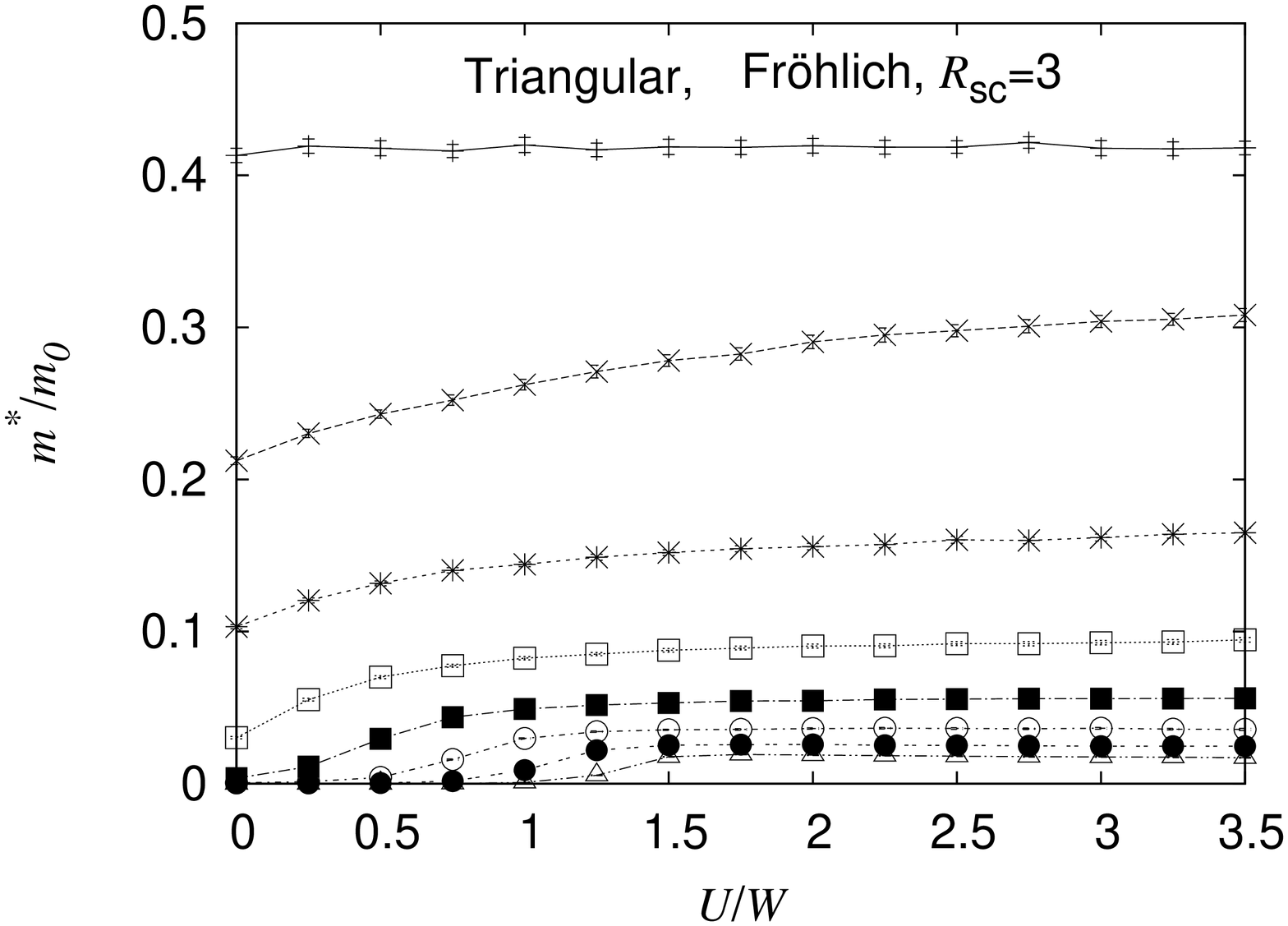}
\includegraphics[height=55mm,angle=270]{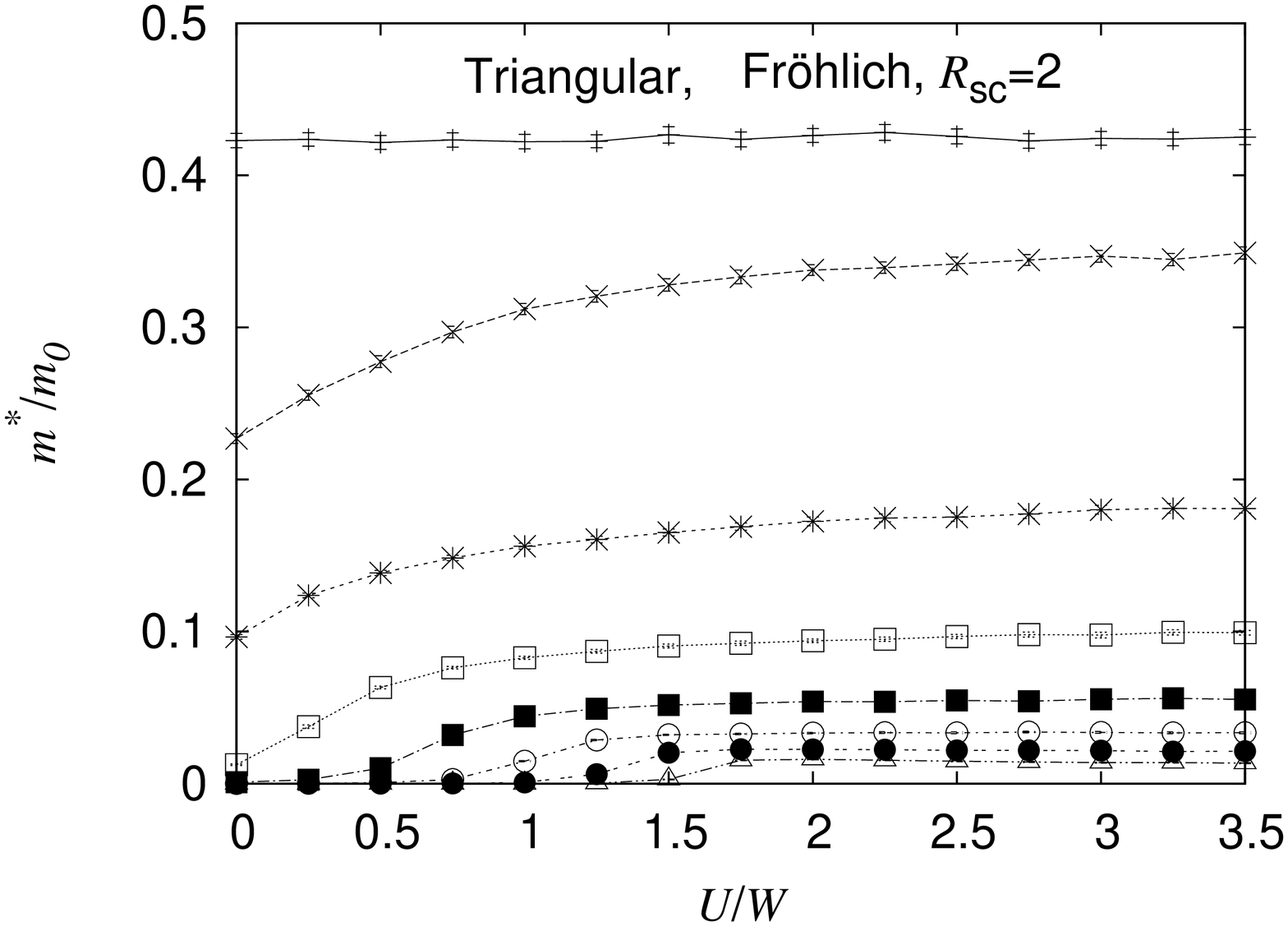}
\includegraphics[height=55mm,angle=270]{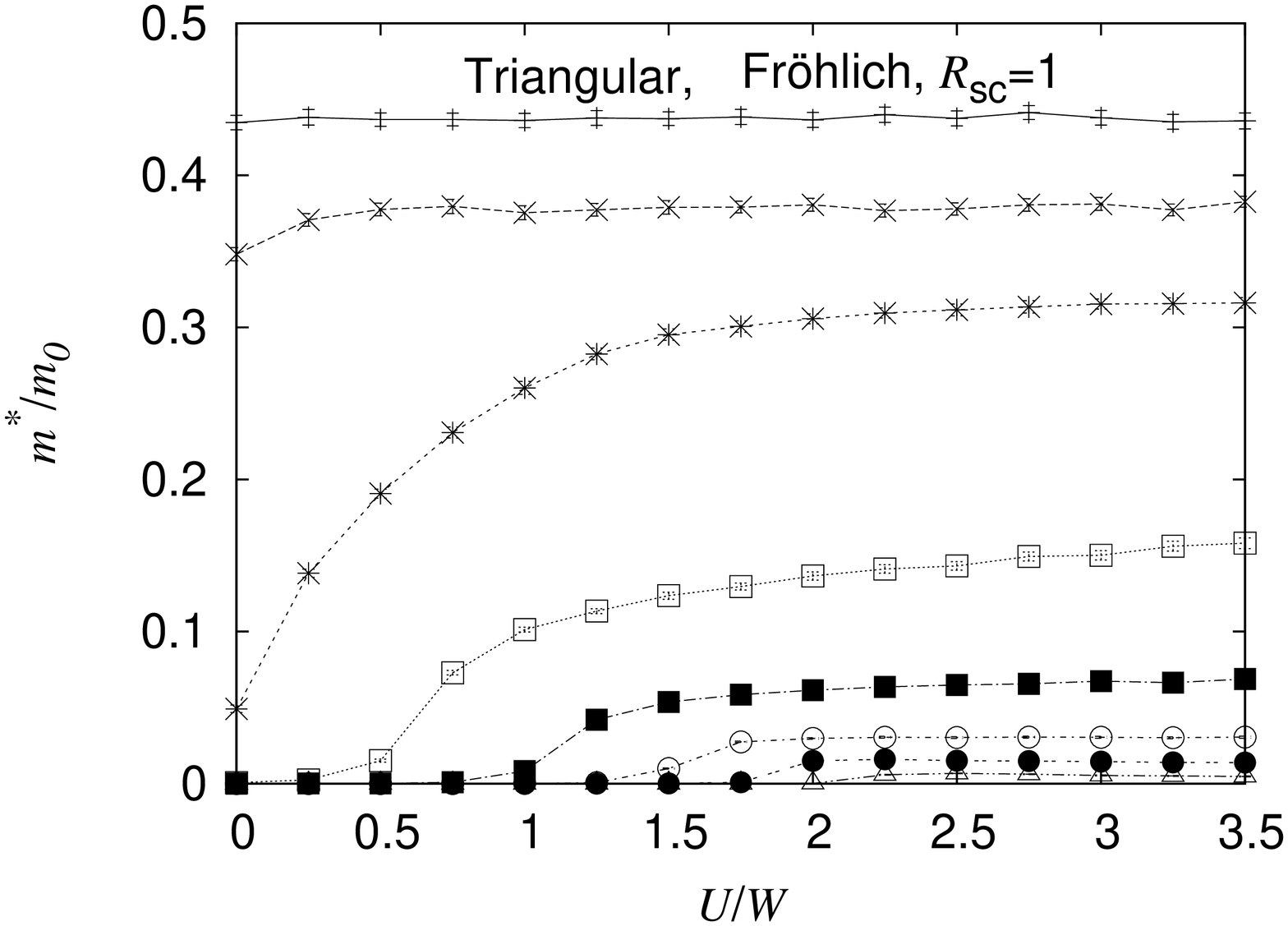}
\includegraphics[height=55mm,angle=270]{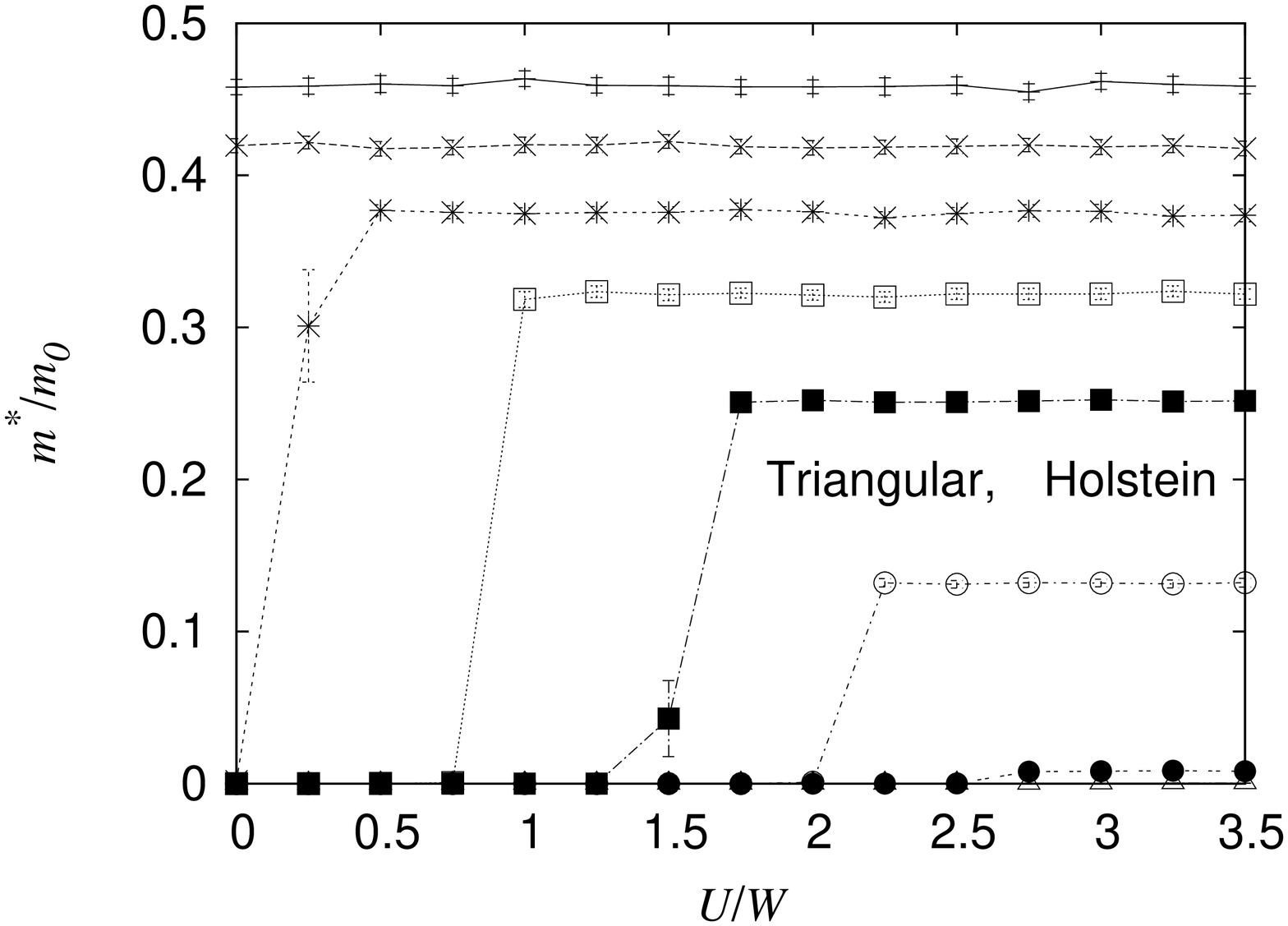}
\includegraphics[height=55mm,angle=270]{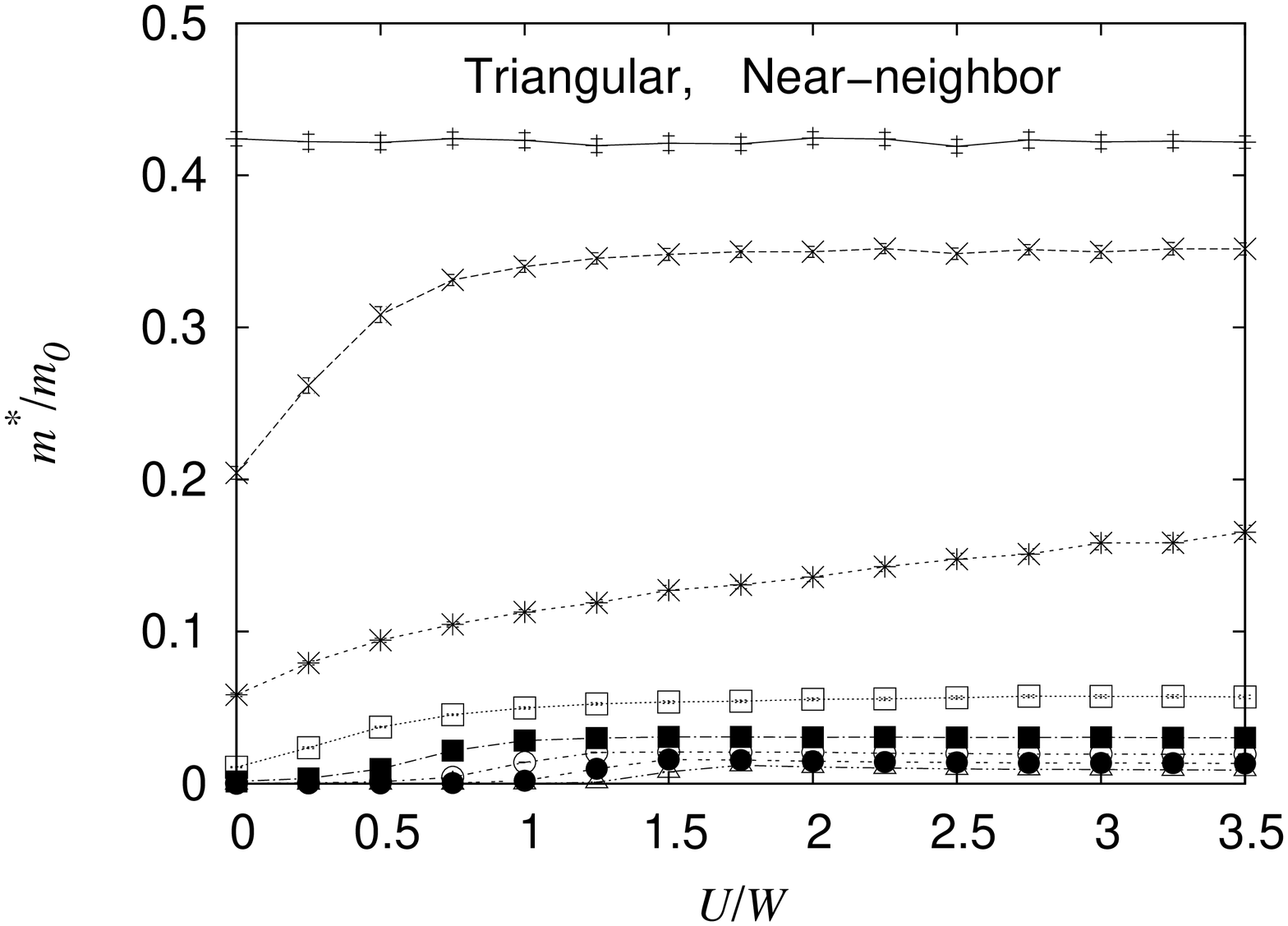}
\caption{Inverse mass of bipolarons on the triangular lattice. Parameters are as in Fig. \ref{fig:tetriangular}. The mass decrease at intermediate $U$ is much smaller than in the case of the square lattice since the intersite bipolarons formed at large $U/t$ are already superlight.}
\label{fig:imtriangular}
\end{figure*}

\begin{figure}
\includegraphics[height=75mm,angle=270]{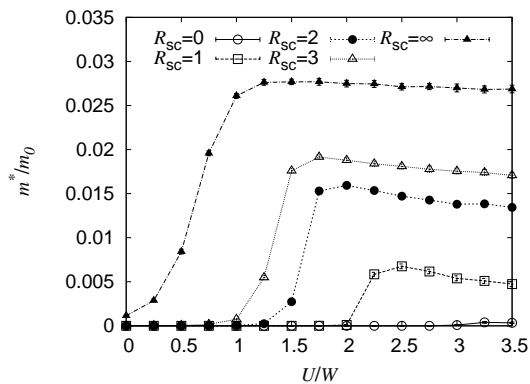}
\caption{Comparison of the inverse mass at $\lambda=1.6$ for bipolarons on the triangular lattice. Parameters are as in Fig. \ref{fig:tetriangular}.}
\label{fig:imcomparelargelambdatri}
\end{figure}

\begin{figure*}
\includegraphics[height=55mm,angle=270]{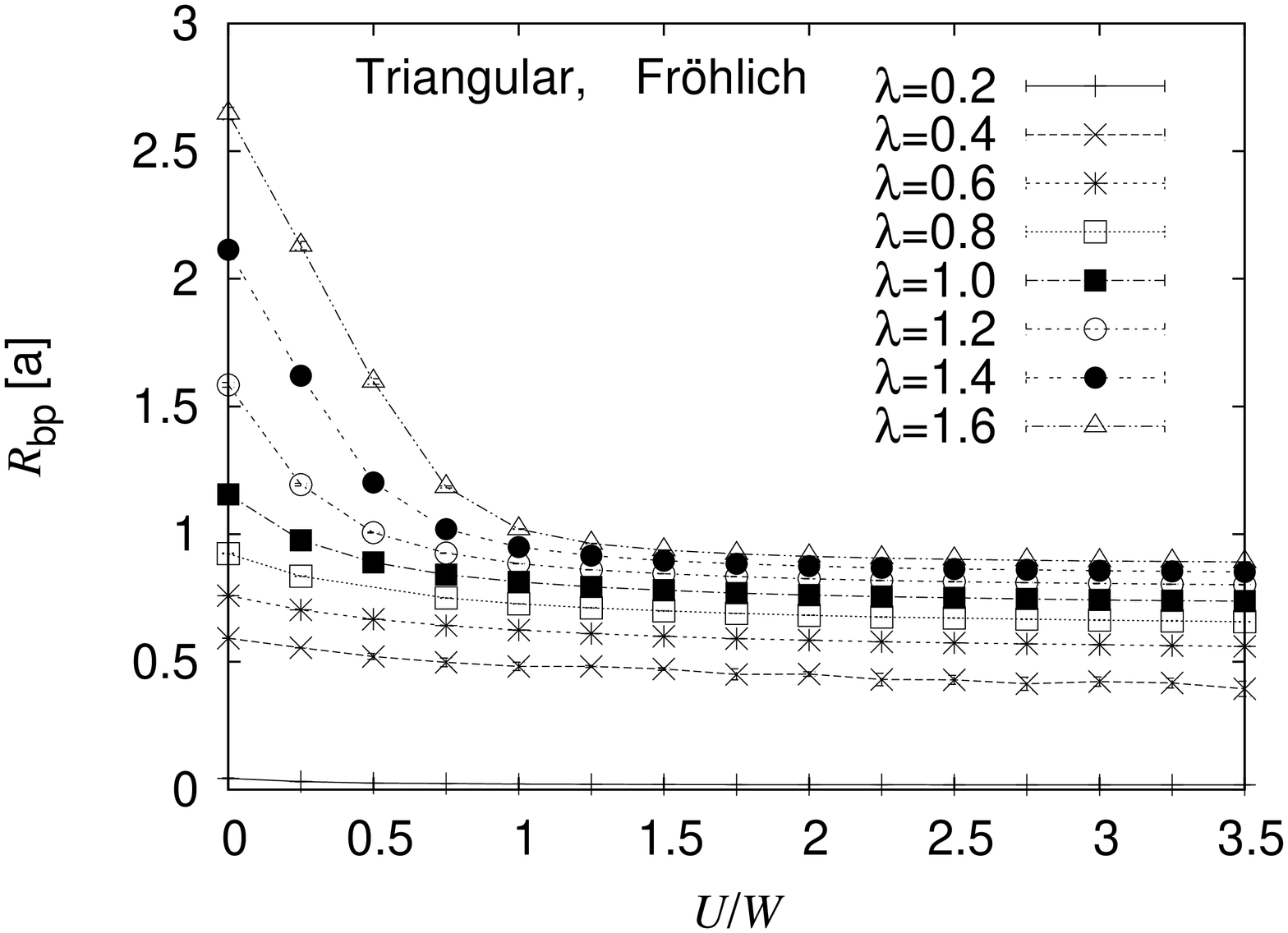}
\includegraphics[height=55mm,angle=270]{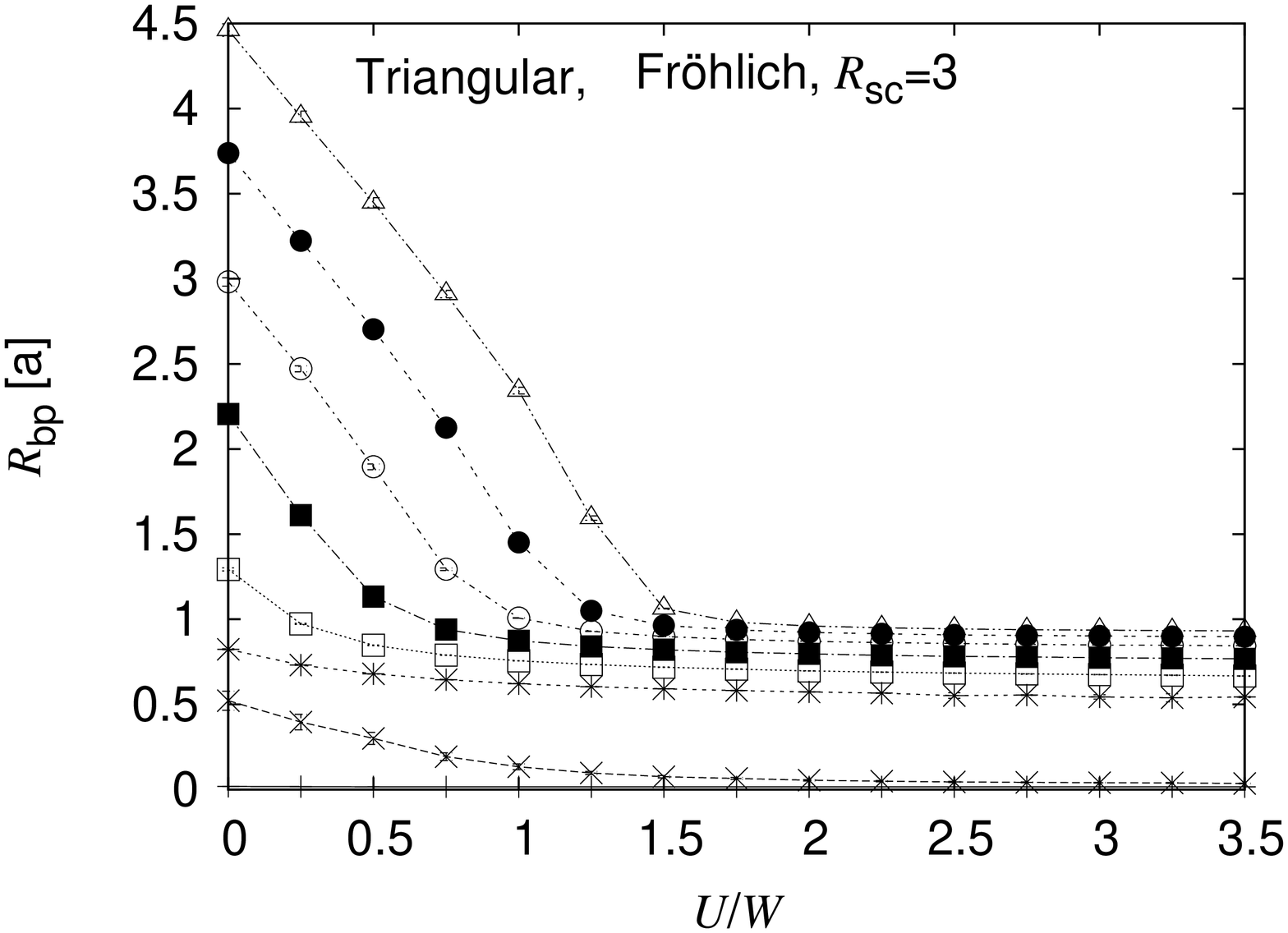}
\includegraphics[height=55mm,angle=270]{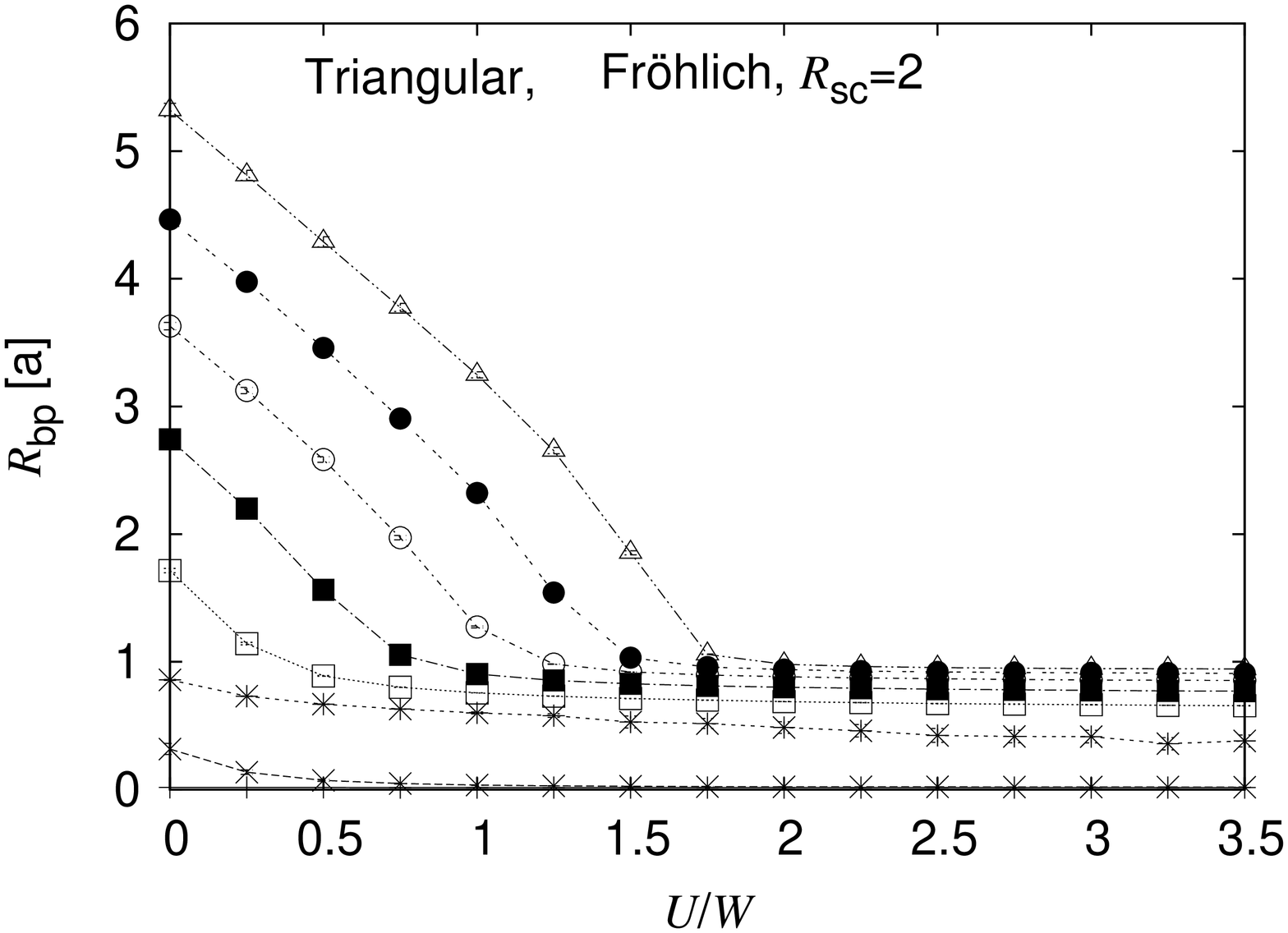}
\includegraphics[height=55mm,angle=270]{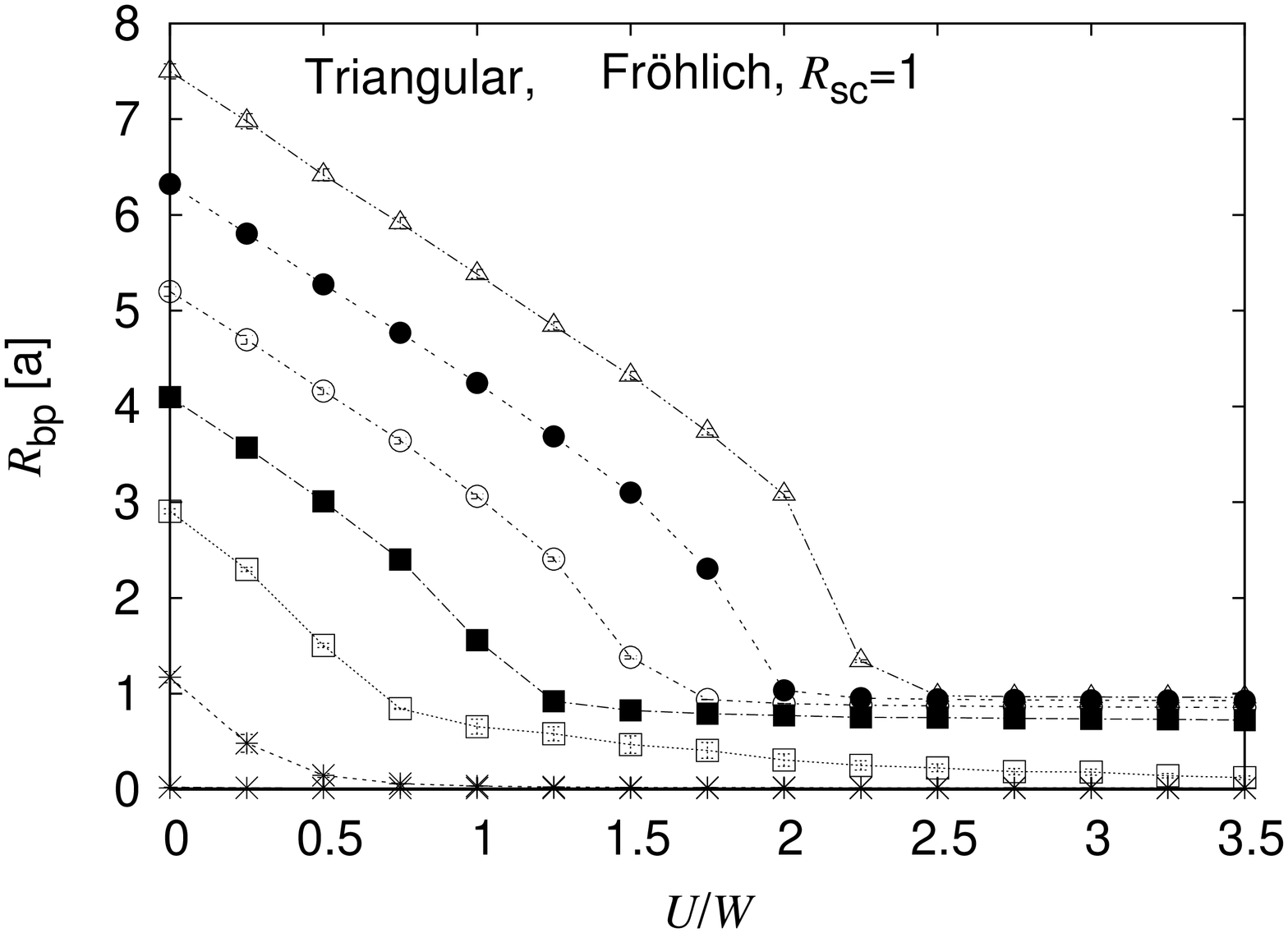}
\includegraphics[height=55mm,angle=270]{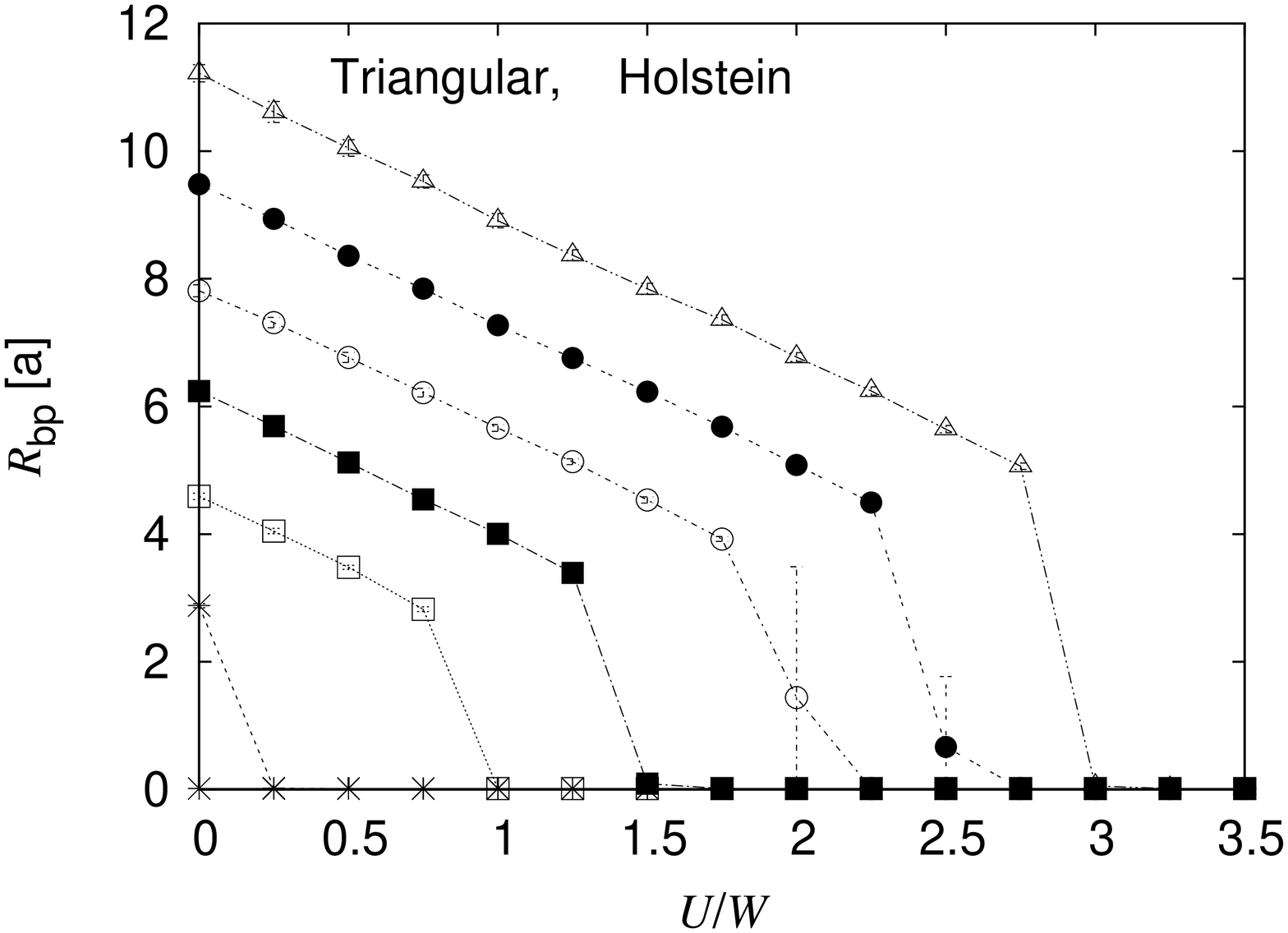}
\includegraphics[height=55mm,angle=270]{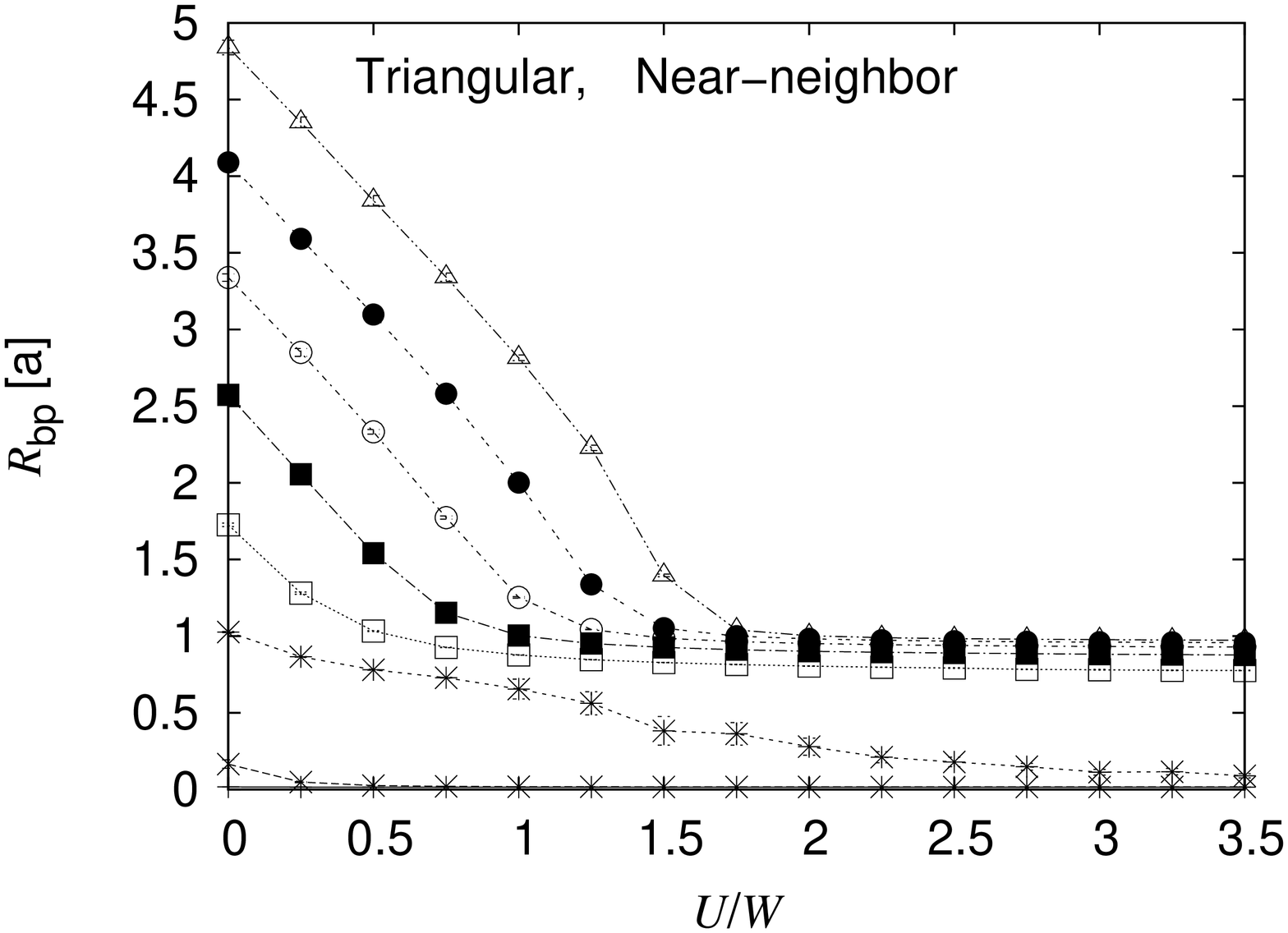}
\caption{Inverse size associated with bipolarons on the triangular lattice. Parameters are as in Fig. \ref{fig:tetriangular}.}
\label{fig:bpstriangular}
\end{figure*}

\section{Singlets on the triangular lattice}
\label{sec:triangular}

In this section, attention is turned to the properties of bipolarons on the triangular lattice, which have already been shown to have unusual properties \cite{hague2007a}. To ensure that results for bipolarons on
square and triangular lattices can be meaningfully compared, we have
kept energy scales fixed in terms of $W$ (the non-interacting kinetic energy of a single particle) rather than $t$. For the square lattice, $W=4t$, while for the triangular lattice, $W=6t$. Similar considerations should be made when comparing plots of total energy between lattice types, which is why we have quoted all energy scales in these sections in terms of $W$. Also, to keep the ratio $\omega/W=1/4$, all properties in this section are computed for $\omega/t=1.5$. Finally, for the same reason, the temperature is also 3/2 higher than that used for computations of the bipolaron on the square lattice, so $\bar{\beta}=28/3$.

Figure \ref{fig:tetriangular} shows the total energy of bipolarons on
the triangular lattice as $U$, $\lambda$ and $R_{sc}$ are
changed. A quick inspection indicated that there is no clear qualitative difference between the total
energy of bipolarons on square and triangular lattices. The same is
true of the number of phonons associated with the bipolaron, which is
shown in Fig. \ref{fig:nptriangular}. It is quite surprising that the
properties are so similar, since there are very different hopping
mechanisms on the triangular lattice compared to the square lattice
(such as the possibility for S1 bipolarons to move with a single hop
on the triangular lattice, while intersite bipolarons on the square
lattice have to tunnel through a high energy intermediate
state to move). However, careful comparison shows that in the strong $U$ limit (where inter-site S1 bipolarons are formed) the
bipolaron formed from near-neighbor interactions has slightly more phonons on
the square lattice than on the triangular lattice. For the very long range Fr\"ohlich interaction,
there are only tiny differences between results for the square and
triangular lattices (there are around 2.5\% more phonons associated
with Fr\"ohlich bipolarons on the triangular lattice). Clearly, the
crossover between S0 and S1 bipolarons occurs at approximately the same value of
$U/W$. The strong similarities between the properties of bipolarons on
square and triangular lattices on the adiabatic side of the parameter
space is quite surprising considering that the properties are very
different in the antiadiabatic limit. However, the differences between
the bipolarons becomes clearer when the inverse mass is examined.

%The unbinding of
%the bipolaron on increasing $U$ occurs at much larger $U/t$ on the
%triangular lattice than on the square lattice. This probably relates
%to the larger bandwidth of the triangular polaron, since at
%$\lambda=0.4$ the bipolaron unbinds at $U/t\sim 4$ on the square
%lattice and $U/t\sim 6$ on the triangular lattice (the ratio of
%half-bandwidths of polarons on triangular and square lattice is
%$3/2$).

It has previously been reported that bipolarons on the triangular
lattice are small and light \cite{hague2007a}. Figure
\ref{fig:imtriangular} shows the inverse mass of bipolarons on the
triangular lattice. Bipolarons formed via the Fr\"ohlich interaction are relatively light, as was observed when the square lattice was considered. The Holstein
bipolaron remains qualitatively different to the bipolarons formed with the screened Fr\"ohlich interaction, with an enormous change in
the mass of several orders of magnitude over a very small range of $U/W$
as the bipolaron binds. The most significant (and surprising)
differences between the bipolarons on square and triangular lattices
can be seen when nearest neighbor interactions are considered and when $\lambda=0.6$ and
$U$ is large. For these parameters, the bipolaron on the square lattice is
lighter than its counterpart on the triangular
lattice (although this does not mean that it is necessarily small). We also note that as $\lambda$ approaches $1.6$, the relative magnitudes of the masses on square and triangular lattices reverses and the bipolaron on
the triangular lattice becomes lighter than the bipolaron on the square lattice (the other parameters are held fixed). This is the part of the parameter space where superlight small bipolarons are expected. We will revisit this point when the size of the
bipolaron is computed.

Again, a hybrid S0-S1 bipolaron is likely when the S0 and S1
configurations become degenerate. As before, this state is visible as
a decrease in the effective mass (increase in inverse effective mass)
at intermediate $U$. A comparison of the inverse masses of bipolarons
as interaction range is changed is shown in
Fig. \ref{fig:imcomparelargelambdatri}. There is only a shallow
maximum in the inverse mass curve for $R_{sc}=\infty$, but there are
more pronounced humps in the curves for smaller values of the
screening, corresponding to a light bipolaron. The peaks are not as
pronounced for the bipolaron on the triangular lattice as they were
when motion was on the square lattice. This is probably because the S1
bipolaron on the triangular lattice at large $\lambda$ and large $U$
is already light and small with a mass that is 1st order in $t$
\cite{hague2007a}, so the presence of the hybrid state does not make
it significantly easier for the bipolaron to move about on the
triangular lattice (whereas on the square lattice, the change from S1
to the hybrid state changes bipolaron hopping processes from 2nd to
1st order in $t$ leading to a reduction in mass).

%Differences in the inverse mass should lead to differences in the mass
%isotope exponent shown in Fig. \ref{fig:iextriangular}. There is no clearly visible dip in the
%isotope exponent for the bipolaron on the triangular lattice formed
%from long range interactions in contrast to the bipolaron that forms on the square lattice. We consider this to
%be related to the smaller peak in the inverse mass of the bipolaron on
%the triangular lattice. Otherwise, there are no significant differences between the isotope exponents of bipolarons formed on square and triangular lattices.

To finish the examination of the differences between square and
triangular lattices, we examine the inverse bipolaron size in figure
\ref{fig:bpstriangular}. For large $\lambda$, the size of the pair is small and S1
bipolarons are clearly formed in the Fr\"ohlich model at large
$U$ (i.e. the radius is approximately one lattice spacing). There are differences between the sizes of bipolarons on
triangular and square lattices. Comparison between the radius of the
bipolaron on the two lattices shows that the bipolaron on the triangular lattice is
significantly smaller than its counterpart on the square lattice at large $U$. Also, the bipolaron on the triangular lattice is slightly bigger than the pair on the square lattice at very
small $U$. The bipolarons formed on square and triangular lattices have similar size at intermediate $U$ where the
pair has hybrid S0-S1 properties. Thus we can see that the slightly
larger (but still light) mass of the S1 bipolaron on the triangular lattice is most likely a result of the stronger binding into the S1 state (indeed for large $U$, the
S1 state is very well bound on the triangular lattice at quite small
$\lambda$ - shown by $R_{bp}^{-1}\approx 1$). It is this
property of small and light bipolarons that could lead to a bipolaron
condensate of S1 bipolarons on a triangular lattice at reasonably high temperatures. However, we have
now identified an additional bipolaron configuration which is
light and small on both triangular and square lattices, and which is formed
at moderate $\lambda$ and $U$: the extra-light hybrid S0-S1 bipolaron.

\section{Triplets}
\label{sec:triplet}

We now direct our attention to triplet properties of the bipolaron. The possibility of triplet superconductivity has received a lot of
interest since the discovery of spin triplet superconductivity in
Sr$_{2}$RuO$_{4}$ \cite{ishida1998a}. Triplet superconductivity has
also been identified in heavy Fermion materials that have a triangular
lattice \cite{ishida2002a}. From the theoretical point of view,
triplet pairs could be of interest for two reasons. First, there is a
wide literature on BCS to BEC crossover
\cite{micnas1990a,noziers1985a,randeria1992a} which is of interest when the parameters governing pairing are intermediate. The presence of stable real-space
triplet pairs would add additional limits to this problem, including a
crossover or transition between singlet and triplet pairing. Second,
the energy difference between singlet and triplet states could be
considered as the energy cost of a spin flip, with a direct
interpretation as a spin gap \cite{mottsashabook}. In the next two
sections of this article, we discuss the possibility and properties of
real space triplet pairs on square and triangular lattices.

\subsection{Hubbard-Holstein model}

\begin{figure}
\includegraphics[height=70mm,angle=270]{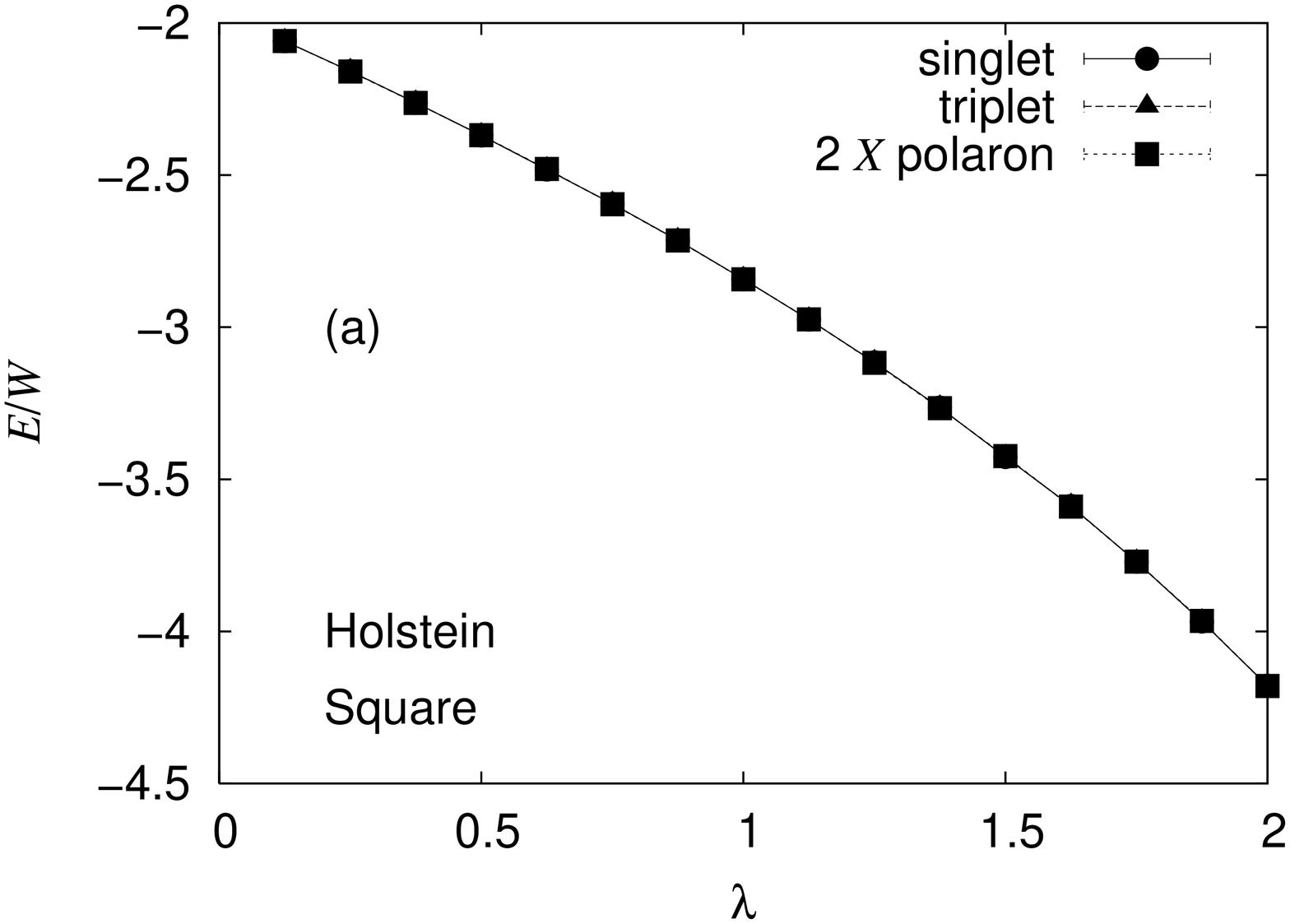}
\includegraphics[height=70mm,angle=270]{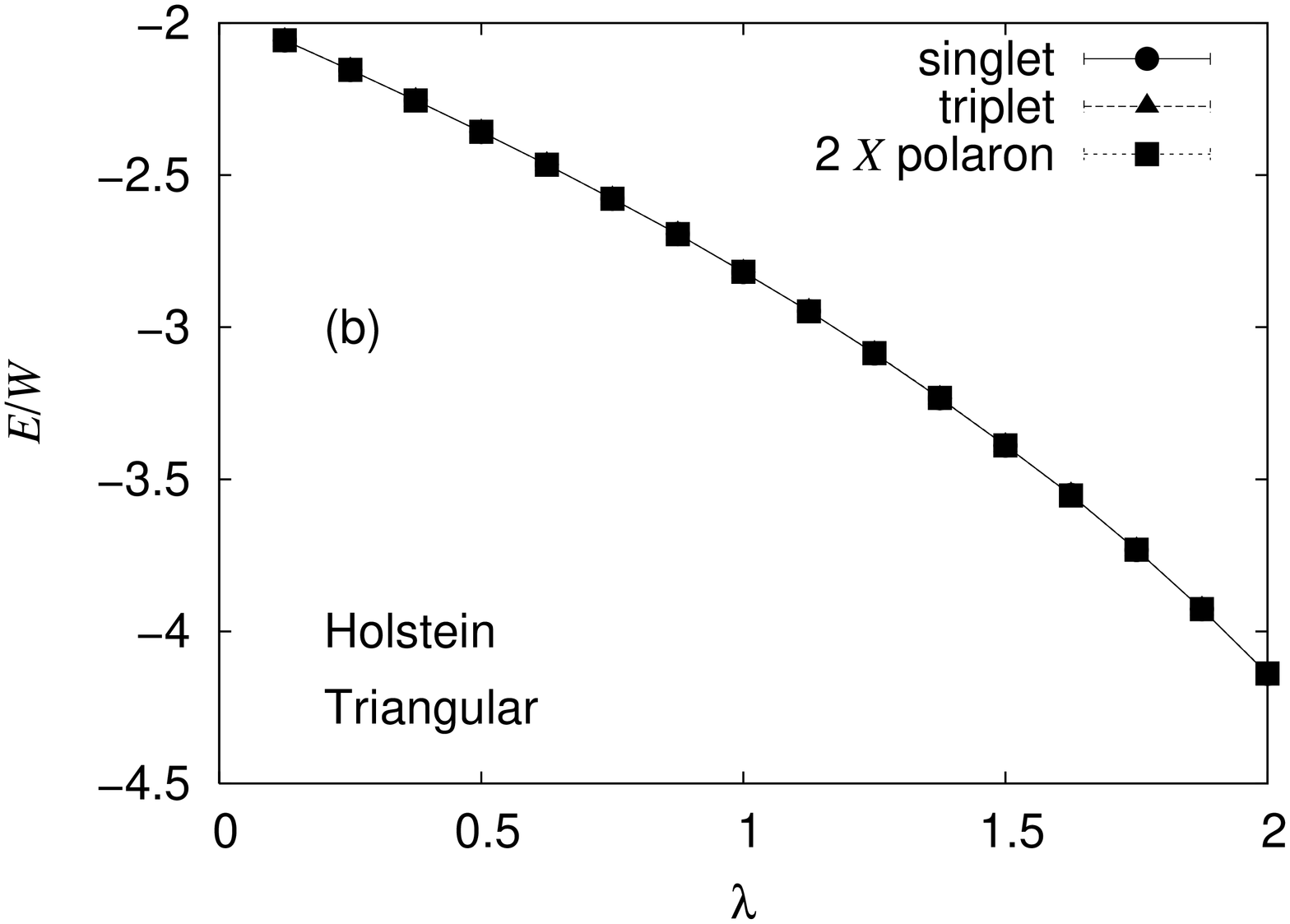}
\caption{Absence of triplet bipolaron in the Holstein model on square
and triangular lattices. (a) Square lattice, $\bar{\beta}=7$,
$\bar{\omega}=1$, $U/t=40$. (b) Triangular lattice,
$\bar{\beta}=14/3$, $\bar{\omega}=1.5$, $U/t=60$. Properties are
computed at various $\lambda$. Note that with the energy expressed in terms of the band-width, lattice type has
very little effect on Holstein polarons. The singlet and triplet
curves lie under the polaron curve, and are identical to within the
statistical error.}
\label{fig:tripletholstein}
\end{figure}

We start our examination by considering the on-site Hubbard-Holstein
interaction. To examine if triplet bipolarons can exist in the 2D
Hubbard-Holstein model, we examine the model with very large $U/W=10$
and compare the energy of two electrons subjected to the Hubbard-Holstein interaction with that of two unbound
polarons (Fig. \ref{fig:tripletholstein}). There are several good
reasons for doing this. First, as we have seen in the previous
sections, the energy of the singlet bipolaron increases monotonically
with $U$, so the binding energy of the singlet bipolaron is minimized,
therefore the singlet triplet splitting is also minimized (since triplet states can not have a higher energy than two free polarons without dissociating). Second, since
triplet states must have equal or higher energy than singlets, an
unbound singlet also implies that the triplet is also not bound. The triplet properties can not depend on $U$, since a node in the
triplet pair wavefunction means that the Hubbard $U$ has no possible
effect on triplet properties, so it is sufficient to analyze only this large $U$ limit. We have carefully computed the energies
of singlets, triplets and two polarons on triangular and square
lattices, and have found no bound triplet states to within the
statistical error.

\subsection{Hubbard-Fr\"ohlich model}

%\begin{figure}
%\includegraphics[height=75mm,angle=270]{Graphs/Triangular/triplet_radius_frohlich.ps}
%\caption{Radius of triplet bipolarons in the Hubbard-Fr\"ohlich model
%with comparison to the singlet bipolaron. $U=\infty$,
%$\bar{\beta}=3.5$, $\bar{\omega}=4$.}
%\end{figure}

%\begin{figure}
%\includegraphics[height=75mm,angle=270]{Graphs/Triangular/triplet_mass_frohlich.ps}
%\caption{Mass of triplet bipolarons in the Hubbard-Fr\"ohlich model
%with comparison to the singlet bipolaron. $U=\infty$,
%$\bar{\beta}=3.5$, $\bar{\omega}=4$.}
%\end{figure}

\begin{figure*}
\includegraphics[height=65mm,angle=270]{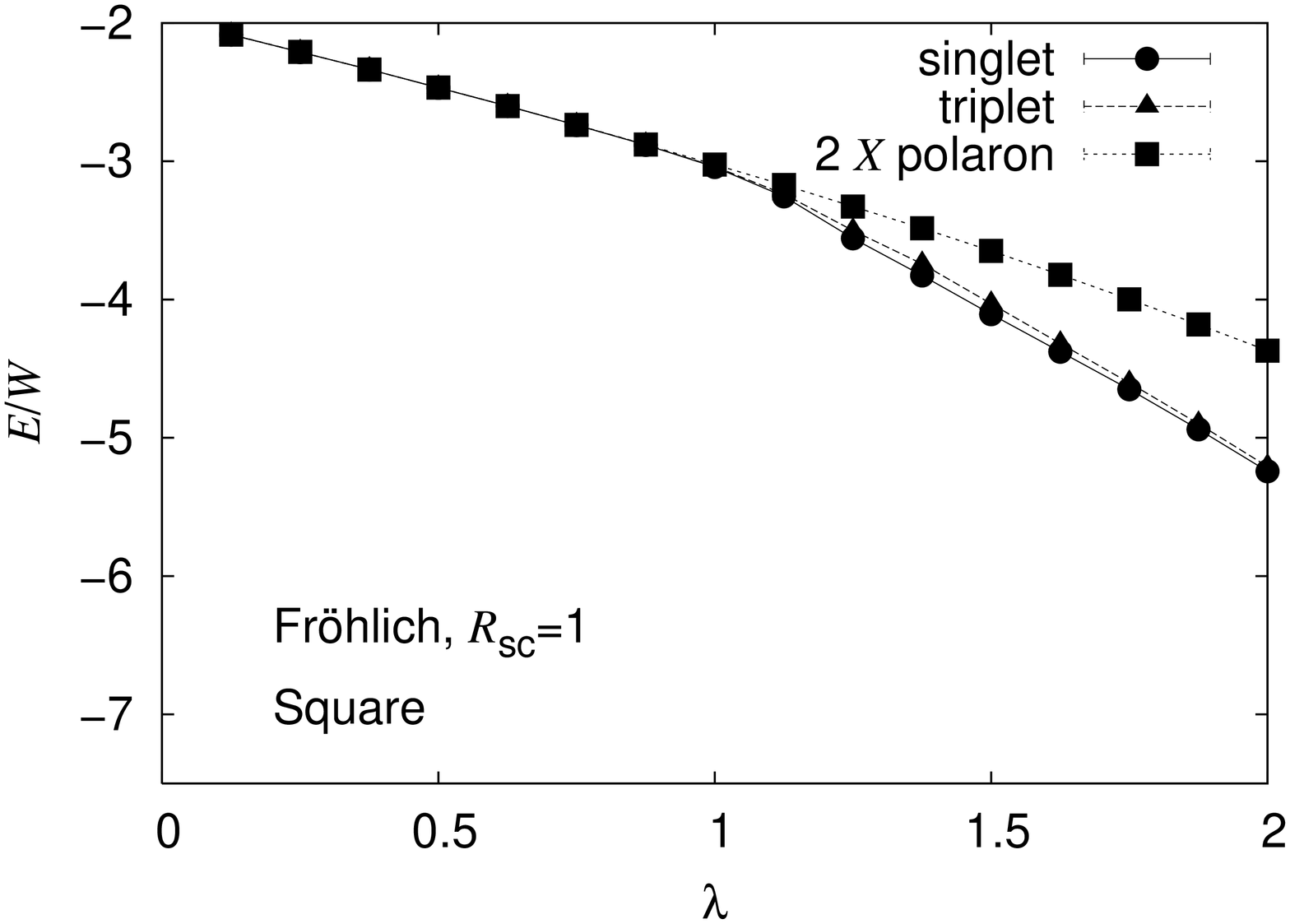}
\includegraphics[height=65mm,angle=270]{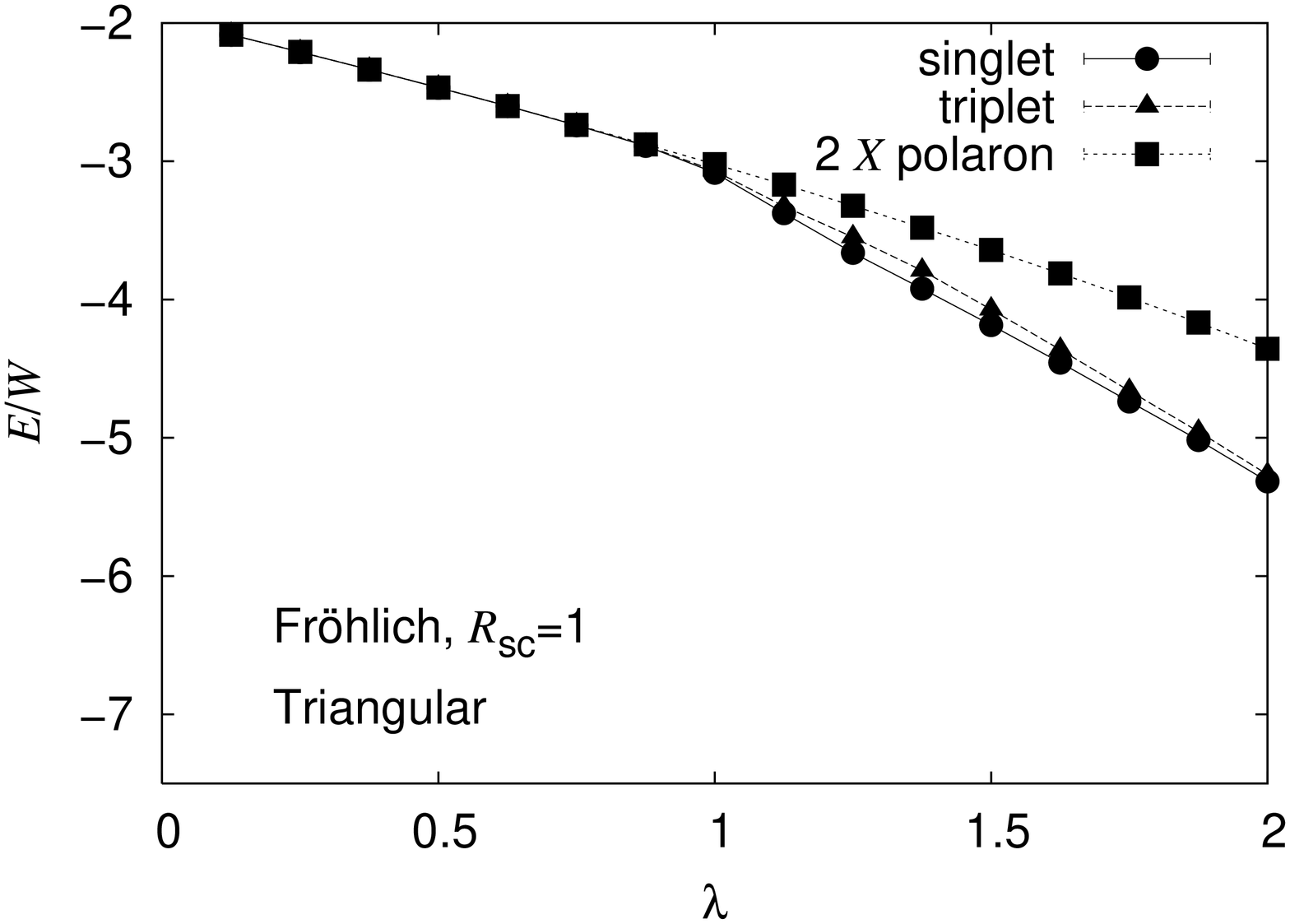}

\includegraphics[height=65mm,angle=270]{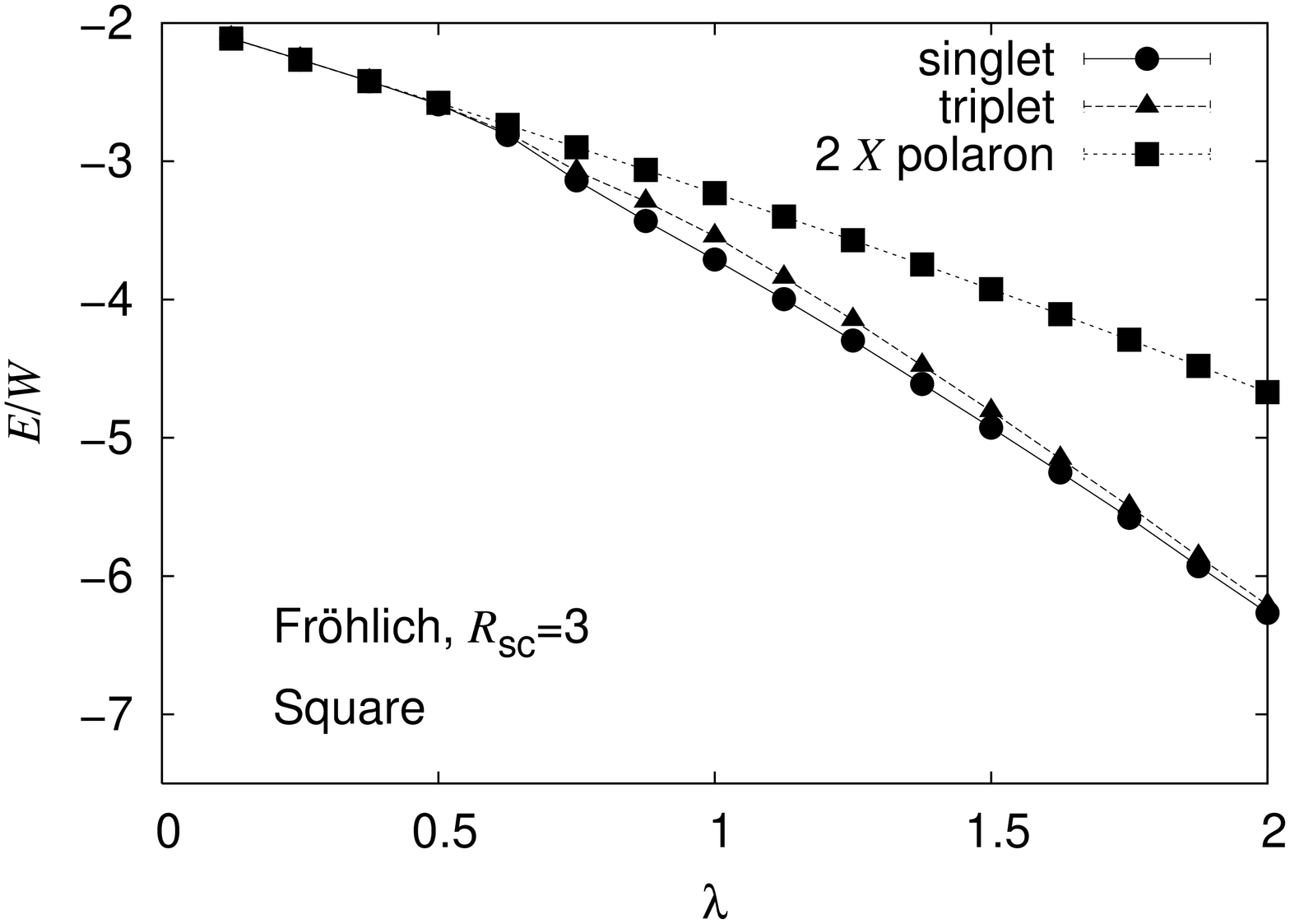}
\includegraphics[height=65mm,angle=270]{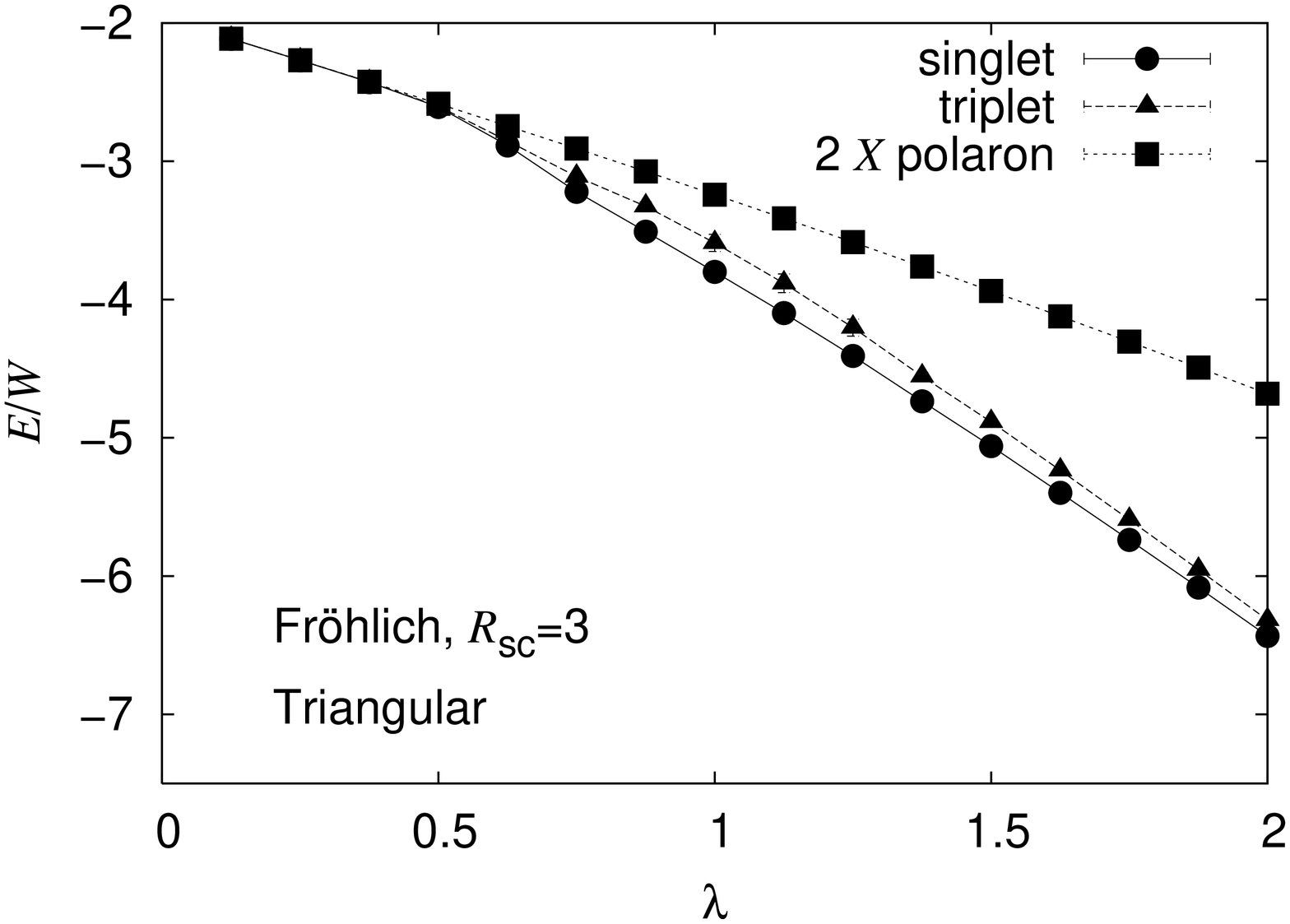}

\includegraphics[height=65mm,angle=270]{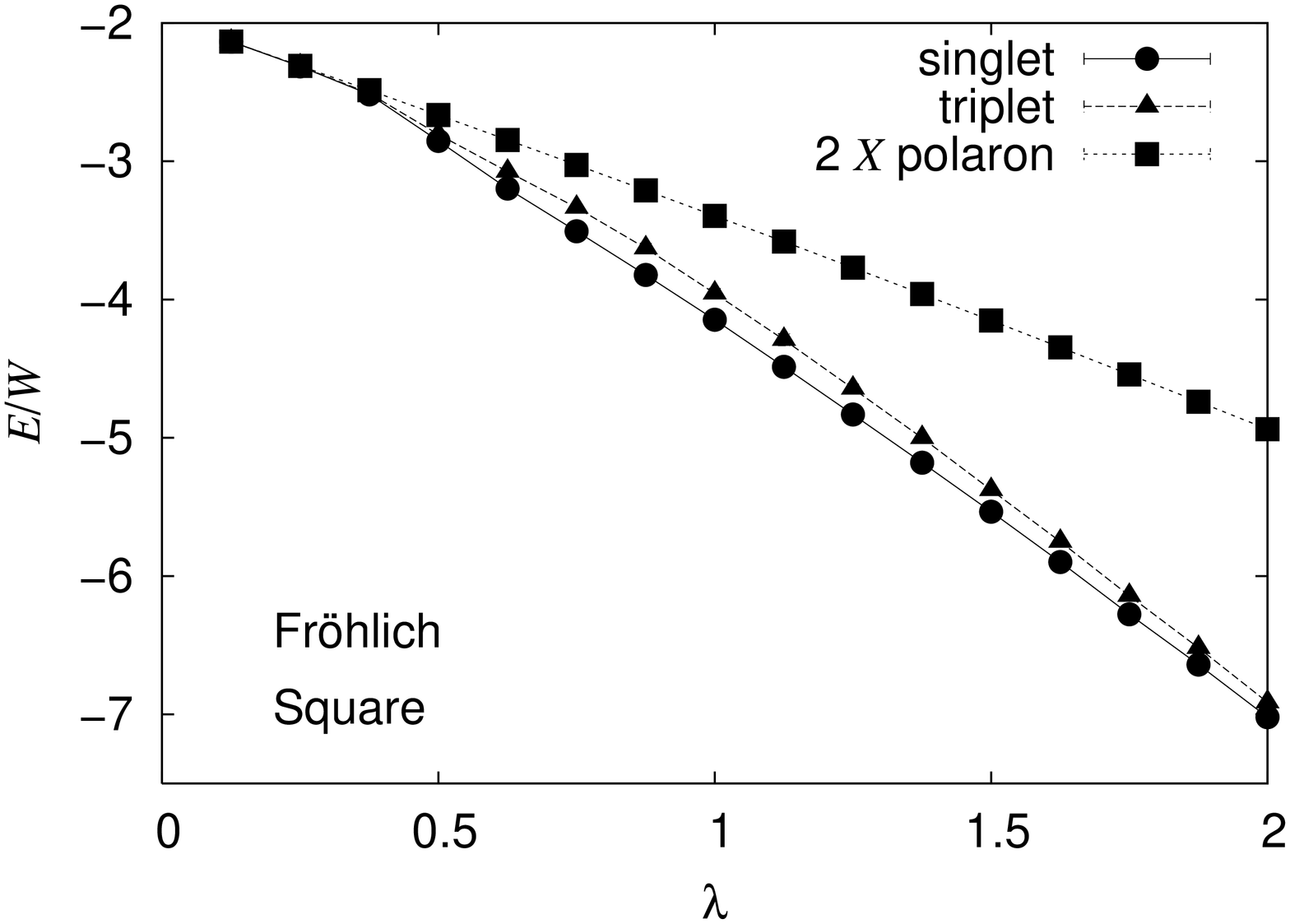}
\includegraphics[height=65mm,angle=270]{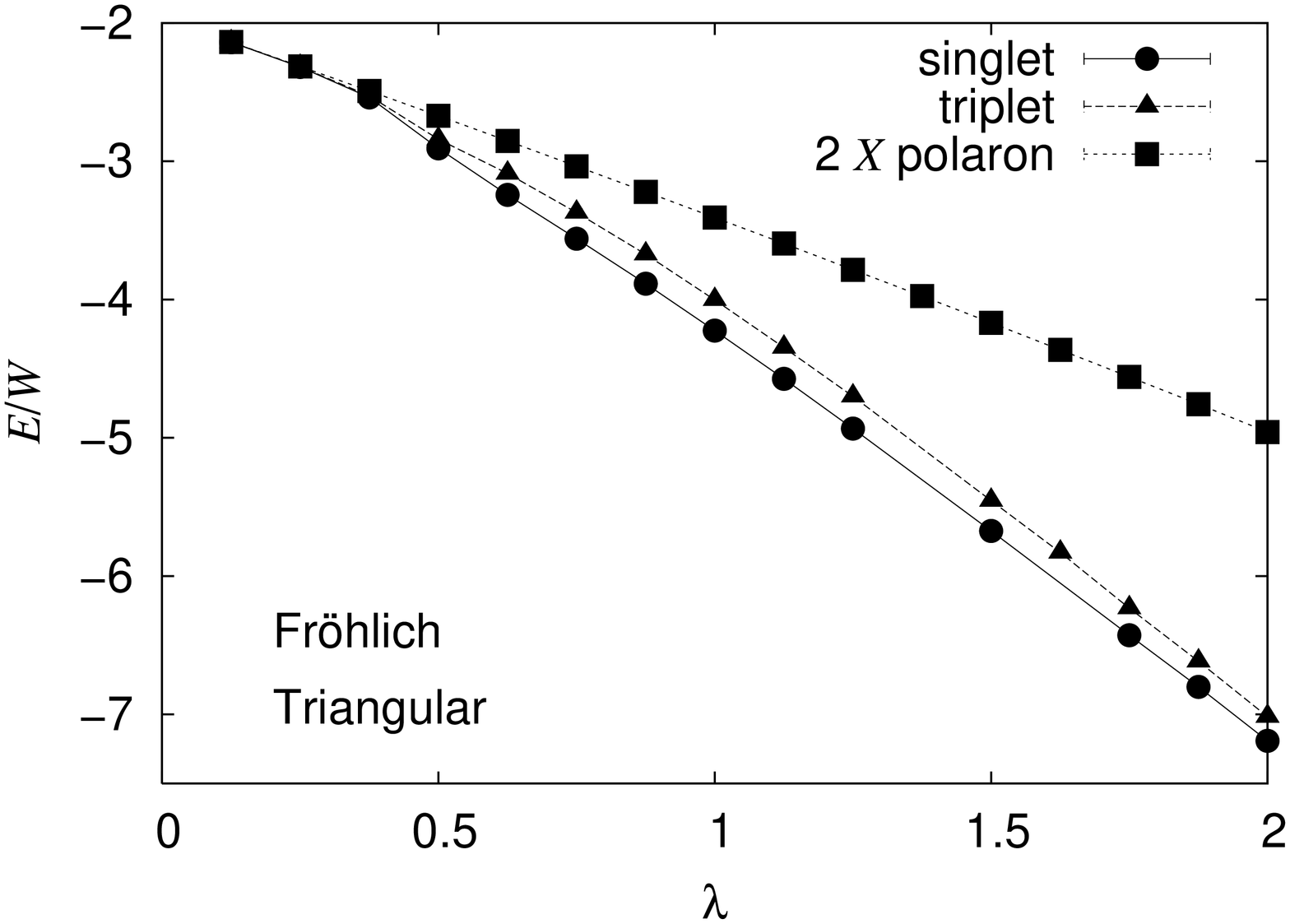}

\includegraphics[height=65mm,angle=270]{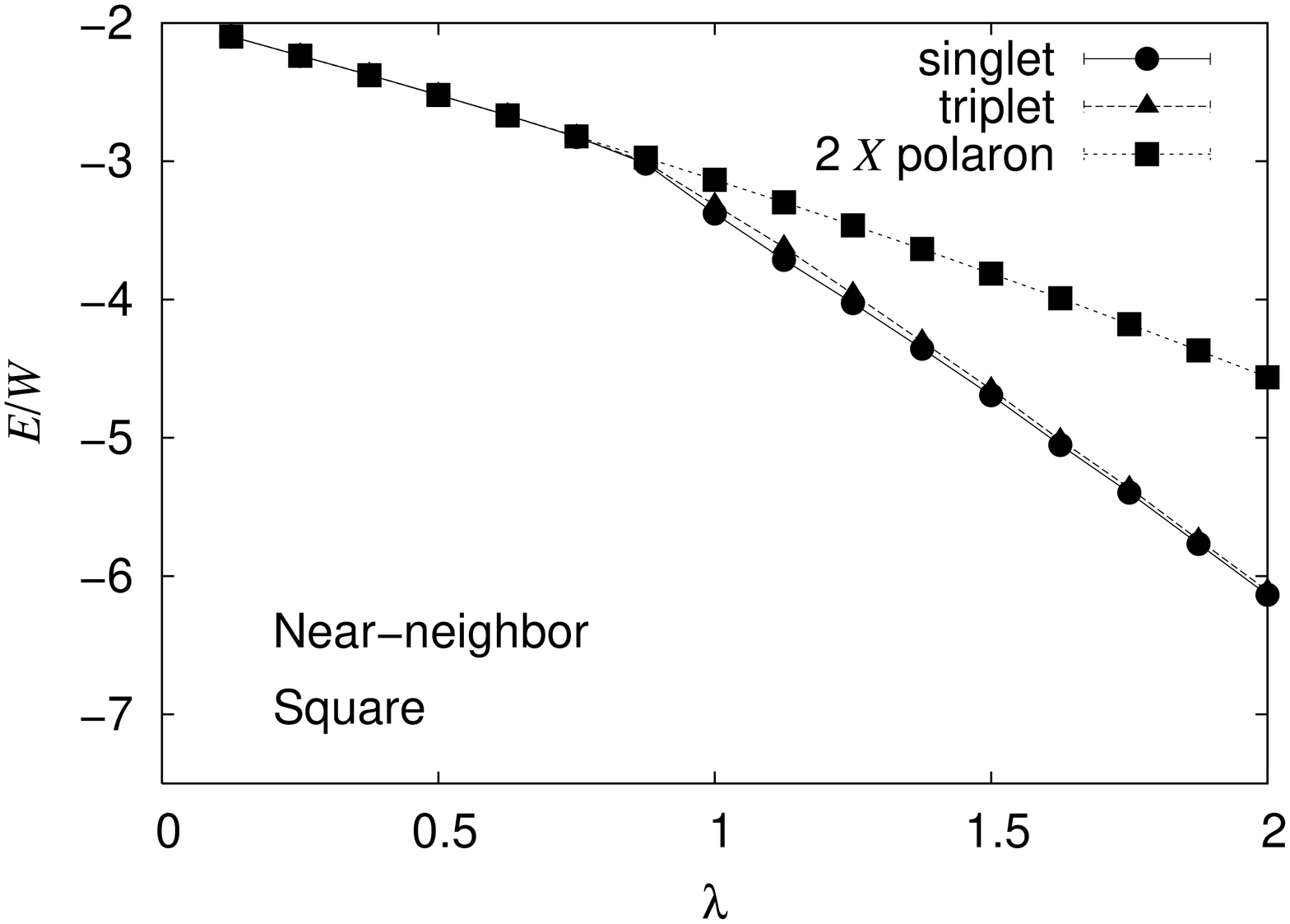}
\includegraphics[height=65mm,angle=270]{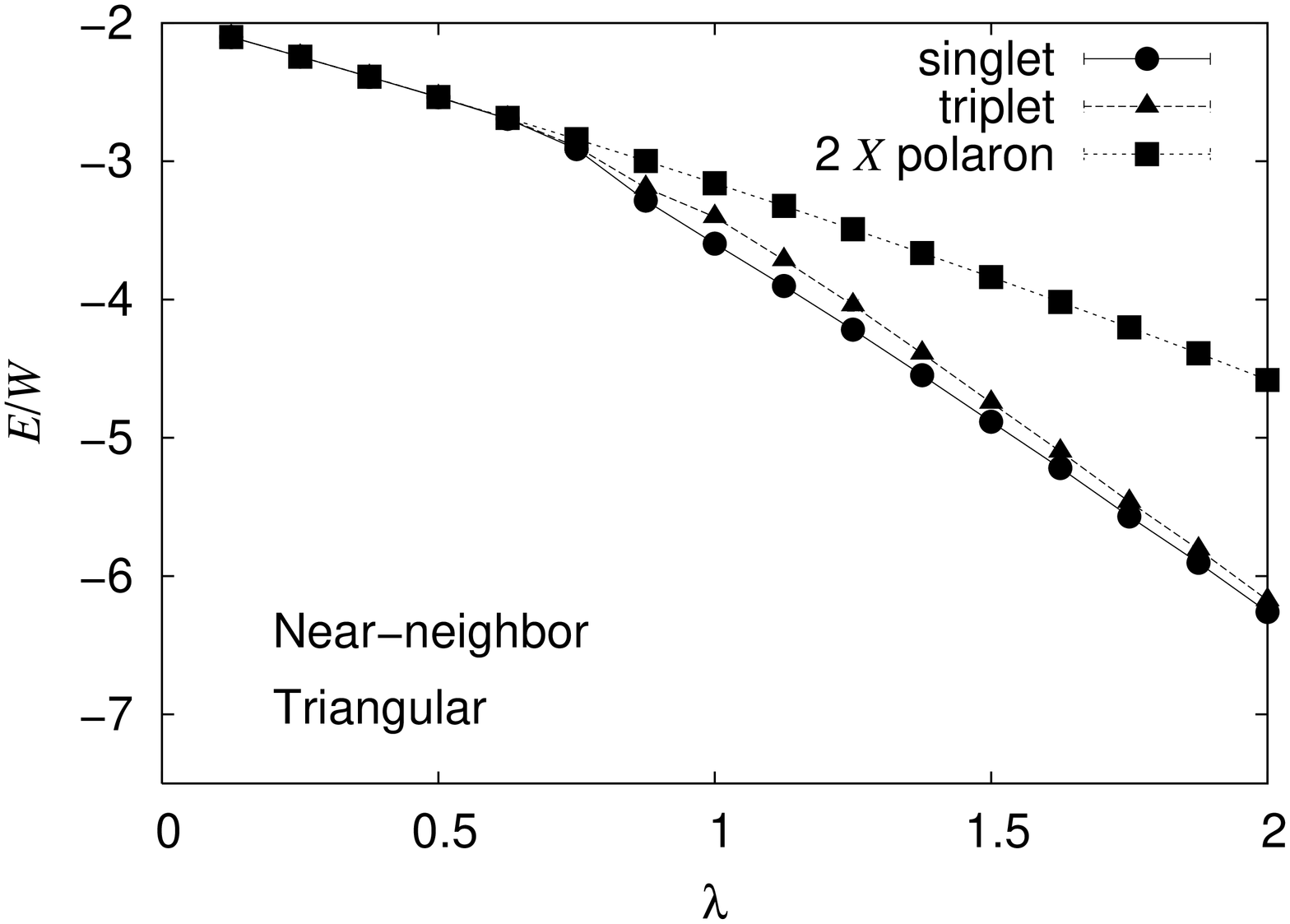}
\caption{Total triplet energy of the Fr\"ohlich bipolaron on the
square lattice (left) and triangular lattice (right). For the square
lattice, $U=40t$, $\bar{\beta}=7$, $\bar{\omega}=2$. On the triangular
lattice, $U/t=60$, $\bar{\omega}=3$, $\bar{\beta}=14/3$. $\lambda$ and
the interaction range are varied. For comparison, we also show the
singlet energy and the energy of two polarons. As the interaction
range increases, the minimum electron-phonon coupling required to bind
both singlet and triplet decreases. The singlet energy is very similar
on both lattices. However, the singlet-triplet splitting on the triangular
lattice is much larger than on the square lattice for all interaction ranges.}
\label{fig:tetriplet}
\end{figure*}

While there are no triplet bipolarons formed in the Hubbard-Holstein
model, the longer range interactions associated with the
Hubbard-Fr\"ohlich model can lead directly to inter-site pairing, which is
likely to lead to triplet states. To determine the existence of triplet
pairs in the model, we compute the singlet and triplet energies, and compare them with the
energy of two polarons, which can be seen in Fig. \ref{fig:tetriplet}. The parameters associated with the computations are $U=40t$, $\bar{\beta}=7$ and $\bar{\omega}=2$ on the square
lattice and $U/t=60$, $\bar{\omega}=3$ and $\bar{\beta}=14/3$ on the
triangular lattice. We use the same parameters throughout this
section. At weak $\lambda$, there are neither singlet nor triplet
bound states. On increasing $\lambda$, the singlet and triplet states bind at a similar value of $\lambda$ (the binding can
be seen as bifurcations of the curves). At very large $\lambda$, the
energies of the singlet and triplet become degenerate. This is a
result of the interaction on the polaron hopping, which becomes
exponentially small. The separation of singlets and triplets in the
$U-V$ model on the square lattice scales as $t^2$, whereas on the
triangular lattice the large $\lambda$ singlet-triplet splitting
scales like $t$. This can be seen in the adiabatic limit as a much
larger singlet triplet splitting in the case of the triangular
lattice. We note that there are strictly two degenerate
triplet states measured here, representing two $p$-pairings.

\begin{figure*}
\includegraphics[height=65mm,angle=270]{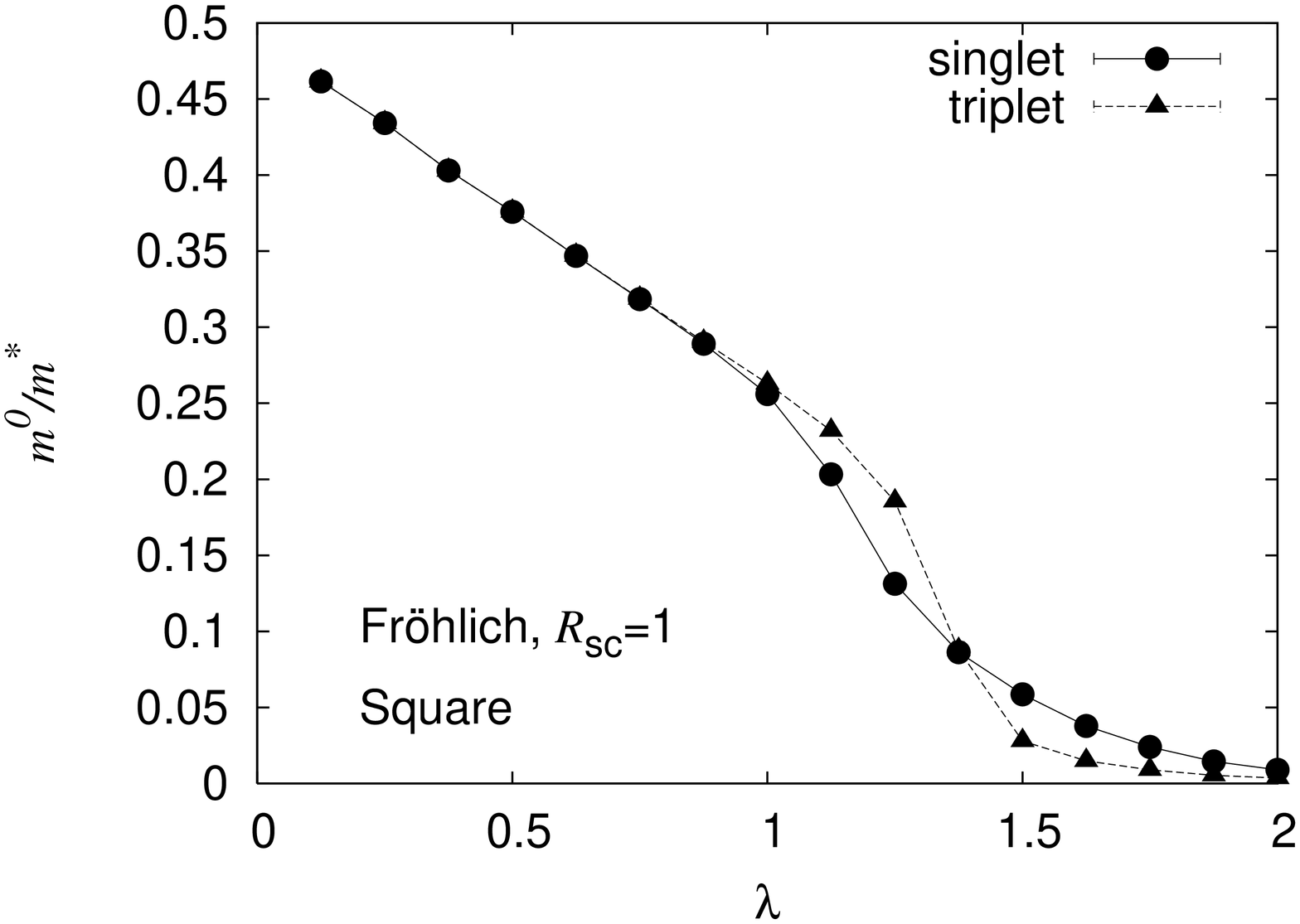}
\includegraphics[height=65mm,angle=270]{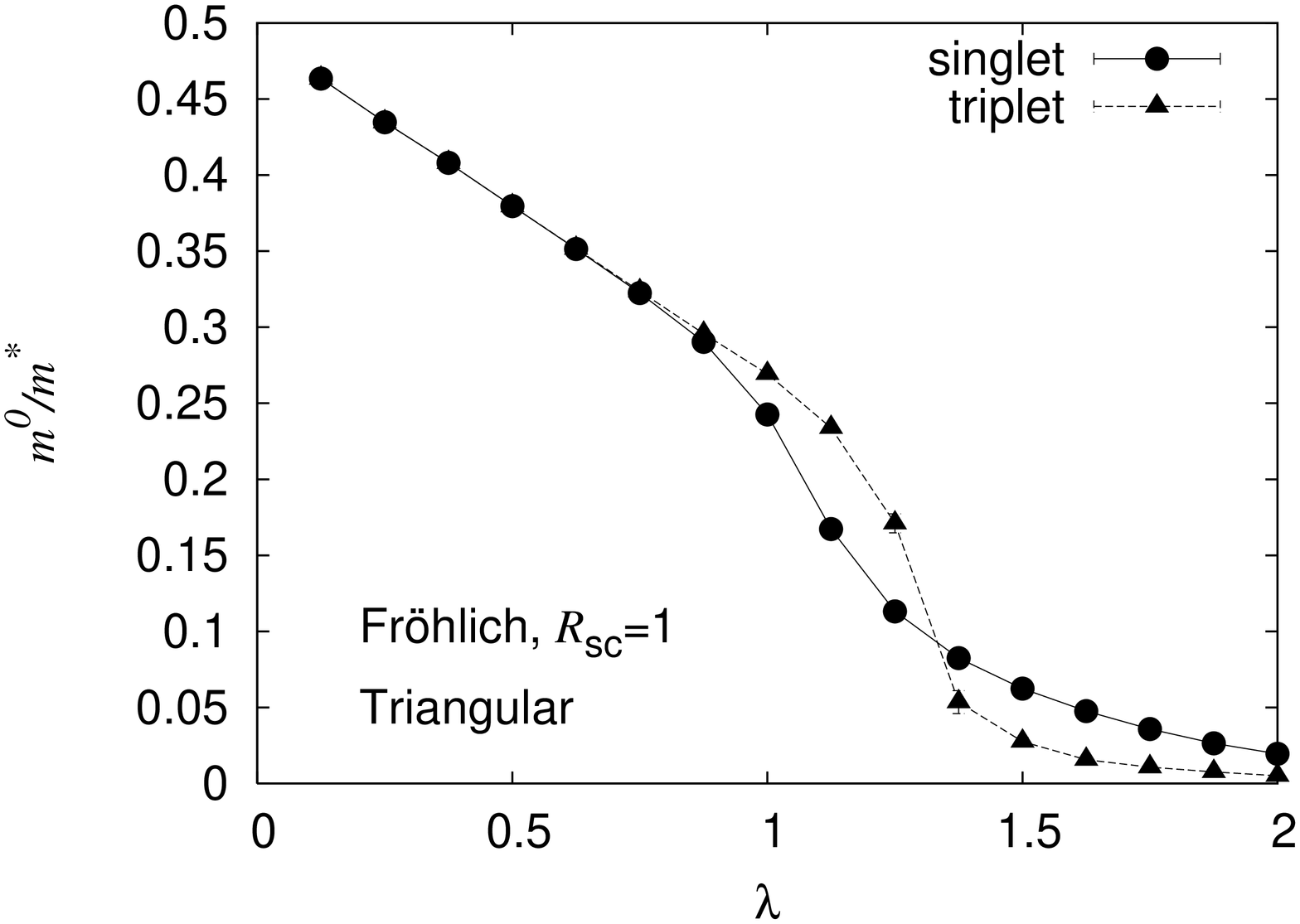}

\includegraphics[height=65mm,angle=270]{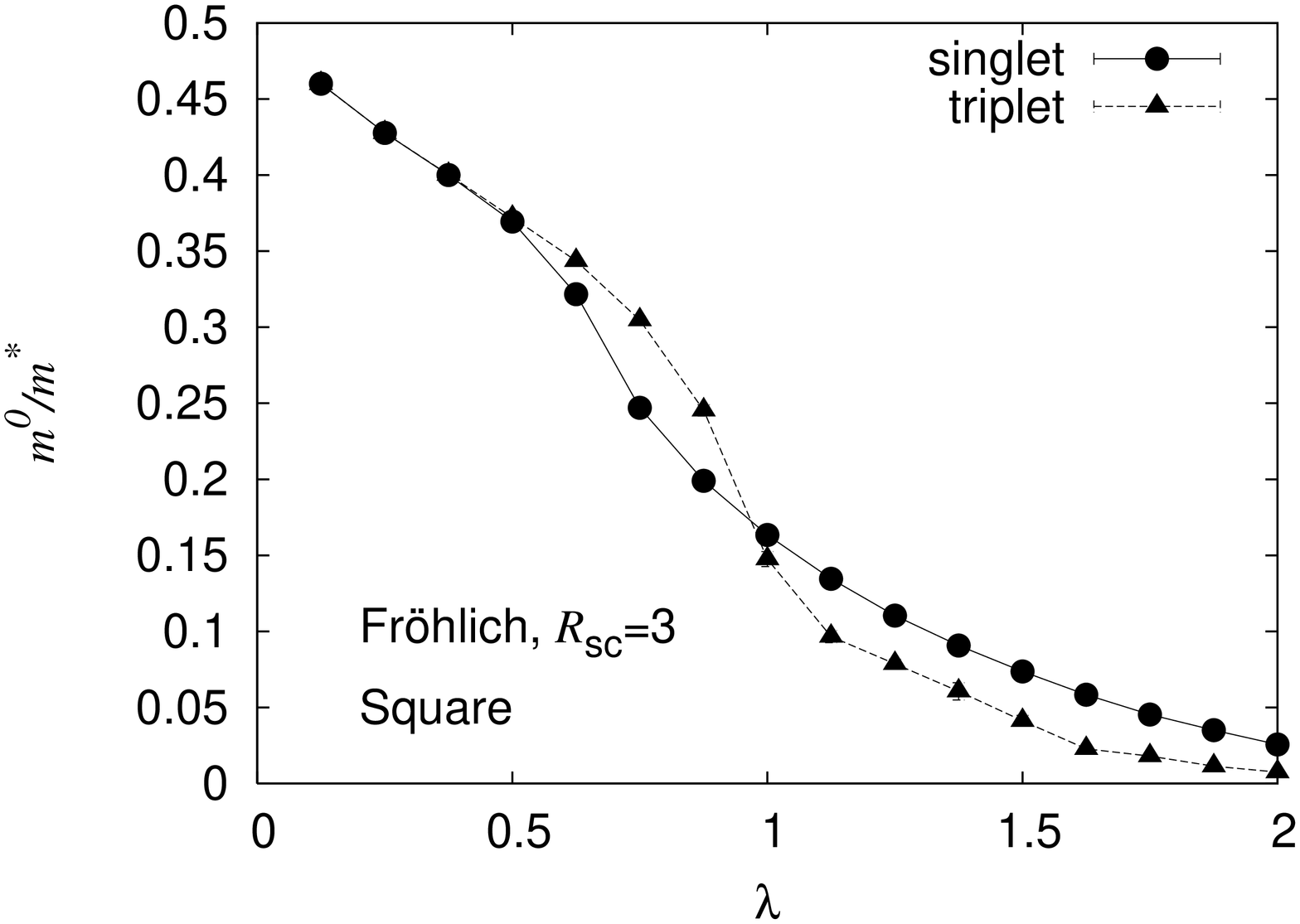}
\includegraphics[height=65mm,angle=270]{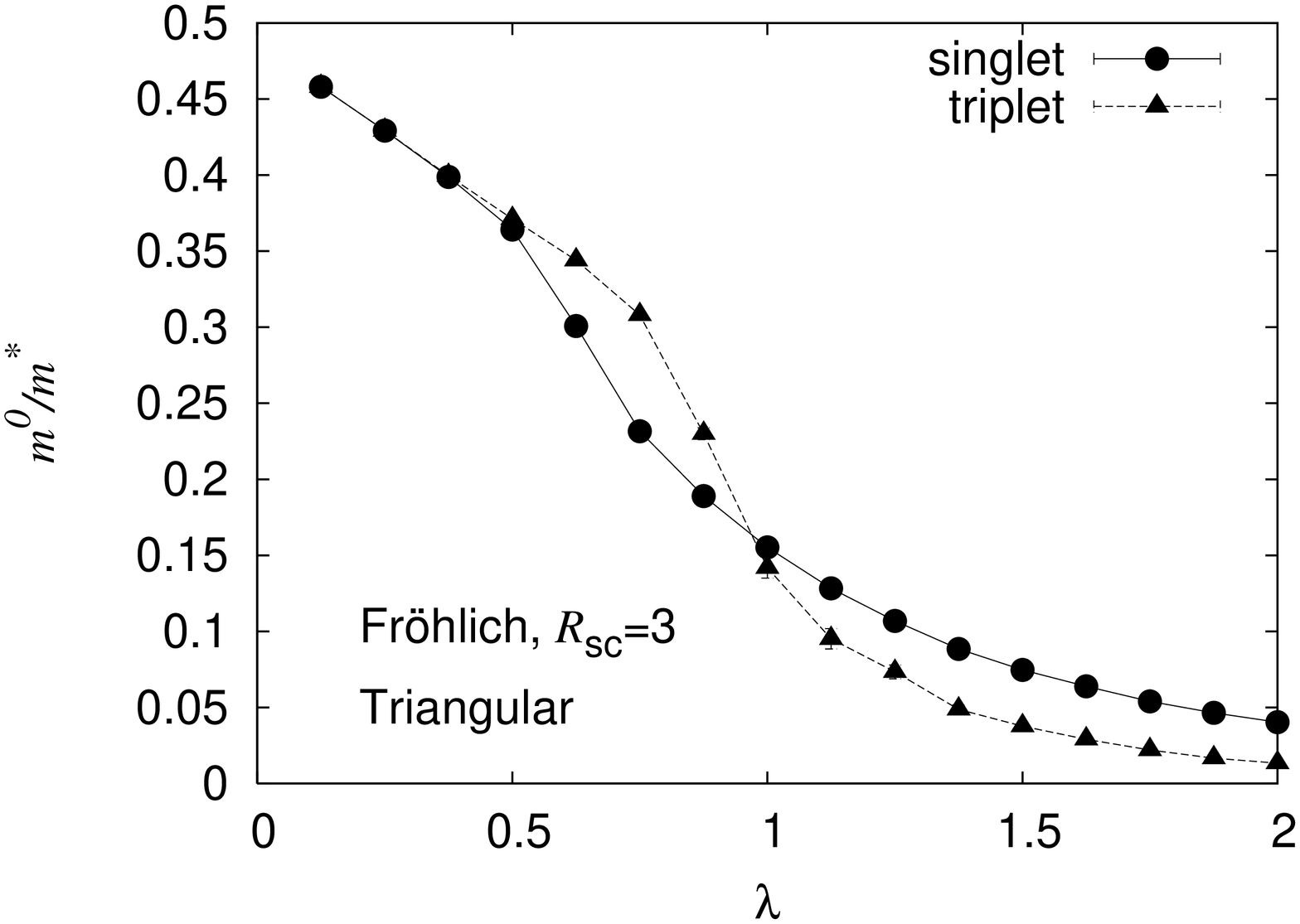}

\includegraphics[height=65mm,angle=270]{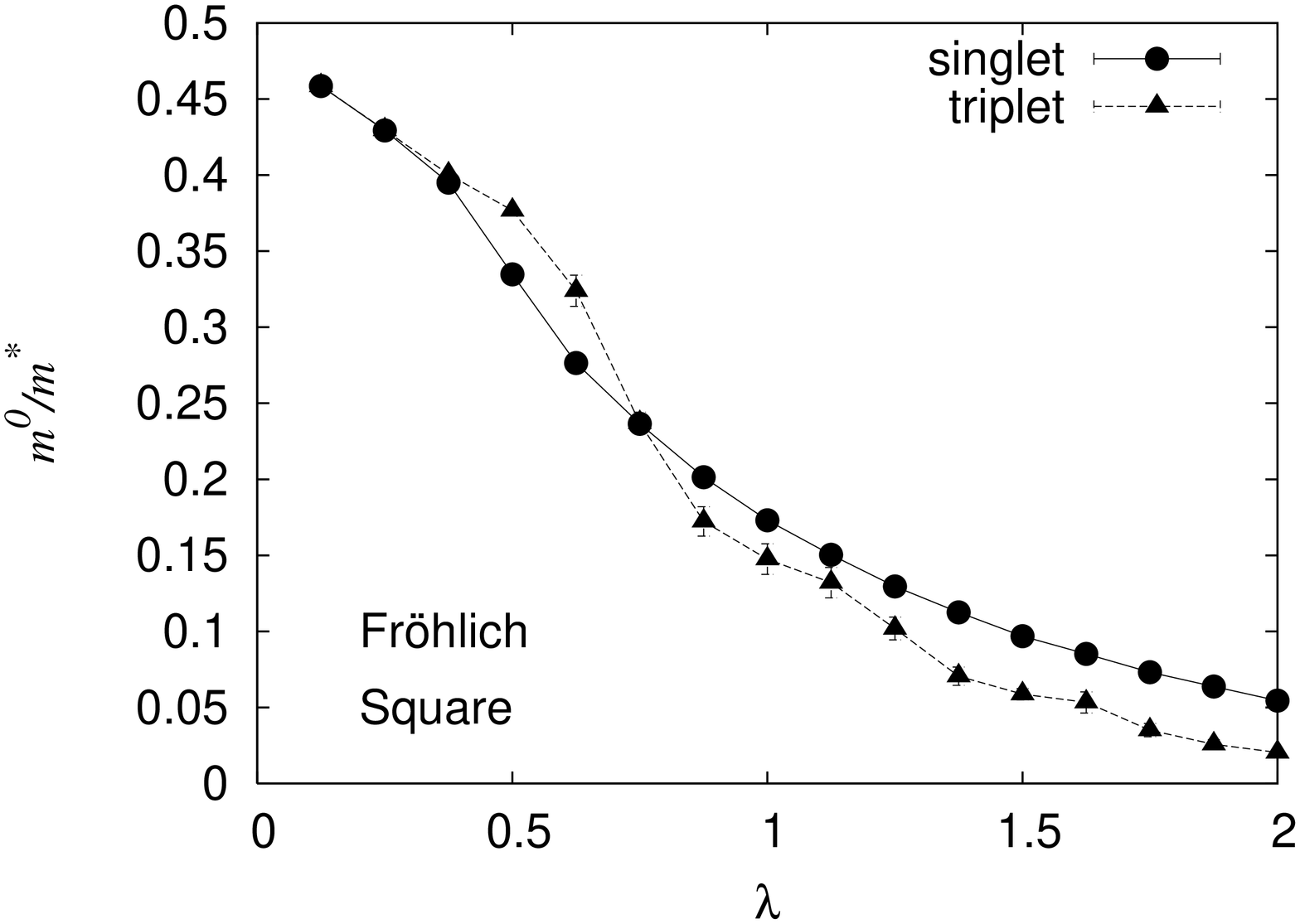}
\includegraphics[height=65mm,angle=270]{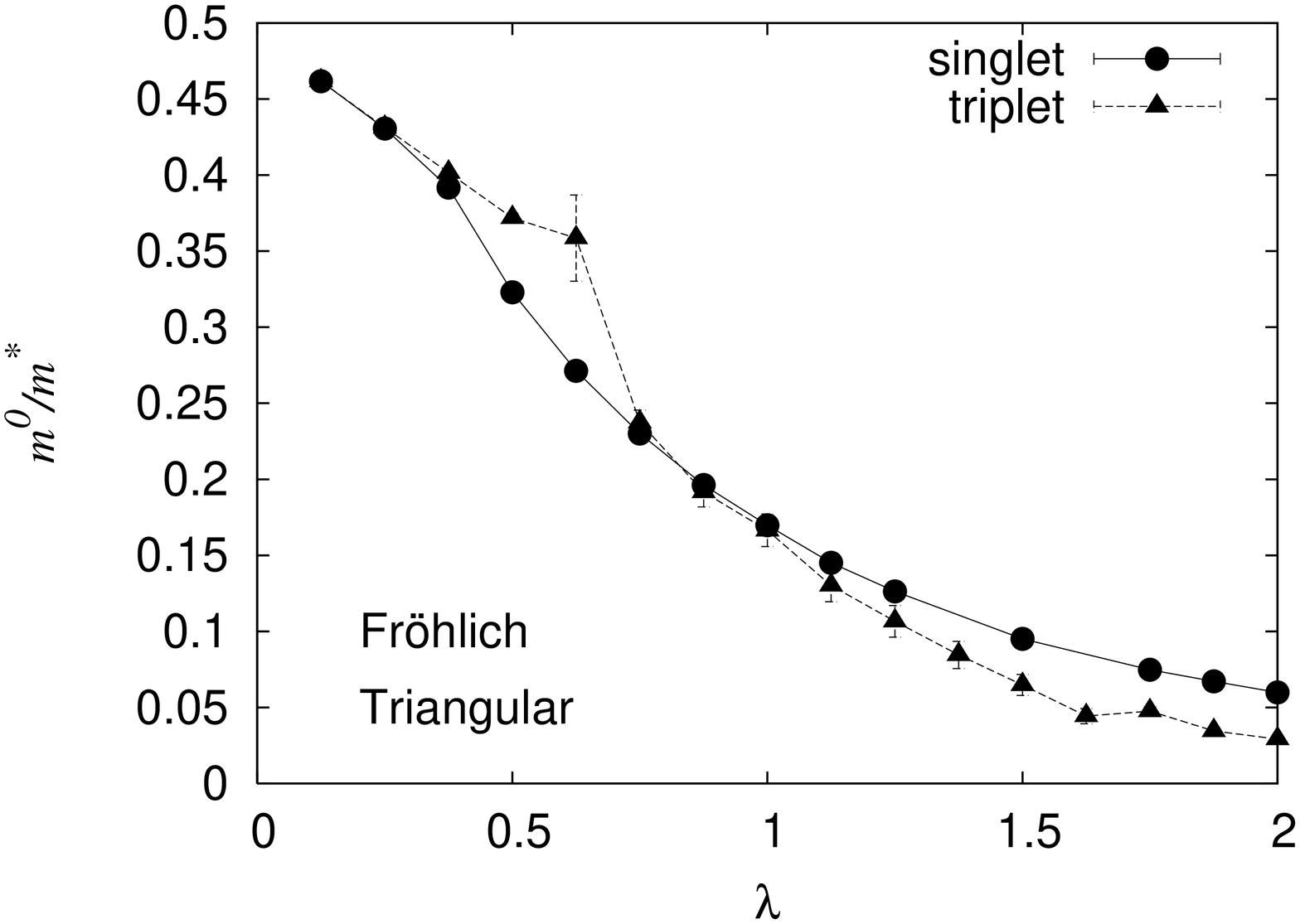}

\includegraphics[height=65mm,angle=270]{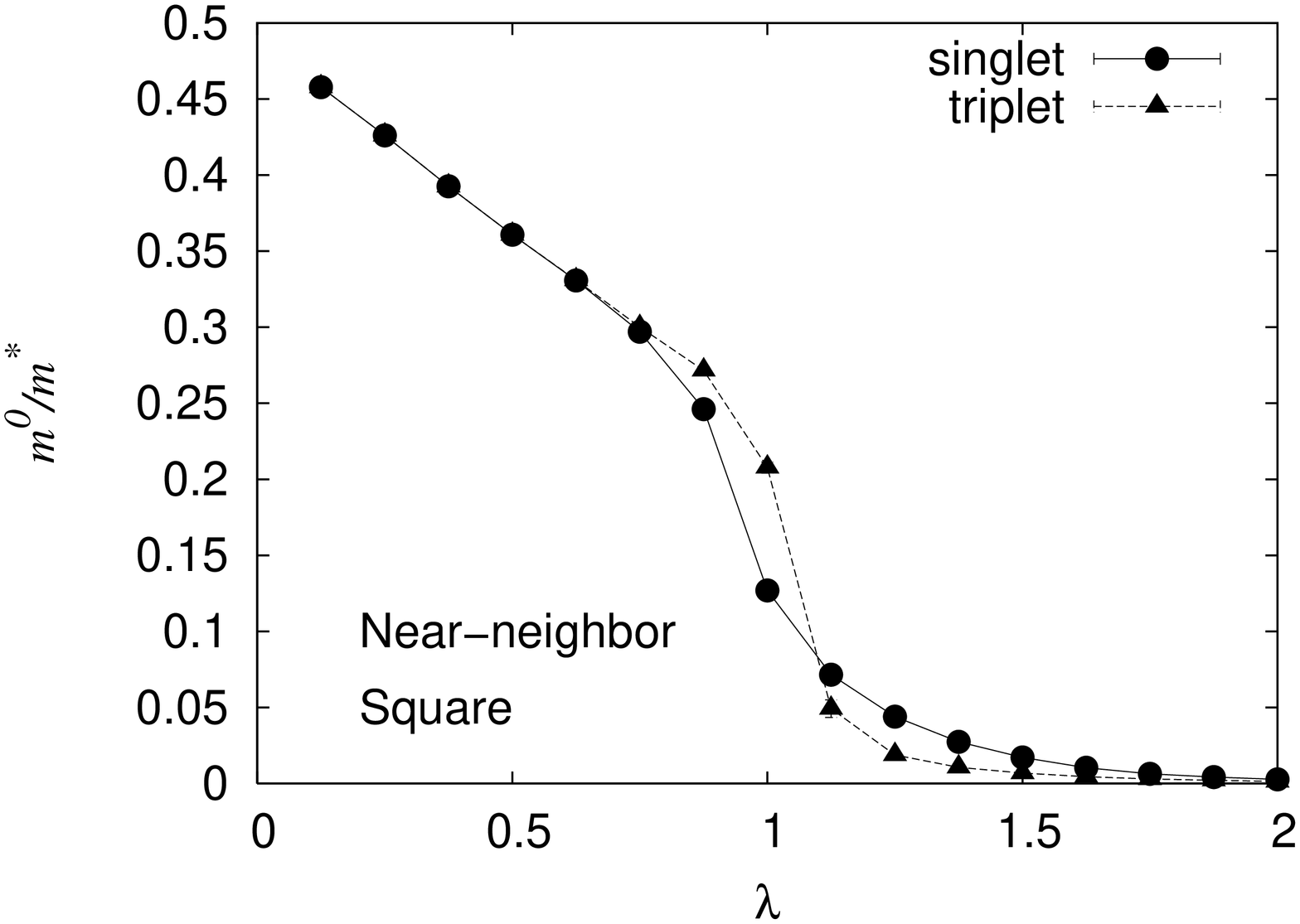}
\includegraphics[height=65mm,angle=270]{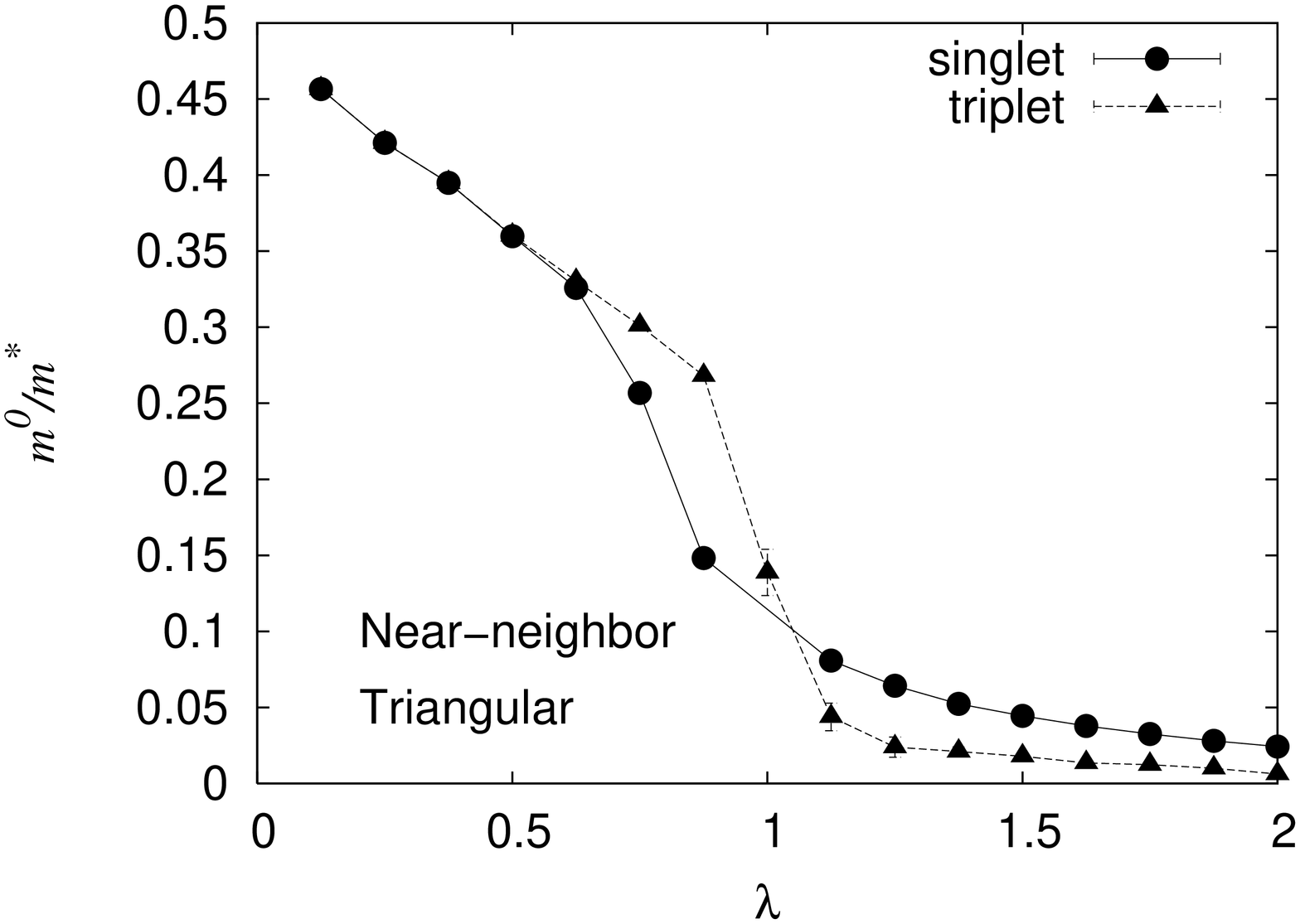}
\caption{Inverse mass of the triplet Fr\"ohlich bipolaron. Parameters are as Fig. \ref{fig:tetriplet}. Error bars represent one standard deviation. At low $\lambda$, the triplet bipolaron is lighter than the singlet bipolaron. Note that since properties of the triplet bipolaron do not depend on $U$, then the singlet will be always be heavier at low $U$.}
\label{fig:imtriplet}
\end{figure*}

The inverse mass of the triplet Fr\"ohlich bipolaron is shown in
Fig. \ref{fig:imtriplet}. At this value of $U/W$ and large electron-phonon coupling, the triplet bipolaron is heavier than the singlet
bipolaron, and at weak coupling the triplet is lighter. As $U$ is
changed, the triplet properties remain constant, whereas the singlet
ones change, notably with an increase in the singlet mass on degrease
in $U$. At very small $U$, we would expect that the singlet is
always heavier. While the strongly coupled singlet bipolaron is
clearly lighter on the triangular lattice, which is due to the
qualitative effects of lattice type, we do not see any qualitative
difference between the mass of the triplets on square and triangular
lattices which are very similar.

\begin{figure*}
\includegraphics[height=65mm,angle=270]{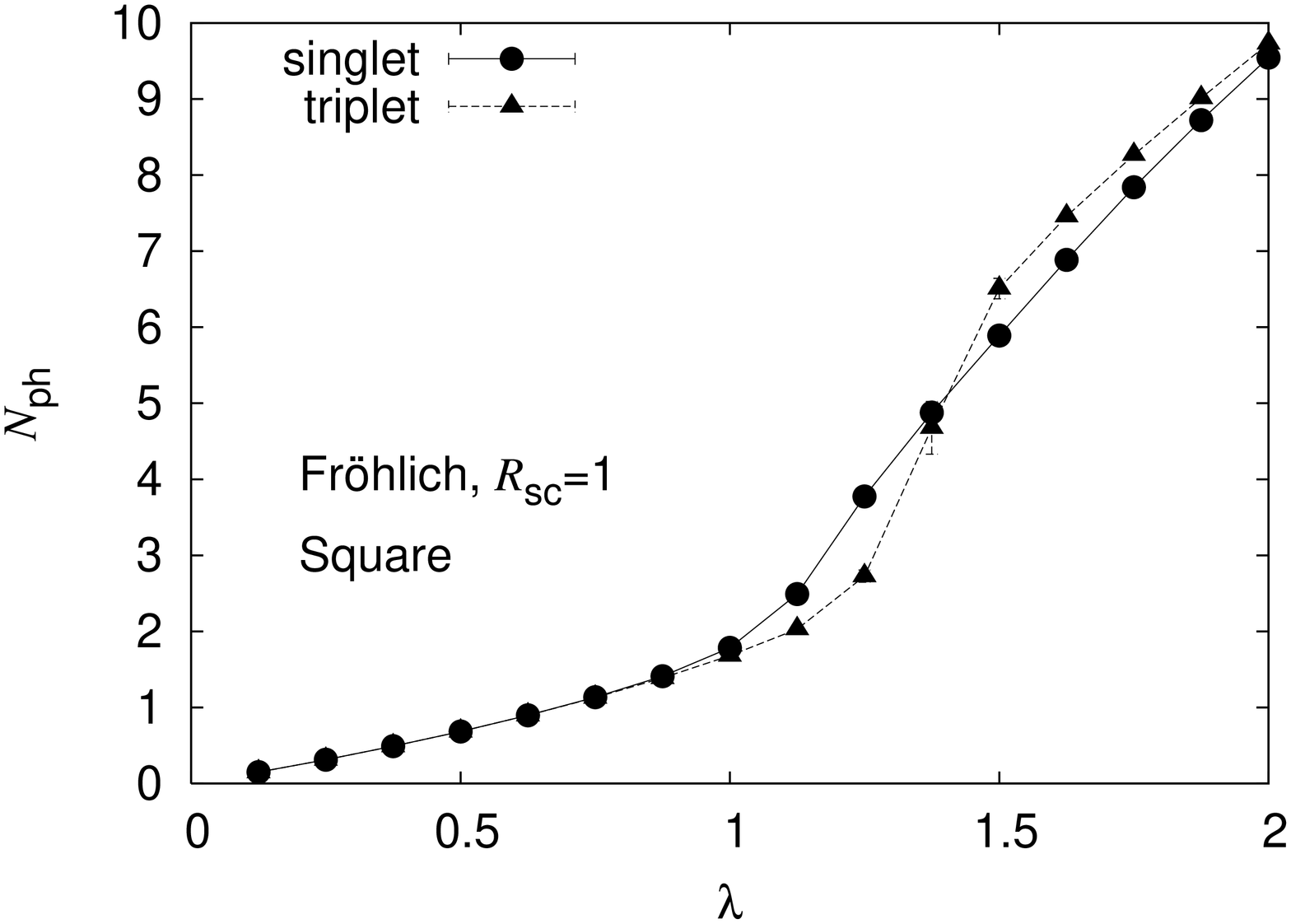}
\includegraphics[height=65mm,angle=270]{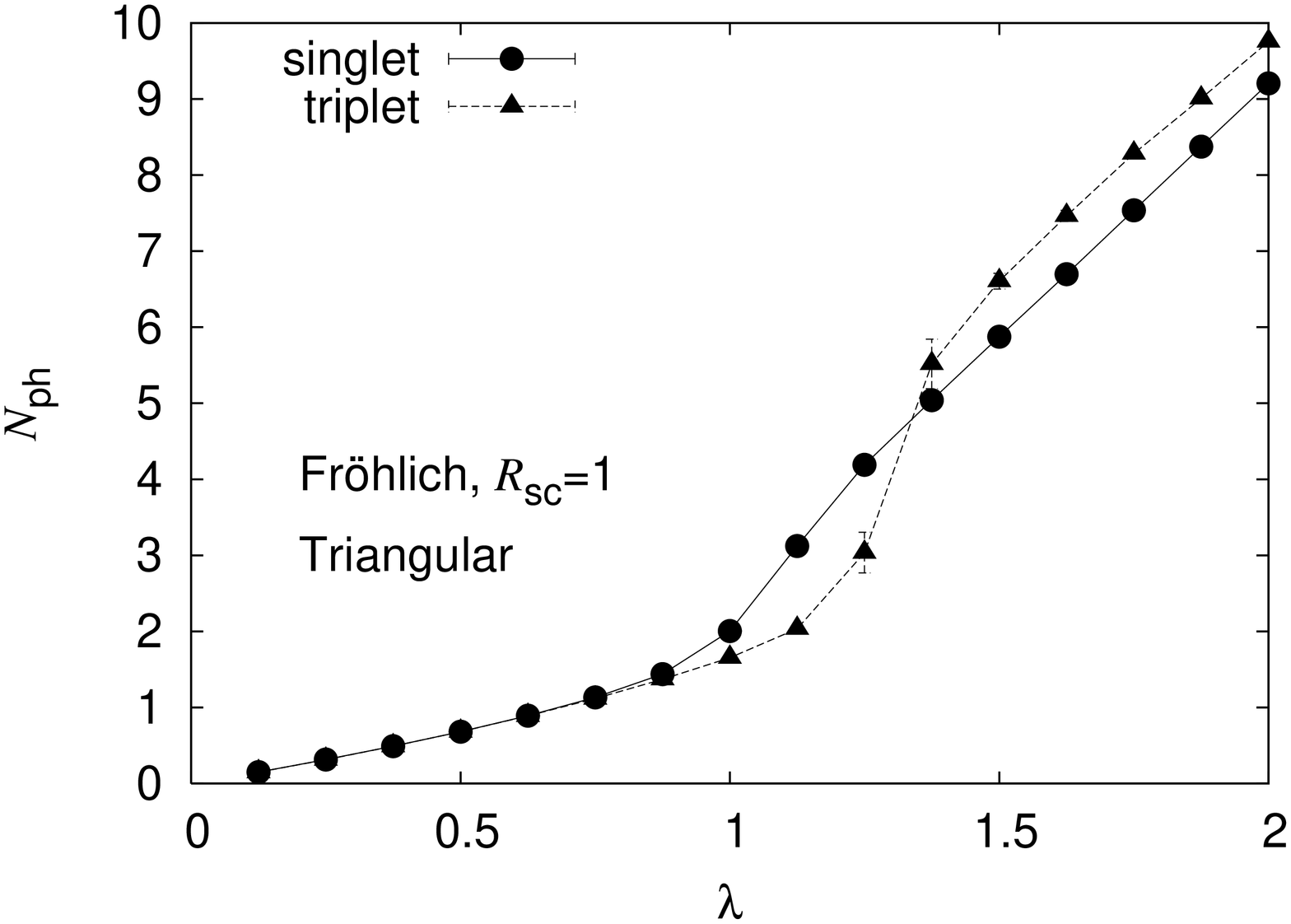}

\includegraphics[height=65mm,angle=270]{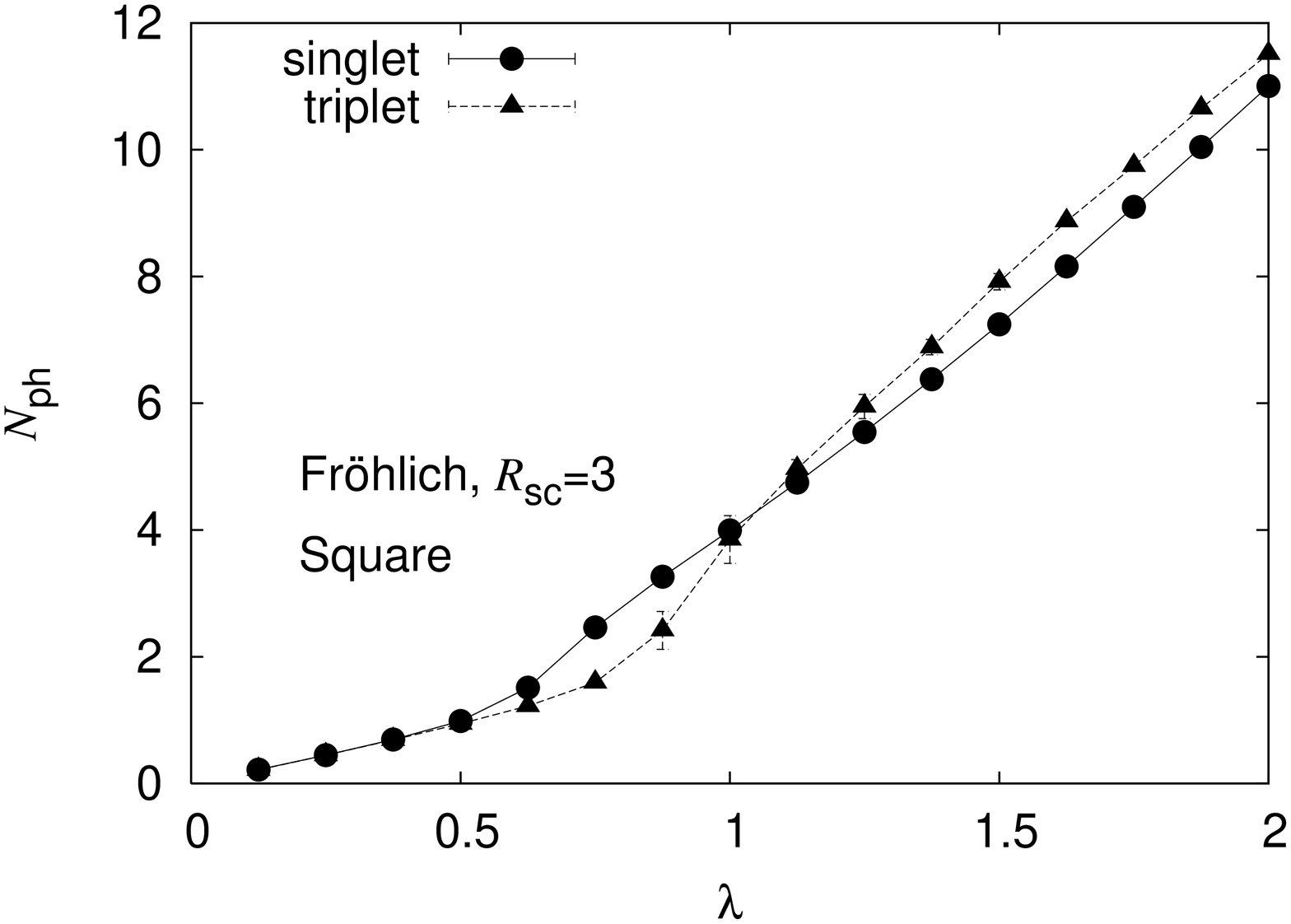}
\includegraphics[height=65mm,angle=270]{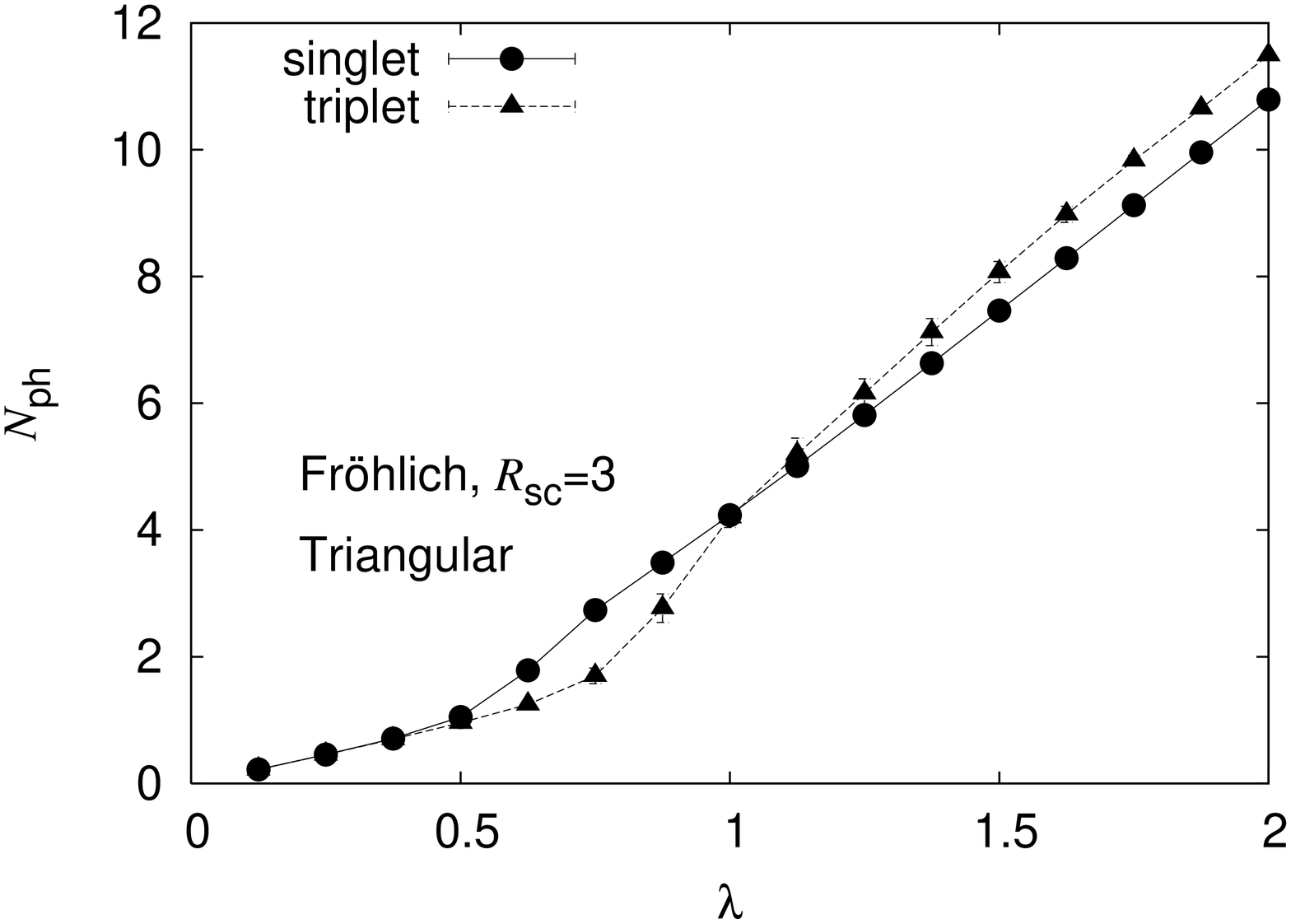}

\includegraphics[height=65mm,angle=270]{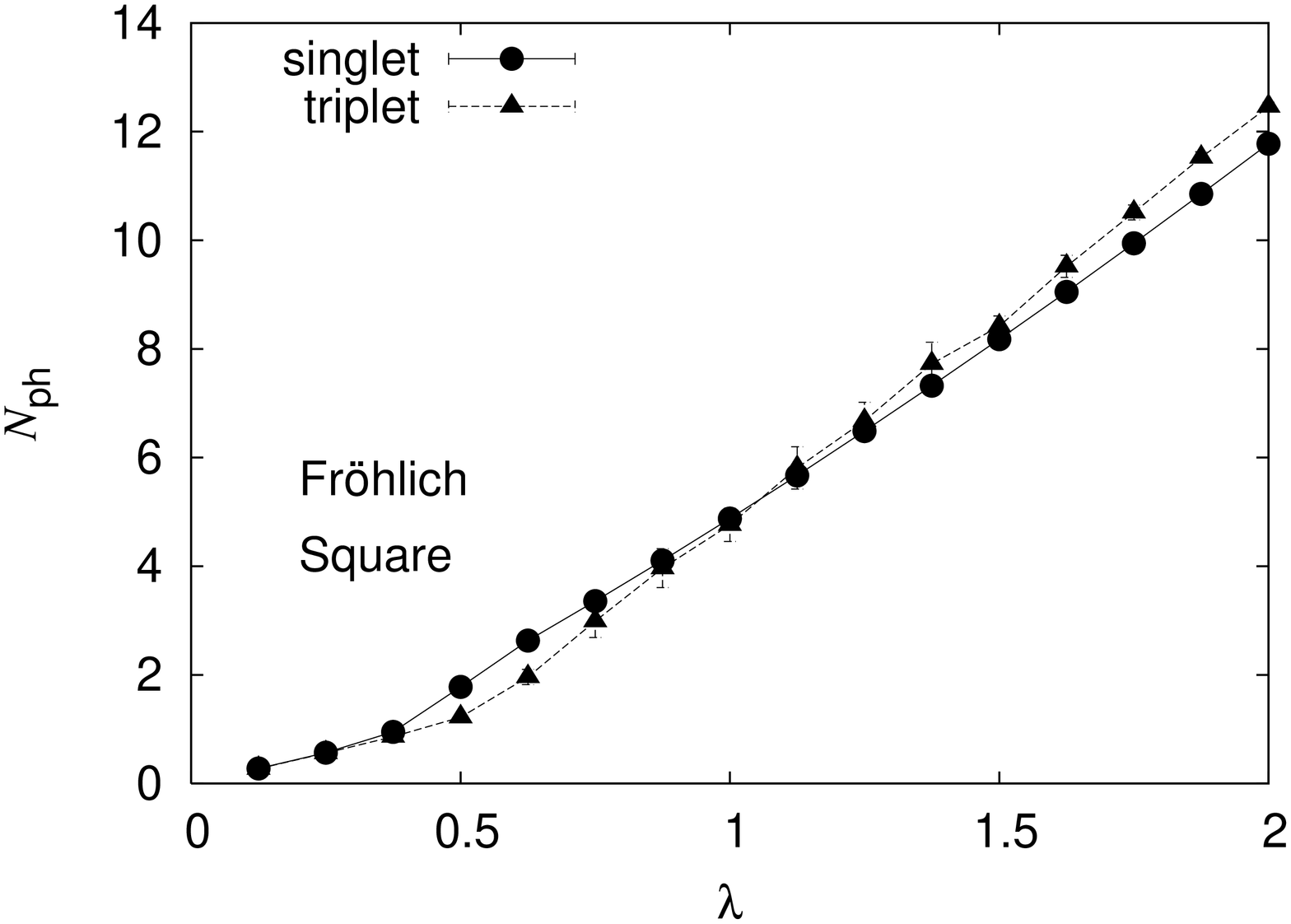}
\includegraphics[height=65mm,angle=270]{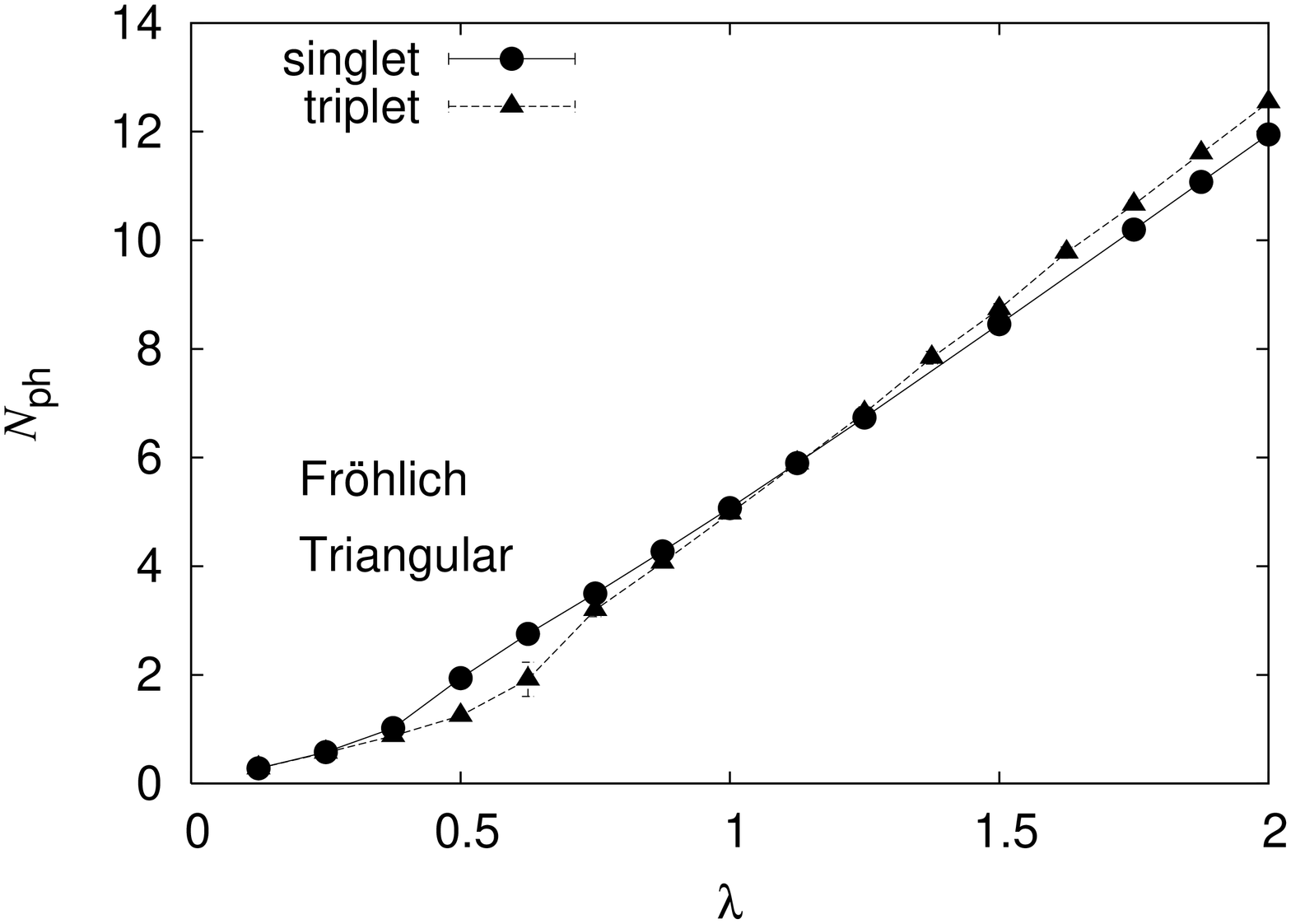}

\includegraphics[height=65mm,angle=270]{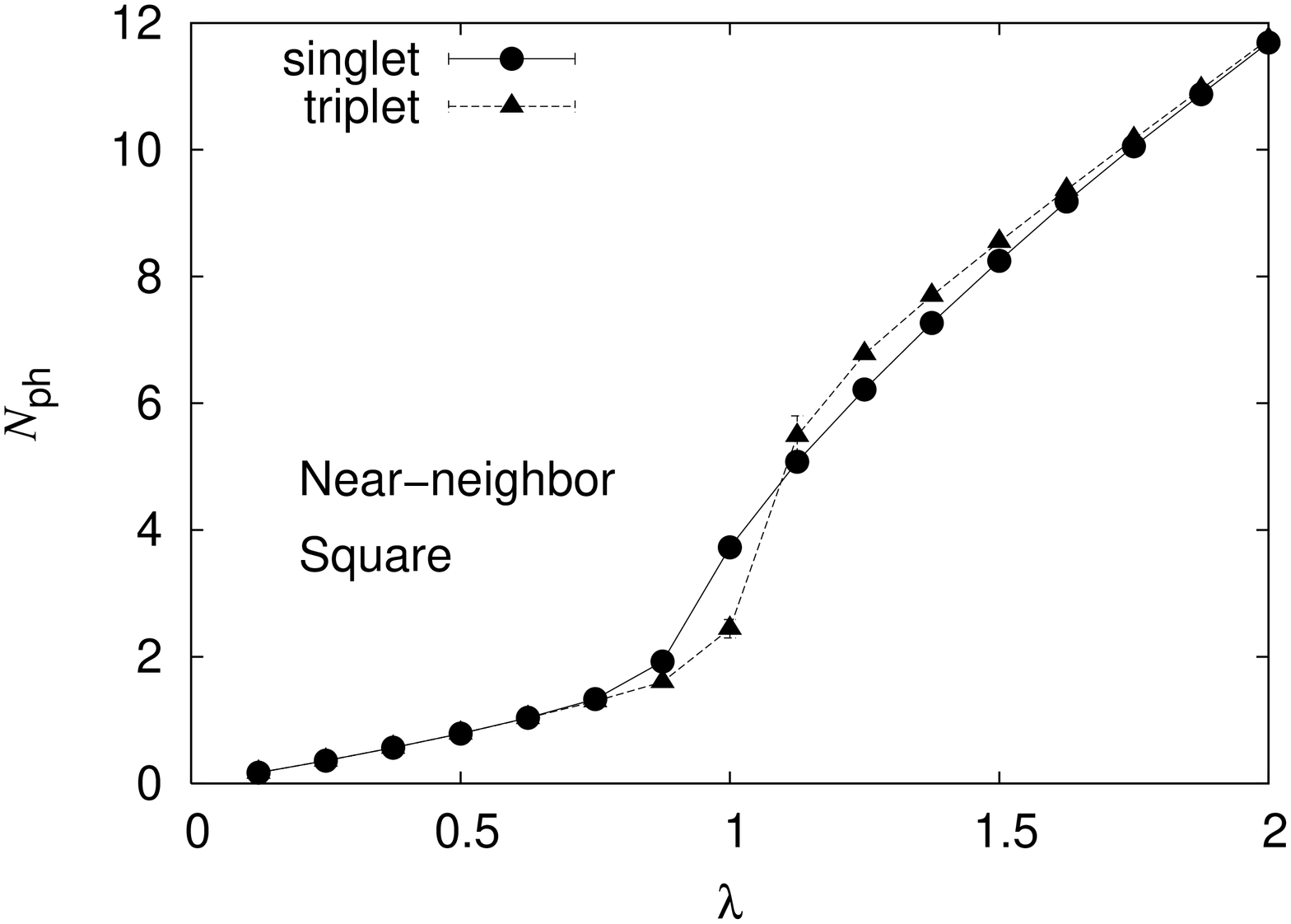}
\includegraphics[height=65mm,angle=270]{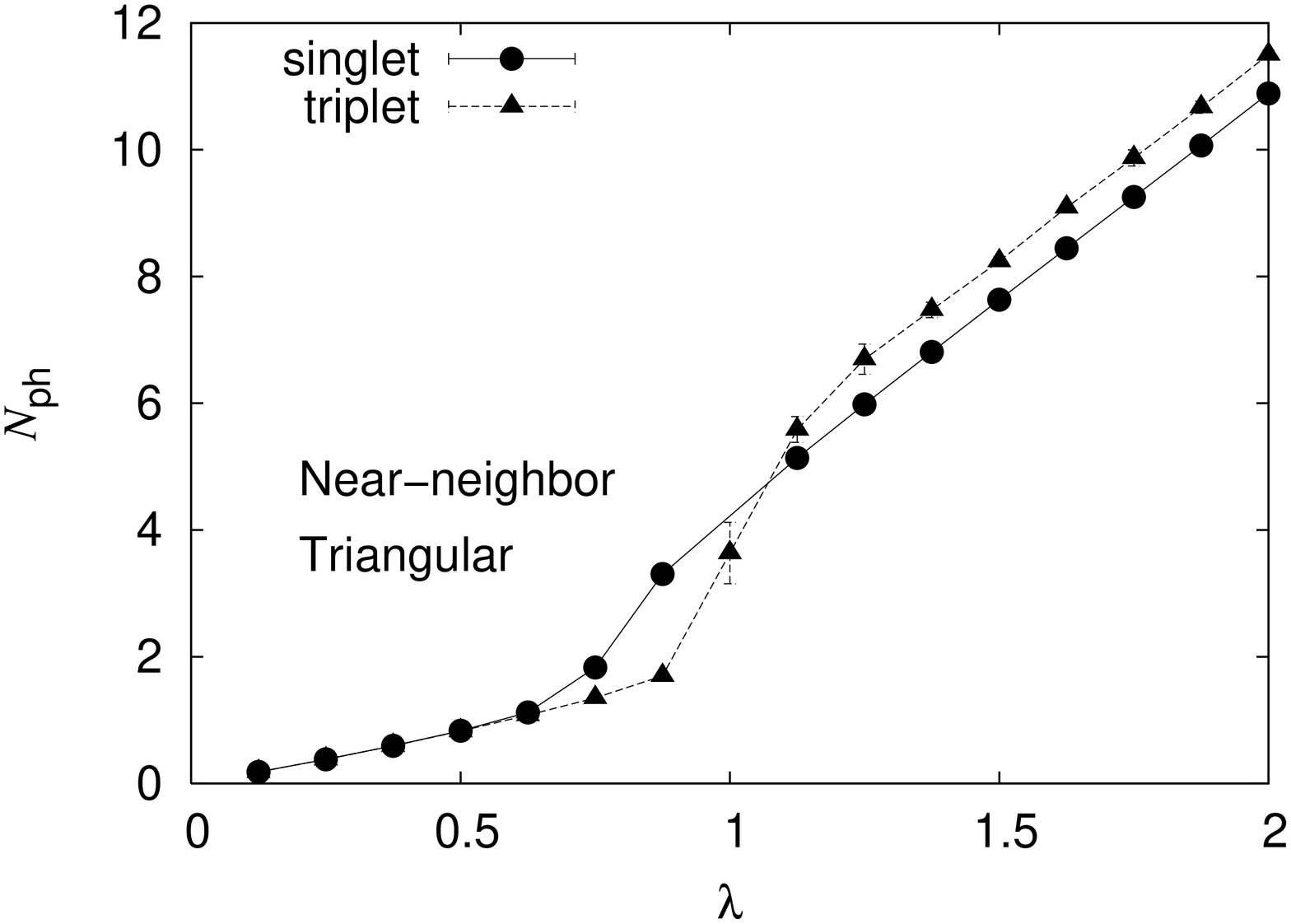}

\caption{Number of phonons associated with the triplet Fr\"ohlich bipolaron. Parameters are as Fig. \ref{fig:tetriplet}.}
\label{fig:nptriplet}
\end{figure*}

Figure \ref{fig:nptriplet} shows the number of phonons associated with
the triplet Fr\"ohlich bipolaron. At
strong coupling on the triangular lattice, there is a clear difference
between the number of phonons associated with singlet and triplet. In
contrast, on the square lattice the number of phonons associated with
the two types of bipolaron converges on a single value at large $\lambda$. We believe that this is related
to the second order hopping on square lattices at strong coupling and
first order hopping on triangular lattices. We also note that
increasing the electron-phonon interaction range reduces this
difference, presumably since the bipolaron has more configurational
freedom when $R_{sc}$ is increased, thus allowing similar hopping processes.

\begin{figure*}
\includegraphics[height=65mm,angle=270]{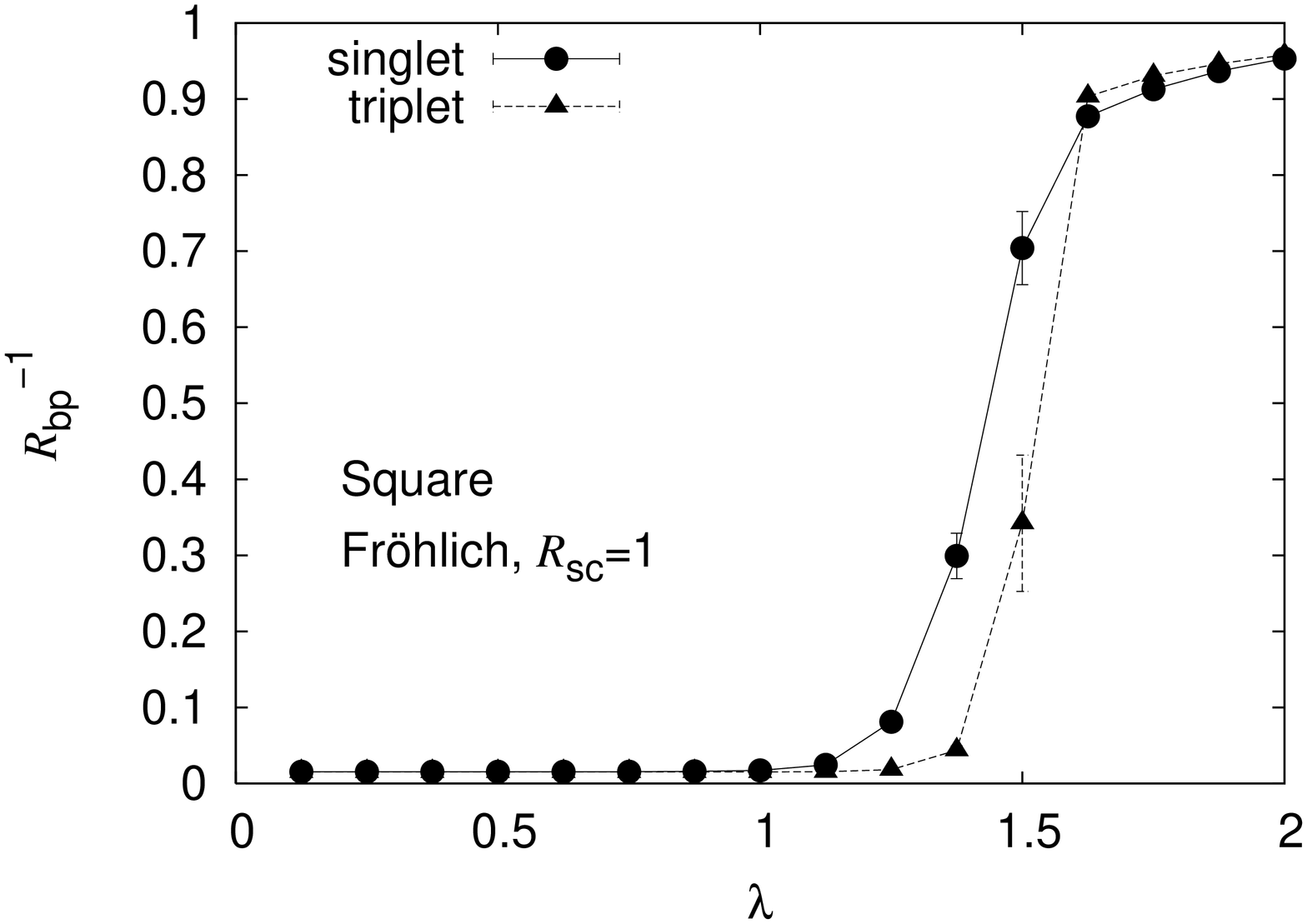}
\includegraphics[height=65mm,angle=270]{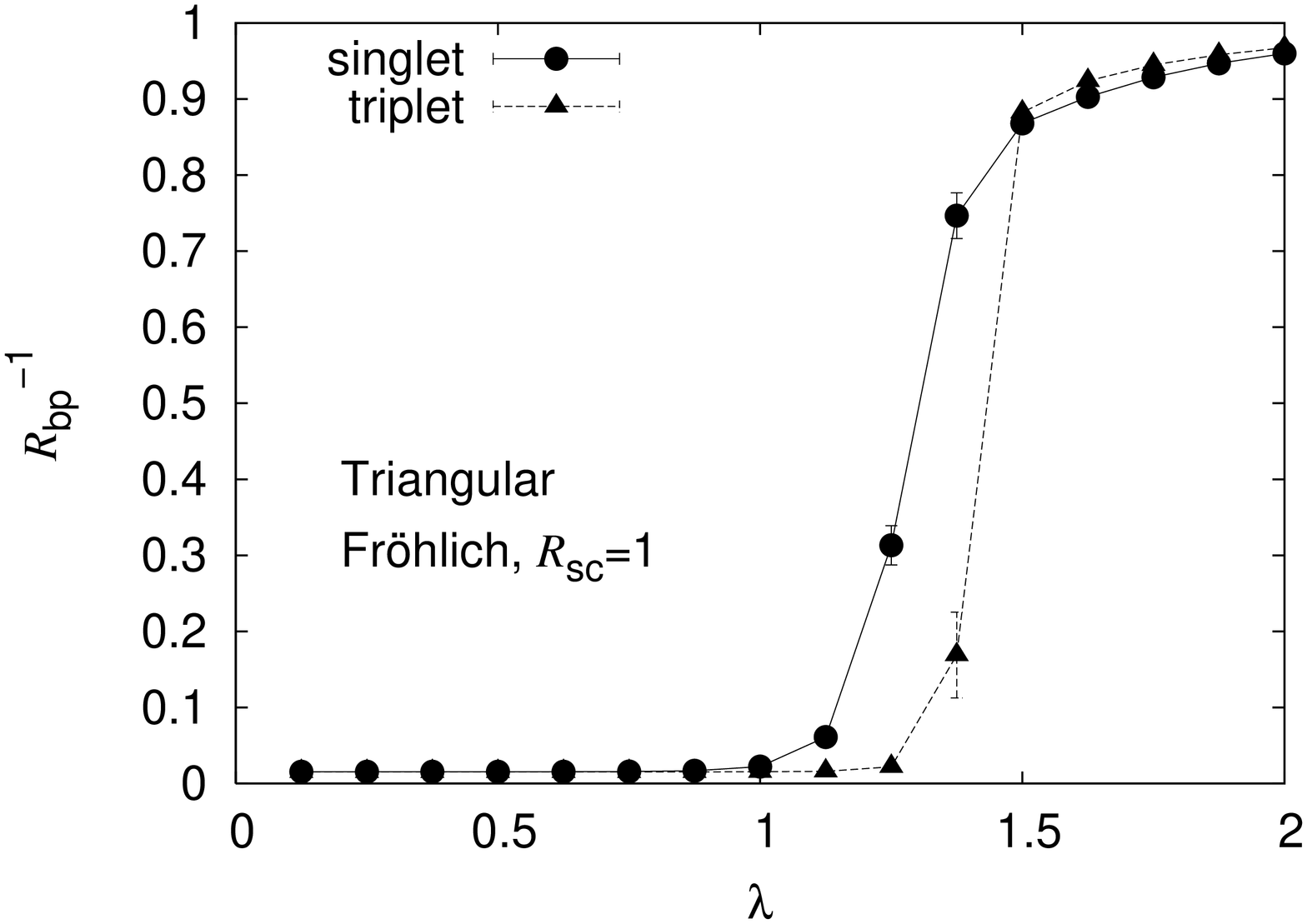}

\includegraphics[height=65mm,angle=270]{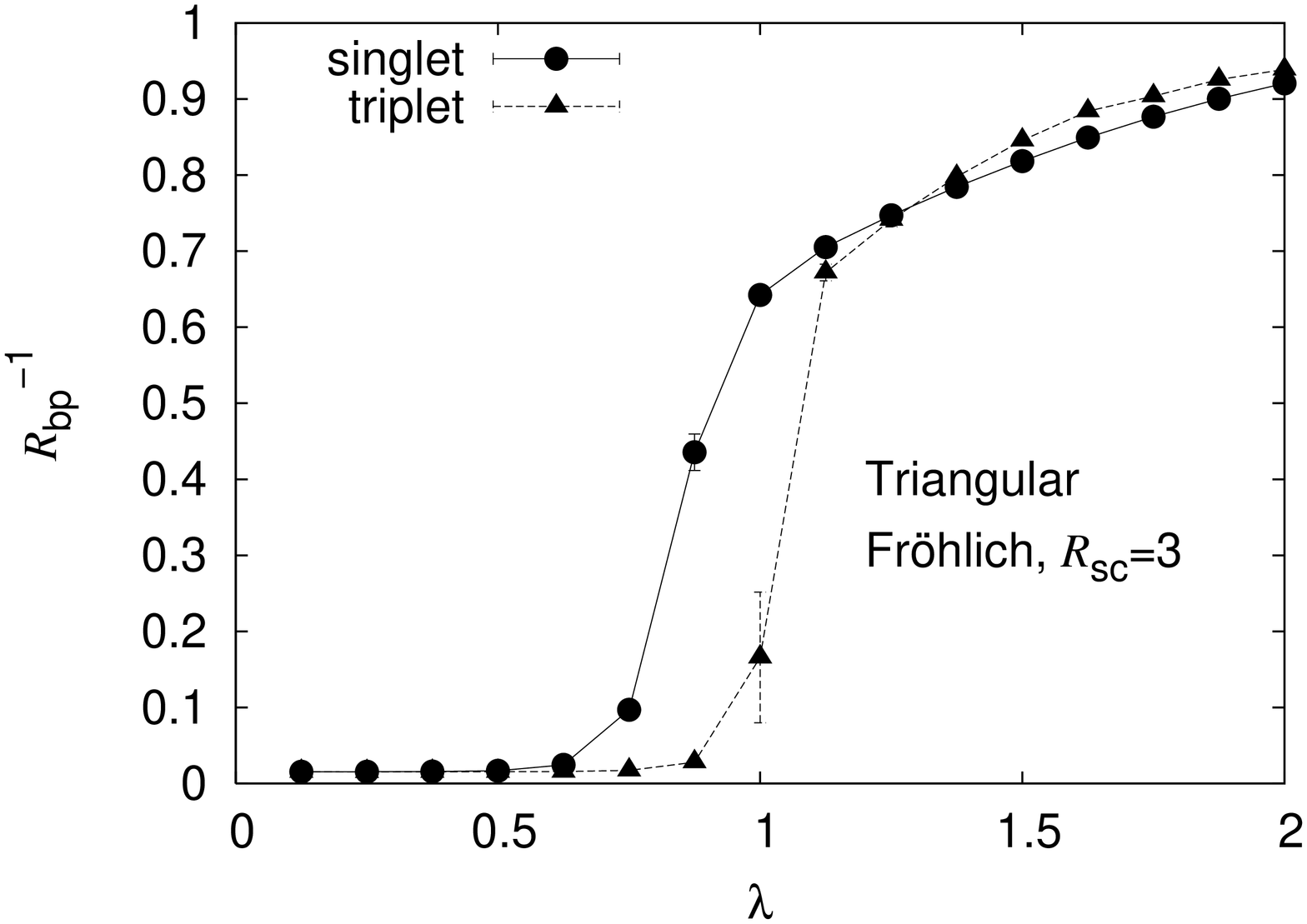}
\includegraphics[height=65mm,angle=270]{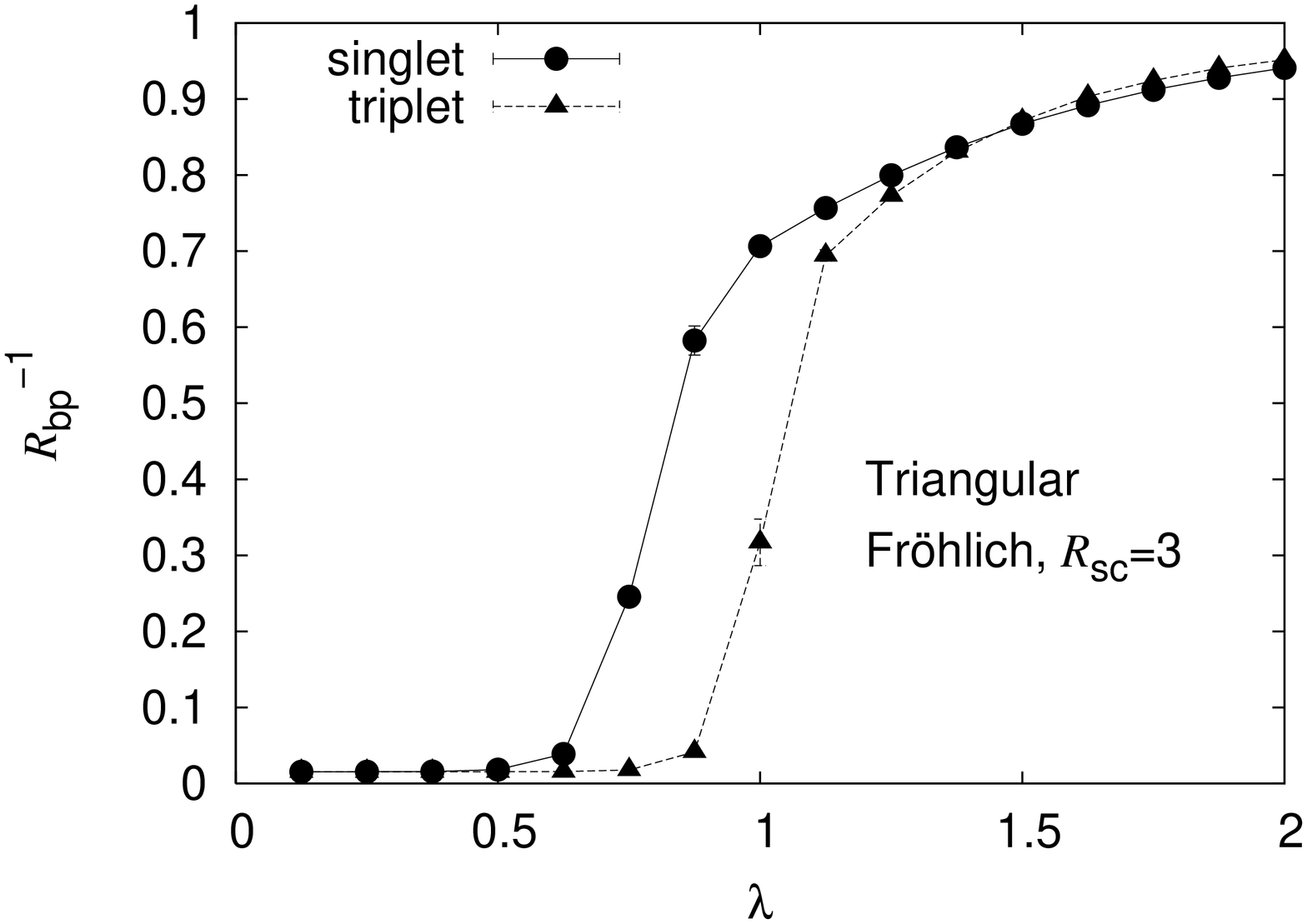}

\includegraphics[height=65mm,angle=270]{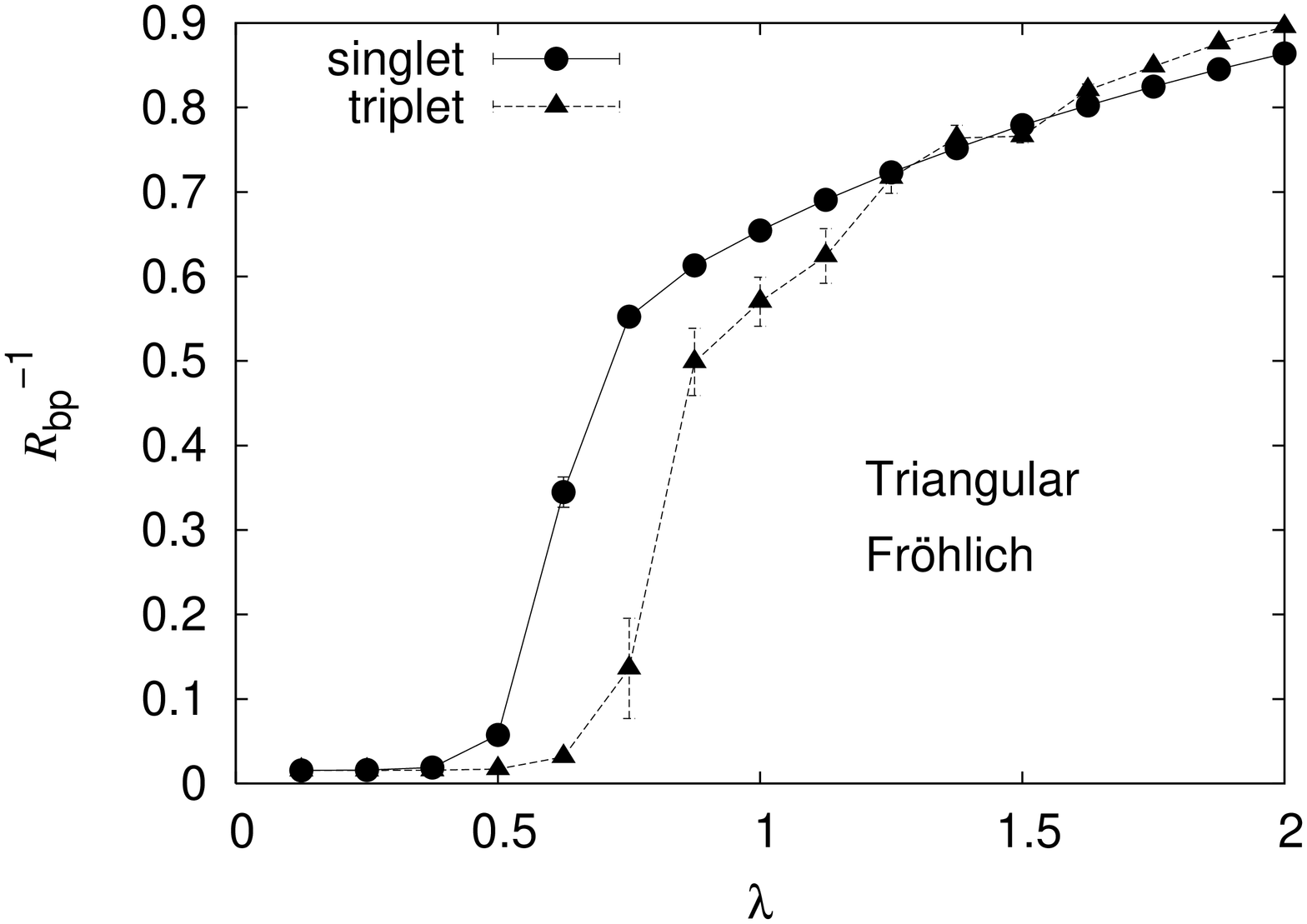}
\includegraphics[height=65mm,angle=270]{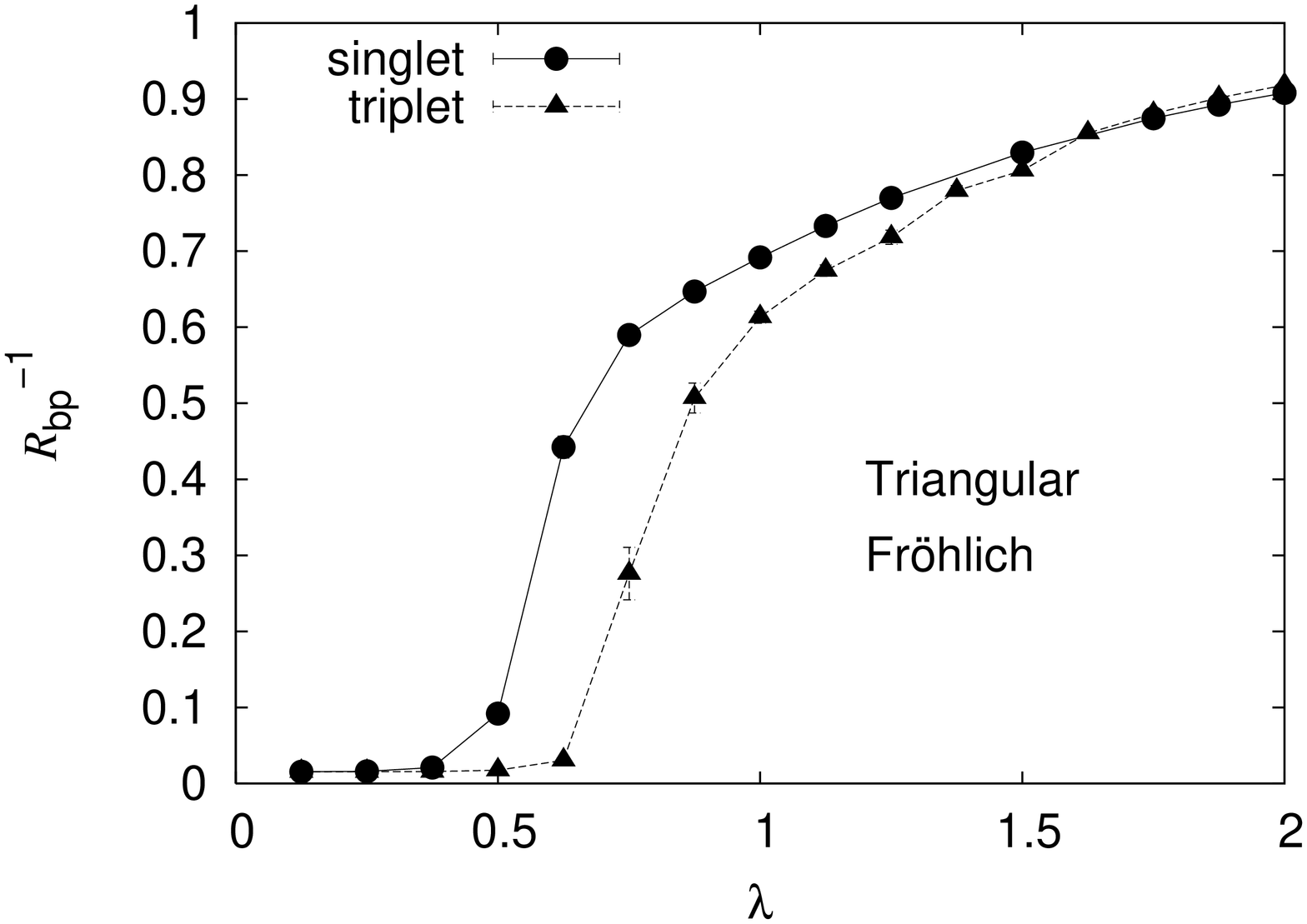}

\includegraphics[height=65mm,angle=270]{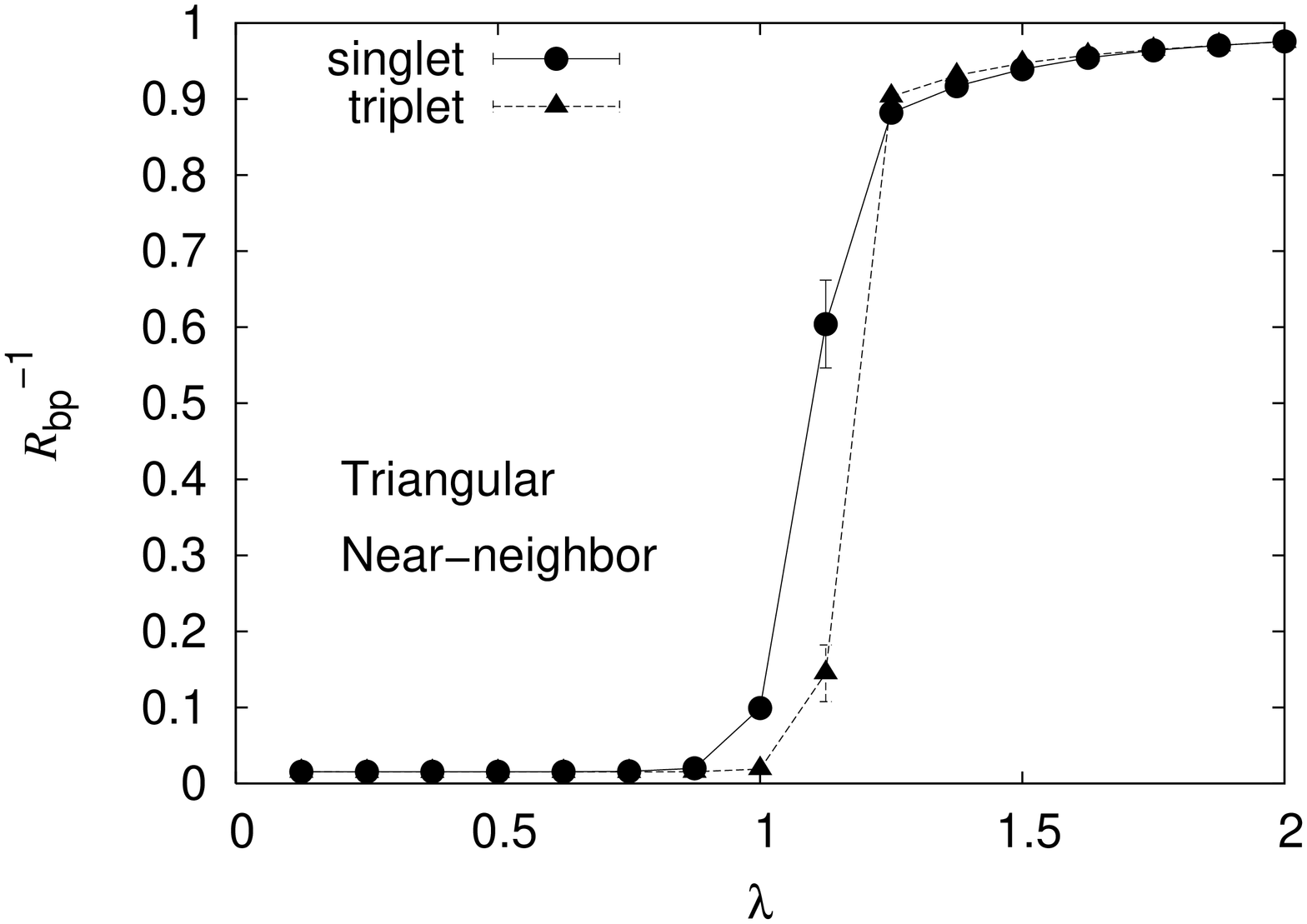}
\includegraphics[height=65mm,angle=270]{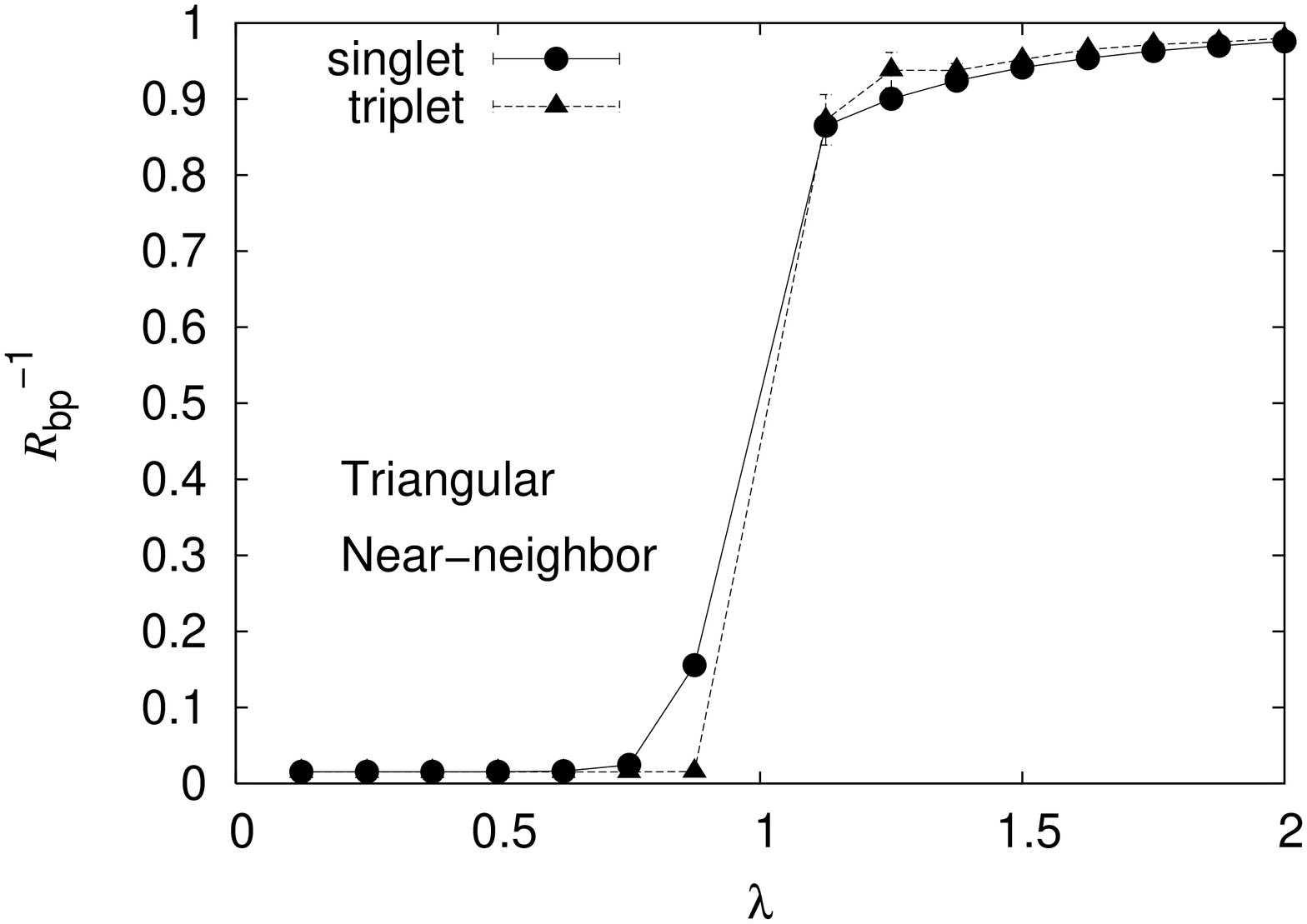}
\caption{Radius of the Fr\"ohlich triplet bipolaron. Parameters are as Fig. \ref{fig:tetriplet}.}
\label{fig:bpstriplet}
\end{figure*}

Finally, we compute the radius of the triplet Fr\"ohlich
bipolaron, which is shown in Fig. \ref{fig:bpstriplet}. There are only
small quantitative changes between the radii of triplet bipolarons on different
lattices. We note a small finite size effect at weak coupling, which
corresponds to a finite inverse radius where the bipolaron is not
bound. On increasing $\lambda$, the singlet bipolaron is the first to bind strongly,
with the triplet radius decreasing slightly, but with the triplet becoming small only for larger $\lambda$. Increased
interaction range decreases the electron-phonon coupling required to
bind both types of bipolaron. The strongly bound triplet is slightly
larger than the singlet.

%\begin{figure*}
%\includegraphics[height=75mm,angle=270]{Graphs/TriSquTriplet_100000/iex_frohlich_rsc1_trip.ps}
%%\includegraphics[height=75mm,angle=270]{Graphs/TriSquTriplet_100000/iex_frohlich_rsc2_trip.ps}
%\includegraphics[height=75mm,angle=270]{Graphs/TriSquTriplet_100000/iex_frohlich_rsc3_trip.ps}
%\includegraphics[height=75mm,angle=270]{Graphs/TriSquTriplet_100000/iex_frohlich_trip.ps}
%\includegraphics[height=75mm,angle=270]{Graphs/TriSquTriplet_100000/iex_bonca_trip.ps}
%\caption{Triplet isotope exponent of the Fr\"ohlich bipolaron on the
%square lattice. Parameters are as Fig. \ref{fig:tetripletsq}.}
%\label{fig:iextripletsq}
%\end{figure*}

%\begin{figure*}
%\includegraphics[height=75mm,angle=270]{Graphs/TriSquTriplet_100000/iex_frohlich_rsc1_trip_tri.ps}
%%\includegraphics[height=75mm,angle=270]{Graphs/TriSquTriplet_100000/iex_frohlich_rsc2_trip_tri.ps}
%\includegraphics[height=75mm,angle=270]{Graphs/TriSquTriplet_100000/iex_frohlich_rsc3_trip_tri.ps}
%\includegraphics[height=75mm,angle=270]{Graphs/TriSquTriplet_100000/iex_frohlich_trip_tri.ps}
%\includegraphics[height=75mm,angle=270]{Graphs/TriSquTriplet_100000/iex_bonca_trip_tri.ps}
%\caption{Triplet isotope exponent, triangular lattice}
%\end{figure*}

\section{Dispersion}
\label{sec:dispersion}

\subsection{Hubbard-Holstein model}

To complete our survey of bipolarons in 2D, we consider dispersions of
the Hubbard-Holstein and Hubbard-Fr\"ohlich bipolarons. The
electron-phonon interaction has the potential to lead to unusual
effects. For example polaron dispersions in the adiabatic regime of
the Holstein model on square and triangular lattices are flattened at
the edge of the Brillouin zone \cite{kornilovitch1998a,hague2006a}. It
is not known if the dispersion remains flattened once bipolarons form,
and also little is known about the dispersions of triplet
bipolarons. Thus, it is of interest to determine how the flat
dispersion evolves as a bipolaron is bound from two polarons.

\begin{figure}
\includegraphics[height=75mm,angle=270]{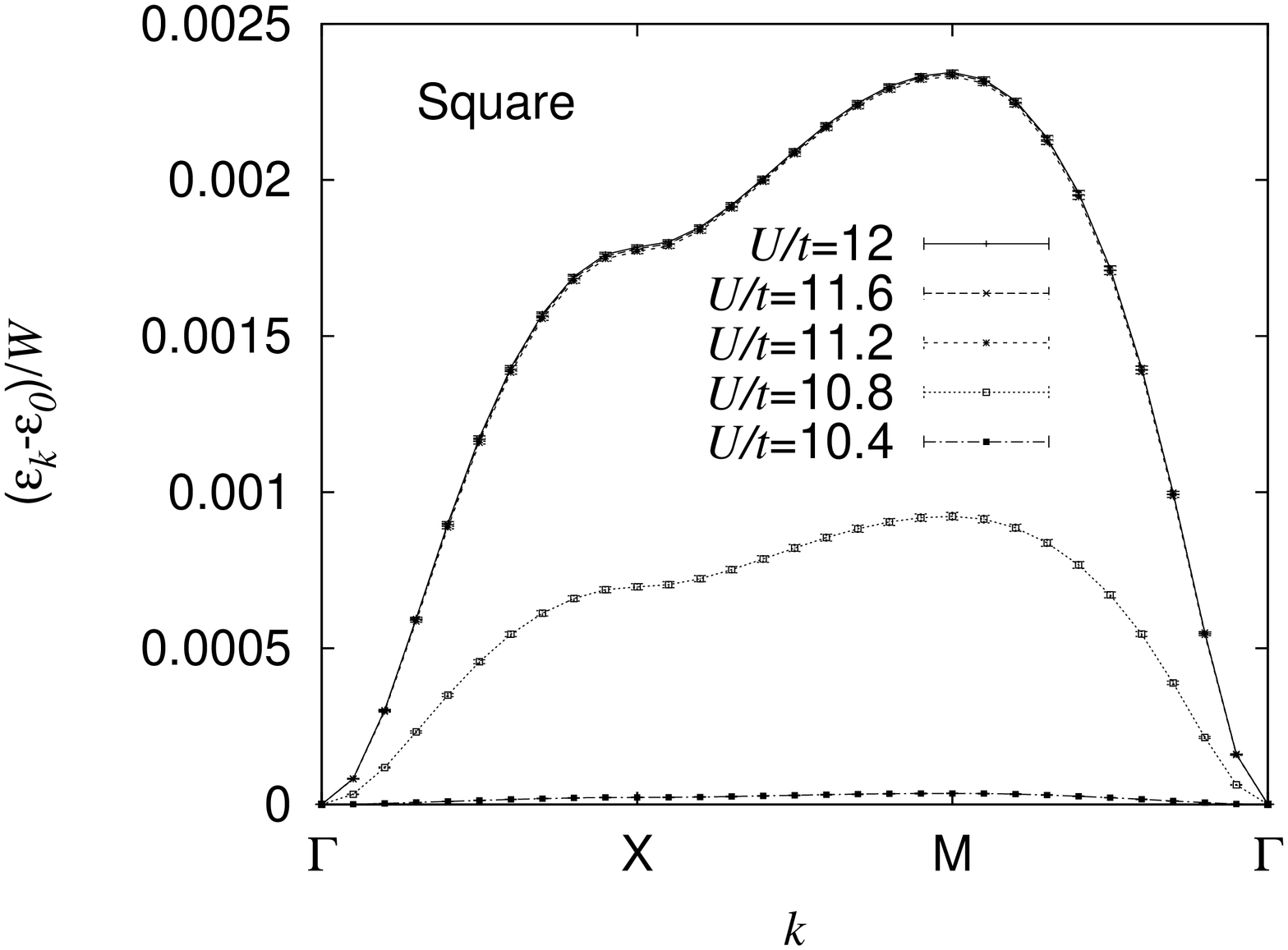}
\includegraphics[height=75mm,angle=270]{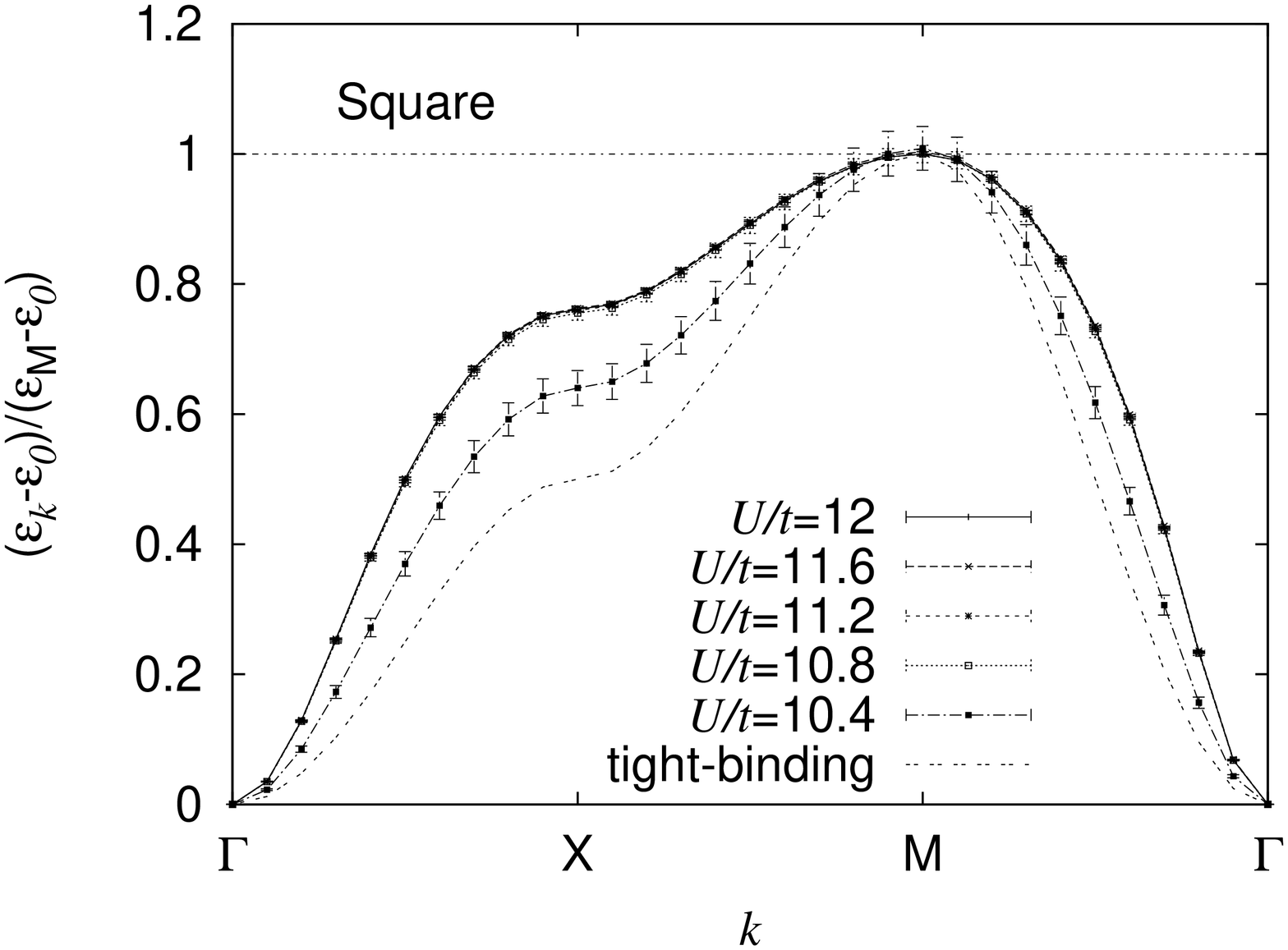}
\caption{Singlet dispersions of the Hubbard-Holstein
bipolaron on the square lattice when $\lambda=1.45$, $\bar{\beta}=14$ and
$\bar{\omega}=1$. $U$ is varied. Note that error bars on the plots of dispersions represent one standard deviation. The lower panel shows the dispersions normalized by $\epsilon_{{\rm M}}-\epsilon_{0}$ compared to the normalized tight binding dispersion. As $U$ decreases, the bipolaron becomes strongly bound on-site and the dispersion develops a form closer to the tight-binding approximation.}
\label{fig:hhdispsq}
\end{figure}

In figure \ref{fig:hhdispsq} we show singlet dispersions of the
Hubbard-Holstein bipolaron on the square lattice when $\lambda=1.45$
and $\bar{\omega}=1$. Computation of the dispersion is carried out
using the same method as in Ref. \onlinecite{hague2009a}. At large
$U$, the bipolaron is unbound and the dispersion represents two
polarons. The dispersions relating to $U/t=12$, $U/t=11.6$ and
$U/t=11.2$ are indistinguishable to the eye (top panel). As the repulsion is
decreased, there is a decrease in bandwidth of several orders of
magnitude at the $U$ corresponding to the binding of the bipolaron. In
the lower panel, we also show dispersions normalized by
$\epsilon_{{\rm M}}-\epsilon_{0}$ ($\epsilon_{{\rm M}}$ is the energy of the bipolaron at the M point) so that the forms of the dispersion
function can be compared to the non-interacting tight-binding
dispersion. At large $U$, the form of the dispersion of the two polarons (and the
weakly bound bipolaron) is quite distorted away from the form of the tight-binding
dispersion. As $U$ decreases further, the electrons become tightly bound
into an on-site bipolaron. Then the form of the dispersion tends
towards that of the tight-binding spectrum. We believe this to be the effect
of the strongly bound bipolaron acting as a single particle, with a
new hopping integral that relates to the overlap of the wavefunction
of the pair between sites.

\begin{figure}
\includegraphics[height=75mm,angle=270]{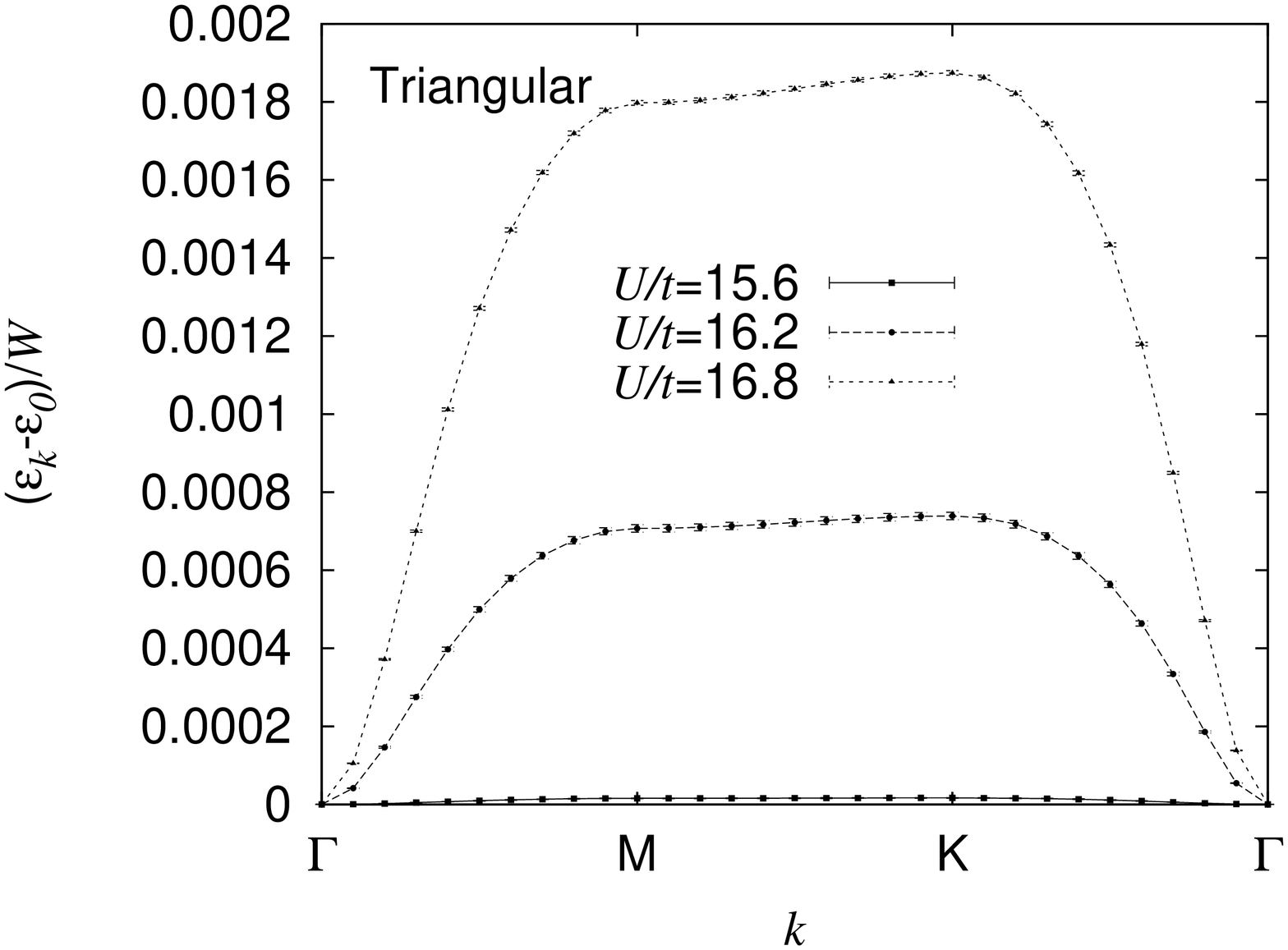}
\includegraphics[height=75mm,angle=270]{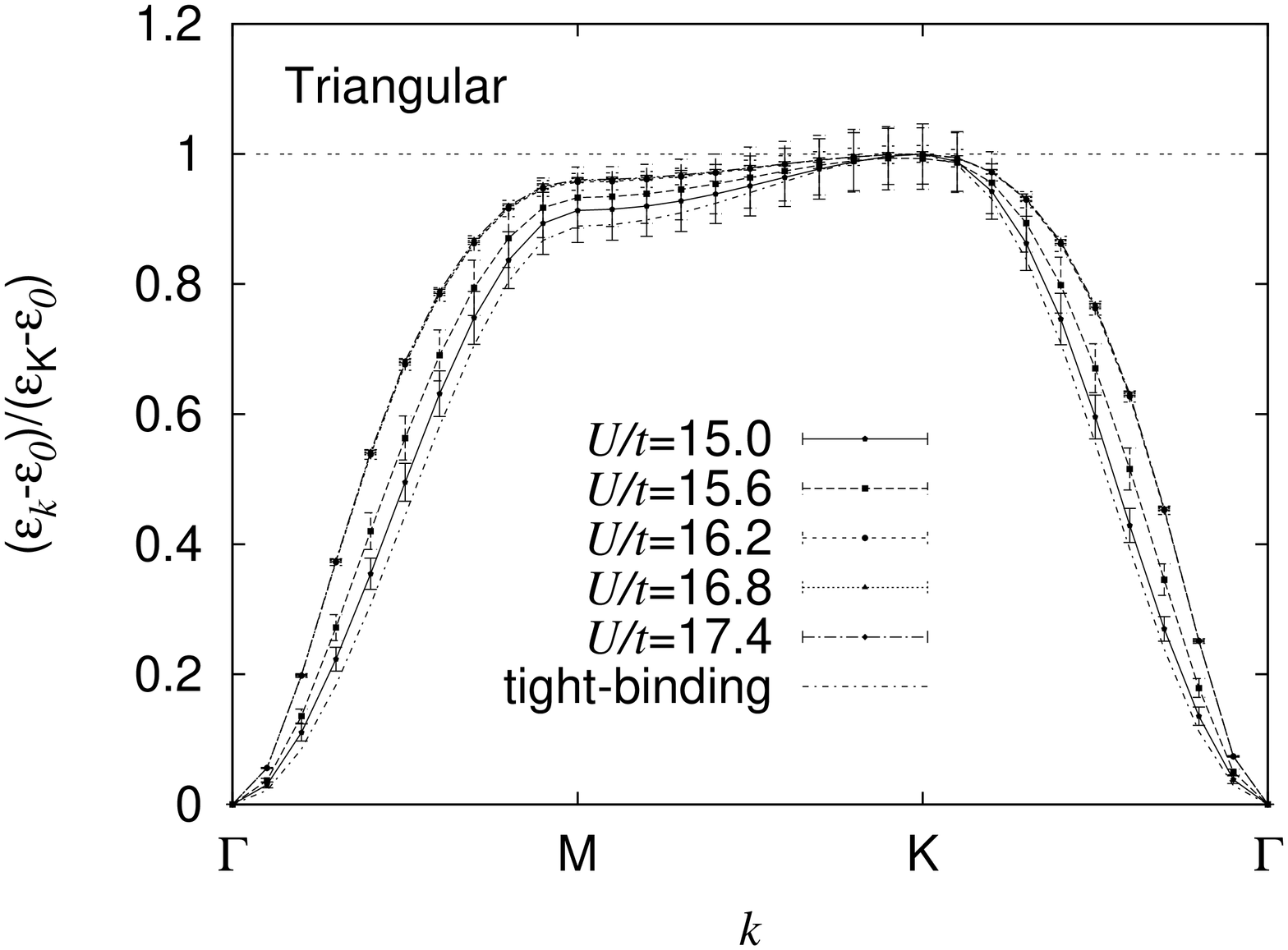}
\caption{Singlet dispersions of the Hubbard-Holstein bipolaron on the
triangular lattice when $\lambda=1.45$, $\bar{\beta}=28/3$ and
$\bar{\omega}=1.5$. $U$ is varied. Error bars represent one standard error. There is a very rapid change in the
bandwidth between $U/t=15.6$ and $U/t=16.8$. This coincides with a
change in the shape of the dispersion from the flattened to the tight binding form.}
\label{fig:hhdisptr}
\end{figure}

In figure \ref{fig:hhdisptr} we show the variation of the singlet
dispersions of the Hubbard-Holstein bipolaron on the triangular
lattice when $\lambda=1.45$, $\bar{\omega}=1.5$ and $U$ is varied. Again, the functional form of the spectrum tends towards the
tight-binding dispersion as pairs become strongly bound, although it
becomes increasingly difficult to collect data since the fractional
variance on the dispersion increases as $\lambda$ increases. Again, we
suggest that the similarity in functional forms is due to the tightly
bound bipolaron acting as a single particle with a single (albeit
small) inter-site hopping parameter. For the Holstein bipolaron,
the pair wavefunction is small, and the lattice distortion is highly
localized.

\subsection{Hubbard-Fr\"ohlich model}

\begin{figure*}
\includegraphics[height=75mm,angle=270]{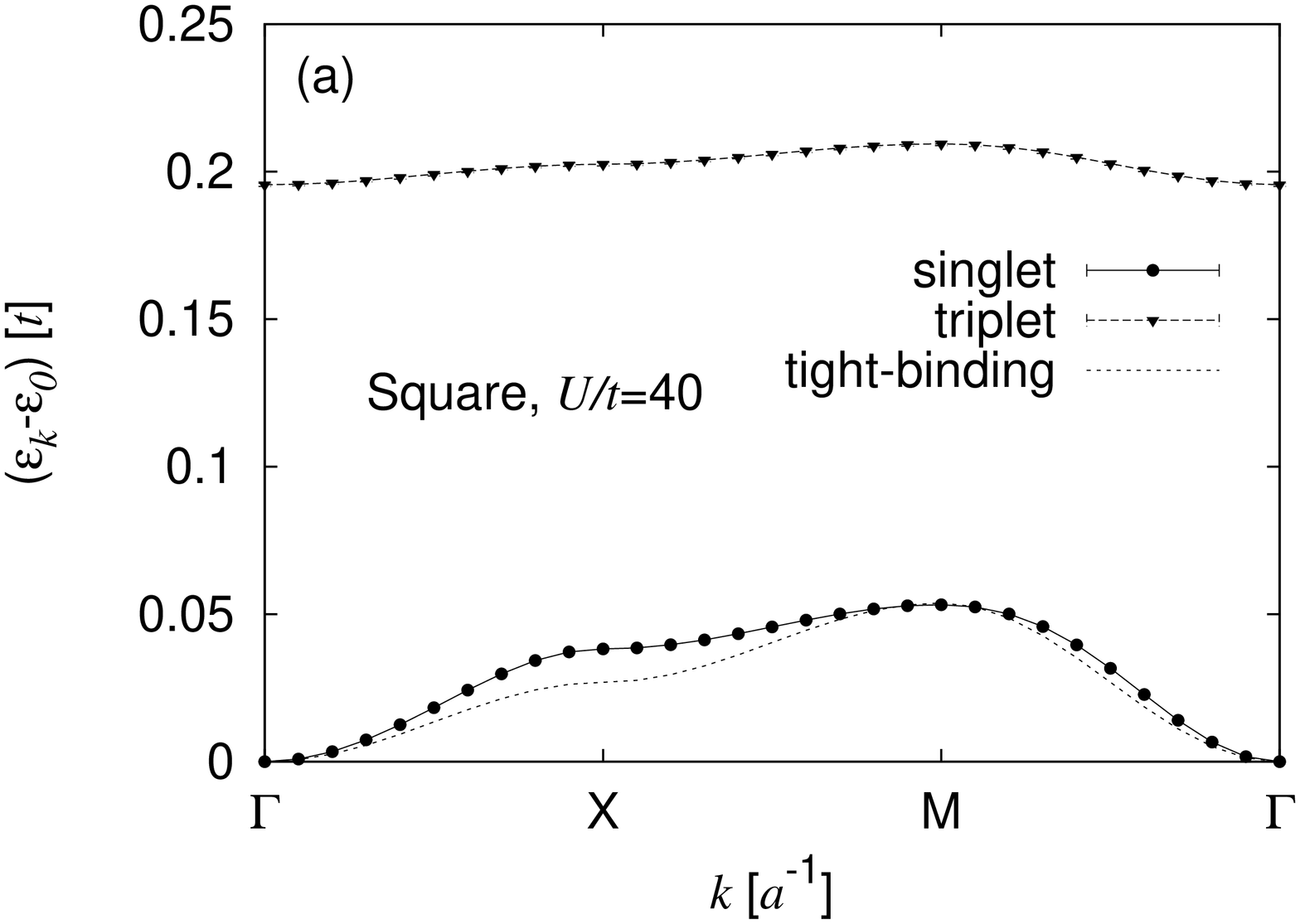}
\includegraphics[height=75mm,angle=270]{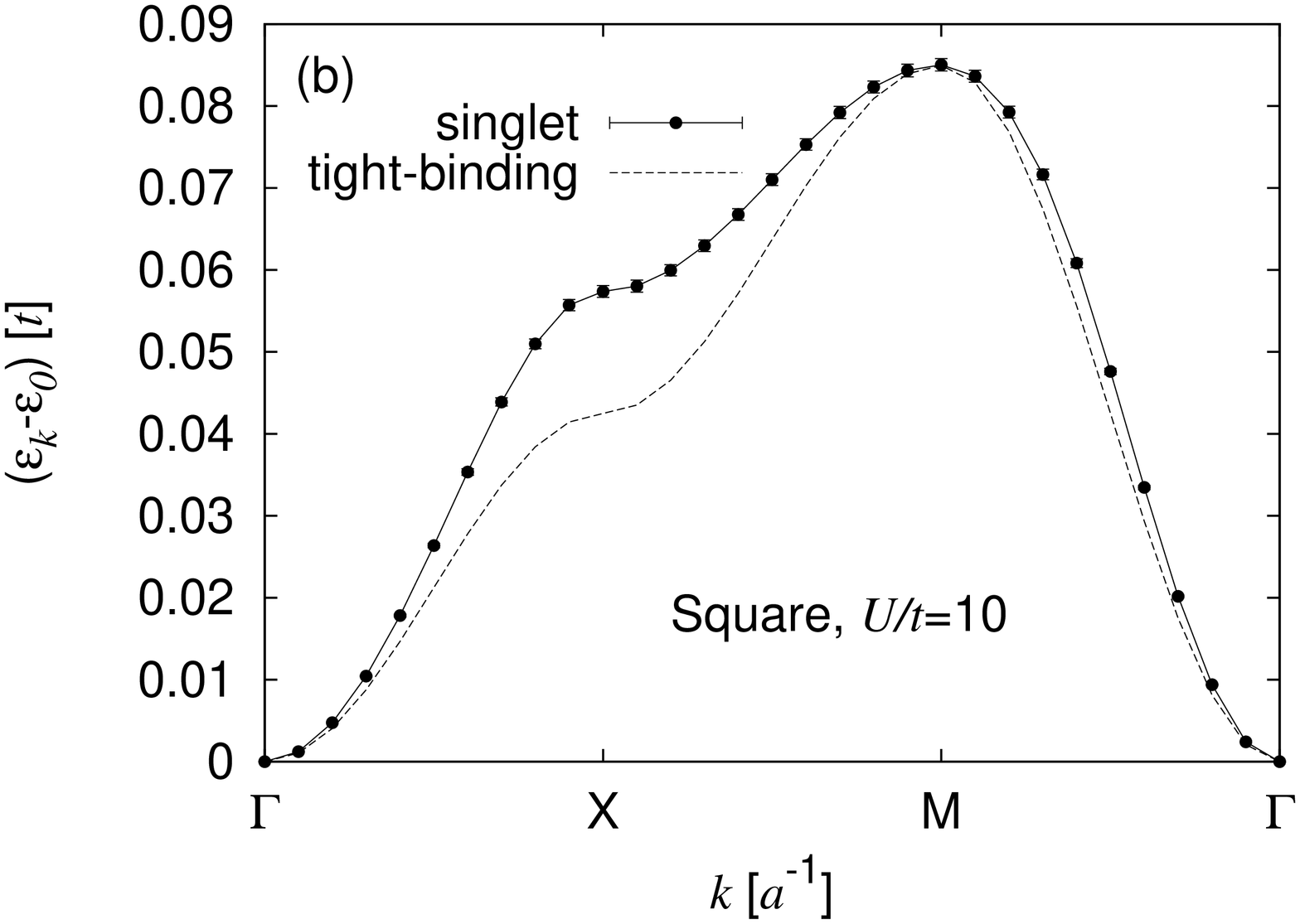}
\includegraphics[height=75mm,angle=270]{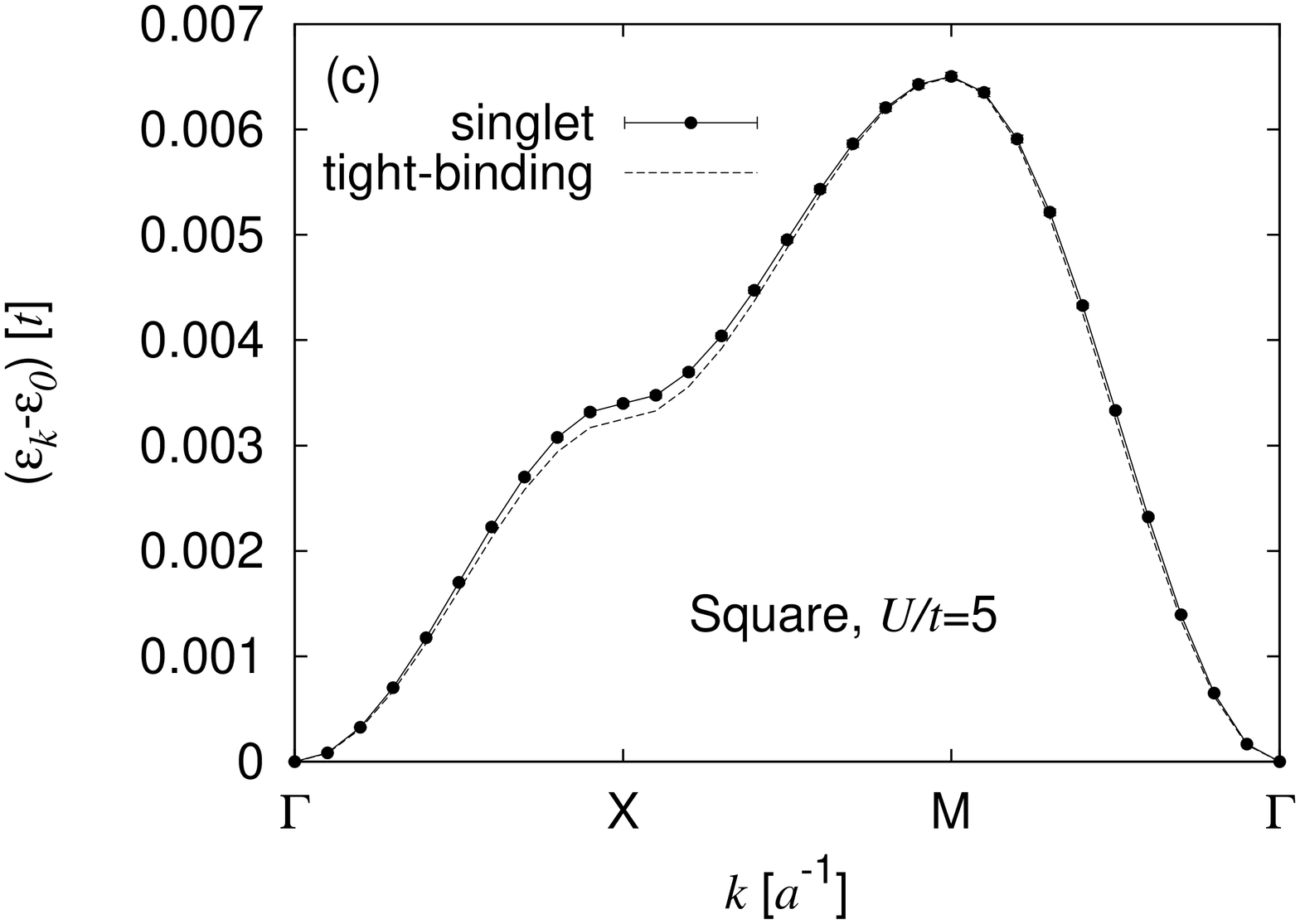}
\includegraphics[height=75mm,angle=270]{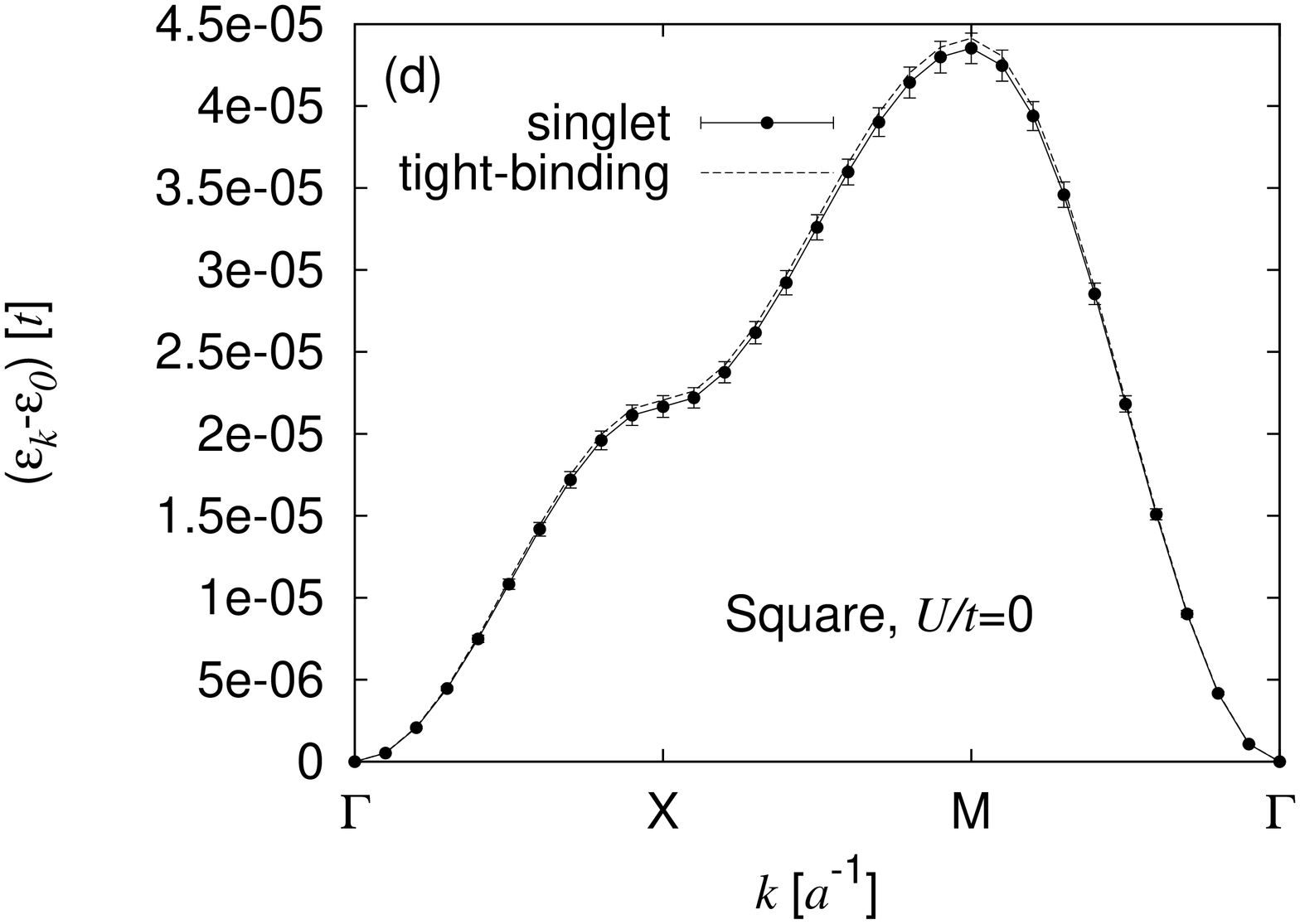}

\caption{Singlet dispersions of the Hubbard-Fr\"ohlich model on the
square lattice. The triplet dispersion is shown in panel (a). For
comparison, the tight-binding dispersion normalized to the bandwidth
is also shown. $\lambda=1.45$, $\bar{\omega}=1$, $R_{sc}=2$ and
$\bar{\beta}=14$. Error bars show one standard deviation. As $U$ decreases and the bipolaron becomes strongly
bound onsite, the shape of the dispersion becomes consistent with that
of a tightly-bound particle. An increase in the band-width can be seen
initially on decreasing $U$ corresponding to the formation of the
hybrid S0-S1 bipolaron. For smaller $U$, the S0 bipolaron is tightly bound and the dispersion is narrow.}
\label{fig:singdispfrohsqu}
\end{figure*}

We also examine the dispersions of the Hubbard-Fr\"ohlich
model. In Fig. \ref{fig:singdispfrohsqu}, we show the dispersion
for the square lattice when the Fr\"ohlich interaction with $R_{sc}=2$ is considered. The
panels show the effects of changing $U$. At large $U$, the bipolaron
is bound into an intersite (S1) configuration. There are two main
features that should be emphasized. The first is remarkable. As the
Coulomb repulsion decreases, the bandwidth \emph{increases}
significantly, i.e. although the bipolaron becomes more strongly
bound, it becomes lighter. This is the same effect as seen in section
\ref{sec:singlet}, where the mass was shown to decrease at
intermediate $U$ as a new type of hybrid S0-S1 crab bipolaron forms. At
small $U$, the degeneracy of the S0 and S1 bipolarons is broken and the bandwidth decreases with $U$.

The second feature in Fig. \ref{fig:singdispfrohsqu} is that the shape
of the bipolaron dispersion changes dramatically as it it bound onsite. For large
$U$, the dispersion is flattened, whereas for the
strongly bound onsite pair ($U/t=5$) the dispersion has the
characteristic tight-binding shape (albeit strongly renormalized by interactions). We
consider this effect to relate to the change between hopping
mechanisms. For a weakly bound bipolaron, the wavefunction is large
and single polaron-like hops dominate the motion of the particle,
whereas for the strongly bound bipolaron, the pair wavefunction is
smaller than a single lattice site and a new hopping integral becomes
relevant, which is relating to the tunneling of the whole bipolaron
between sites. The dispersions of triplet states on the square lattice are also shown in the panel relating to $U/t=40$. The triplet state is higher in energy and has a narrower bandwidth. Singlet and triplet dispersions do not cross.

\begin{figure*}
\includegraphics[height=75mm,angle=270]{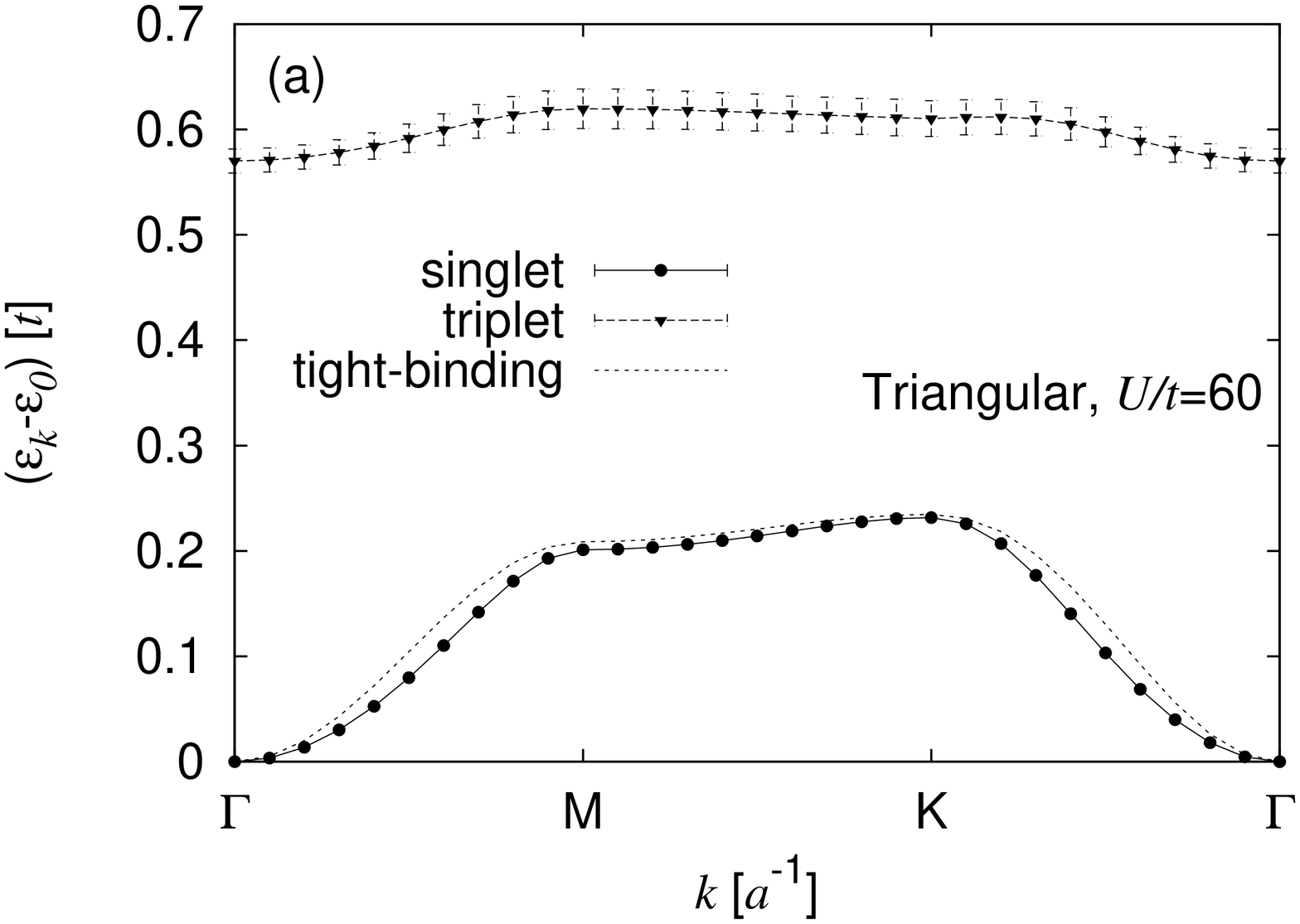}
\includegraphics[height=75mm,angle=270]{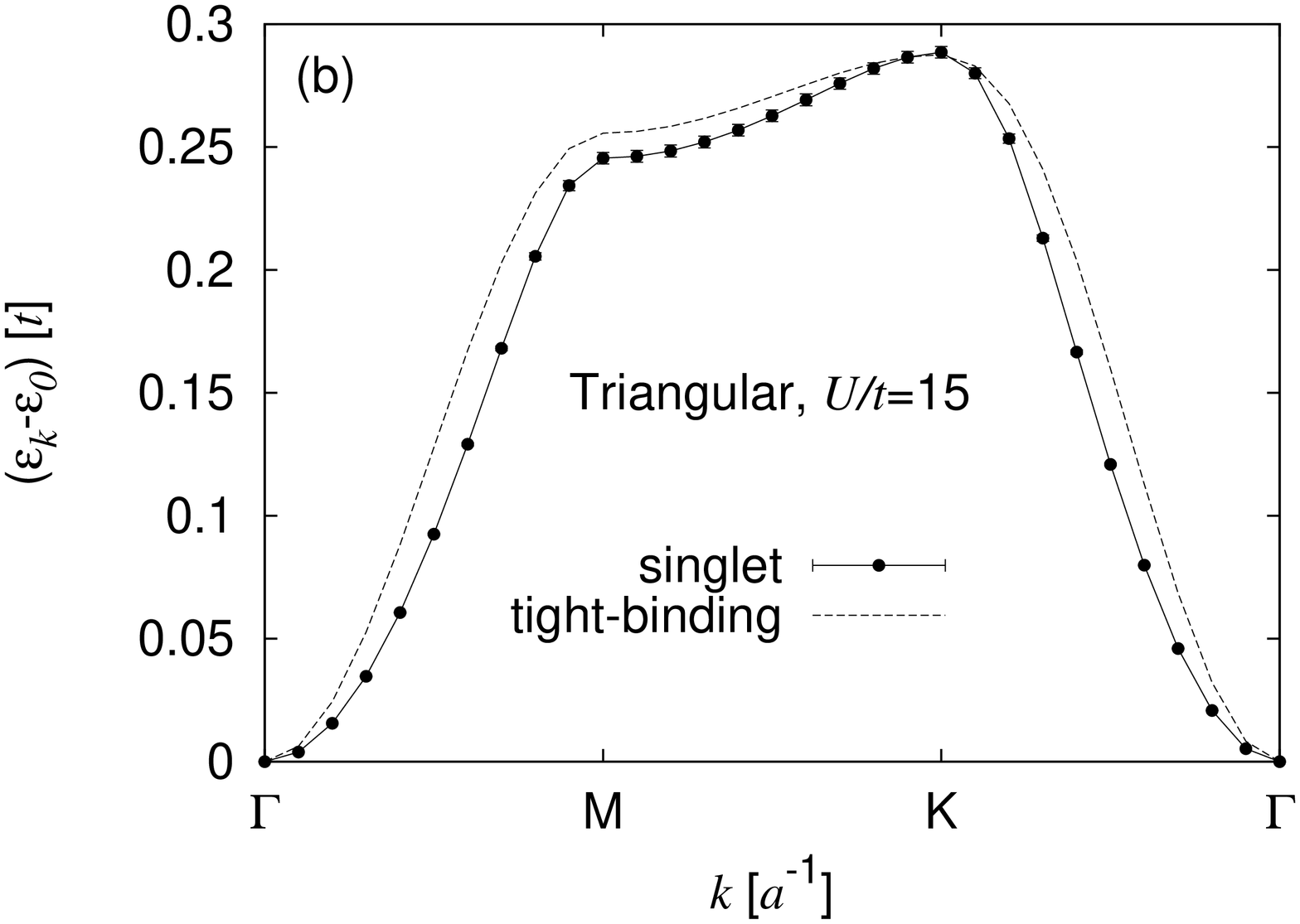}
\includegraphics[height=75mm,angle=270]{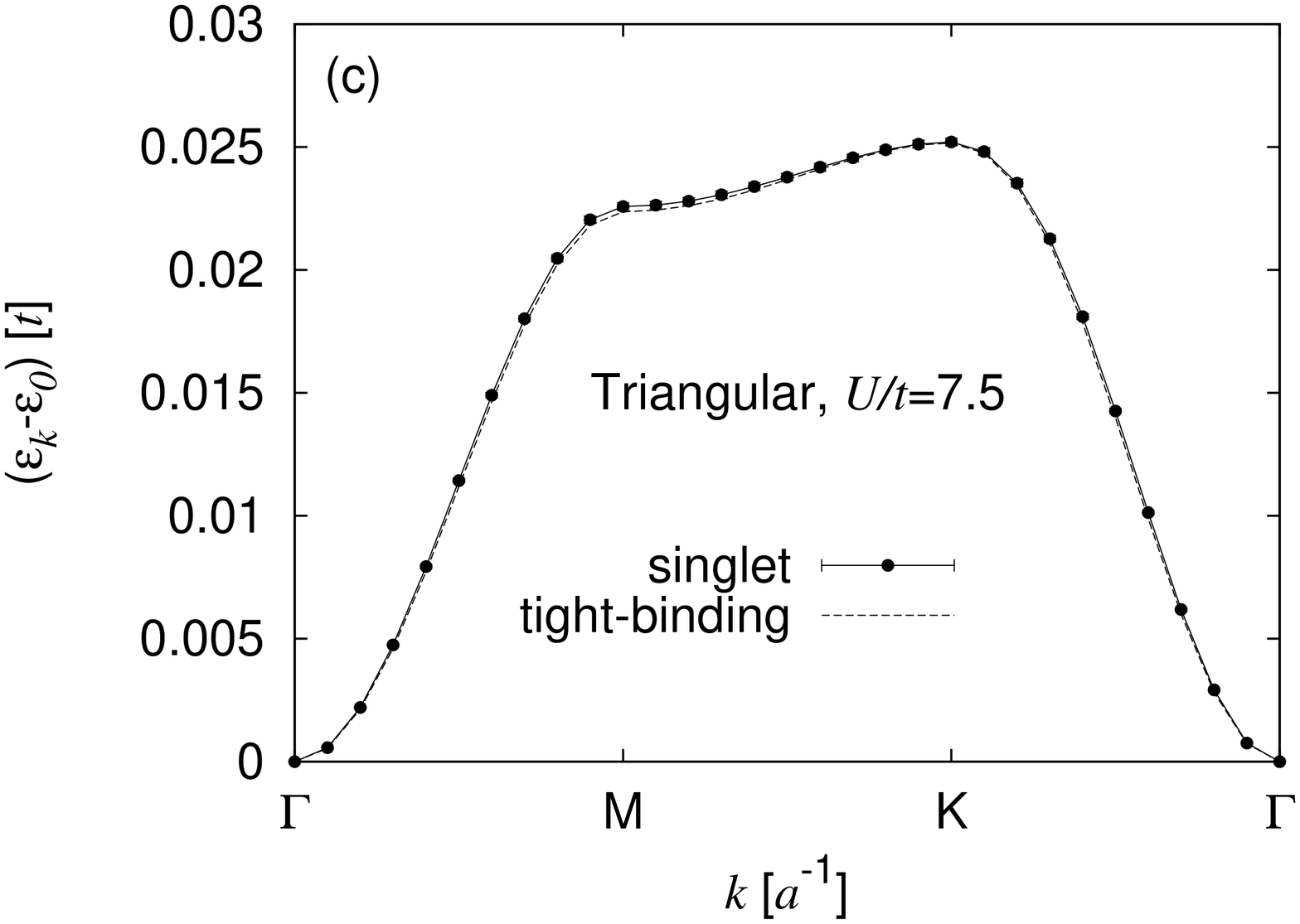}
\includegraphics[height=75mm,angle=270]{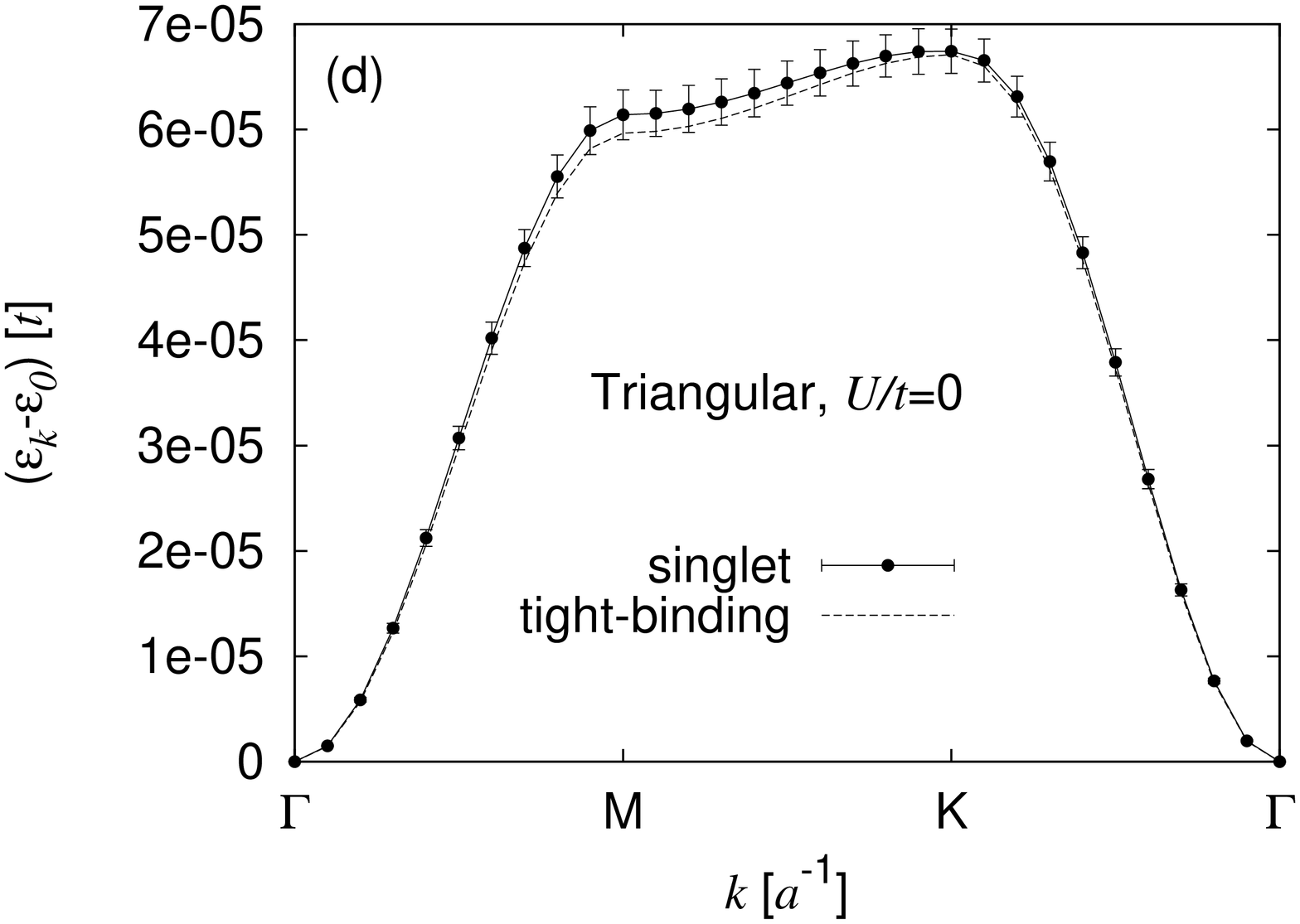}
\caption{Singlet dispersions of the bipolaron formed in the screened
Hubbard-Fr\"ohlich model on the triangular lattice. $R_{sc}=2$,
$\lambda=1.45$, $\bar{\omega}=1.5$, $\bar{\beta}=28/3$ and $U$ is varied. Error bars represent one standard deviation. As $U$ is
decreased, the bandwidth initially \emph{increases} as the S0-S1
hybrid bipolaron is approached from the S1 bipolaron (until around
$U/t=15$). For small $U$, the S0 bipolaron is approached, and the
bandwidth decreases significantly. For reference the tight-binding dispersion
normalized to the same bandwidth is also plotted. The triplet dispersion is shown in panel (a).}
\label{fig:singdispfrohtri}
\end{figure*}

We also examine the dispersions of the Coulomb-Fr\"ohlich bipolaron on
a triangular lattice, which are shown in
Fig. \ref{fig:singdispfrohtri}. The first significant difference
between bipolarons on square and triangular lattices is the existence
of superlight bipolarons at large $U\rightarrow\infty$. These
contribute to a factor of $\sim 4$ difference between the bandwidths
of the bipolaron on square and triangular lattices. Again, as $U$
decreases there is a remarkable increase in the bandwidth. This effect
is not as dramatic as in the case of the square lattice, presumably
because the bipolaron already moves with a crab like motion at large
$U$. However the degeneracy between $S0$ and $S1$ bipolarons accounts
for an additional $\sim$20\% increase in the bandwidth, making the
bipolarons extra-light. This shows how bipolarons can be light over an
extremely large regime of the parameter space on triangular lattices.

\begin{figure}
\includegraphics[height=75mm,angle=270]{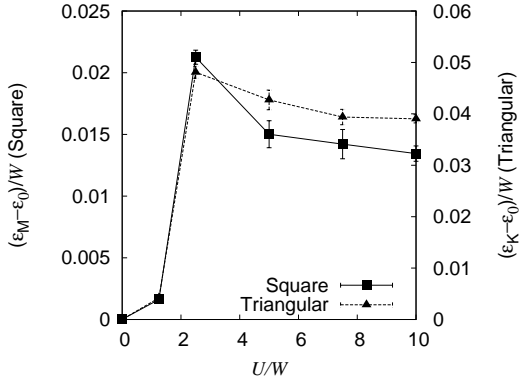}
\caption{Comparison of the variation of the bandwidth with $U$ for the square and triangular lattices. The widening of the bandwidth at intermediate $U$ is more significant for the square lattice, but is also visible for the triangular lattice.}
\label{fig:cmpbandwidth}
\end{figure}

To highlight the effect of the formation of the hybrid S0-S1 bipolaron
on the dispersion, Fig. \ref{fig:cmpbandwidth} shows a comparison of
the variation of the bandwidth as $U$ is changed. The bandwidths of
bipolarons on both square and triangular lattices are shown, although
it should be noted that the scales are different. There is an increase
in bandwidth at $U/W\sim 3$. The widening of the bandwidth is more
significant for the square lattice, but is also visible for the
triangular lattice. The broadening of the bandwidth on the square
lattice at intermediate to large $U$ is important, since it shows that
light bipolarons are not confined to lattices constructed from
triangular plaquettes, but can also exist on simple square, linear and
presumably cubic lattices.

\begin{figure*}
\includegraphics[height=75mm,angle=270]{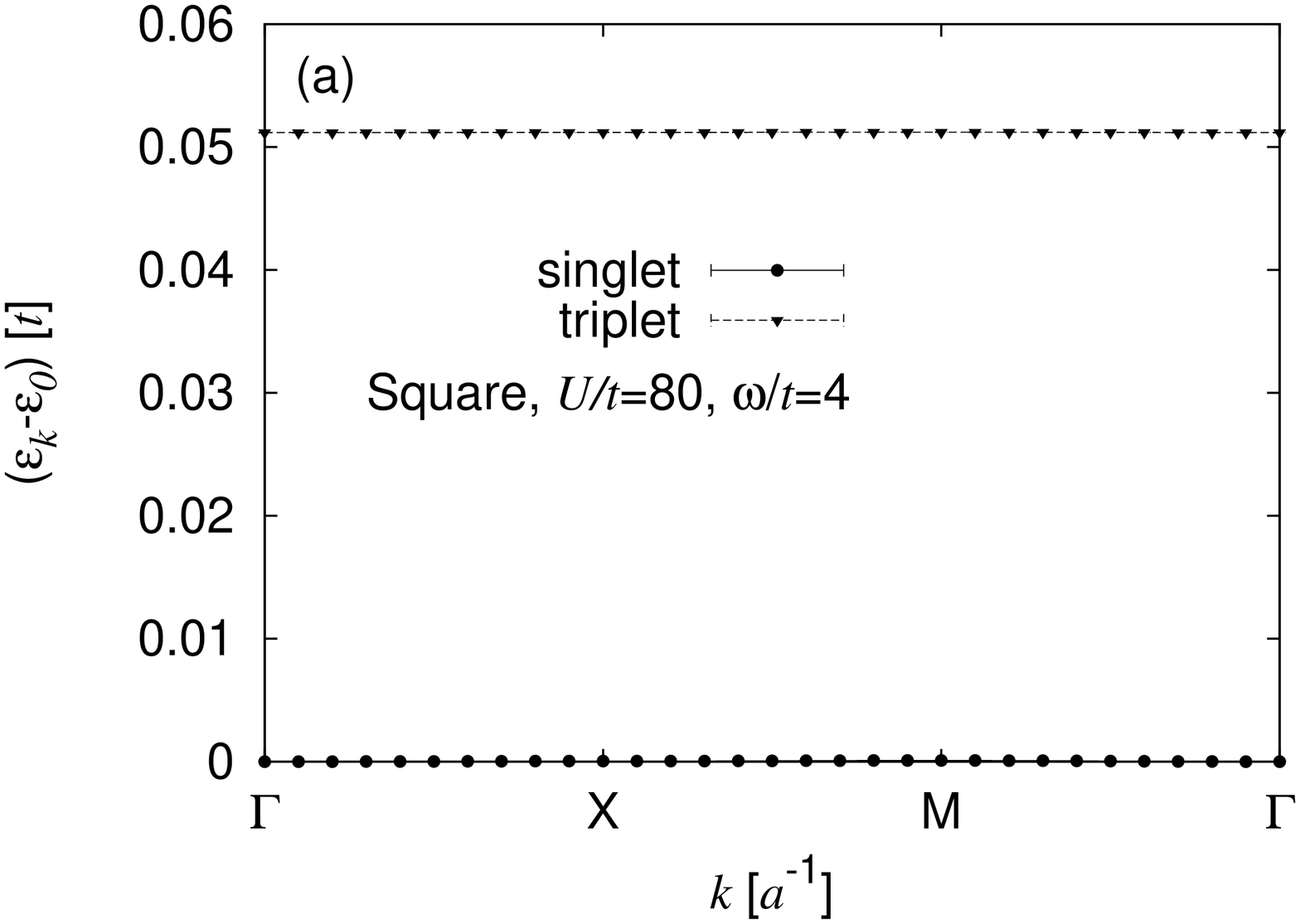}
\includegraphics[height=75mm,angle=270]{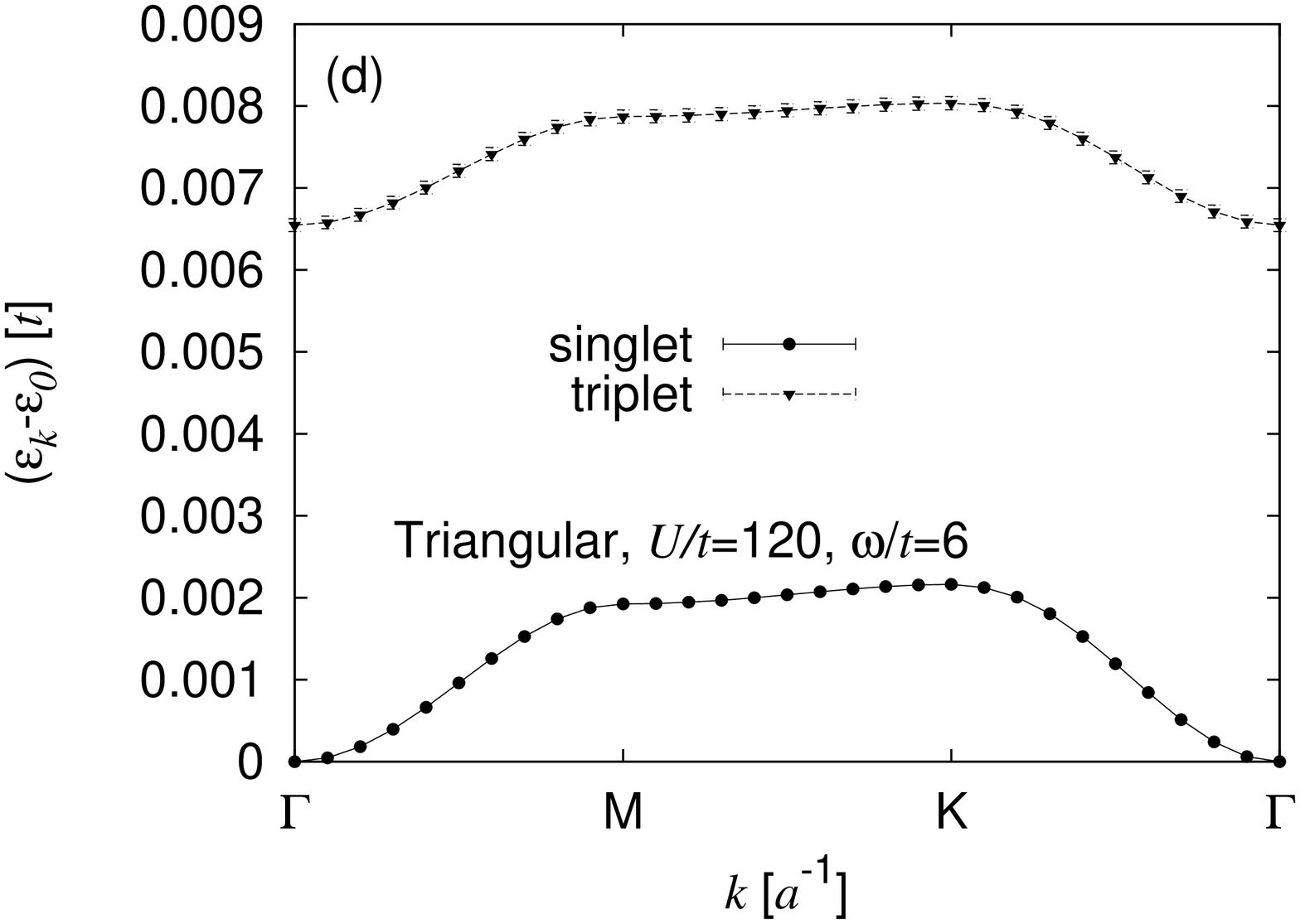}
\includegraphics[height=75mm,angle=270]{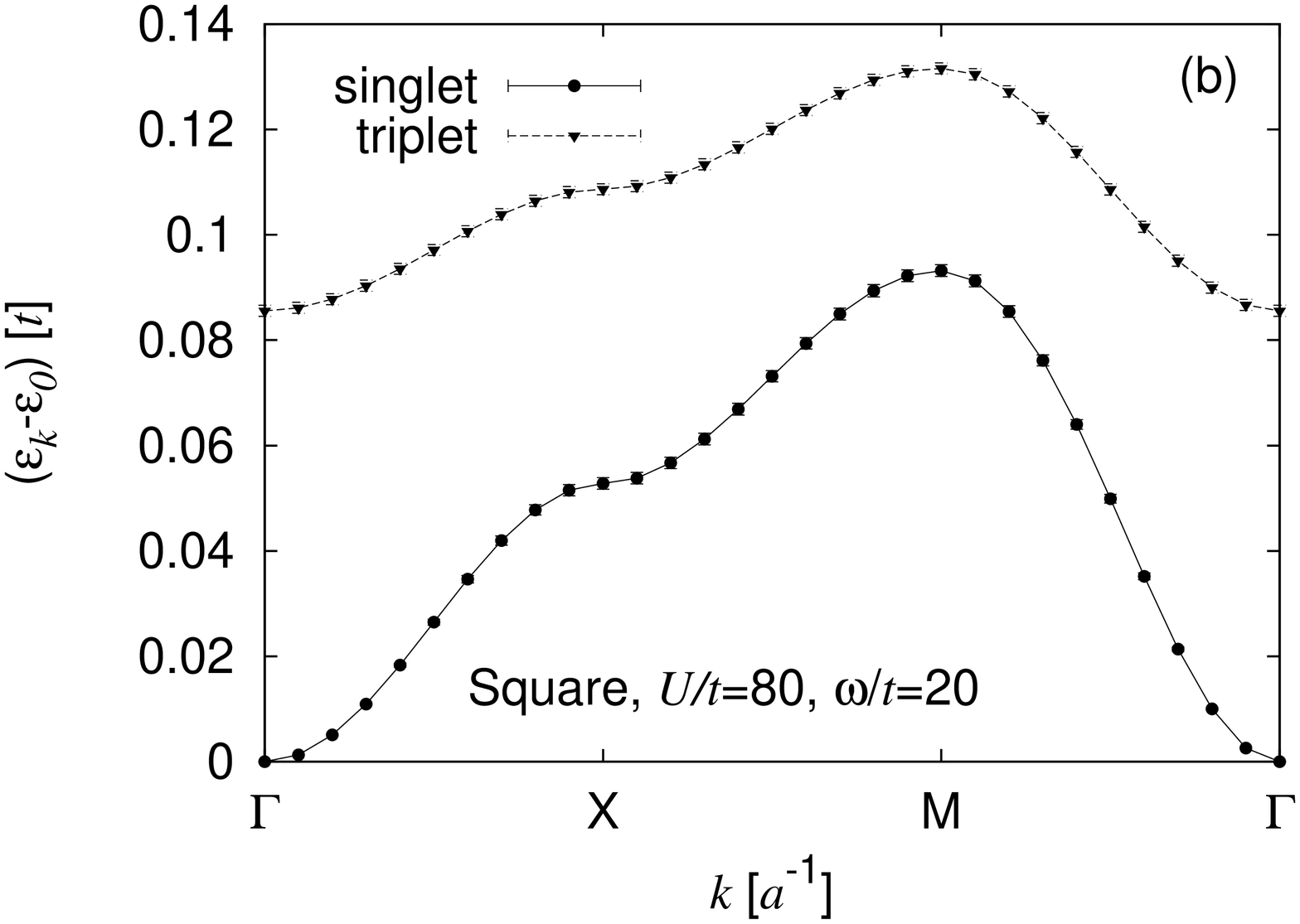}
\includegraphics[height=75mm,angle=270]{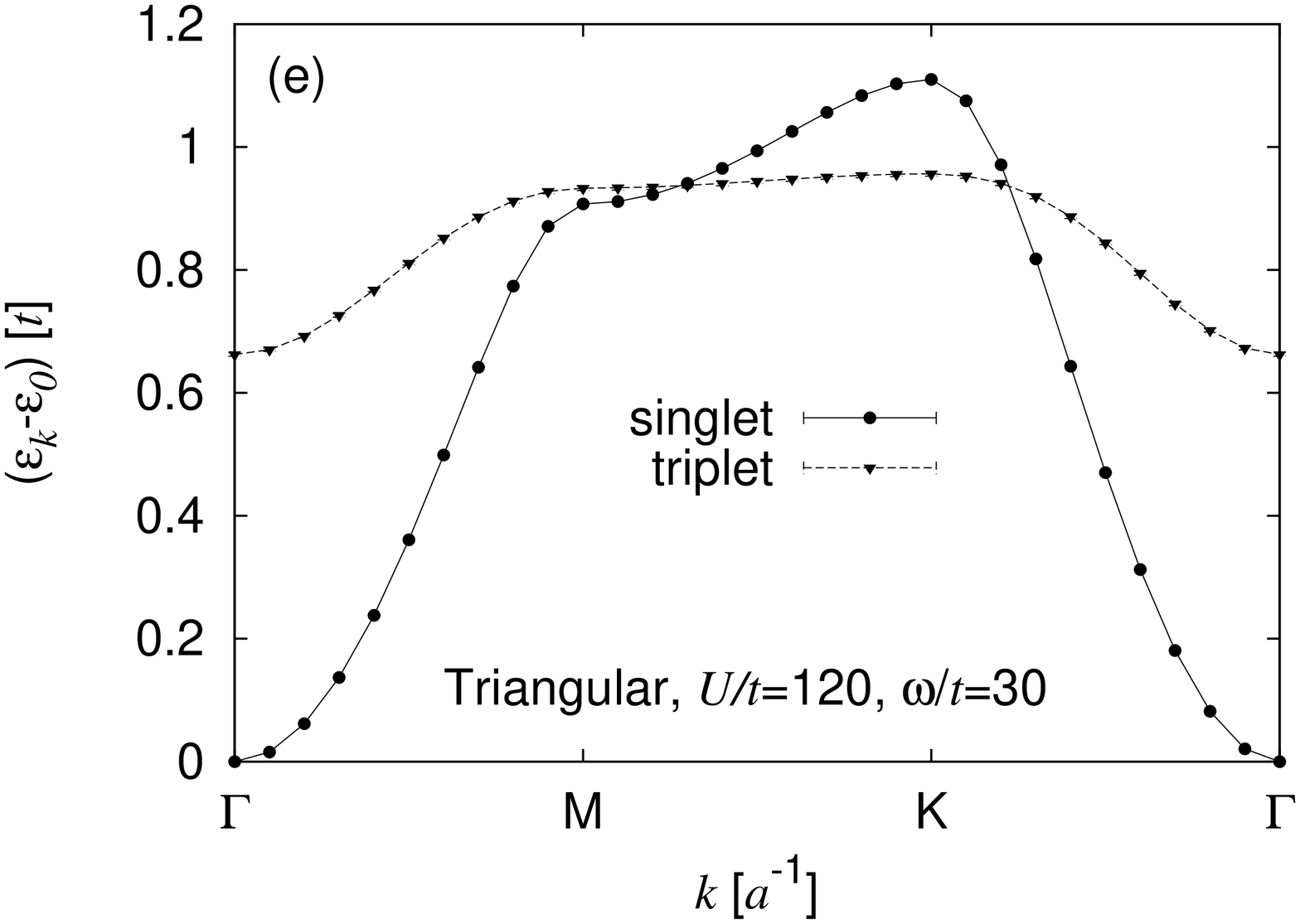}
\includegraphics[height=75mm,angle=270]{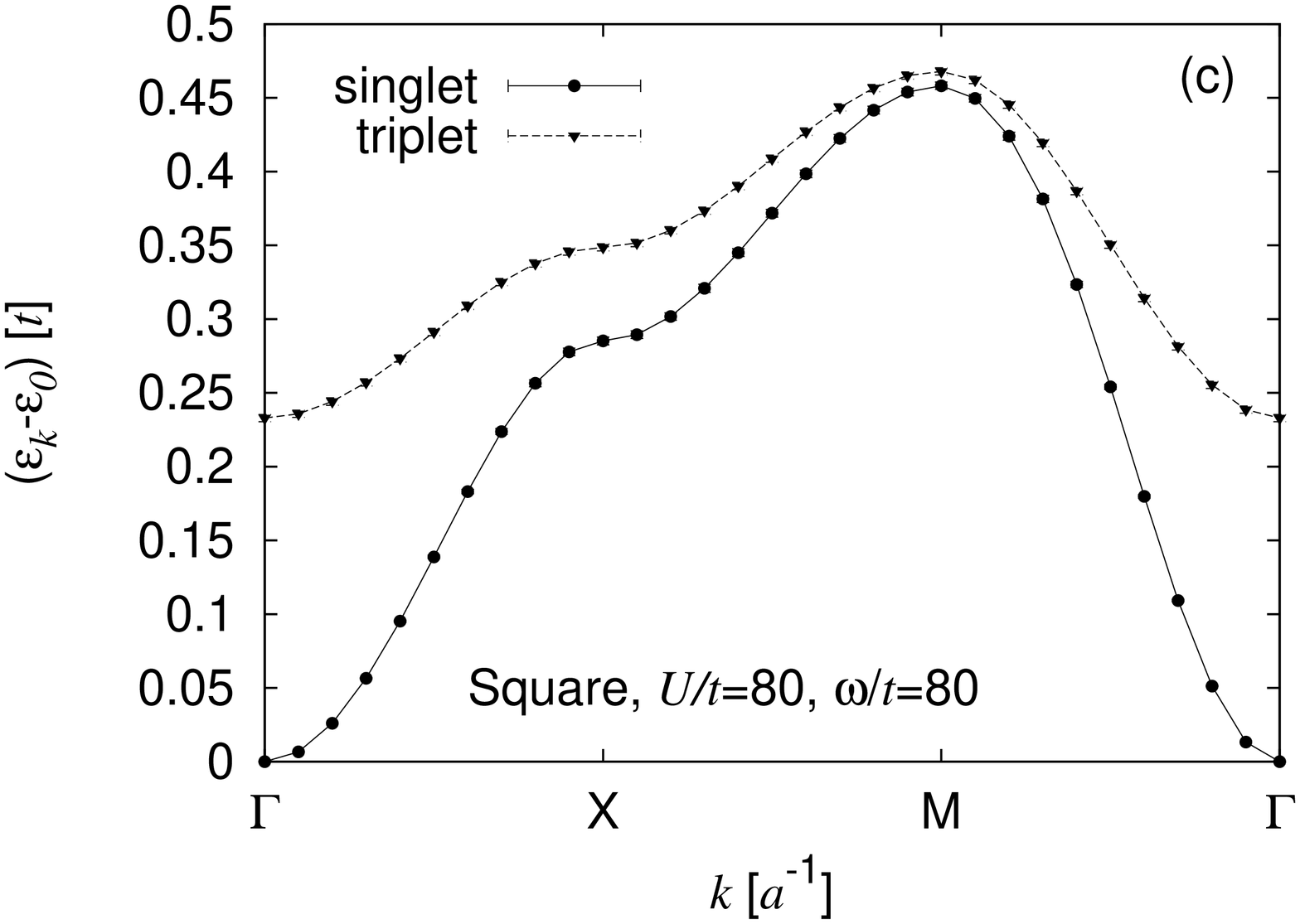}
\includegraphics[height=75mm,angle=270]{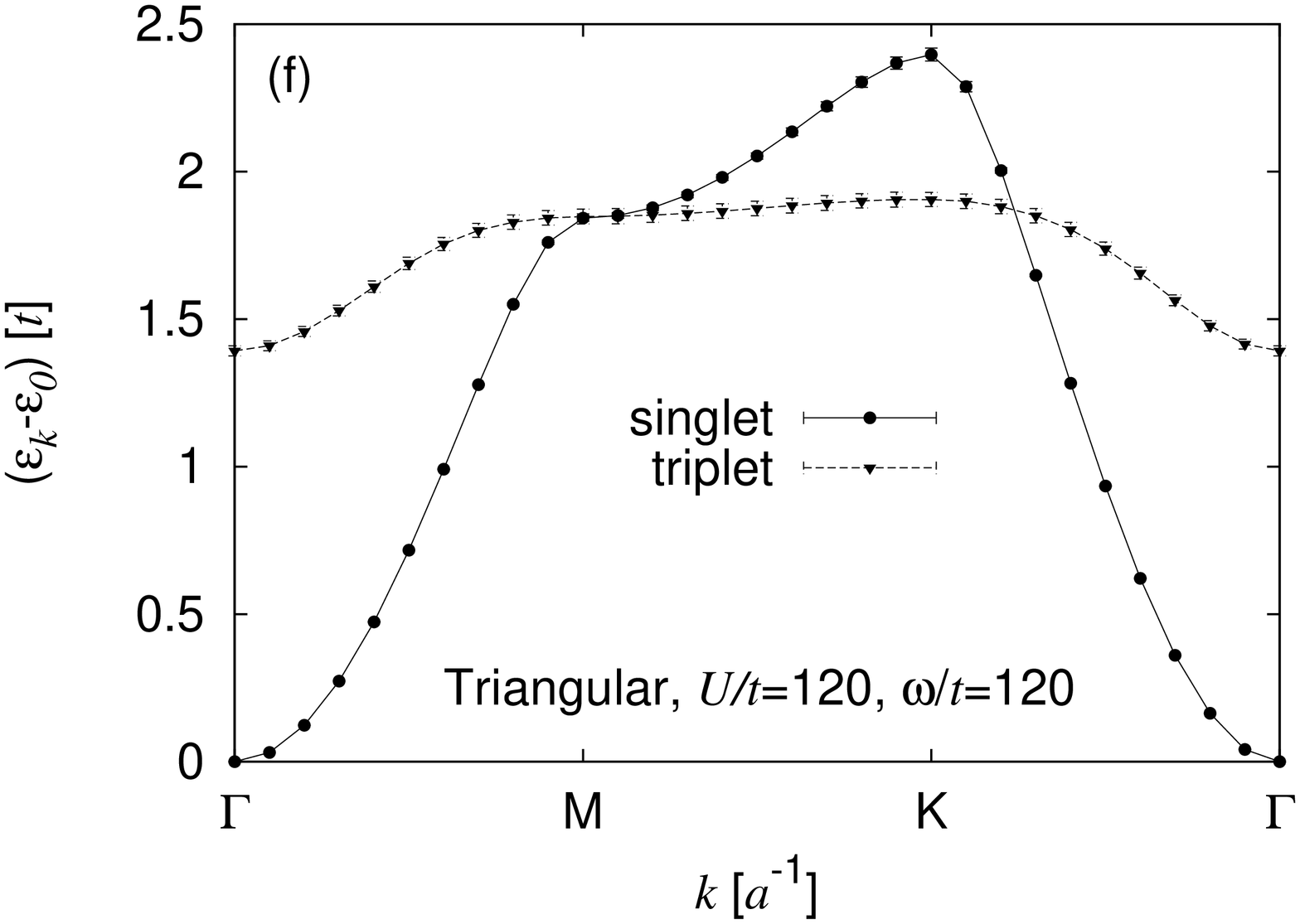}
\caption{Graphs showing the variation of the singlet and triplet
dispersions in the near-neighbor model with $\omega$ with $\lambda=8$,
$\omega=120t$ ($\omega=80t$), $U=120t$ ($U=80t$) to match with a $U-V$
model with $\bar{U}=24t$ ($\bar{U}=16t$) and $\bar{V}=48t$
($\bar{V}=32t$). Also $\bar{\omega}=30$ and $\bar{\omega}=6$ (and
equivalent for the square lattice, $\bar{\omega}=20$ and
$\bar{\omega}=4$).}
\label{fig:dispnn}
\end{figure*}

As noted in the introduction, we can also consider the properties of
bipolarons formed from a simplified nearest-neighbor interaction as a
means to compare with analytical results. Fig. \ref{fig:dispnn} shows
the variation of the singlet and triplet dispersions in the
near-neighbor model as $\omega$ is changed . Bipolarons are considered on both
square and triangular lattices with a large electron-phonon coupling
of $\lambda=8$. To make comparisons with the $UV$-model, we choose a
large phonon frequency $\omega=120t$ on the triangular lattice (and
$\omega=80t$ on the square lattice). A large coulomb repulsion of
$U=120t$ ($U=80t$) stops on-site pairing, which leads to a $U-V$ model
with $\tilde{U}=24t$ ($\tilde{U}=16t$) and $\tilde{V}=48t$
($\tilde{V}=32t$). We also compute the dispersions of bipolarons on
the triangular lattice when $\bar{\omega}=30$ and $\bar{\omega}=6$
($\bar{\omega}=20$ and $\bar{\omega}=4$ on the square lattice).

The dispersion of the near-neighbor bipolaron on the triangular
lattice at large phonon frequency yields an unusual result. The
singlet and triplet bands cross. We note again that the QMC algorithm
picks out only the lowest energies associated with singlet and triplet
states. In the plot, the $s$ band is visible, along with the
lowest $p$ band. A crossing is very unexpected, since in all
situations that we have previously studied, the triplet band has
always sat above the singlet one. This indicates that there will be a
transition from singlet to triplet states if sufficient momentum can
be imparted to the bipolarons. The crossing remains down to lower
phonon frequencies $\hbar\omega\sim 5W$, but is not visible in the
adiabatic regime.

In the appendix, we summarize the analytical approach to solving the
$UV$-model on a triangular lattice when pairs have finite momentum,
and the full dispersion can be seen in Fig. \ref{fig:aone}. The energy
level crossing is clearly visible, and by comparison it is clear that a single
$s$ band and a single $p$ band are visible in
Fig. \ref{fig:dispnn}. Note that there are are a total of 6 possible
bands associated with the bipolaron, although it is unlikely that the
higher energy bands would be excited in any normal transport process.

The band crossing can not be seen for bipolarons on the square lattice, rather the
singlet and triplet pairs become degenerate at the zone edge as the
phonon frequency increases. This feature is also visible in the
dispersion of bipolarons on the chain \cite{hague2009a}, and is
probably a generic feature of bipolarons on lattices with cubic
symmetry.

\section{Summary}
\label{sec:summary}

In this paper, we computed the properties of bipolarons on
two-dimensional lattices using a continuous time quantum Monte Carlo
algorithm. Properties of the bipolaron including the total energy,
inverse mass, bipolaron radius and number of phonons associated with the
bipolaron demonstrated the qualitative
difference between models of electron phonon interaction with
long-range interaction (screened Fr\"ohlich) and those with purely
local (Holstein) interaction, with only small long range tails needed
to completely change the properties of the bipolaron. A major result
of our survey of the parameter space is the existence of extra-light
hybrid bipolarons consisting of an on-site and an off-site component
when on-site and inter-site bipolaron configurations are degenerate,
similar to the previously reported superlight bipolarons, but slightly
lighter on the triangular lattice, and significantly lighter than the
intersite bipolarons on the square lattice. We also compute triplet
properties of the bipolarons. A major surprise is that triplet pairs
with large momentum (close to the K point) are more stable than
singlet states on the triangular lattice.

The potential for the existence of local triplet pairs that are
preferentially bound has the potential to open new avenues in the
theory of superconductivity. At low momenta, triplet pairs are
necessarily higher in energy than their singlet counterparts since an
exact theorem requires the ground state wavefunction of a pair to have
no nodes. No such theorem exists for states with high pair momentum,
however the existence of stable triplet states is unexpected. Thus, in
addition to the potential for BCS-BEC crossover as the attractive potential is tuned, there is an
additional possibility of triplet BCS to triplet BEC crossover,
or even a transition between singlet and triplet pairing. We
believe that such a possibility should be investigated further.

Our other finding of light hybrid bipolarons on the square lattice at
intermediate coupling and Coulomb repulsion is also significant. While
the effect only exists over a very narrow range of coupling constants,
it shows that exotic lattice types are not necessary for small
light pairs that could have potential to form a high temperature Bose
condensate. As in the case of the triangular lattice, light
pairs are only a prerequisite for a high temperature BEC, as other factors
such as an absence of clustering need to be confirmed. Work to discuss
this clustering will form the basis of a future publication, but
clearly there are still some surprises left in the bipolaron problem.

\section{Acknowledgements}

We would like to thank Sasha Alexandrov and John Samson for useful discussions. JPH would like to acknowledge support from EPSRC grant no. EP/H015655/1.

\appendix

\section{UV model on the triangular lattice}

The quantum-mechanical problem of two interacting particles on a lattice can be solved
exactly as long as the radius of the interaction is finite. Of special interest here
is the spectrum of bound pairs that form when at least part of the potential is attractive. 
The general solution of the bound state problem results in a spectral equation 
\cite{kornilovitch1995a, kornilovitch1997, kornilovitch2004a} in the form of a determinant with a size equal to the
number of lattice sites within the interaction range. For the non-retarded $UV$ model 
with on-site Hubbard repulsion $U$ and nearest-neighbor attraction $-V$, the 
determinant is $3 \times 3$ for the one-dimensional chain, $5 \times 5$ for the square
lattice, $7 \times 7$ for the triangular lattice, and so on. The subsequent analysis 
can be simplified by considering the symmetric and anti-symmetric states (that is 
the singlets and triplets) separately. This effectively halves the number of interaction 
sites, and halves the size of the determinants as a result. For the $UV$-model on the
triangular lattice, the size of the singlet determinant reduces to 4 (symmetric 
combinations of the nearest neighbors plus the central site), and that of the triplet 
determinant to 3 (anti-symmetric combinations of the nearest neighbors).     
  
Omitting the straightforward but lengthy derivation, the spectral equation for the 
singlet bound pair of the lattice momentum $\Kvec$ is    
\begin{equation}
\left\vert 
\begin{array}{cccc}
U S_{00} - 1 & -V S_{01}     & -V S_{02}     & -V S_{03}     \\
  -V S_{10}  & -V S_{11} - 1 & -V S_{12}     & -V S_{13}     \\
  -V S_{20}  & -V S_{21}     & -V S_{22} - 1 & -V S_{23}     \\
  -V S_{30}  & -V S_{31}     & -V S_{32}     & -V S_{33} - 1  
\end{array} \right\vert = 0 \: ,
\label{eq:aone}
\end{equation}
where
\begin{equation}
S_{j0} (\Kvec) = \frac{1}{N} \sum_{\qvec} \frac{e^{i \qvec \nvec_j}}
                                               { E + \Delta(\Kvec,\qvec) } \: ,
\label{eq:atwo}
\end{equation}
\begin{equation}
S_{jk} (\Kvec) = \frac{2}{N} \sum_{\qvec} \frac{e^{i \qvec \nvec_j} \cos{(\qvec \nvec_k)} }
                                               { E + \Delta(\Kvec,\qvec) } \: ,
\label{eq:athree}
\end{equation}
\begin{eqnarray}
\Delta(\Kvec,\qvec) \!\! & = & 4t \cos{ \frac{K_x}{2} } \cos{q_x}  \nonumber \\ 
 & + & 8t \cos{\frac{K_x}{4}} \cos{\frac{\sqrt{3}K_y}{4}} 
          \cos{\frac{q_x}{2}} \cos{\frac{\sqrt{3}q_y}{2}}     \nonumber \\
 & + & 8t \sin{\frac{K_x}{4}} \sin{\frac{\sqrt{3}K_y}{4}} 
          \sin{\frac{q_x}{2}} \sin{\frac{\sqrt{3}q_y}{2}}     .
\label{eq:afour}
\end{eqnarray}
Here the vectors $\nvec_{j,k}$ denote the nearest neighbor sites of the triangular lattice:
$\nvec_{0} = (0,0)$; $\nvec_{1} = (1,0)$; $\nvec_{2} = (\frac{1}{2},\frac{\sqrt{3}}{2})$; 
$\nvec_{3} = (- \frac{1}{2},\frac{\sqrt{3}}{2})$. 

\begin{figure}
\includegraphics[height=0.48\textwidth,angle=270]{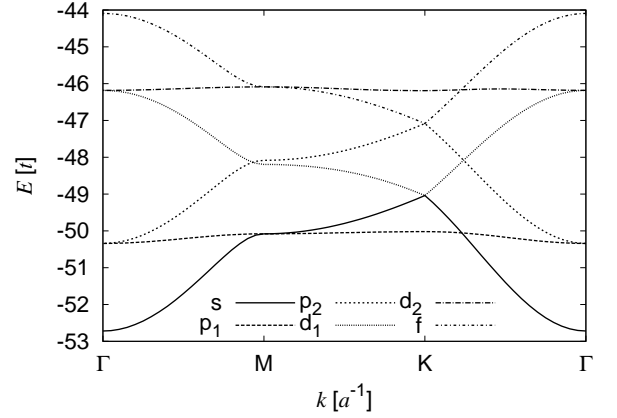}
\caption{The spectrum of bound pair in the $UV$ model on the triangular lattice. 
$U = 24 \, t$ and $V = 48 \, t$, which corresponds to $\lambda=8$, $U=20W$ and $\omega\rightarrow\infty$. 
Notice the crossing of the lowest singlet and triplet states near the edge of the
Brillouin zone at the $M$ point.}
\label{fig:aone}
\end{figure}

The equation (\ref{eq:aone}) can be solved numerically with respect to the pair energy $E$
for the given pair momentum $\Kvec$. For a positive on-site repulsion $U$ and a large enough 
intersite attraction $V$, the equation has three roots that we denote $s$, $d_1$ and $d_2$
states. The evolution of the energies as a function of $\Kvec$ is shown in figure~\ref{fig:aone}
At the high-symmetry points of the Brillouin zone, the system of equations (\ref{eq:aone})
can be further diagonalized into smaller blocks by a proper linear transformation.   
Thus, at the $\Gamma$ point $\Kvec = (0,0)$, the spectrum splits into the standalone 
ground state $s$ and a degenerate doublet $(d_1,d_2)$.  In the corner of the Brillouin zone,
the states $s$ and $d_1$ become degenerate, as can be seen from figure~\ref{fig:aone}.

The triplet bound states can be treated similarly. The exact spectrum equation is
a $3 \times 3$ determinant:
\begin{equation}
\left\vert 
\begin{array}{ccc}
  -V T_{11} - 1 & -V T_{12}     & -V T_{13}     \\
  -V T_{21}     & -V T_{22} - 1 & -V T_{23}     \\
  -V T_{31}     & -V T_{32}     & -V T_{33} - 1  
\end{array} \right\vert = 0 \: ,
\label{eq:asix}
\end{equation}
\begin{equation}
T_{jk} (\Kvec) = \frac{(-2i)}{N} \sum_{\qvec} \frac{e^{i \qvec \nvec_j} \sin{(\qvec \nvec_k)} }
                                               { E + \Delta(\Kvec,\qvec) } \: ,
\label{eq:aseven}
\end{equation}
with the same meaning of $\nvec_{j,k}$ and $\Delta$. The equation does not depend on the
on-site repulsion at all, as expected. At the $\Gamma$ point, the spectrum splits into
a low-energy doublet $(p_1,p_2)$, and a standalone high-energy $f$-state, 
(see figure~\ref{fig:aone}).      
At the $K$-point, $p_2$ and $f$ are degenerate.

Two comments are in order. (i) For strongly-coupled pairs, the
classification of the different orbital states is basically the one of
a single particle on a six-site tight-binding ring. (ii) Notice the
general property of level crossing between the lowest singlet and
triplet states. While the singlet is always the ground state at small
momenta, the triplet becomes the lowest state at large momenta. The
same property has been observed for the bipolaron, as discussed in the
main text of the paper.

\bibliography{coulombfrohlichtwo}

\end{document}